%% file: paper.tex
\documentclass[creativecommons]{eptcs}
\providecommand{\event}{MSFP 2022}
\usepackage{breakurl}
\usepackage{underscore}

\usepackage[utf8]{inputenc}
\usepackage{agda}
\usepackage{subcaption}
\usepackage{tikz}
\usepackage{wrapfig}
\usetikzlibrary{positioning}
\usetikzlibrary{arrows}
\usetikzlibrary{cd}

\usepackage{prftree}
\global\prfinterspace=1.5em

\usepackage{enumitem}
\setlist{leftmargin=4mm}

\input{commands.tex}
\input{unicode.tex}

\title{Sikkel: Multimode Simple Type Theory as an Agda Library}
\author{Joris Ceulemans
  \qquad\qquad Andreas Nuyts
  \qquad\qquad Dominique Devriese
  \institute{Department of Computer Science, KU Leuven}}
\def\titlerunning{Sikkel: Multimode Simple Type Theory as an Agda Library}
\def\authorrunning{Joris Ceulemans, Andreas Nuyts \& Dominique Devriese}

\begin{document}
\maketitle

\begin{abstract}
  Many variants of type theory extend a basic theory with additional primitives or properties like univalence, guarded recursion or parametricity, to enable constructions or proofs that would be harder or impossible to do in the original theory.
  However, implementing such extended type theories (either from scratch or by modifying an existing implementation) is a big hurdle for their wider adoption.
  In this paper we present Sikkel, a library in the dependently typed programming language Agda that allows users to program in extended type theories.
  It uses a deeply embedded language
  that can be easily extended with additional type and term constructors, thus supporting a wide variety of type theories.
  Moreover, Sikkel has a type checker that is sound by construction in the sense that all well-typed programs are automatically translated to their semantics in a shallow embedding based on presheaf models.
  Additionally, our model supports combining different base categories by using modalities to transport definitions between them.
  This enables in particular a general approach for extracting definitions to the meta-level, so that we can use the extended type theories to define regular Agda functions and prove properties of them.
  In this paper, we demonstrate Sikkel theories with guarded recursion and parametricity, but other extensions can be easily plugged in.
  For now, Sikkel supports only simple type theories but its model
  already anticipates the future addition of dependent types and a universe.
\end{abstract}

\input{content-tex/intro}
\input{content-lagda/deep-embedding}

\input{content-lagda/guarded-recursion}
\input{content-lagda/presheaves}
\input{content-lagda/sound-typechecker}
\input{content-lagda/extraction}
\input{content-lagda/parametricity}

\input{content-tex/discussion-relatedwork.tex}

\bibliographystyle{eptcs}
\bibliography{bibliography}

\appendix

\input{content-lagda/appendix-presheaves}

\end{document}

%% file: commands.tex
\usepackage{ifthen}
\usepackage{newtxmath}

\DeclareRobustCommand{\AgdaFormat}[2]{%
    \ifthenelse{\equal{#1}{δx}}{$\delta_{\textit{x}}$}{%
    \ifthenelse{\equal{#1}{δy}}{$\delta_{\textit{y}}$}{%
    \ifthenelse{\equal{#1}{γx}}{$\gamma_{\textit{x}}$}{%
    \ifthenelse{\equal{#1}{γy}}{$\gamma_{\textit{y}}$}{%
    \ifthenelse{\equal{#1}{γz}}{$\gamma_{\textit{z}}$}{%
    \ifthenelse{\equal{#1}{≅ᵗʸ}}{$\cong^{\text{ty}}$}{%
    \ifthenelse{\equal{#1}{≅ᵐ}}{$\cong^{\text{m}}$}{%
    \ifthenelse{\equal{#1}{≅ᵗᵐ}}{$\cong^{\text{tm}}$}{%
    \ifthenelse{\equal{#1}{≅ˢ}}{$\cong^{\text{s}}$}{%
    \ifthenelse{\equal{#1}{≃ᵗʸ?}}{$\simeq^{\text{ty}}\text{?}$}{%
    \ifthenelse{\equal{#1}{≃ᵐ?}}{$\simeq^{\text{m}}\text{?}$}{%
    \ifthenelse{\equal{#1}{lamι}}{lam$\iotaup$}{%
    \ifthenelse{\equal{#1}{lamι[}}{lam$\iotaup$[}{%
    \ifthenelse{\equal{#1}{löbι}}{löb$\iotaup$}{%
    \ifthenelse{\equal{#1}{löbι[}}{löb$\iotaup$[}{%
    \ifthenelse{\equal{#1}{varι}}{var$\iotaup$}{\mbox{#2}}}}}}}}}}}}}}}}}}

\newcommand{\later}[1]{\AF{▻}\AS #1}

\newcommand{\tyeq}{\simeq^\mathsf{ty}}
\newcommand{\modeq}{\simeq^\mathsf{m}}
\newcommand{\rulename}[1]{\textsc{\scriptsize #1}}
\newcommand{\inlinerulename}[1]{\textsc{\footnotesize #1}}
\newcommand{\twocell}[3]{#1 $\in$ #2 $\Rightarrow$ #3}

\newcommand{\omode}{\AIC{$\omega$}}
\newcommand{\stmode}{\AIC{$\star$}}
\newcommand{\wsmode}{\AIC{$\wedge$}}

\newcommand{\AB}[1]{\AgdaBound{#1}}
\newcommand{\ADT}[1]{\AgdaDatatype{#1}}
\newcommand{\AF}[1]{\AgdaFunction{#1}}
\newcommand{\AFi}[1]{\AgdaField{#1}}
\newcommand{\AIC}[1]{\AgdaInductiveConstructor{#1}}
\newcommand{\AK}[1]{\AgdaKeyword{#1}}

\newcommand{\APt}[1]{\AgdaPrimitive{#1}}
\newcommand{\AR}[1]{\AgdaRecord{#1}}
\newcommand{\AS}{\AgdaSpace{}}
\newcommand{\ASt}[1]{\AgdaString{#1}}
\newcommand{\ASy}[1]{\AgdaSymbol{#1}}
\newcommand{\AU}{\AgdaUnderscore{}}

\newcommand{\adapted}[1]{#1}

\newcommand{\todo}[1]{}
\newcommand{\remark}[1]{}
\newcommand{\question}[1]{}
\newcommand{\domi}[1]{}
\newcommand{\joris}[1]{}
\newcommand{\andreas}[1]{}


%% file: unicode.tex
\usepackage{newunicodechar}
\usepackage{stmaryrd}
\usepackage{bbold}

\newunicodechar{ℕ}{\ensuremath{\mathnormal{\mathbb{N}}}}
\newunicodechar{→}{\ensuremath{\mathnormal\to}}
\newunicodechar{∀}{\ensuremath{\mathnormal\forall}}
\newunicodechar{λ}{\ensuremath{\mathnormal\lambda}}
\newunicodechar{ω}{\ensuremath{\mathnormal\omega}}
\newunicodechar{◇}{\ensuremath{\mathnormal\diamond}}
\newunicodechar{⇛}{\ensuremath{\mathnormal\Rrightarrow}}
\newunicodechar{∈}{\ensuremath{\mathnormal\in}}
\newunicodechar{∋}{\ensuremath{\mathnormal\ni}}
\newunicodechar{ι}{\ensuremath{\mathnormal\iota}}
\newunicodechar{▻}{\ensuremath{\mathnormal\vartriangleright}}
\newunicodechar{⊛}{\ensuremath{\mathnormal\circledast}}
\newunicodechar{Γ}{$\Gammaup$}
\newunicodechar{₁}{\ensuremath{\mathnormal{}_1}}
\newunicodechar{Δ}{$\Deltaup$}
\newunicodechar{⇒}{\ensuremath{\mathnormal\Rightarrow}}
\newunicodechar{⊚}{\ensuremath{\mathnormal\circledcirc}}
\newunicodechar{Θ}{$\Theta$}
\newunicodechar{σ}{$\sigma$}
\newunicodechar{π}{$\pi$}
\newunicodechar{⊠}{\ensuremath{\mathnormal\boxtimes}}
\newunicodechar{τ}{$\tau$}
\newunicodechar{◄}{\ensuremath{\mathnormal\blacktriangleleft}}
\newunicodechar{Φ}{$\Phi$}
\newunicodechar{β}{$\beta$}
\newunicodechar{η}{$\eta$}
\newunicodechar{μ}{$\mu$}
\newunicodechar{★}{\ensuremath{\mathnormal\star}}
\newunicodechar{ℓ}{\ensuremath{\mathnormal\ell}}
\newunicodechar{⋀}{\ensuremath{\mathnormal\wedge}}
\newunicodechar{ℤ}{\ensuremath{\mathnormal{\mathbb{Z}}}}
\newunicodechar{∼}{\ensuremath{\mathnormal\sim}}
\newunicodechar{∙}{\ensuremath{\mathnormal\cdot}}
\newunicodechar{≡}{\ensuremath{\mathnormal\equiv}}
\newunicodechar{ʳ}{\ensuremath{\mathnormal{}^r}}
\newunicodechar{ˡ}{\ensuremath{\mathnormal{}^l}}
\newunicodechar{⟪}{\ensuremath{\langle\!\langle}}
\newunicodechar{⟫}{\ensuremath{\rangle\!\rangle}}
\newunicodechar{γ}{$\gamma$}
\newunicodechar{⊤}{\ensuremath{\top}}
\newunicodechar{δ}{$\delta$}
\newunicodechar{ρ}{$\rho$}
\newunicodechar{≤}{\ensuremath{\mathnormal{\leq}}}
\newunicodechar{Σ}{$\Sigma$}
\newunicodechar{Ψ}{$\Psi$}
\newunicodechar{⟦}{\ensuremath{\llbracket}}
\newunicodechar{⟧}{\ensuremath{\rrbracket}}
\newunicodechar{₂}{\ensuremath{\mathnormal{}_2}}
\newunicodechar{⊘}{\ensuremath{\mathnormal{\varnothing}}}
\newunicodechar{κ}{$\kappa$}
\newunicodechar{∣}{\ensuremath{\mid}}
\newunicodechar{⁻}{\ensuremath{\mathnormal{}^-}}
\newunicodechar{¹}{\ensuremath{\mathnormal{}^1}}
\newunicodechar{ⓜ}{\textcircled{{\raisebox{.08ex}{\footnotesize m}}}}
\newunicodechar{𝟙}{\ensuremath{\mathbb{1}}}
\newunicodechar{₃}{\ensuremath{\mathnormal{}_3}}
\newunicodechar{α}{$\alpha$}
\newunicodechar{≟}{\ensuremath{\overset{\scriptscriptstyle?}{=}}}
\newunicodechar{≅}{\ensuremath{\mathnormal\cong}}
\newunicodechar{∘}{\ensuremath{\mathnormal\circ}}
\newunicodechar{≃}{\ensuremath{\mathnormal\simeq}}
\newunicodechar{ᵐ}{\ensuremath{\mathnormal{}^{\text{m}}}}
\newunicodechar{ᵗ}{\ensuremath{\mathnormal{}^{\text{ty}}}}
\newunicodechar{ʸ}{} 
\newunicodechar{ⓣ}{\textcircled{\footnotesize t}}
\newunicodechar{⟨}{\ensuremath{\langle}}
\newunicodechar{⟩}{\ensuremath{\rangle}}
\newunicodechar{←}{\ensuremath{\mathnormal\leftarrow}}


%% file: content-tex/intro.tex
\section{Introduction}

Dependently typed programming languages like Agda or Coq are based on mathematical formal systems such as Martin-L\"of Type Theory (MLTT \cite{ML84}) or the Calculus of Inductive Constructions (CIC \cite{coc,colog88}).
Over the past years many authors have proposed new type systems that extend a theory like MLTT or CIC with new primitives which allow users to prove more theorems or write more programs.
Examples of such extensions include guarded recursion \cite{clock-cat,gratzer20-multimodal}, parametricity \cite{fomega-parametricity,dtt-parametricity,moulin-param3,nuyts17-parametric,nuyts18-degrees,cavallo-harper-paramhott-pub}, univalence \cite{model-cubical,cubical-unifying,cubical}, directed type theory \cite{shulman-riehl-directed,weaver-licata-dua} and nominal reasoning \cite{freshmltt}.

However, the question remains how these extended formal systems can be used in existing languages like Agda or Coq.
In this paper we present Sikkel\footnote{Available at \url{https://github.com/JorisCeulemans/sikkel/releases/tag/v1.0}.}, an Agda library that allows users to work in a class of extended type theories called multimode or multimodal type theories \cite{adjoint-logic,gratzer20-multimodal}.
Such type theories are parametrized by a mode theory which specifies new primitive type constructors called modalities.

\begin{wrapfigure}[10]{R}{.45\textwidth}
  \centering
  \begin{tikzpicture}[node distance=0.7cm]
    \draw node[draw, rounded corners] (syn) {Multimode Type Theory};
    \draw node[draw, rounded corners,below=of syn] (sem) {Presheaf Models};
    \draw node[draw, rounded corners,below=of sem] (agda) {Agda};

    \draw[->] (syn) to node[midway,right] {Sound Type-checker} (sem);
    \draw[->] (sem) to node[midway,right] {Extraction} (agda);
  \end{tikzpicture}%
  \caption{Sikkel's architecture.}
  \label{fig:sikkel-architecture}
\end{wrapfigure}
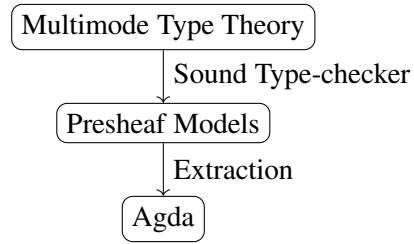

The Sikkel library consists of 3 layers, depicted in Figure~\ref{fig:sikkel-architecture}.
The first is an extrinsically typed syntax of multimode simple type theory (MSTT) defined in Agda.
Once a mode theory has been fixed%
, a user can write programs in a modal setting at this layer.
Moreover, the syntax can be easily extended with additional (non-modal) type and term formers.

Programs written at the first layer cannot be directly interpreted as Agda programs because that would require Agda interpretations for the newly added primitives.
Since our goal is to work within off-the-shelf Agda, this is impossible for many of the features we want to support.
However, many extensions of type theory \adapted{(including all examples mentioned above)} can be interpreted in a class of mathematical models called presheaf models.
Sikkel's second, semantic layer therefore consists of a formalization of presheaf models in Agda, essentially a shallow embedding of MSTT \cite{wildmoser-certifying-2004}.
The bridge between the first and the second layer is formed by a type checker, implemented in Agda, for the deeply embedded syntax. It is sound by construction in the sense that it will either refute a judgment or accept it \emph{and} interpret it in the presheaf model.

\adapted{However, this is insufficient if we want to incorporate programs defined in Sikkel into an ``ordinary'' Agda project.}
For that reason, Sikkel considers a third layer: plain Agda, and provides an extraction mechanism that translates terms at the semantic layer to ordinary Agda terms.
We only support this for terms in the trivial mode (modeled in the category of sets), but modalities allow us to carry over terms from other modes (modeled in other presheaf categories) to the trivial mode, whence they can be extracted to Agda.

We demonstrate the use of Sikkel in two applications: guarded recursive type theory and a restricted form of parametricity.
With guarded recursion, we can write definitions of infinite streams that would not be accepted by Agda's termination checker.
The extraction mechanism allows us to interpret these definitions as ordinary Agda streams.
For parametricity, we demonstrate how a function definition can be interpreted with two different representations for an abstract type, and how parametricity allows us to relate the two resulting definitions.

Sikkel is implemented in a standard off-the-shelf version of Agda (in which uniqueness of identity proofs is by default enabled).
We postulate function extensionality but do not use other postulates or unsafe Agda features like pragmas or sized types \cite{sized-types-unsafe}, which could (potentially) break soundness of the meta-theory.
Consequently, our library should also be implementable in other dependently typed languages than Agda.

\paragraph{Contributions}
Our central contribution is Sikkel, an Agda library for multimode simple type theory (MSTT) based on presheaf models.
It combines some novel and some existing ideas to obtain a system with novel characteristics:
\begin{itemize}
\item Sikkel's semantic layer is fully general w.r.t.\ the base category.
  As such, it can in principle represent many different type theory extensions like guarded recursion \cite{clock-cat,gratzer20-multimodal}, parametricity \cite{fomega-parametricity,dtt-parametricity,moulin-param3,nuyts17-parametric,nuyts18-degrees}, univalence \cite{model-cubical,cubical-unifying,cubical}, nominal \cite{freshmltt} and directed \cite{shulman-riehl-directed} type theory, etc., as well as combinations of these \cite{cavallo-harper-paramhott-pub,weaver-licata-dua}.
  
\item Sikkel is built on multimode type theory, which allows programs to move between modes (presheaf models over different base categories), along modalities (adjunctions between those presheaf models).
  Thanks to this, the extraction mechanism does not expose the user to extension-specific models (like the topos of trees or reflexive graphs).
  Essentially, conversion from the richer object theory to the meta-theory is done within Sikkel (using a modality to the trivial mode) rather than outside it.

\item \adapted{Sikkel is easy to extend.
  The layer 2 model has an open mode theory and modes and modalities can be added simply by implementing an additional base category or a dependent right adjoint on presheaf categories, respectively.
  In the syntactic layer, the mode theory as well as primitive types and their typing rules can be changed as long as they can be checked in a way that is sound with respect to the Sikkel model.}

\item To accomodate future Sikkel support for dependent types, the presheaf model already implements the structural rules of an internal category with families (CwF) \cite{dybjer96-internal}: contexts, context-dependent types and terms, equality judgments,
  and type and term substitution.
  
\item We demonstrate two concrete examples of multimode simple type theories in Sikkel: one with support for guarded recursion, where we are able to implement all examples involving streams that were presented on paper by Clouston et al.~\cite[pp.\ 8--12]{clouston17-guarded}, and one with support for proving representation independence results using parametricity.

\end{itemize}

\paragraph{Overview of the paper}
We explain Sikkel's syntactic layer in Section~\ref{sec:deep-embedding}, followed by a discussion of guarded recursion in Section~\ref{sec:appl-guard-recurs}.
This application will be used as a motivating example throughout the rest of the paper.
Section~\ref{sec:presheaves} presents the semantic layer of Sikkel.
The type checker which translates between the first two layers is presented in Section~\ref{sec:sound-typechecker} and the extraction from the semantic layer to Agda can be found in Section~\ref{sec:extraction}.
To demonstrate Sikkel's generality, Section~\ref{sec:parametricity} discusses a second application: the use of parametricity to obtain representation independence results.
We conclude by discussing the current, related and future work in Section~\ref{sec:discussion}.

%% file: content-lagda/deep-embedding.tex
\section{Multimode Simple Type Theory (MSTT)}
\label{sec:deep-embedding}

\begin{code}[hide]%
\>[0]\AgdaSymbol{\{-\#}\AgdaSpace{}%
\AgdaKeyword{OPTIONS}\AgdaSpace{}%
\AgdaPragma{--allow-unsolved-metas}\AgdaSpace{}%
\AgdaSymbol{\#-\}}\<%
\\
\\[\AgdaEmptyExtraSkip]%
\>[0]\AgdaKeyword{module}\AgdaSpace{}%
\AgdaModule{deep-embedding}\AgdaSpace{}%
\AgdaKeyword{where}\<%
\\
\\[\AgdaEmptyExtraSkip]%
\>[0]\AgdaKeyword{open}\AgdaSpace{}%
\AgdaKeyword{import}\AgdaSpace{}%
\AgdaModule{Data.Nat}\<%
\\
\>[0]\AgdaKeyword{open}\AgdaSpace{}%
\AgdaKeyword{import}\AgdaSpace{}%
\AgdaModule{Data.String}\<%
\\
\>[0]\AgdaKeyword{open}\AgdaSpace{}%
\AgdaKeyword{import}\AgdaSpace{}%
\AgdaModule{Data.Unit}\<%
\\
\>[0]\AgdaKeyword{open}\AgdaSpace{}%
\AgdaKeyword{import}\AgdaSpace{}%
\AgdaModule{Level}\AgdaSpace{}%
\AgdaKeyword{using}\AgdaSpace{}%
\AgdaSymbol{(}\AgdaPostulate{Level}\AgdaSymbol{)}\<%
\\
\>[0]\AgdaKeyword{open}\AgdaSpace{}%
\AgdaKeyword{import}\AgdaSpace{}%
\AgdaModule{Relation.Binary.PropositionalEquality}\<%
\end{code}

\begin{wrapfigure}[19]{R}{.35\textwidth}
\vspace{-25pt}
\begin{AgdaMultiCode}
\begin{subfigure}{\linewidth}
\begin{code} \label{code:mode-theory}%
\>[0]\AgdaKeyword{record}\AgdaSpace{}%
\AgdaRecord{ModeTheory}\AgdaSpace{}%
\AgdaSymbol{:}\AgdaSpace{}%
\AgdaPrimitive{Set₁}\AgdaSpace{}%
\AgdaKeyword{where}\<%
\\
\>[0][@{}l@{\AgdaIndent{0}}]%
\>[2]\AgdaKeyword{field}\<%
\\
\>[2][@{}l@{\AgdaIndent{0}}]%
\>[4]\AgdaField{ModeExpr}\AgdaSpace{}%
\AgdaSymbol{:}\AgdaSpace{}%
\AgdaPrimitive{Set}\<%
\\
\>[4]\AgdaField{ModalityExpr}\AgdaSpace{}%
\AgdaSymbol{:}\AgdaSpace{}%
\AgdaField{ModeExpr}\AgdaSpace{}%
\AgdaSymbol{→}\<%
\\
\>[4][@{}l@{\AgdaIndent{0}}]%
\>[6]\AgdaField{ModeExpr}\AgdaSpace{}%
\AgdaSymbol{→}\AgdaSpace{}%
\AgdaPrimitive{Set}\<%
\\
\>[4]\AgdaOperator{\AgdaField{\AgdaUnderscore{}ⓜ\AgdaUnderscore{}}}\AgdaSpace{}%
\AgdaSymbol{:}\AgdaSpace{}%
\>[35I]\AgdaField{ModalityExpr}\AgdaSpace{}%
\AgdaBound{n}\AgdaSpace{}%
\AgdaBound{o}\AgdaSpace{}%
\AgdaSymbol{→}\<%
\\
\>[35I][@{}l@{\AgdaIndent{0}}]%
\>[34]\AgdaField{ModalityExpr}\AgdaSpace{}%
\AgdaBound{m}\AgdaSpace{}%
\AgdaBound{n}\AgdaSpace{}%
\AgdaSymbol{→}\<%
\\
\>[34]\AgdaField{ModalityExpr}\AgdaSpace{}%
\AgdaBound{m}\AgdaSpace{}%
\AgdaBound{o}\<%
\\
\>[4]\AgdaField{𝟙}\AgdaSpace{}%
\AgdaSymbol{:}\AgdaSpace{}%
\AgdaField{ModalityExpr}\AgdaSpace{}%
\AgdaBound{m}\AgdaSpace{}%
\AgdaBound{m}\<%
\end{code}
\caption{Fields introduced in Section~\ref{sec:mode-theory}.}
\label{fig:modetheory}
\end{subfigure}
\begin{subfigure}{\linewidth}
  \begin{code}%
\>[0]\<%
\\
\>[4]\AgdaField{TwoCellExpr}\AgdaSpace{}%
\AgdaSymbol{:}\AgdaSpace{}%
\AgdaPrimitive{Set}\<%
\\
\>[4]\AgdaField{id-cell}\AgdaSpace{}%
\AgdaSymbol{:}\AgdaSpace{}%
\AgdaField{TwoCellExpr}\<%
\end{code}
\caption{Fields introduced in Section~\ref{sec:typing-rules}.}
\label{fig:modetheory-cells}
\end{subfigure}
\end{AgdaMultiCode}
\caption{Type signature of the mode theory parameter to Sikkel. More fields are introduced in Figs.~\ref{fig:interp-mode-theory},~\ref{fig:interpret-2-cell} and \ref{fig:mode-theory-eq}.}
\label{fig:modetheorycomplete}
\end{wrapfigure}

We first present Sikkel's syntactic layer: Multimode Simple Type Theory (MSTT), which is essentially Multimode Type Theory (MTT) by Gratzer et al.~\cite{gratzer20-multimodal} restricted to simple (non-dependent) types.
Just like MTT, MSTT is parametrized by a mode theory specifying the available modalities, discussed in Section~\ref{sec:mode-theory}.
In Section~\ref{sec:types-contexts}, we present MSTT's type formers and contexts.
Finally, the typing rules are explained in Section~\ref{sec:typing-rules}.
We note that MSTT does not have an equational theory or evaluation function for terms.
This is unnecessary because MSTT is not dependently typed and all computational aspects will be handled by the model after interpretation of the syntax.

\subsection{Mode Theory}
\label{sec:mode-theory}
In a multimode type theory, every context, type or term lives at a particular mode and the available modes are specified by the mode theory.
Moreover, a mode theory fixes for every two modes $m$ and $n$ a collection of modalities going from $m$ to $n$ together with a composition operation for modalities and a unit modality for every mode.
As such, a mode theory can be any small category with the objects functioning as modes and the morphisms as modalities.

A Sikkel mode theory consists of an Agda type of modes and a type family of modalities together with the composition operation and unit modality (Fig.~\ref{fig:modetheory}).%
\footnote{Here \APt{Set} and \APt{Set₁} are the first two types in Agda's universe hierarchy.
  Note that the seemingly unbound variables $m$, $n$ and $o$ are implicitly universally quantified.
  \AFi{\AU{}\textcircled{\scriptsize m}\AU{}} is Agda syntax for defining a mixfix operator \AFi{\textcircled{\raisebox{.08ex}{\scriptsize m}}}. The underscore characters indicate where arguments are expected.
}
The \AR{ModeTheory} record actually contains a few other fields that we will introduce in Sections~\ref{sec:typing-rules} and~\ref{sec:sound-typechecker}.
In particular, a mode theory also specifies an equivalence relation $\modeq$ on
\begin{code}[hide]%
\>[0]\AgdaKeyword{module}\AgdaSpace{}%
\AgdaModule{\AgdaUnderscore{}}\AgdaSpace{}%
\AgdaSymbol{(}\AgdaBound{mt}\AgdaSpace{}%
\AgdaSymbol{:}\AgdaSpace{}%
\AgdaRecord{ModeTheory}\AgdaSymbol{)}\AgdaSpace{}%
\AgdaKeyword{where}\<%
\\
\>[0][@{}l@{\AgdaIndent{0}}]%
\>[2]\AgdaKeyword{open}\AgdaSpace{}%
\AgdaModule{ModeTheory}\AgdaSpace{}%
\AgdaBound{mt}\<%
\\
\>[2]\AgdaFunction{\AgdaUnderscore{}}\AgdaSpace{}%
\AgdaSymbol{:}\AgdaSpace{}%
\AgdaField{ModeExpr}\AgdaSpace{}%
\AgdaSymbol{→}\AgdaSpace{}%
\AgdaField{ModeExpr}\AgdaSpace{}%
\AgdaSymbol{→}\AgdaSpace{}%
\AgdaPrimitive{Set}\<%
\\
\>[2]\AgdaSymbol{\AgdaUnderscore{}}\AgdaSpace{}%
\AgdaSymbol{=}\AgdaSpace{}%
\AgdaSymbol{λ}%
\>[70I]\AgdaBound{m}\AgdaSpace{}%
\AgdaBound{n}\AgdaSpace{}%
\AgdaSymbol{→}\<%
\end{code}
\begin{code}[inline*]%
\>[.][@{}l@{}]\<[70I]%
\>[8]\AgdaField{ModalityExpr}\AgdaSpace{}%
\AgdaBound{m}\AgdaSpace{}%
\AgdaBound{n}\<%
\end{code}
for \adapted{every two} modes $m$ and $n$.
Associativity and unit laws for \AFi{ⓜ} are expected to hold up to $\modeq$ in order for the \adapted{resulting type theory to be well-behaved}, but there are no corresponding proof obligations when constructing a \AR{ModeTheory} record.
In the remainder of this section, we assume a mode theory and field names of the record type will refer to this theory.
\begin{code}[hide]%
\>[0]\AgdaKeyword{module}\AgdaSpace{}%
\AgdaModule{\AgdaUnderscore{}}\AgdaSpace{}%
\AgdaSymbol{(}\AgdaBound{mt}\AgdaSpace{}%
\AgdaSymbol{:}\AgdaSpace{}%
\AgdaRecord{ModeTheory}\AgdaSymbol{)}\AgdaSpace{}%
\AgdaKeyword{where}\<%
\\
\>[0][@{}l@{\AgdaIndent{0}}]%
\>[2]\AgdaKeyword{open}\AgdaSpace{}%
\AgdaModule{ModeTheory}\AgdaSpace{}%
\AgdaBound{mt}\<%
\\
\\[\AgdaEmptyExtraSkip]%
\>[2]\AgdaKeyword{private}\AgdaSpace{}%
\AgdaKeyword{variable}\<%
\\
\>[2][@{}l@{\AgdaIndent{0}}]%
\>[4]\AgdaGeneralizable{m}\AgdaSpace{}%
\AgdaGeneralizable{n}\AgdaSpace{}%
\AgdaSymbol{:}\AgdaSpace{}%
\AgdaField{ModeExpr}\<%
\end{code}

\subsection{Types and Contexts}
\label{sec:types-contexts}
We define a type \ADT{TyExpr} of MSTT type expressions as an inductive family indexed by a \AFi{ModeExpr} (Fig.~\ref{fig:tyexpr}).
\begin{code}[hide]%
\>[2]\AgdaKeyword{infixr}\AgdaSpace{}%
\AgdaNumber{6}\AgdaSpace{}%
\AgdaOperator{\AgdaInductiveConstructor{\AgdaUnderscore{}⇛\AgdaUnderscore{}}}\<%
\\
\>[2]\AgdaKeyword{infixl}\AgdaSpace{}%
\AgdaNumber{5}\AgdaSpace{}%
\AgdaOperator{\AgdaInductiveConstructor{\AgdaUnderscore{}⊠\AgdaUnderscore{}}}\<%
\end{code}
\begin{wrapfigure}[14]{R}{.5\textwidth}
\vspace{-28pt}
\begin{AgdaMultiCode}
\begin{code}%
\>[2]\AgdaKeyword{data}\AgdaSpace{}%
\AgdaDatatype{TyExpr}\AgdaSpace{}%
\AgdaSymbol{:}\AgdaSpace{}%
\AgdaField{ModeExpr}\AgdaSpace{}%
\AgdaSymbol{→}\AgdaSpace{}%
\AgdaPrimitive{Set}\AgdaSpace{}%
\AgdaKeyword{where}\<%
\\
\>[2][@{}l@{\AgdaIndent{0}}]%
\>[4]\AgdaInductiveConstructor{Nat}\AgdaSpace{}%
\AgdaInductiveConstructor{Bool}\AgdaSpace{}%
\AgdaSymbol{:}\AgdaSpace{}%
\AgdaDatatype{TyExpr}\AgdaSpace{}%
\AgdaGeneralizable{m}\<%
\\
\>[4]\AgdaOperator{\AgdaInductiveConstructor{\AgdaUnderscore{}⇛\AgdaUnderscore{}}}\AgdaSpace{}%
\AgdaOperator{\AgdaInductiveConstructor{\AgdaUnderscore{}⊠\AgdaUnderscore{}}}\AgdaSpace{}%
\AgdaSymbol{:}\AgdaSpace{}%
\AgdaDatatype{TyExpr}\AgdaSpace{}%
\AgdaGeneralizable{m}\AgdaSpace{}%
\AgdaSymbol{→}\AgdaSpace{}%
\AgdaDatatype{TyExpr}\AgdaSpace{}%
\AgdaGeneralizable{m}\AgdaSpace{}%
\AgdaSymbol{→}\AgdaSpace{}%
\AgdaDatatype{TyExpr}\AgdaSpace{}%
\AgdaGeneralizable{m}\<%
\\
\>[4]\AgdaOperator{\AgdaInductiveConstructor{⟨\AgdaUnderscore{}∣\AgdaUnderscore{}⟩}}\AgdaSpace{}%
\AgdaSymbol{:}\AgdaSpace{}%
\AgdaField{ModalityExpr}\AgdaSpace{}%
\AgdaGeneralizable{m}\AgdaSpace{}%
\AgdaGeneralizable{n}\AgdaSpace{}%
\AgdaSymbol{→}\AgdaSpace{}%
\AgdaDatatype{TyExpr}\AgdaSpace{}%
\AgdaGeneralizable{m}\AgdaSpace{}%
\AgdaSymbol{→}\AgdaSpace{}%
\AgdaDatatype{TyExpr}\AgdaSpace{}%
\AgdaGeneralizable{n}\<%
\\
\>[0]\<%
\end{code}
\begin{code}%
\>[0][@{}l@{\AgdaIndent{1}}]%
\>[2]\AgdaKeyword{data}\AgdaSpace{}%
\AgdaDatatype{CtxExpr}\AgdaSpace{}%
\AgdaSymbol{:}\AgdaSpace{}%
\AgdaField{ModeExpr}\AgdaSpace{}%
\AgdaSymbol{→}\AgdaSpace{}%
\AgdaPrimitive{Set}\AgdaSpace{}%
\AgdaKeyword{where}\<%
\\
\>[2][@{}l@{\AgdaIndent{0}}]%
\>[4]\AgdaInductiveConstructor{◇}\AgdaSpace{}%
\AgdaSymbol{:}\AgdaSpace{}%
\AgdaDatatype{CtxExpr}\AgdaSpace{}%
\AgdaGeneralizable{m}\<%
\\
\>[4]\AgdaOperator{\AgdaInductiveConstructor{\AgdaUnderscore{},\AgdaUnderscore{}∣\AgdaUnderscore{}∈\AgdaUnderscore{}}}\AgdaSpace{}%
\AgdaSymbol{:}\AgdaSpace{}%
\AgdaDatatype{CtxExpr}\AgdaSpace{}%
\AgdaGeneralizable{m}\AgdaSpace{}%
\AgdaSymbol{→}\AgdaSpace{}%
\AgdaField{ModalityExpr}\AgdaSpace{}%
\AgdaGeneralizable{n}\AgdaSpace{}%
\AgdaGeneralizable{m}\AgdaSpace{}%
\AgdaSymbol{→}\<%
\\
\>[4][@{}l@{\AgdaIndent{0}}]%
\>[6]\AgdaPostulate{String}\AgdaSpace{}%
\AgdaSymbol{→}\AgdaSpace{}%
\AgdaDatatype{TyExpr}\AgdaSpace{}%
\AgdaGeneralizable{n}\AgdaSpace{}%
\AgdaSymbol{→}\AgdaSpace{}%
\AgdaDatatype{CtxExpr}\AgdaSpace{}%
\AgdaGeneralizable{m}\<%
\\
\>[4]\AgdaOperator{\AgdaInductiveConstructor{\AgdaUnderscore{},lock⟨\AgdaUnderscore{}⟩}}\AgdaSpace{}%
\AgdaSymbol{:}\AgdaSpace{}%
\AgdaDatatype{CtxExpr}\AgdaSpace{}%
\AgdaGeneralizable{n}\AgdaSpace{}%
\AgdaSymbol{→}\AgdaSpace{}%
\AgdaField{ModalityExpr}\AgdaSpace{}%
\AgdaGeneralizable{m}\AgdaSpace{}%
\AgdaGeneralizable{n}\AgdaSpace{}%
\AgdaSymbol{→}\<%
\\
\>[4][@{}l@{\AgdaIndent{0}}]%
\>[6]\AgdaDatatype{CtxExpr}\AgdaSpace{}%
\AgdaGeneralizable{m}\<%
\end{code}
\end{AgdaMultiCode}
\caption{Syntax for MSTT types and contexts.}
\label{fig:ctxexpr} \label{fig:tyexpr}
\end{wrapfigure}
MSTT has built-in types for natural numbers, booleans, functions (\AIC{⇛}) and products (\AIC{⊠}) at any mode.
The most interesting type former takes a modality \AB{μ} from mode $m$ to $n$ and a type \AB{T} at mode $m$ to produce the modal type
\begin{code}[hide]%
\>[2]\AgdaFunction{\AgdaUnderscore{}}\AgdaSpace{}%
\AgdaSymbol{:}\AgdaSpace{}%
\AgdaField{ModalityExpr}\AgdaSpace{}%
\AgdaGeneralizable{m}\AgdaSpace{}%
\AgdaGeneralizable{n}\AgdaSpace{}%
\AgdaSymbol{→}\AgdaSpace{}%
\AgdaDatatype{TyExpr}\AgdaSpace{}%
\AgdaGeneralizable{m}\AgdaSpace{}%
\AgdaSymbol{→}\AgdaSpace{}%
\AgdaDatatype{TyExpr}\AgdaSpace{}%
\AgdaGeneralizable{n}\<%
\\
\>[2]\AgdaSymbol{\AgdaUnderscore{}}\AgdaSpace{}%
\AgdaSymbol{=}%
\>[163I]\AgdaSymbol{λ}\AgdaSpace{}%
\AgdaBound{μ}\AgdaSpace{}%
\AgdaBound{T}\AgdaSpace{}%
\AgdaSymbol{→}\<%
\end{code}%
\begin{code}[inline*]%
\>[.][@{}l@{}]\<[163I]%
\>[6]\AgdaOperator{\AgdaInductiveConstructor{⟨}}\AgdaSpace{}%
\AgdaBound{μ}\AgdaSpace{}%
\AgdaOperator{\AgdaInductiveConstructor{∣}}\AgdaSpace{}%
\AgdaBound{T}\AgdaSpace{}%
\AgdaOperator{\AgdaInductiveConstructor{⟩}}\<%
\end{code}
at mode $n$.

The typing rules for MSTT will use an equivalence relation $\tyeq$ on types as a form of judgmental equality.
This relation $\tyeq$ is the smallest congruence on types such that the modal type former respects the relation $\modeq$ on modalities.

MSTT contexts also live at a particular mode (Fig.~\ref{fig:ctxexpr}).
Every mode has an empty context \AIC{◇}.
Variables in MSTT are represented by strings%
\footnote{A de~Bruijn approach is inherent to the CwF-structure of the semantic layer. As such, the strings will be resolved to de~Bruijn indices during type-checking/interpretation and hence before any equational or computational reasoning can occur \adapted{(recall that MSTT does not have an equational theory or evaluation function for terms)}.}
and are annotated with a modality. As such, we can extend any context \AB{Γ} at mode $m$ with a variable \AB{x} of type \AB{T} (at mode $n$) under a modality \AB{μ} from $n$ to $m$, to obtain a new context
\begin{code}[hide]%
\>[2]\AgdaFunction{\AgdaUnderscore{}}\AgdaSpace{}%
\AgdaSymbol{:}\AgdaSpace{}%
\AgdaDatatype{CtxExpr}\AgdaSpace{}%
\AgdaGeneralizable{m}\AgdaSpace{}%
\AgdaSymbol{→}\AgdaSpace{}%
\AgdaField{ModalityExpr}\AgdaSpace{}%
\AgdaGeneralizable{n}\AgdaSpace{}%
\AgdaGeneralizable{m}\AgdaSpace{}%
\AgdaSymbol{→}\AgdaSpace{}%
\AgdaPostulate{String}\AgdaSpace{}%
\AgdaSymbol{→}\AgdaSpace{}%
\AgdaDatatype{TyExpr}\AgdaSpace{}%
\AgdaGeneralizable{n}\AgdaSpace{}%
\AgdaSymbol{→}\AgdaSpace{}%
\AgdaDatatype{CtxExpr}\AgdaSpace{}%
\AgdaGeneralizable{m}\<%
\\
\>[2]\AgdaSymbol{\AgdaUnderscore{}}\AgdaSpace{}%
\AgdaSymbol{=}%
\>[187I]\AgdaSymbol{λ}\AgdaSpace{}%
\AgdaBound{Γ}\AgdaSpace{}%
\AgdaBound{μ}\AgdaSpace{}%
\AgdaBound{x}\AgdaSpace{}%
\AgdaBound{T}\AgdaSpace{}%
\AgdaSymbol{→}\<%
\end{code}
\begin{code}[inline]%
\>[187I][@{}l@{\AgdaIndent{1}}]%
\>[7]\AgdaBound{Γ}\AgdaSpace{}%
\AgdaOperator{\AgdaInductiveConstructor{,}}\AgdaSpace{}%
\AgdaBound{μ}\AgdaSpace{}%
\AgdaOperator{\AgdaInductiveConstructor{∣}}\AgdaSpace{}%
\AgdaBound{x}\AgdaSpace{}%
\AgdaOperator{\AgdaInductiveConstructor{∈}}\AgdaSpace{}%
\AgdaBound{T}\<%
\end{code}.
Finally, a context \AB{Γ} at mode $n$ can be locked with a modality $\mu$ from mode $m$ to $n$ to produce a new context
\begin{code}[hide]%
\>[2]\AgdaFunction{\AgdaUnderscore{}}\AgdaSpace{}%
\AgdaSymbol{:}\AgdaSpace{}%
\AgdaDatatype{CtxExpr}\AgdaSpace{}%
\AgdaGeneralizable{n}\AgdaSpace{}%
\AgdaSymbol{→}\AgdaSpace{}%
\AgdaField{ModalityExpr}\AgdaSpace{}%
\AgdaGeneralizable{m}\AgdaSpace{}%
\AgdaGeneralizable{n}\AgdaSpace{}%
\AgdaSymbol{→}\AgdaSpace{}%
\AgdaDatatype{CtxExpr}\AgdaSpace{}%
\AgdaGeneralizable{m}\<%
\\
\>[2]\AgdaSymbol{\AgdaUnderscore{}}\AgdaSpace{}%
\AgdaSymbol{=}%
\>[210I]\AgdaSymbol{λ}\AgdaSpace{}%
\AgdaBound{Γ}\AgdaSpace{}%
\AgdaBound{μ}\AgdaSpace{}%
\AgdaSymbol{→}\<%
\end{code}
\begin{code}[inline*]%
\>[210I][@{}l@{\AgdaIndent{1}}]%
\>[7]\AgdaBound{Γ}\AgdaSpace{}%
\AgdaOperator{\AgdaInductiveConstructor{,lock⟨}}\AgdaSpace{}%
\AgdaBound{μ}\AgdaSpace{}%
\AgdaOperator{\AgdaInductiveConstructor{⟩}}\<%
\end{code}
at mode $m$.
This operation will play an important role in the introduction rule for modal types and in the variable rule.

\subsection{The Typing Rules}
\label{sec:typing-rules}

\begin{code}[hide]%
\>[2]\AgdaKeyword{infixl}\AgdaSpace{}%
\AgdaNumber{50}\AgdaSpace{}%
\AgdaOperator{\AgdaInductiveConstructor{\AgdaUnderscore{}∙\AgdaUnderscore{}}}\<%
\\
\>[2]\AgdaKeyword{infixr}\AgdaSpace{}%
\AgdaNumber{4}\AgdaSpace{}%
\AgdaOperator{\AgdaInductiveConstructor{lam[\AgdaUnderscore{}∈\AgdaUnderscore{}]\AgdaUnderscore{}}}\AgdaSpace{}%
\AgdaOperator{\AgdaInductiveConstructor{mod⟨\AgdaUnderscore{}⟩\AgdaUnderscore{}}}\<%
\\
\>[2]\AgdaKeyword{data}\AgdaSpace{}%
\AgdaDatatype{TmExpr}\AgdaSpace{}%
\AgdaSymbol{:}\AgdaSpace{}%
\AgdaField{ModeExpr}\AgdaSpace{}%
\AgdaSymbol{→}\AgdaSpace{}%
\AgdaPrimitive{Set}\AgdaSpace{}%
\AgdaKeyword{where}\<%
\\
\>[2][@{}l@{\AgdaIndent{0}}]%
\>[4]\AgdaOperator{\AgdaInductiveConstructor{ann\AgdaUnderscore{}∈\AgdaUnderscore{}}}\AgdaSpace{}%
\AgdaSymbol{:}\AgdaSpace{}%
\AgdaDatatype{TmExpr}\AgdaSpace{}%
\AgdaGeneralizable{m}\AgdaSpace{}%
\AgdaSymbol{→}\AgdaSpace{}%
\AgdaDatatype{TyExpr}\AgdaSpace{}%
\AgdaGeneralizable{m}\AgdaSpace{}%
\AgdaSymbol{→}\AgdaSpace{}%
\AgdaDatatype{TmExpr}\AgdaSpace{}%
\AgdaGeneralizable{m}\<%
\\
\>[4]\AgdaInductiveConstructor{var}\AgdaSpace{}%
\AgdaSymbol{:}\AgdaSpace{}%
\AgdaPostulate{String}\AgdaSpace{}%
\AgdaSymbol{→}\AgdaSpace{}%
\AgdaField{TwoCellExpr}\AgdaSpace{}%
\AgdaSymbol{→}\AgdaSpace{}%
\AgdaDatatype{TmExpr}\AgdaSpace{}%
\AgdaGeneralizable{m}\<%
\\
\>[4]\AgdaOperator{\AgdaInductiveConstructor{lam[\AgdaUnderscore{}∈\AgdaUnderscore{}]\AgdaUnderscore{}}}\AgdaSpace{}%
\AgdaSymbol{:}\AgdaSpace{}%
\AgdaPostulate{String}\AgdaSpace{}%
\AgdaSymbol{→}\AgdaSpace{}%
\AgdaDatatype{TyExpr}\AgdaSpace{}%
\AgdaGeneralizable{m}\AgdaSpace{}%
\AgdaSymbol{→}\AgdaSpace{}%
\AgdaDatatype{TmExpr}\AgdaSpace{}%
\AgdaGeneralizable{m}\AgdaSpace{}%
\AgdaSymbol{→}\AgdaSpace{}%
\AgdaDatatype{TmExpr}\AgdaSpace{}%
\AgdaGeneralizable{m}\<%
\\
\>[4]\AgdaOperator{\AgdaInductiveConstructor{\AgdaUnderscore{}∙\AgdaUnderscore{}}}\AgdaSpace{}%
\AgdaSymbol{:}\AgdaSpace{}%
\AgdaDatatype{TmExpr}\AgdaSpace{}%
\AgdaGeneralizable{m}\AgdaSpace{}%
\AgdaSymbol{→}\AgdaSpace{}%
\AgdaDatatype{TmExpr}\AgdaSpace{}%
\AgdaGeneralizable{m}\AgdaSpace{}%
\AgdaSymbol{→}\AgdaSpace{}%
\AgdaDatatype{TmExpr}\AgdaSpace{}%
\AgdaGeneralizable{m}\<%
\\
\>[4]\AgdaInductiveConstructor{lit}\AgdaSpace{}%
\AgdaSymbol{:}\AgdaSpace{}%
\AgdaDatatype{ℕ}\AgdaSpace{}%
\AgdaSymbol{→}\AgdaSpace{}%
\AgdaDatatype{TmExpr}\AgdaSpace{}%
\AgdaGeneralizable{m}\<%
\\
\>[4]\AgdaInductiveConstructor{suc}\AgdaSpace{}%
\AgdaInductiveConstructor{plus}\AgdaSpace{}%
\AgdaSymbol{:}\AgdaSpace{}%
\AgdaDatatype{TmExpr}\AgdaSpace{}%
\AgdaGeneralizable{m}\<%
\\
\>[4]\AgdaInductiveConstructor{nat-elim}\AgdaSpace{}%
\AgdaSymbol{:}\AgdaSpace{}%
\AgdaDatatype{TmExpr}\AgdaSpace{}%
\AgdaGeneralizable{m}\AgdaSpace{}%
\AgdaSymbol{→}\AgdaSpace{}%
\AgdaDatatype{TmExpr}\AgdaSpace{}%
\AgdaGeneralizable{m}\AgdaSpace{}%
\AgdaSymbol{→}\AgdaSpace{}%
\AgdaDatatype{TmExpr}\AgdaSpace{}%
\AgdaGeneralizable{m}\<%
\\
\>[4]\AgdaInductiveConstructor{true}\AgdaSpace{}%
\AgdaInductiveConstructor{false}\AgdaSpace{}%
\AgdaSymbol{:}\AgdaSpace{}%
\AgdaDatatype{TmExpr}\AgdaSpace{}%
\AgdaGeneralizable{m}\<%
\\
\>[4]\AgdaInductiveConstructor{if}\AgdaSpace{}%
\AgdaSymbol{:}\AgdaSpace{}%
\AgdaDatatype{TmExpr}\AgdaSpace{}%
\AgdaGeneralizable{m}\AgdaSpace{}%
\AgdaSymbol{→}\AgdaSpace{}%
\AgdaDatatype{TmExpr}\AgdaSpace{}%
\AgdaGeneralizable{m}\AgdaSpace{}%
\AgdaSymbol{→}\AgdaSpace{}%
\AgdaDatatype{TmExpr}\AgdaSpace{}%
\AgdaGeneralizable{m}\AgdaSpace{}%
\AgdaSymbol{→}\AgdaSpace{}%
\AgdaDatatype{TmExpr}\AgdaSpace{}%
\AgdaGeneralizable{m}\<%
\\
\>[4]\AgdaInductiveConstructor{pair}\AgdaSpace{}%
\AgdaSymbol{:}\AgdaSpace{}%
\AgdaDatatype{TmExpr}\AgdaSpace{}%
\AgdaGeneralizable{m}\AgdaSpace{}%
\AgdaSymbol{→}\AgdaSpace{}%
\AgdaDatatype{TmExpr}\AgdaSpace{}%
\AgdaGeneralizable{m}\AgdaSpace{}%
\AgdaSymbol{→}\AgdaSpace{}%
\AgdaDatatype{TmExpr}\AgdaSpace{}%
\AgdaGeneralizable{m}\<%
\\
\>[4]\AgdaInductiveConstructor{fst}\AgdaSpace{}%
\AgdaInductiveConstructor{snd}\AgdaSpace{}%
\AgdaSymbol{:}\AgdaSpace{}%
\AgdaDatatype{TmExpr}\AgdaSpace{}%
\AgdaGeneralizable{m}\AgdaSpace{}%
\AgdaSymbol{→}\AgdaSpace{}%
\AgdaDatatype{TmExpr}\AgdaSpace{}%
\AgdaGeneralizable{m}\<%
\\
\>[4]\AgdaOperator{\AgdaInductiveConstructor{mod⟨\AgdaUnderscore{}⟩\AgdaUnderscore{}}}\AgdaSpace{}%
\AgdaSymbol{:}\AgdaSpace{}%
\AgdaField{ModalityExpr}\AgdaSpace{}%
\AgdaGeneralizable{m}\AgdaSpace{}%
\AgdaGeneralizable{n}\AgdaSpace{}%
\AgdaSymbol{→}\AgdaSpace{}%
\AgdaDatatype{TmExpr}\AgdaSpace{}%
\AgdaGeneralizable{m}\AgdaSpace{}%
\AgdaSymbol{→}\AgdaSpace{}%
\AgdaDatatype{TmExpr}\AgdaSpace{}%
\AgdaGeneralizable{n}\<%
\\
\>[4]\AgdaInductiveConstructor{mod-elim}\AgdaSpace{}%
\AgdaSymbol{:}\AgdaSpace{}%
\AgdaSymbol{∀}\AgdaSpace{}%
\AgdaSymbol{\{}\AgdaBound{m}\AgdaSpace{}%
\AgdaBound{m'}\AgdaSpace{}%
\AgdaBound{m''}\AgdaSymbol{\}}\AgdaSpace{}%
\AgdaSymbol{→}\AgdaSpace{}%
\AgdaField{ModalityExpr}\AgdaSpace{}%
\AgdaBound{m'}\AgdaSpace{}%
\AgdaBound{m}\AgdaSpace{}%
\AgdaSymbol{→}\AgdaSpace{}%
\AgdaField{ModalityExpr}\AgdaSpace{}%
\AgdaBound{m''}\AgdaSpace{}%
\AgdaBound{m'}\AgdaSpace{}%
\AgdaSymbol{→}\AgdaSpace{}%
\AgdaPostulate{String}\AgdaSpace{}%
\AgdaSymbol{→}\AgdaSpace{}%
\AgdaDatatype{TmExpr}\AgdaSpace{}%
\AgdaBound{m'}\AgdaSpace{}%
\AgdaSymbol{→}\AgdaSpace{}%
\AgdaDatatype{TmExpr}\AgdaSpace{}%
\AgdaBound{m}\AgdaSpace{}%
\AgdaSymbol{→}\AgdaSpace{}%
\AgdaDatatype{TmExpr}\AgdaSpace{}%
\AgdaBound{m}\<%
\\
\\[\AgdaEmptyExtraSkip]%
\>[2]\AgdaKeyword{syntax}\AgdaSpace{}%
\AgdaInductiveConstructor{mod-elim}\AgdaSpace{}%
\AgdaBound{ρ}\AgdaSpace{}%
\AgdaBound{μ}\AgdaSpace{}%
\AgdaBound{x}\AgdaSpace{}%
\AgdaBound{t}\AgdaSpace{}%
\AgdaBound{s}\AgdaSpace{}%
\AgdaSymbol{=}\AgdaSpace{}%
\AgdaInductiveConstructor{let⟨}\AgdaSpace{}%
\AgdaBound{ρ}\AgdaSpace{}%
\AgdaInductiveConstructor{⟩}\AgdaSpace{}%
\AgdaInductiveConstructor{mod⟨}\AgdaSpace{}%
\AgdaBound{μ}\AgdaSpace{}%
\AgdaInductiveConstructor{⟩}\AgdaSpace{}%
\AgdaBound{x}\AgdaSpace{}%
\AgdaInductiveConstructor{←}\AgdaSpace{}%
\AgdaBound{t}\AgdaSpace{}%
\AgdaInductiveConstructor{in'}\AgdaSpace{}%
\AgdaBound{s}\<%
\\
\\[\AgdaEmptyExtraSkip]%
\>[2]\AgdaFunction{mod-elim'}\AgdaSpace{}%
\AgdaSymbol{:}\AgdaSpace{}%
\AgdaSymbol{∀}\AgdaSpace{}%
\AgdaSymbol{\{}\AgdaBound{m'}\AgdaSpace{}%
\AgdaBound{m}\AgdaSymbol{\}}\AgdaSpace{}%
\AgdaSymbol{→}\AgdaSpace{}%
\AgdaField{ModalityExpr}\AgdaSpace{}%
\AgdaBound{m'}\AgdaSpace{}%
\AgdaBound{m}\AgdaSpace{}%
\AgdaSymbol{→}\AgdaSpace{}%
\AgdaPostulate{String}\AgdaSpace{}%
\AgdaSymbol{→}\AgdaSpace{}%
\AgdaDatatype{TmExpr}\AgdaSpace{}%
\AgdaBound{m}\AgdaSpace{}%
\AgdaSymbol{→}\AgdaSpace{}%
\AgdaDatatype{TmExpr}\AgdaSpace{}%
\AgdaBound{m}\AgdaSpace{}%
\AgdaSymbol{→}\AgdaSpace{}%
\AgdaDatatype{TmExpr}\AgdaSpace{}%
\AgdaBound{m}\<%
\\
\>[2]\AgdaFunction{mod-elim'}\AgdaSpace{}%
\AgdaSymbol{=}\AgdaSpace{}%
\AgdaInductiveConstructor{mod-elim}\AgdaSpace{}%
\AgdaField{𝟙}\<%
\\
\\[\AgdaEmptyExtraSkip]%
\>[2]\AgdaKeyword{syntax}\AgdaSpace{}%
\AgdaFunction{mod-elim'}\AgdaSpace{}%
\AgdaBound{μ}\AgdaSpace{}%
\AgdaBound{x}\AgdaSpace{}%
\AgdaBound{t}\AgdaSpace{}%
\AgdaBound{s}\AgdaSpace{}%
\AgdaSymbol{=}\AgdaSpace{}%
\AgdaFunction{let'}\AgdaSpace{}%
\AgdaFunction{mod⟨}\AgdaSpace{}%
\AgdaBound{μ}\AgdaSpace{}%
\AgdaFunction{⟩}\AgdaSpace{}%
\AgdaBound{x}\AgdaSpace{}%
\AgdaFunction{←}\AgdaSpace{}%
\AgdaBound{t}\AgdaSpace{}%
\AgdaFunction{in'}\AgdaSpace{}%
\AgdaBound{s}\<%
\\
\\[\AgdaEmptyExtraSkip]%
\>[2]\AgdaKeyword{infixr}\AgdaSpace{}%
\AgdaNumber{4}\AgdaSpace{}%
\AgdaOperator{\AgdaFunction{lam[\AgdaUnderscore{}∣\AgdaUnderscore{}∈\AgdaUnderscore{}]\AgdaUnderscore{}}}\<%
\\
\>[2]\AgdaOperator{\AgdaFunction{lam[\AgdaUnderscore{}∣\AgdaUnderscore{}∈\AgdaUnderscore{}]\AgdaUnderscore{}}}\AgdaSpace{}%
\AgdaSymbol{:}\AgdaSpace{}%
\AgdaField{ModalityExpr}\AgdaSpace{}%
\AgdaGeneralizable{n}\AgdaSpace{}%
\AgdaGeneralizable{m}\AgdaSpace{}%
\AgdaSymbol{→}\AgdaSpace{}%
\AgdaPostulate{String}\AgdaSpace{}%
\AgdaSymbol{→}\AgdaSpace{}%
\AgdaDatatype{TyExpr}\AgdaSpace{}%
\AgdaGeneralizable{n}\AgdaSpace{}%
\AgdaSymbol{→}\AgdaSpace{}%
\AgdaDatatype{TmExpr}\AgdaSpace{}%
\AgdaGeneralizable{m}\AgdaSpace{}%
\AgdaSymbol{→}\AgdaSpace{}%
\AgdaDatatype{TmExpr}\AgdaSpace{}%
\AgdaGeneralizable{m}\<%
\\
\>[2]\AgdaOperator{\AgdaFunction{lam[}}\AgdaSpace{}%
\AgdaBound{μ}\AgdaSpace{}%
\AgdaOperator{\AgdaFunction{∣}}\AgdaSpace{}%
\AgdaBound{x}\AgdaSpace{}%
\AgdaOperator{\AgdaFunction{∈}}\AgdaSpace{}%
\AgdaBound{T}\AgdaSpace{}%
\AgdaOperator{\AgdaFunction{]}}\AgdaSpace{}%
\AgdaBound{b}\AgdaSpace{}%
\AgdaSymbol{=}\AgdaSpace{}%
\AgdaOperator{\AgdaInductiveConstructor{lam[}}\AgdaSpace{}%
\AgdaBound{x}\AgdaSpace{}%
\AgdaOperator{\AgdaInductiveConstructor{∈}}\AgdaSpace{}%
\AgdaOperator{\AgdaInductiveConstructor{⟨}}\AgdaSpace{}%
\AgdaBound{μ}\AgdaSpace{}%
\AgdaOperator{\AgdaInductiveConstructor{∣}}\AgdaSpace{}%
\AgdaBound{T}\AgdaSpace{}%
\AgdaOperator{\AgdaInductiveConstructor{⟩}}\AgdaSpace{}%
\AgdaOperator{\AgdaInductiveConstructor{]}}\AgdaSpace{}%
\AgdaSymbol{(}\AgdaFunction{let'}\AgdaSpace{}%
\AgdaFunction{mod⟨}\AgdaSpace{}%
\AgdaBound{μ}\AgdaSpace{}%
\AgdaFunction{⟩}\AgdaSpace{}%
\AgdaBound{x}\AgdaSpace{}%
\AgdaFunction{←}\AgdaSpace{}%
\AgdaInductiveConstructor{var}\AgdaSpace{}%
\AgdaBound{x}\AgdaSpace{}%
\AgdaField{id-cell}\AgdaSpace{}%
\AgdaFunction{in'}\AgdaSpace{}%
\AgdaBound{b}\AgdaSymbol{)}\<%
\\
\\[\AgdaEmptyExtraSkip]%
\>[2]\AgdaKeyword{infixl}\AgdaSpace{}%
\AgdaNumber{50}\AgdaSpace{}%
\AgdaOperator{\AgdaFunction{\AgdaUnderscore{}∙⟨\AgdaUnderscore{}⟩\AgdaUnderscore{}}}\<%
\\
\>[2]\AgdaOperator{\AgdaFunction{\AgdaUnderscore{}∙⟨\AgdaUnderscore{}⟩\AgdaUnderscore{}}}\AgdaSpace{}%
\AgdaSymbol{:}\AgdaSpace{}%
\AgdaDatatype{TmExpr}\AgdaSpace{}%
\AgdaGeneralizable{m}\AgdaSpace{}%
\AgdaSymbol{→}\AgdaSpace{}%
\AgdaField{ModalityExpr}\AgdaSpace{}%
\AgdaGeneralizable{n}\AgdaSpace{}%
\AgdaGeneralizable{m}\AgdaSpace{}%
\AgdaSymbol{→}\AgdaSpace{}%
\AgdaDatatype{TmExpr}\AgdaSpace{}%
\AgdaGeneralizable{n}\AgdaSpace{}%
\AgdaSymbol{→}\AgdaSpace{}%
\AgdaDatatype{TmExpr}\AgdaSpace{}%
\AgdaGeneralizable{m}\<%
\\
\>[2]\AgdaBound{f}\AgdaSpace{}%
\AgdaOperator{\AgdaFunction{∙⟨}}\AgdaSpace{}%
\AgdaBound{μ}\AgdaSpace{}%
\AgdaOperator{\AgdaFunction{⟩}}\AgdaSpace{}%
\AgdaBound{t}\AgdaSpace{}%
\AgdaSymbol{=}\AgdaSpace{}%
\AgdaBound{f}\AgdaSpace{}%
\AgdaOperator{\AgdaInductiveConstructor{∙}}\AgdaSpace{}%
\AgdaSymbol{(}\AgdaOperator{\AgdaInductiveConstructor{mod⟨}}\AgdaSpace{}%
\AgdaBound{μ}\AgdaSpace{}%
\AgdaOperator{\AgdaInductiveConstructor{⟩}}\AgdaSpace{}%
\AgdaBound{t}\AgdaSymbol{)}\<%
\end{code}

\newcommand{\tmVarCtx}{%
  \begin{code}[hide]%
\>[2]\AgdaFunction{\AgdaUnderscore{}}\AgdaSpace{}%
\AgdaSymbol{:}\AgdaSpace{}%
\AgdaDatatype{CtxExpr}\AgdaSpace{}%
\AgdaGeneralizable{m}\AgdaSpace{}%
\AgdaSymbol{→}\AgdaSpace{}%
\AgdaField{ModalityExpr}\AgdaSpace{}%
\AgdaGeneralizable{n}\AgdaSpace{}%
\AgdaGeneralizable{m}\AgdaSpace{}%
\AgdaSymbol{→}\AgdaSpace{}%
\AgdaPostulate{String}\AgdaSpace{}%
\AgdaSymbol{→}\AgdaSpace{}%
\AgdaDatatype{TyExpr}\AgdaSpace{}%
\AgdaGeneralizable{n}\AgdaSpace{}%
\AgdaSymbol{→}\AgdaSpace{}%
\AgdaDatatype{CtxExpr}\AgdaSpace{}%
\AgdaGeneralizable{m}\<%
\\
\>[2]\AgdaSymbol{\AgdaUnderscore{}}\AgdaSpace{}%
\AgdaSymbol{=}%
\>[490I]\AgdaSymbol{λ}\AgdaSpace{}%
\AgdaBound{Γ}\AgdaSpace{}%
\AgdaBound{μ}\AgdaSpace{}%
\AgdaBound{x}\AgdaSpace{}%
\AgdaBound{T}\AgdaSpace{}%
\AgdaSymbol{→}\<%
\end{code}%
  \begin{code}[inline*]%
\>[.][@{}l@{}]\<[490I]%
\>[6]\AgdaBound{Γ}\AgdaSpace{}%
\AgdaOperator{\AgdaInductiveConstructor{,}}\AgdaSpace{}%
\AgdaBound{μ}\AgdaSpace{}%
\AgdaOperator{\AgdaInductiveConstructor{∣}}\AgdaSpace{}%
\AgdaBound{x}\AgdaSpace{}%
\AgdaOperator{\AgdaInductiveConstructor{∈}}\AgdaSpace{}%
\AgdaBound{T}\<%
\end{code}
}
\newcommand{\tmVarTm}{%
  \begin{code}[hide]%
\>[2]\AgdaFunction{\AgdaUnderscore{}}\AgdaSpace{}%
\AgdaSymbol{:}\AgdaSpace{}%
\AgdaPostulate{String}\AgdaSpace{}%
\AgdaSymbol{→}\AgdaSpace{}%
\AgdaField{TwoCellExpr}\AgdaSpace{}%
\AgdaSymbol{→}\AgdaSpace{}%
\AgdaDatatype{TmExpr}\AgdaSpace{}%
\AgdaGeneralizable{m}\<%
\\
\>[2]\AgdaSymbol{\AgdaUnderscore{}}\AgdaSpace{}%
\AgdaSymbol{=}%
\>[510I]\AgdaSymbol{λ}\AgdaSpace{}%
\AgdaBound{x}\AgdaSpace{}%
\AgdaBound{α}\AgdaSpace{}%
\AgdaSymbol{→}\<%
\end{code}%
  \begin{code}[inline]%
\>[.][@{}l@{}]\<[510I]%
\>[6]\AgdaInductiveConstructor{var}\AgdaSpace{}%
\AgdaBound{x}\AgdaSpace{}%
\AgdaBound{α}\<%
\end{code}
}
\newcommand{\tmLamCtx}{%
  \begin{code}[hide]%
\>[2]\AgdaFunction{\AgdaUnderscore{}}\AgdaSpace{}%
\AgdaSymbol{:}\AgdaSpace{}%
\AgdaDatatype{CtxExpr}\AgdaSpace{}%
\AgdaGeneralizable{m}\AgdaSpace{}%
\AgdaSymbol{→}\AgdaSpace{}%
\AgdaPostulate{String}\AgdaSpace{}%
\AgdaSymbol{→}\AgdaSpace{}%
\AgdaDatatype{TyExpr}\AgdaSpace{}%
\AgdaGeneralizable{m}\AgdaSpace{}%
\AgdaSymbol{→}\AgdaSpace{}%
\AgdaDatatype{CtxExpr}\AgdaSpace{}%
\AgdaGeneralizable{m}\<%
\\
\>[2]\AgdaSymbol{\AgdaUnderscore{}}\AgdaSpace{}%
\AgdaSymbol{=}%
\>[528I]\AgdaSymbol{λ}\AgdaSpace{}%
\AgdaBound{Γ}\AgdaSpace{}%
\AgdaBound{x}\AgdaSpace{}%
\AgdaBound{T}\AgdaSpace{}%
\AgdaSymbol{→}\<%
\end{code}%
  \begin{code}[inline*]%
\>[.][@{}l@{}]\<[528I]%
\>[6]\AgdaBound{Γ}\AgdaSpace{}%
\AgdaOperator{\AgdaInductiveConstructor{,}}\AgdaSpace{}%
\AgdaField{𝟙}\AgdaSpace{}%
\AgdaOperator{\AgdaInductiveConstructor{∣}}\AgdaSpace{}%
\AgdaBound{x}\AgdaSpace{}%
\AgdaOperator{\AgdaInductiveConstructor{∈}}\AgdaSpace{}%
\AgdaBound{T}\<%
\end{code}
}
\newcommand{\tmLamTm}{%
  \begin{code}[hide]%
\>[2]\AgdaFunction{\AgdaUnderscore{}}\AgdaSpace{}%
\AgdaSymbol{:}\AgdaSpace{}%
\AgdaPostulate{String}\AgdaSpace{}%
\AgdaSymbol{→}\AgdaSpace{}%
\AgdaDatatype{TyExpr}\AgdaSpace{}%
\AgdaGeneralizable{m}\AgdaSpace{}%
\AgdaSymbol{→}\AgdaSpace{}%
\AgdaDatatype{TmExpr}\AgdaSpace{}%
\AgdaGeneralizable{m}\AgdaSpace{}%
\AgdaSymbol{→}\AgdaSpace{}%
\AgdaDatatype{TmExpr}\AgdaSpace{}%
\AgdaGeneralizable{m}\<%
\\
\>[2]\AgdaSymbol{\AgdaUnderscore{}}\AgdaSpace{}%
\AgdaSymbol{=}%
\>[551I]\AgdaSymbol{λ}\AgdaSpace{}%
\AgdaBound{x}\AgdaSpace{}%
\AgdaBound{T}\AgdaSpace{}%
\AgdaBound{s}\AgdaSpace{}%
\AgdaSymbol{→}\<%
\end{code}%
  \begin{code}[inline]%
\>[.][@{}l@{}]\<[551I]%
\>[6]\AgdaOperator{\AgdaInductiveConstructor{lam[}}\AgdaSpace{}%
\AgdaBound{x}\AgdaSpace{}%
\AgdaOperator{\AgdaInductiveConstructor{∈}}\AgdaSpace{}%
\AgdaBound{T}\AgdaSpace{}%
\AgdaOperator{\AgdaInductiveConstructor{]}}\AgdaSpace{}%
\AgdaBound{s}\<%
\end{code}
}
\newcommand{\tmLamTy}{%
  \begin{code}[hide]%
\>[2]\AgdaFunction{\AgdaUnderscore{}}\AgdaSpace{}%
\AgdaSymbol{:}\AgdaSpace{}%
\AgdaDatatype{TyExpr}\AgdaSpace{}%
\AgdaGeneralizable{m}\AgdaSpace{}%
\AgdaSymbol{→}\AgdaSpace{}%
\AgdaDatatype{TyExpr}\AgdaSpace{}%
\AgdaGeneralizable{m}\AgdaSpace{}%
\AgdaSymbol{→}\AgdaSpace{}%
\AgdaDatatype{TyExpr}\AgdaSpace{}%
\AgdaGeneralizable{m}\<%
\\
\>[2]\AgdaSymbol{\AgdaUnderscore{}}\AgdaSpace{}%
\AgdaSymbol{=}%
\>[571I]\AgdaSymbol{λ}\AgdaSpace{}%
\AgdaBound{T}\AgdaSpace{}%
\AgdaBound{S}\AgdaSpace{}%
\AgdaSymbol{→}\<%
\end{code}%
  \begin{code}[inline]%
\>[.][@{}l@{}]\<[571I]%
\>[6]\AgdaBound{T}\AgdaSpace{}%
\AgdaOperator{\AgdaInductiveConstructor{⇛}}\AgdaSpace{}%
\AgdaBound{S}\<%
\end{code}
}
\newcommand{\tmAppTy}{%
  \begin{code}[hide]%
\>[2]\AgdaFunction{\AgdaUnderscore{}}\AgdaSpace{}%
\AgdaSymbol{:}\AgdaSpace{}%
\AgdaDatatype{TyExpr}\AgdaSpace{}%
\AgdaGeneralizable{m}\AgdaSpace{}%
\AgdaSymbol{→}\AgdaSpace{}%
\AgdaDatatype{TyExpr}\AgdaSpace{}%
\AgdaGeneralizable{m}\AgdaSpace{}%
\AgdaSymbol{→}\AgdaSpace{}%
\AgdaDatatype{TyExpr}\AgdaSpace{}%
\AgdaGeneralizable{m}\<%
\\
\>[2]\AgdaSymbol{\AgdaUnderscore{}}\AgdaSpace{}%
\AgdaSymbol{=}%
\>[587I]\AgdaSymbol{λ}\AgdaSpace{}%
\AgdaBound{R}\AgdaSpace{}%
\AgdaBound{S}\AgdaSpace{}%
\AgdaSymbol{→}\<%
\end{code}%
  \begin{code}[inline]%
\>[.][@{}l@{}]\<[587I]%
\>[6]\AgdaBound{R}\AgdaSpace{}%
\AgdaOperator{\AgdaInductiveConstructor{⇛}}\AgdaSpace{}%
\AgdaBound{S}\<%
\end{code}
}
\newcommand{\tmAppTm}{%
  \begin{code}[hide]%
\>[2]\AgdaFunction{\AgdaUnderscore{}}\AgdaSpace{}%
\AgdaSymbol{:}\AgdaSpace{}%
\AgdaDatatype{TmExpr}\AgdaSpace{}%
\AgdaGeneralizable{m}\AgdaSpace{}%
\AgdaSymbol{→}\AgdaSpace{}%
\AgdaDatatype{TmExpr}\AgdaSpace{}%
\AgdaGeneralizable{m}\AgdaSpace{}%
\AgdaSymbol{→}\AgdaSpace{}%
\AgdaDatatype{TmExpr}\AgdaSpace{}%
\AgdaGeneralizable{m}\<%
\\
\>[2]\AgdaSymbol{\AgdaUnderscore{}}\AgdaSpace{}%
\AgdaSymbol{=}%
\>[603I]\AgdaSymbol{λ}\AgdaSpace{}%
\AgdaBound{f}\AgdaSpace{}%
\AgdaBound{t}\AgdaSpace{}%
\AgdaSymbol{→}\<%
\end{code}%
  \begin{code}[inline]%
\>[.][@{}l@{}]\<[603I]%
\>[6]\AgdaBound{f}\AgdaSpace{}%
\AgdaOperator{\AgdaInductiveConstructor{∙}}\AgdaSpace{}%
\AgdaBound{t}\<%
\end{code}
}
\newcommand{\tmModIntroCtx}{%
  \begin{code}[hide]%
\>[2]\AgdaFunction{\AgdaUnderscore{}}\AgdaSpace{}%
\AgdaSymbol{:}\AgdaSpace{}%
\AgdaDatatype{CtxExpr}\AgdaSpace{}%
\AgdaGeneralizable{n}\AgdaSpace{}%
\AgdaSymbol{→}\AgdaSpace{}%
\AgdaField{ModalityExpr}\AgdaSpace{}%
\AgdaGeneralizable{m}\AgdaSpace{}%
\AgdaGeneralizable{n}\AgdaSpace{}%
\AgdaSymbol{→}\AgdaSpace{}%
\AgdaDatatype{CtxExpr}\AgdaSpace{}%
\AgdaGeneralizable{m}\<%
\\
\>[2]\AgdaSymbol{\AgdaUnderscore{}}\AgdaSpace{}%
\AgdaSymbol{=}%
\>[620I]\AgdaSymbol{λ}\AgdaSpace{}%
\AgdaBound{Γ}\AgdaSpace{}%
\AgdaBound{μ}\AgdaSpace{}%
\AgdaSymbol{→}\<%
\end{code}%
  \begin{code}[inline]%
\>[.][@{}l@{}]\<[620I]%
\>[6]\AgdaBound{Γ}\AgdaSpace{}%
\AgdaOperator{\AgdaInductiveConstructor{,lock⟨}}\AgdaSpace{}%
\AgdaBound{μ}\AgdaSpace{}%
\AgdaOperator{\AgdaInductiveConstructor{⟩}}\<%
\end{code}
}
\newcommand{\tmModIntroTy}{%
  \begin{code}[hide]%
\>[2]\AgdaFunction{\AgdaUnderscore{}}\AgdaSpace{}%
\AgdaSymbol{:}\AgdaSpace{}%
\AgdaField{ModalityExpr}\AgdaSpace{}%
\AgdaGeneralizable{m}\AgdaSpace{}%
\AgdaGeneralizable{n}\AgdaSpace{}%
\AgdaSymbol{→}\AgdaSpace{}%
\AgdaDatatype{TyExpr}\AgdaSpace{}%
\AgdaGeneralizable{m}\AgdaSpace{}%
\AgdaSymbol{→}\AgdaSpace{}%
\AgdaDatatype{TyExpr}\AgdaSpace{}%
\AgdaGeneralizable{n}\<%
\\
\>[2]\AgdaSymbol{\AgdaUnderscore{}}\AgdaSpace{}%
\AgdaSymbol{=}%
\>[638I]\AgdaSymbol{λ}\AgdaSpace{}%
\AgdaBound{μ}\AgdaSpace{}%
\AgdaBound{T}\AgdaSpace{}%
\AgdaSymbol{→}\<%
\end{code}%
  \begin{code}[inline]%
\>[.][@{}l@{}]\<[638I]%
\>[6]\AgdaOperator{\AgdaInductiveConstructor{⟨}}\AgdaSpace{}%
\AgdaBound{μ}\AgdaSpace{}%
\AgdaOperator{\AgdaInductiveConstructor{∣}}\AgdaSpace{}%
\AgdaBound{T}\AgdaSpace{}%
\AgdaOperator{\AgdaInductiveConstructor{⟩}}\<%
\end{code}
}
\newcommand{\tmModIntroTm}{%
  \begin{code}[hide]%
\>[2]\AgdaFunction{\AgdaUnderscore{}}\AgdaSpace{}%
\AgdaSymbol{:}\AgdaSpace{}%
\AgdaField{ModalityExpr}\AgdaSpace{}%
\AgdaGeneralizable{m}\AgdaSpace{}%
\AgdaGeneralizable{n}\AgdaSpace{}%
\AgdaSymbol{→}\AgdaSpace{}%
\AgdaDatatype{TmExpr}\AgdaSpace{}%
\AgdaGeneralizable{m}\AgdaSpace{}%
\AgdaSymbol{→}\AgdaSpace{}%
\AgdaDatatype{TmExpr}\AgdaSpace{}%
\AgdaGeneralizable{n}\<%
\\
\>[2]\AgdaSymbol{\AgdaUnderscore{}}\AgdaSpace{}%
\AgdaSymbol{=}%
\>[657I]\AgdaSymbol{λ}\AgdaSpace{}%
\AgdaBound{μ}\AgdaSpace{}%
\AgdaBound{t}\AgdaSpace{}%
\AgdaSymbol{→}\<%
\end{code}%
  \begin{code}[inline]%
\>[.][@{}l@{}]\<[657I]%
\>[6]\AgdaOperator{\AgdaInductiveConstructor{mod⟨}}\AgdaSpace{}%
\AgdaBound{μ}\AgdaSpace{}%
\AgdaOperator{\AgdaInductiveConstructor{⟩}}\AgdaSpace{}%
\AgdaBound{t}\<%
\end{code}
}
\newcommand{\tmModElimCtx}{%
  \begin{code}[hide]%
\>[2]\AgdaFunction{\AgdaUnderscore{}}\AgdaSpace{}%
\AgdaSymbol{:}\AgdaSpace{}%
\AgdaDatatype{CtxExpr}\AgdaSpace{}%
\AgdaGeneralizable{n}\AgdaSpace{}%
\AgdaSymbol{→}\AgdaSpace{}%
\AgdaField{ModalityExpr}\AgdaSpace{}%
\AgdaGeneralizable{m}\AgdaSpace{}%
\AgdaGeneralizable{n}\AgdaSpace{}%
\AgdaSymbol{→}\AgdaSpace{}%
\AgdaPostulate{String}\AgdaSpace{}%
\AgdaSymbol{→}\AgdaSpace{}%
\AgdaDatatype{TyExpr}\AgdaSpace{}%
\AgdaGeneralizable{m}\AgdaSpace{}%
\AgdaSymbol{→}\AgdaSpace{}%
\AgdaDatatype{CtxExpr}\AgdaSpace{}%
\AgdaGeneralizable{n}\<%
\\
\>[2]\AgdaSymbol{\AgdaUnderscore{}}\AgdaSpace{}%
\AgdaSymbol{=}%
\>[680I]\AgdaSymbol{λ}\AgdaSpace{}%
\AgdaBound{Γ}\AgdaSpace{}%
\AgdaBound{μ}\AgdaSpace{}%
\AgdaBound{x}\AgdaSpace{}%
\AgdaBound{T}\AgdaSpace{}%
\AgdaSymbol{→}\<%
\end{code}%
  \begin{code}[inline*]%
\>[.][@{}l@{}]\<[680I]%
\>[6]\AgdaBound{Γ}\AgdaSpace{}%
\AgdaOperator{\AgdaInductiveConstructor{,}}\AgdaSpace{}%
\AgdaBound{μ}\AgdaSpace{}%
\AgdaOperator{\AgdaInductiveConstructor{∣}}\AgdaSpace{}%
\AgdaBound{x}\AgdaSpace{}%
\AgdaOperator{\AgdaInductiveConstructor{∈}}\AgdaSpace{}%
\AgdaBound{T}\<%
\end{code}
}
\newcommand{\tmModElimTy}{%
  \begin{code}[hide]%
\>[2]\AgdaFunction{\AgdaUnderscore{}}\AgdaSpace{}%
\AgdaSymbol{:}\AgdaSpace{}%
\AgdaField{ModalityExpr}\AgdaSpace{}%
\AgdaGeneralizable{m}\AgdaSpace{}%
\AgdaGeneralizable{n}\AgdaSpace{}%
\AgdaSymbol{→}\AgdaSpace{}%
\AgdaDatatype{TyExpr}\AgdaSpace{}%
\AgdaGeneralizable{m}\AgdaSpace{}%
\AgdaSymbol{→}\AgdaSpace{}%
\AgdaDatatype{TyExpr}\AgdaSpace{}%
\AgdaGeneralizable{n}\<%
\\
\>[2]\AgdaSymbol{\AgdaUnderscore{}}\AgdaSpace{}%
\AgdaSymbol{=}%
\>[703I]\AgdaSymbol{λ}\AgdaSpace{}%
\AgdaBound{ρ}\AgdaSpace{}%
\AgdaBound{T}\AgdaSpace{}%
\AgdaSymbol{→}\<%
\end{code}%
  \begin{code}[inline]%
\>[.][@{}l@{}]\<[703I]%
\>[6]\AgdaOperator{\AgdaInductiveConstructor{⟨}}\AgdaSpace{}%
\AgdaBound{ρ}\AgdaSpace{}%
\AgdaOperator{\AgdaInductiveConstructor{∣}}\AgdaSpace{}%
\AgdaBound{T}\AgdaSpace{}%
\AgdaOperator{\AgdaInductiveConstructor{⟩}}\<%
\end{code}
}
\newcommand{\tmModElimTm}{%
  \begin{code}[hide]%
\>[2]\AgdaFunction{\AgdaUnderscore{}}\AgdaSpace{}%
\AgdaSymbol{:}\AgdaSpace{}%
\AgdaField{ModalityExpr}\AgdaSpace{}%
\AgdaGeneralizable{m}\AgdaSpace{}%
\AgdaGeneralizable{n}\AgdaSpace{}%
\AgdaSymbol{→}\AgdaSpace{}%
\AgdaPostulate{String}\AgdaSpace{}%
\AgdaSymbol{→}\AgdaSpace{}%
\AgdaDatatype{TmExpr}\AgdaSpace{}%
\AgdaGeneralizable{n}\AgdaSpace{}%
\AgdaSymbol{→}\AgdaSpace{}%
\AgdaDatatype{TmExpr}\AgdaSpace{}%
\AgdaGeneralizable{n}\AgdaSpace{}%
\AgdaSymbol{→}\AgdaSpace{}%
\AgdaDatatype{TmExpr}\AgdaSpace{}%
\AgdaGeneralizable{n}\<%
\\
\>[2]\AgdaSymbol{\AgdaUnderscore{}}\AgdaSpace{}%
\AgdaSymbol{=}%
\>[727I]\AgdaSymbol{λ}\AgdaSpace{}%
\AgdaBound{μ}\AgdaSpace{}%
\AgdaBound{x}\AgdaSpace{}%
\AgdaBound{t}\AgdaSpace{}%
\AgdaBound{s}\AgdaSpace{}%
\AgdaSymbol{→}\<%
\end{code}%
  \begin{code}[inline]%
\>[.][@{}l@{}]\<[727I]%
\>[6]\AgdaFunction{let'}\AgdaSpace{}%
\AgdaFunction{mod⟨}\AgdaSpace{}%
\AgdaBound{μ}\AgdaSpace{}%
\AgdaFunction{⟩}\AgdaSpace{}%
\AgdaBound{x}\AgdaSpace{}%
\AgdaFunction{←}\AgdaSpace{}%
\AgdaBound{t}\AgdaSpace{}%
\AgdaFunction{in'}\AgdaSpace{}%
\AgdaBound{s}\<%
\end{code}
}
\newcommand{\tmAnn}{%
  \begin{code}[hide]%
\>[2]\AgdaFunction{\AgdaUnderscore{}}\AgdaSpace{}%
\AgdaSymbol{:}\AgdaSpace{}%
\AgdaDatatype{TmExpr}\AgdaSpace{}%
\AgdaGeneralizable{m}\AgdaSpace{}%
\AgdaSymbol{→}\AgdaSpace{}%
\AgdaDatatype{TyExpr}\AgdaSpace{}%
\AgdaGeneralizable{m}\AgdaSpace{}%
\AgdaSymbol{→}\AgdaSpace{}%
\AgdaDatatype{TmExpr}\AgdaSpace{}%
\AgdaGeneralizable{m}\<%
\\
\>[2]\AgdaSymbol{\AgdaUnderscore{}}\AgdaSpace{}%
\AgdaSymbol{=}%
\>[751I]\AgdaSymbol{λ}\AgdaSpace{}%
\AgdaBound{t}\AgdaSpace{}%
\AgdaBound{S}\AgdaSpace{}%
\AgdaSymbol{→}\<%
\end{code}%
  \begin{code}[inline]%
\>[.][@{}l@{}]\<[751I]%
\>[6]\AgdaOperator{\AgdaInductiveConstructor{ann}}\AgdaSpace{}%
\AgdaBound{t}\AgdaSpace{}%
\AgdaOperator{\AgdaInductiveConstructor{∈}}\AgdaSpace{}%
\AgdaBound{S}\<%
\end{code}
}
\begin{figure*}[tb]
  \begin{minipage}{\textwidth}
    \begin{equation*}
      \prftree[r]{\rulename{Tm-Var}}{
        \prfassumption{\AB{α} \in \AB{μ} \Rightarrow \mathsf{locks}(\AB{Δ})}}{
        \prfassumption{\AB{x} \notin \AB{Δ}}}{
        \text{\tmVarCtx} , \AB{Δ} \vdash \text{\tmVarTm{}} : \AB{T}}
      \qquad
      \prftree[r]{\rulename{Tm-Ann}}{
        \prfassumption{\AB{Γ} \vdash \AB{t}\AS{} : \AB{T}}}{
        \prfassumption{\AB{T} \tyeq \AB{S}}}{
        \AB{Γ} \vdash \text{\tmAnn{}} : \AB{S}}
    \end{equation*}
  \end{minipage} \vspace{7pt} \\
  \begin{minipage}{\textwidth}
    \begin{equation*}
      \prftree[r]{\rulename{Tm-Lam}}{
        \prfassumption{\text{\tmLamCtx{}} \vdash \AB{s} : \AB{S}}}{
        \AB{Γ} \vdash \text{\tmLamTm{}} : \text{\tmLamTy{}}}
      \qquad
      \prftree[r]{\rulename{Tm-App}}{
        \prfassumption{\AB{Γ} \vdash \AB{f} : \text{\tmAppTy{}}}}{
        \prfassumption{\AB{Γ} \vdash \AB{t} : \AB{T}}}{
        \prfassumption{\AB{R} \tyeq \AB{T}}}{
        \AB{Γ} \vdash \text{\tmAppTm{}} : \AB{S}}
    \end{equation*}
  \end{minipage} \\ 
  \begin{minipage}{\textwidth}
    \begin{equation*}
      \prftree[r]{\rulename{Tm-ModIntro}}{
        \prfassumption{\text{\tmModIntroCtx{}} \vdash \AB{t} : \AB{T}}}{
        \AB{Γ} \vdash \text{\tmModIntroTm{}} : \text{\tmModIntroTy{}}}
      \qquad
      \prftree[r]{\rulename{Tm-ModElim}}{
        \prfassumption{\AB{Γ} \vdash \AB{t} : \text{\tmModElimTy{}}}}{
        \prfassumption{\rho \modeq \mu}}{
        \prfassumption{\text{\tmModElimCtx{}} \vdash \AB{s} : \AB{S}}}{
        \text{\AB{Γ}} \vdash \text{\tmModElimTm} : \AB{S}}
    \end{equation*}
  \end{minipage}
  \caption{Some typing rules of MSTT.}
  \label{fig:typing-rules}
\end{figure*}

The most interesting typing rules of MSTT can be found in Figure~\ref{fig:typing-rules}.
The rules and term formers for constructing and destructing product values, booleans, and natural numbers have been omitted.
Note that we are formulating MSTT as an \emph{algorithmic} system for type \emph{inference}, i.e.\ the typing judgment can be seen as taking the context and term as inputs and returning -- if the term is well-typed -- an inferred type as output.
This explains why in some places we reuse a type or modality symbol whereas in other places we use different symbols and check for equivalence.
In particular, we never have control over the type inferred by a premise (although of course we can raise a type error if it does not match a certain pattern) so we can only assign it a new symbol and check for equivalence if necessary.


\begin{code}[hide]%
\>[2]\AgdaOperator{\AgdaFunction{\AgdaUnderscore{}ⓣ-vert\AgdaUnderscore{}}}\AgdaSpace{}%
\AgdaOperator{\AgdaFunction{\AgdaUnderscore{}ⓣ-hor\AgdaUnderscore{}}}\AgdaSpace{}%
\AgdaSymbol{:}\AgdaSpace{}%
\AgdaField{TwoCellExpr}\AgdaSpace{}%
\AgdaSymbol{→}\AgdaSpace{}%
\AgdaField{TwoCellExpr}\AgdaSpace{}%
\AgdaSymbol{→}\AgdaSpace{}%
\AgdaField{TwoCellExpr}\<%
\\
\>[2]\AgdaOperator{\AgdaFunction{\AgdaUnderscore{}ⓣ-vert\AgdaUnderscore{}}}\AgdaSpace{}%
\AgdaSymbol{=}\AgdaSpace{}%
\AgdaHole{\{!!\}}\<%
\\
\>[2]\AgdaOperator{\AgdaFunction{\AgdaUnderscore{}ⓣ-hor\AgdaUnderscore{}}}\AgdaSpace{}%
\AgdaSymbol{=}\AgdaSpace{}%
\AgdaHole{\{!!\}}\<%
\end{code}
\newcommand{\alphaBetaVert}{%
\begin{code}[hide]
\>[2]\AgdaFunction{\AgdaUnderscore{}}\AgdaSpace{}%
\AgdaSymbol{:}\AgdaSpace{}%
\AgdaField{TwoCellExpr}\AgdaSpace{}%
\AgdaSymbol{→}\AgdaSpace{}%
\AgdaField{TwoCellExpr}\AgdaSpace{}%
\AgdaSymbol{→}\AgdaSpace{}%
\AgdaField{TwoCellExpr}\<%
\\
\>[2]\AgdaSymbol{\AgdaUnderscore{}}%
\>[775I]\AgdaSymbol{=}\AgdaSpace{}%
\AgdaSymbol{λ}\AgdaSpace{}%
\AgdaBound{α}\AgdaSpace{}%
\AgdaBound{β}\AgdaSpace{}%
\AgdaSymbol{→}\<%
\end{code}%
\begin{code}[inline]
\>[.][@{}l@{}]\<[775I]%
\>[4]\AgdaBound{α}\AgdaSpace{}%
\AgdaOperator{\AgdaFunction{ⓣ-vert}}\AgdaSpace{}%
\AgdaBound{β}\<%
\end{code}%
}%
\newcommand{\alphaHor}{%
\begin{code}[hide]
\>[2]\AgdaFunction{\AgdaUnderscore{}}\AgdaSpace{}%
\AgdaSymbol{:}\AgdaSpace{}%
\AgdaField{TwoCellExpr}\AgdaSpace{}%
\AgdaSymbol{→}\AgdaSpace{}%
\AgdaField{TwoCellExpr}\AgdaSpace{}%
\AgdaSymbol{→}\AgdaSpace{}%
\AgdaField{TwoCellExpr}\<%
\\
\>[2]\AgdaSymbol{\AgdaUnderscore{}}%
\>[788I]\AgdaSymbol{=}\AgdaSpace{}%
\AgdaSymbol{λ}\AgdaSpace{}%
\AgdaBound{α₁}\AgdaSpace{}%
\AgdaBound{α₂}\AgdaSpace{}%
\AgdaSymbol{→}\<%
\end{code}%
\begin{code}[inline]
\>[.][@{}l@{}]\<[788I]%
\>[4]\AgdaBound{α₁}\AgdaSpace{}%
\AgdaOperator{\AgdaFunction{ⓣ-hor}}\AgdaSpace{}%
\AgdaBound{α₂}\<%
\end{code}%
}%
\newcommand{\domainCompTwoCell}{%
\begin{code}[hide]
\>[2]\AgdaFunction{\AgdaUnderscore{}}\AgdaSpace{}%
\AgdaSymbol{:}\AgdaSpace{}%
\AgdaSymbol{∀}\AgdaSpace{}%
\AgdaSymbol{\{}\AgdaBound{m}\AgdaSpace{}%
\AgdaBound{n}\AgdaSpace{}%
\AgdaBound{o}\AgdaSymbol{\}}\AgdaSpace{}%
\AgdaSymbol{→}\AgdaSpace{}%
\AgdaField{ModalityExpr}\AgdaSpace{}%
\AgdaBound{n}\AgdaSpace{}%
\AgdaBound{o}\AgdaSpace{}%
\AgdaSymbol{→}\AgdaSpace{}%
\AgdaField{ModalityExpr}\AgdaSpace{}%
\AgdaBound{m}\AgdaSpace{}%
\AgdaBound{n}\AgdaSpace{}%
\AgdaSymbol{→}\AgdaSpace{}%
\AgdaField{ModalityExpr}\AgdaSpace{}%
\AgdaBound{m}\AgdaSpace{}%
\AgdaBound{o}\<%
\\
\>[2]\AgdaSymbol{\AgdaUnderscore{}}%
\>[812I]\AgdaSymbol{=}\AgdaSpace{}%
\AgdaSymbol{λ}\AgdaSpace{}%
\AgdaBound{μ₁}\AgdaSpace{}%
\AgdaBound{μ₂}\AgdaSpace{}%
\AgdaSymbol{→}\<%
\end{code}%
\begin{code}[inline]
\>[.][@{}l@{}]\<[812I]%
\>[4]\AgdaBound{μ₁}\AgdaSpace{}%
\AgdaOperator{\AgdaField{ⓜ}}\AgdaSpace{}%
\AgdaBound{μ₂}\<%
\end{code}%
}%
\newcommand{\codomainCompTwoCell}{%
\begin{code}[hide]
\>[2]\AgdaFunction{\AgdaUnderscore{}}\AgdaSpace{}%
\AgdaSymbol{:}\AgdaSpace{}%
\AgdaSymbol{∀}\AgdaSpace{}%
\AgdaSymbol{\{}\AgdaBound{m}\AgdaSpace{}%
\AgdaBound{n}\AgdaSpace{}%
\AgdaBound{o}\AgdaSymbol{\}}\AgdaSpace{}%
\AgdaSymbol{→}\AgdaSpace{}%
\AgdaField{ModalityExpr}\AgdaSpace{}%
\AgdaBound{n}\AgdaSpace{}%
\AgdaBound{o}\AgdaSpace{}%
\AgdaSymbol{→}\AgdaSpace{}%
\AgdaField{ModalityExpr}\AgdaSpace{}%
\AgdaBound{m}\AgdaSpace{}%
\AgdaBound{n}\AgdaSpace{}%
\AgdaSymbol{→}\AgdaSpace{}%
\AgdaField{ModalityExpr}\AgdaSpace{}%
\AgdaBound{m}\AgdaSpace{}%
\AgdaBound{o}\<%
\\
\>[2]\AgdaSymbol{\AgdaUnderscore{}}%
\>[836I]\AgdaSymbol{=}\AgdaSpace{}%
\AgdaSymbol{λ}\AgdaSpace{}%
\AgdaBound{ρ₁}\AgdaSpace{}%
\AgdaBound{ρ₂}\AgdaSpace{}%
\AgdaSymbol{→}\<%
\end{code}%
\begin{code}[inline]
\>[.][@{}l@{}]\<[836I]%
\>[4]\AgdaBound{ρ₁}\AgdaSpace{}%
\AgdaOperator{\AgdaField{ⓜ}}\AgdaSpace{}%
\AgdaBound{ρ₂}\<%
\end{code}
}%

The variable rule \inlinerulename{Tm-Var} mentions a new concept of the mode theory that we have not yet introduced.
Apart from modes and modalities, a \AR{ModeTheory} also specifies an Agda type \AFi{TwoCellExpr} of 2-cell expressions together with a particular so-called trivial 2-cell \AFi{id-cell} (Fig.~\ref{fig:modetheory-cells}).
\adapted{Such 2-cells can be seen as morphisms between modalities and they will control when a variable in the context can be used in the construction of a term.}
We write \twocell{$\alpha$}{$\mu$}{$\rho$} to mean that the 2-cell \AB{α} can have the modality \AB{μ} as its domain and the modality \AB{ρ} as its codomain and a \AR{ModeTheory} will additionally consist of a function to check such statements (Fig.~\ref{fig:interpret-2-cell}).
For the trivial 2-cell, it is required that \twocell{\AFi{id-cell}}{$\mu$}{$\rho$} for any \AB{μ} and \AB{ρ} as long as $\mu \modeq \rho$.%
\footnote{\adapted{
  In MTT \cite{gratzer20-multimodal}, a mode theory additionally provides 2 composition operations for 2-cells: vertical composition lets us combine \twocell{$\alpha$}{$\mu$}{$\rho$} and \twocell{$\beta$}{$\kappa$}{$\mu$} to obtain \twocell{\alphaBetaVert{}}{$\kappa$}{$\rho$} and horizontal composition allows to combine \twocell{$\alpha_1$}{$\mu_1$}{$\rho_1$} and \twocell{$\alpha_2$}{$\mu_2$}{$\rho_2$} to obtain \twocell{\alphaHor{}}{\domainCompTwoCell{}}{\codomainCompTwoCell{}} (the latter is of course only possible when the modalities involved can be composed).
  These operations make an MTT mode theory a strict 2-category instead of just a 1-category, but they are not required when constructing a record of type \AR{ModeTheory} because they are not necessary in the formulation of the MSTT syntax or type checker.
  It is however recommendable to provide them to the programmer from a user-friendliness perspective.}} 
The MSTT variable rule expresses that we can use a variable \AB{x} which is in the context under a modality \AB{μ}, provided that there is a 2-cell going from \AB{μ} to the composite of all modalities that appear in the locks to the right of \AB{x}.
In many cases, these two modalities are equivalent so Sikkel has the abbreviation
\begin{code}[hide]%
\>[2]\AgdaFunction{svar}\AgdaSpace{}%
\AgdaSymbol{:}\AgdaSpace{}%
\AgdaPostulate{String}\AgdaSpace{}%
\AgdaSymbol{→}\AgdaSpace{}%
\AgdaDatatype{TmExpr}\AgdaSpace{}%
\AgdaGeneralizable{m}\<%
\end{code}
\begin{code}[inline]%
\>[2]\AgdaFunction{svar}\AgdaSpace{}%
\AgdaBound{x}\AgdaSpace{}%
\AgdaSymbol{=}\AgdaSpace{}%
\AgdaInductiveConstructor{var}\AgdaSpace{}%
\AgdaBound{x}\AgdaSpace{}%
\AgdaField{id-cell}\<%
\end{code}.

The rule \inlinerulename{Tm-Ann} allows to cast a term to an equivalent type.
There are standard typing rules for lambda abstraction (\inlinerulename{Tm-Lam}, introducing a variable under the trivial modality \AFi{𝟙}, \adapted{see further below for details about modal functions}) and function application (\inlinerulename{Tm-App}).
The modal constructor
\begin{code}[hide]%
\>[2]\AgdaFunction{\AgdaUnderscore{}}\AgdaSpace{}%
\AgdaSymbol{:}\AgdaSpace{}%
\AgdaField{ModalityExpr}\AgdaSpace{}%
\AgdaGeneralizable{m}\AgdaSpace{}%
\AgdaGeneralizable{n}\AgdaSpace{}%
\AgdaSymbol{→}\AgdaSpace{}%
\AgdaDatatype{TmExpr}\AgdaSpace{}%
\AgdaGeneralizable{m}\AgdaSpace{}%
\AgdaSymbol{→}\AgdaSpace{}%
\AgdaDatatype{TmExpr}\AgdaSpace{}%
\AgdaGeneralizable{n}\<%
\\
\>[2]\AgdaSymbol{\AgdaUnderscore{}}%
\>[863I]\AgdaSymbol{=}\AgdaSpace{}%
\AgdaSymbol{λ}\AgdaSpace{}%
\AgdaBound{μ}\AgdaSpace{}%
\AgdaBound{t}\AgdaSpace{}%
\AgdaSymbol{→}\<%
\end{code}
\begin{code}[inline*]%
\>[.][@{}l@{}]\<[863I]%
\>[4]\AgdaOperator{\AgdaInductiveConstructor{mod⟨}}\AgdaSpace{}%
\AgdaBound{μ}\AgdaSpace{}%
\AgdaOperator{\AgdaInductiveConstructor{⟩}}\<%
\end{code}
\begin{code}[hide]%
\>[4][@{}l@{\AgdaIndent{1}}]%
\>[6]\AgdaBound{t}\<%
\end{code}
allows us to construct a term of the modal type
\begin{code}[hide]%
\>[2]\AgdaFunction{\AgdaUnderscore{}}\AgdaSpace{}%
\AgdaSymbol{:}\AgdaSpace{}%
\AgdaField{ModalityExpr}\AgdaSpace{}%
\AgdaGeneralizable{m}\AgdaSpace{}%
\AgdaGeneralizable{n}\AgdaSpace{}%
\AgdaSymbol{→}\AgdaSpace{}%
\AgdaDatatype{TyExpr}\AgdaSpace{}%
\AgdaGeneralizable{m}\AgdaSpace{}%
\AgdaSymbol{→}\AgdaSpace{}%
\AgdaDatatype{TyExpr}\AgdaSpace{}%
\AgdaGeneralizable{n}\<%
\\
\>[2]\AgdaSymbol{\AgdaUnderscore{}}%
\>[880I]\AgdaSymbol{=}\AgdaSpace{}%
\AgdaSymbol{λ}\AgdaSpace{}%
\AgdaBound{μ}\AgdaSpace{}%
\AgdaBound{T}\AgdaSpace{}%
\AgdaSymbol{→}\<%
\end{code}
\begin{code}[inline*]%
\>[880I][@{}l@{\AgdaIndent{1}}]%
\>[5]\AgdaOperator{\AgdaInductiveConstructor{⟨}}\AgdaSpace{}%
\AgdaBound{μ}\AgdaSpace{}%
\AgdaOperator{\AgdaInductiveConstructor{∣}}\AgdaSpace{}%
\AgdaBound{T}\AgdaSpace{}%
\AgdaOperator{\AgdaInductiveConstructor{⟩}}\<%
\end{code}
if we can produce a term of type \AB{T} after locking the context with modality \AB{μ} (\inlinerulename{Tm-ModIntro}).
This lock ensures that in both the premise and the conclusion of \inlinerulename{Tm-ModIntro}, context, term and type live at the same mode.
Indeed, if \AB{μ} goes from $m$ to $n$, then \AB{t} and \AB{T} will live at mode $m$ but \AB{Γ} will live at mode $n$.
The modal elimination rule \inlinerulename{Tm-ModElim} allows us to pattern match in some sense on a term of a modal type.%
\footnote{The primes in \AF{let'} and \AF{in'} avoid conflicts with Agda's keywords \AK{let} and \AK{in}.} 
More concretely, it allows to view any term \AB{t} of type
\begin{code}[hide]%
\>[2]\AgdaFunction{\AgdaUnderscore{}}\AgdaSpace{}%
\AgdaSymbol{:}\AgdaSpace{}%
\AgdaField{ModalityExpr}\AgdaSpace{}%
\AgdaGeneralizable{m}\AgdaSpace{}%
\AgdaGeneralizable{n}\AgdaSpace{}%
\AgdaSymbol{→}\AgdaSpace{}%
\AgdaDatatype{TyExpr}\AgdaSpace{}%
\AgdaGeneralizable{m}\AgdaSpace{}%
\AgdaSymbol{→}\AgdaSpace{}%
\AgdaDatatype{TyExpr}\AgdaSpace{}%
\AgdaGeneralizable{n}\<%
\\
\>[2]\AgdaSymbol{\AgdaUnderscore{}}%
\>[899I]\AgdaSymbol{=}\AgdaSpace{}%
\AgdaSymbol{λ}\AgdaSpace{}%
\AgdaBound{μ}\AgdaSpace{}%
\AgdaBound{T}\AgdaSpace{}%
\AgdaSymbol{→}\<%
\end{code}
\begin{code}[inline*]%
\>[899I][@{}l@{\AgdaIndent{1}}]%
\>[5]\AgdaOperator{\AgdaInductiveConstructor{⟨}}\AgdaSpace{}%
\AgdaBound{μ}\AgdaSpace{}%
\AgdaOperator{\AgdaInductiveConstructor{∣}}\AgdaSpace{}%
\AgdaBound{T}\AgdaSpace{}%
\AgdaOperator{\AgdaInductiveConstructor{⟩}}\<%
\end{code}
as a term of the form
\begin{code}[hide]%
\>[2]\AgdaFunction{\AgdaUnderscore{}}\AgdaSpace{}%
\AgdaSymbol{:}\AgdaSpace{}%
\AgdaField{ModalityExpr}\AgdaSpace{}%
\AgdaGeneralizable{m}\AgdaSpace{}%
\AgdaGeneralizable{n}\AgdaSpace{}%
\AgdaSymbol{→}\AgdaSpace{}%
\AgdaDatatype{TmExpr}\AgdaSpace{}%
\AgdaGeneralizable{m}\AgdaSpace{}%
\AgdaSymbol{→}\AgdaSpace{}%
\AgdaDatatype{TmExpr}\AgdaSpace{}%
\AgdaGeneralizable{n}\<%
\\
\>[2]\AgdaSymbol{\AgdaUnderscore{}}%
\>[918I]\AgdaSymbol{=}\AgdaSpace{}%
\AgdaSymbol{λ}\AgdaSpace{}%
\AgdaBound{μ}\AgdaSpace{}%
\AgdaBound{x}\AgdaSpace{}%
\AgdaSymbol{→}\<%
\end{code}
\begin{code}[inline*]%
\>[918I][@{}l@{\AgdaIndent{1}}]%
\>[5]\AgdaOperator{\AgdaInductiveConstructor{mod⟨}}\AgdaSpace{}%
\AgdaBound{μ}\AgdaSpace{}%
\AgdaOperator{\AgdaInductiveConstructor{⟩}}\AgdaSpace{}%
\AgdaBound{x}\<%
\end{code}
for some variable \AB{x} of type \AB{T} that appears in the context under modality \AB{μ}.
Just like in MTT, the modal eliminator is actually more expressive and may take an additional modality as a parameter, but this will not be needed in the paper.

MSTT has no \emph{primitive} term formers for modal functions, but they are provided as syntactic sugar and implemented using the primitives introduced in Figure~\ref{fig:typing-rules}.
More concretely, one can write
\begin{code}[hide]%
\>[2]\AgdaFunction{\AgdaUnderscore{}}\AgdaSpace{}%
\AgdaSymbol{:}\AgdaSpace{}%
\AgdaField{ModalityExpr}\AgdaSpace{}%
\AgdaGeneralizable{m}\AgdaSpace{}%
\AgdaGeneralizable{n}\AgdaSpace{}%
\AgdaSymbol{→}\AgdaSpace{}%
\AgdaPostulate{String}\AgdaSpace{}%
\AgdaSymbol{→}\AgdaSpace{}%
\AgdaDatatype{TyExpr}\AgdaSpace{}%
\AgdaGeneralizable{m}\AgdaSpace{}%
\AgdaSymbol{→}\AgdaSpace{}%
\AgdaDatatype{TmExpr}\AgdaSpace{}%
\AgdaGeneralizable{n}\AgdaSpace{}%
\AgdaSymbol{→}\AgdaSpace{}%
\AgdaDatatype{TmExpr}\AgdaSpace{}%
\AgdaGeneralizable{n}\<%
\\
\>[2]\AgdaSymbol{\AgdaUnderscore{}}%
\>[941I]\AgdaSymbol{=}\AgdaSpace{}%
\AgdaSymbol{λ}\AgdaSpace{}%
\AgdaBound{μ}\AgdaSpace{}%
\AgdaBound{x}\AgdaSpace{}%
\AgdaBound{T}\AgdaSpace{}%
\AgdaBound{s}\AgdaSpace{}%
\AgdaSymbol{→}\<%
\end{code}
\begin{code}[inline*]%
\>[.][@{}l@{}]\<[941I]%
\>[4]\AgdaOperator{\AgdaFunction{lam[}}\AgdaSpace{}%
\AgdaBound{μ}\AgdaSpace{}%
\AgdaOperator{\AgdaFunction{∣}}\AgdaSpace{}%
\AgdaBound{x}\AgdaSpace{}%
\AgdaOperator{\AgdaFunction{∈}}\AgdaSpace{}%
\AgdaBound{T}\AgdaSpace{}%
\AgdaOperator{\AgdaFunction{]}}\AgdaSpace{}%
\AgdaBound{s}\<%
\end{code}
to construct a function of type
\begin{code}[hide]%
\>[2]\AgdaFunction{\AgdaUnderscore{}}\AgdaSpace{}%
\AgdaSymbol{:}\AgdaSpace{}%
\AgdaField{ModalityExpr}\AgdaSpace{}%
\AgdaGeneralizable{m}\AgdaSpace{}%
\AgdaGeneralizable{n}\AgdaSpace{}%
\AgdaSymbol{→}\AgdaSpace{}%
\AgdaDatatype{TyExpr}\AgdaSpace{}%
\AgdaGeneralizable{m}\AgdaSpace{}%
\AgdaSymbol{→}\AgdaSpace{}%
\AgdaDatatype{TyExpr}\AgdaSpace{}%
\AgdaGeneralizable{n}\AgdaSpace{}%
\AgdaSymbol{→}\AgdaSpace{}%
\AgdaDatatype{TyExpr}\AgdaSpace{}%
\AgdaGeneralizable{n}\<%
\\
\>[2]\AgdaSymbol{\AgdaUnderscore{}}%
\>[968I]\AgdaSymbol{=}\AgdaSpace{}%
\AgdaSymbol{λ}\AgdaSpace{}%
\AgdaBound{μ}\AgdaSpace{}%
\AgdaBound{T}\AgdaSpace{}%
\AgdaBound{S}\AgdaSpace{}%
\AgdaSymbol{→}\<%
\end{code}
\begin{code}[inline]%
\>[.][@{}l@{}]\<[968I]%
\>[4]\AgdaOperator{\AgdaInductiveConstructor{⟨}}\AgdaSpace{}%
\AgdaBound{μ}\AgdaSpace{}%
\AgdaOperator{\AgdaInductiveConstructor{∣}}\AgdaSpace{}%
\AgdaBound{T}\AgdaSpace{}%
\AgdaOperator{\AgdaInductiveConstructor{⟩}}\AgdaSpace{}%
\AgdaOperator{\AgdaInductiveConstructor{⇛}}\AgdaSpace{}%
\AgdaBound{S}\<%
\end{code}.
In this case, when type-checking the term \AB{s}, the variable \AB{x} will appear in the context under the modality \AB{μ} instead of under \AFi{𝟙} as in \inlinerulename{Tm-Lam}.
We can also use modal function application
\begin{code}[hide]%
\>[2]\AgdaFunction{\AgdaUnderscore{}}\AgdaSpace{}%
\AgdaSymbol{:}\AgdaSpace{}%
\AgdaDatatype{TmExpr}\AgdaSpace{}%
\AgdaGeneralizable{n}\AgdaSpace{}%
\AgdaSymbol{→}\AgdaSpace{}%
\AgdaField{ModalityExpr}\AgdaSpace{}%
\AgdaGeneralizable{m}\AgdaSpace{}%
\AgdaGeneralizable{n}\AgdaSpace{}%
\AgdaSymbol{→}\AgdaSpace{}%
\AgdaDatatype{TmExpr}\AgdaSpace{}%
\AgdaGeneralizable{m}\AgdaSpace{}%
\AgdaSymbol{→}\AgdaSpace{}%
\AgdaDatatype{TmExpr}\AgdaSpace{}%
\AgdaGeneralizable{n}\<%
\\
\>[2]\AgdaSymbol{\AgdaUnderscore{}}\AgdaSpace{}%
\AgdaSymbol{=}%
\>[994I]\AgdaSymbol{λ}\AgdaSpace{}%
\AgdaBound{f}\AgdaSpace{}%
\AgdaBound{μ}\AgdaSpace{}%
\AgdaBound{t}\AgdaSpace{}%
\AgdaSymbol{→}\<%
\end{code}
\begin{code}[inline*]%
\>[.][@{}l@{}]\<[994I]%
\>[6]\AgdaBound{f}\AgdaSpace{}%
\AgdaOperator{\AgdaFunction{∙⟨}}\AgdaSpace{}%
\AgdaBound{μ}\AgdaSpace{}%
\AgdaOperator{\AgdaFunction{⟩}}\AgdaSpace{}%
\AgdaBound{t}\<%
\end{code}
to apply a modal function \AB{f} of type
\begin{code}[hide]%
\>[2]\AgdaFunction{\AgdaUnderscore{}}\AgdaSpace{}%
\AgdaSymbol{:}\AgdaSpace{}%
\AgdaField{ModalityExpr}\AgdaSpace{}%
\AgdaGeneralizable{m}\AgdaSpace{}%
\AgdaGeneralizable{n}\AgdaSpace{}%
\AgdaSymbol{→}\AgdaSpace{}%
\AgdaDatatype{TyExpr}\AgdaSpace{}%
\AgdaGeneralizable{m}\AgdaSpace{}%
\AgdaSymbol{→}\AgdaSpace{}%
\AgdaDatatype{TyExpr}\AgdaSpace{}%
\AgdaGeneralizable{n}\AgdaSpace{}%
\AgdaSymbol{→}\AgdaSpace{}%
\AgdaDatatype{TyExpr}\AgdaSpace{}%
\AgdaGeneralizable{n}\<%
\\
\>[2]\AgdaSymbol{\AgdaUnderscore{}}%
\>[1016I]\AgdaSymbol{=}\AgdaSpace{}%
\AgdaSymbol{λ}\AgdaSpace{}%
\AgdaBound{μ}\AgdaSpace{}%
\AgdaBound{T}\AgdaSpace{}%
\AgdaBound{S}\AgdaSpace{}%
\AgdaSymbol{→}\<%
\end{code}
\begin{code}[inline*]%
\>[.][@{}l@{}]\<[1016I]%
\>[4]\AgdaOperator{\AgdaInductiveConstructor{⟨}}\AgdaSpace{}%
\AgdaBound{μ}\AgdaSpace{}%
\AgdaOperator{\AgdaInductiveConstructor{∣}}\AgdaSpace{}%
\AgdaBound{T}\AgdaSpace{}%
\AgdaOperator{\AgdaInductiveConstructor{⟩}}\AgdaSpace{}%
\AgdaOperator{\AgdaInductiveConstructor{⇛}}\AgdaSpace{}%
\AgdaBound{S}\<%
\end{code}
to a term \AB{t} of type \AB{T}.
To check such a term in a context \AB{Γ}, the function \AB{f} will also be checked in \AB{Γ} but the argument \AB{t} will be checked in the locked context
\begin{code}[hide]%
\>[2]\AgdaFunction{\AgdaUnderscore{}}\AgdaSpace{}%
\AgdaSymbol{:}\AgdaSpace{}%
\AgdaDatatype{CtxExpr}\AgdaSpace{}%
\AgdaGeneralizable{n}\AgdaSpace{}%
\AgdaSymbol{→}\AgdaSpace{}%
\AgdaField{ModalityExpr}\AgdaSpace{}%
\AgdaGeneralizable{m}\AgdaSpace{}%
\AgdaGeneralizable{n}\AgdaSpace{}%
\AgdaSymbol{→}\AgdaSpace{}%
\AgdaDatatype{CtxExpr}\AgdaSpace{}%
\AgdaGeneralizable{m}\<%
\\
\>[2]\AgdaSymbol{\AgdaUnderscore{}}%
\>[1038I]\AgdaSymbol{=}\AgdaSpace{}%
\AgdaSymbol{λ}\AgdaSpace{}%
\AgdaBound{Γ}\AgdaSpace{}%
\AgdaBound{μ}\AgdaSpace{}%
\AgdaSymbol{→}\<%
\end{code}
\begin{code}[inline]%
\>[1038I][@{}l@{\AgdaIndent{1}}]%
\>[5]\AgdaBound{Γ}\AgdaSpace{}%
\AgdaOperator{\AgdaInductiveConstructor{,lock⟨}}\AgdaSpace{}%
\AgdaBound{μ}\AgdaSpace{}%
\AgdaOperator{\AgdaInductiveConstructor{⟩}}\<%
\end{code}.%
\footnote{This is \emph{almost} as good as a native modal function type like in MTT, except that 1) our modal function type of modality \AFi{𝟙} is not equivalent to the non-modal function type and 2) since $\eta$-equality cannot in general be expressed for modal types in MTT, if we were to add an equational theory to MSTT, then we cannot have $\eta$-equality for these `sugary' functions. These equalities however do hold in the presheaf model.}

%% file: content-lagda/guarded-recursion.tex
\section{Application 1: Guarded Recursive Type Theory}
\label{sec:appl-guard-recurs}

\begin{code}[hide]%
\>[0]\AgdaKeyword{module}\AgdaSpace{}%
\AgdaModule{guarded-recursion}\AgdaSpace{}%
\AgdaKeyword{where}\<%
\\
\\[\AgdaEmptyExtraSkip]%
\>[0]\AgdaKeyword{open}\AgdaSpace{}%
\AgdaKeyword{import}\AgdaSpace{}%
\AgdaModule{Data.String}\AgdaSpace{}%
\AgdaKeyword{using}\AgdaSpace{}%
\AgdaSymbol{(}\AgdaPostulate{String}\AgdaSymbol{)}\<%
\end{code}


\begin{code}[hide]%
\>[0]\AgdaKeyword{open}\AgdaSpace{}%
\AgdaKeyword{import}\AgdaSpace{}%
\AgdaModule{Applications.GuardedRecursion.MSTT.ModeTheory}\AgdaSpace{}%
\AgdaKeyword{renaming}\AgdaSpace{}%
\AgdaSymbol{(}\AgdaInductiveConstructor{𝟙≤later}\AgdaSpace{}%
\AgdaSymbol{to}\AgdaSpace{}%
\AgdaInductiveConstructor{𝟙-to-later}\AgdaSymbol{;}\AgdaSpace{}%
\AgdaInductiveConstructor{constantly∘forever≤𝟙}\AgdaSpace{}%
\AgdaSymbol{to}\AgdaSpace{}%
\AgdaInductiveConstructor{const∘forev-to-𝟙}\AgdaSymbol{)}\<%
\\
\>[0]\AgdaKeyword{open}\AgdaSpace{}%
\AgdaKeyword{import}\AgdaSpace{}%
\AgdaModule{Applications.GuardedRecursion.MSTT.Syntax.Type}\AgdaSpace{}%
\AgdaKeyword{renaming}\AgdaSpace{}%
\AgdaSymbol{(}\AgdaInductiveConstructor{Nat'}\AgdaSpace{}%
\AgdaSymbol{to}\AgdaSpace{}%
\AgdaInductiveConstructor{Nat}\AgdaSymbol{;}\AgdaSpace{}%
\AgdaInductiveConstructor{Bool'}\AgdaSpace{}%
\AgdaSymbol{to}\AgdaSpace{}%
\AgdaInductiveConstructor{Bool}\AgdaSymbol{)}\<%
\\
\>[0]\AgdaKeyword{open}\AgdaSpace{}%
\AgdaKeyword{import}\AgdaSpace{}%
\AgdaModule{Applications.GuardedRecursion.MSTT.Syntax.Term}\<%
\\
\>[0]\AgdaKeyword{open}\AgdaSpace{}%
\AgdaKeyword{import}\AgdaSpace{}%
\AgdaModule{Applications.GuardedRecursion.MSTT.TypeExtension}\<%
\\
\>[0]\AgdaKeyword{open}\AgdaSpace{}%
\AgdaKeyword{import}\AgdaSpace{}%
\AgdaModule{Applications.GuardedRecursion.MSTT.TermExtension}\<%
\\
\>[0]\AgdaKeyword{open}\AgdaSpace{}%
\AgdaKeyword{import}\AgdaSpace{}%
\AgdaModule{MSTT.Syntax.Context}\AgdaSpace{}%
\AgdaFunction{GR-mode-theory}\AgdaSpace{}%
\AgdaFunction{GR-ty-ext}\<%
\\
\>[0]\AgdaKeyword{open}\AgdaSpace{}%
\AgdaKeyword{import}\AgdaSpace{}%
\AgdaModule{MSTT.BasicOperations}\AgdaSpace{}%
\AgdaFunction{GR-mode-theory}\AgdaSpace{}%
\AgdaFunction{GR-ty-ext}\AgdaSpace{}%
\AgdaFunction{GR-tm-ext}\<%
\end{code}

\newcommand{\tmLobCtx}{%
  \begin{code}[hide]%
\>[0]\AgdaFunction{\AgdaUnderscore{}}\AgdaSpace{}%
\AgdaSymbol{:}\AgdaSpace{}%
\AgdaDatatype{CtxExpr}\AgdaSpace{}%
\AgdaInductiveConstructor{ω}\AgdaSpace{}%
\AgdaSymbol{→}\AgdaSpace{}%
\AgdaPostulate{String}\AgdaSpace{}%
\AgdaSymbol{→}\AgdaSpace{}%
\AgdaDatatype{TyExpr}\AgdaSpace{}%
\AgdaInductiveConstructor{ω}\AgdaSpace{}%
\AgdaSymbol{→}\AgdaSpace{}%
\AgdaDatatype{CtxExpr}\AgdaSpace{}%
\AgdaInductiveConstructor{ω}\<%
\\
\>[0]\AgdaSymbol{\AgdaUnderscore{}}\AgdaSpace{}%
\AgdaSymbol{=}\AgdaSpace{}%
\AgdaSymbol{λ}%
\>[52I]\AgdaBound{Γ}\AgdaSpace{}%
\AgdaBound{x}\AgdaSpace{}%
\AgdaBound{T}\AgdaSpace{}%
\AgdaSymbol{→}\<%
\end{code}%
  \begin{code}[inline]%
\>[.][@{}l@{}]\<[52I]%
\>[6]\AgdaBound{Γ}\AgdaSpace{}%
\AgdaOperator{\AgdaInductiveConstructor{,}}\AgdaSpace{}%
\AgdaInductiveConstructor{later}\AgdaSpace{}%
\AgdaOperator{\AgdaInductiveConstructor{∣}}\AgdaSpace{}%
\AgdaBound{x}\AgdaSpace{}%
\AgdaOperator{\AgdaInductiveConstructor{∈}}\AgdaSpace{}%
\AgdaBound{T}\<%
\end{code}
}
\newcommand{\tmLobTm}{%
  \begin{code}[hide]%
\>[0]\AgdaFunction{\AgdaUnderscore{}}\AgdaSpace{}%
\AgdaSymbol{:}\AgdaSpace{}%
\AgdaPostulate{String}\AgdaSpace{}%
\AgdaSymbol{→}\AgdaSpace{}%
\AgdaDatatype{TyExpr}\AgdaSpace{}%
\AgdaInductiveConstructor{ω}\AgdaSpace{}%
\AgdaSymbol{→}\AgdaSpace{}%
\AgdaDatatype{TmExpr}\AgdaSpace{}%
\AgdaInductiveConstructor{ω}\AgdaSpace{}%
\AgdaSymbol{→}\AgdaSpace{}%
\AgdaDatatype{TmExpr}\AgdaSpace{}%
\AgdaInductiveConstructor{ω}\<%
\\
\>[0]\AgdaSymbol{\AgdaUnderscore{}}\AgdaSpace{}%
\AgdaSymbol{=}\AgdaSpace{}%
\AgdaSymbol{λ}%
\>[75I]\AgdaBound{x}\AgdaSpace{}%
\AgdaBound{T}\AgdaSpace{}%
\AgdaBound{t}\AgdaSpace{}%
\AgdaSymbol{→}\<%
\end{code}%
  \begin{code}[inline]%
\>[.][@{}l@{}]\<[75I]%
\>[6]\AgdaOperator{\AgdaInductiveConstructor{löb[later∣}}\AgdaSpace{}%
\AgdaBound{x}\AgdaSpace{}%
\AgdaOperator{\AgdaInductiveConstructor{∈}}\AgdaSpace{}%
\AgdaBound{T}\AgdaSpace{}%
\AgdaOperator{\AgdaInductiveConstructor{]}}\AgdaSpace{}%
\AgdaBound{t}\<%
\end{code}
}
\newcommand{\tmGHeadTy}{%
  \begin{code}[hide]%
\>[0]\AgdaFunction{\AgdaUnderscore{}}\AgdaSpace{}%
\AgdaSymbol{:}\AgdaSpace{}%
\AgdaDatatype{TyExpr}\AgdaSpace{}%
\AgdaInductiveConstructor{★}\AgdaSpace{}%
\AgdaSymbol{→}\AgdaSpace{}%
\AgdaDatatype{TyExpr}\AgdaSpace{}%
\AgdaInductiveConstructor{ω}\<%
\\
\>[0]\AgdaSymbol{\AgdaUnderscore{}}\AgdaSpace{}%
\AgdaSymbol{=}\AgdaSpace{}%
\AgdaSymbol{λ}%
\>[92I]\AgdaBound{A}\AgdaSpace{}%
\AgdaSymbol{→}\<%
\end{code}%
  \begin{code}[inline]%
\>[.][@{}l@{}]\<[92I]%
\>[6]\AgdaInductiveConstructor{GStream}\AgdaSpace{}%
\AgdaBound{A}\AgdaSpace{}%
\AgdaOperator{\AgdaInductiveConstructor{⇛}}\AgdaSpace{}%
\AgdaOperator{\AgdaInductiveConstructor{⟨}}\AgdaSpace{}%
\AgdaInductiveConstructor{constantly}\AgdaSpace{}%
\AgdaOperator{\AgdaInductiveConstructor{∣}}\AgdaSpace{}%
\AgdaBound{A}\AgdaSpace{}%
\AgdaOperator{\AgdaInductiveConstructor{⟩}}\<%
\end{code}
}
\newcommand{\tmGTailTy}{%
  \begin{code}[hide]%
\>[0]\AgdaFunction{\AgdaUnderscore{}}\AgdaSpace{}%
\AgdaSymbol{:}\AgdaSpace{}%
\AgdaDatatype{TyExpr}\AgdaSpace{}%
\AgdaInductiveConstructor{★}\AgdaSpace{}%
\AgdaSymbol{→}\AgdaSpace{}%
\AgdaDatatype{TyExpr}\AgdaSpace{}%
\AgdaInductiveConstructor{ω}\<%
\\
\>[0]\AgdaSymbol{\AgdaUnderscore{}}\AgdaSpace{}%
\AgdaSymbol{=}\AgdaSpace{}%
\AgdaSymbol{λ}%
\>[109I]\AgdaBound{A}\AgdaSpace{}%
\AgdaSymbol{→}\<%
\end{code}%
  \begin{code}[inline]%
\>[.][@{}l@{}]\<[109I]%
\>[6]\AgdaInductiveConstructor{GStream}\AgdaSpace{}%
\AgdaBound{A}\AgdaSpace{}%
\AgdaOperator{\AgdaInductiveConstructor{⇛}}\AgdaSpace{}%
\AgdaFunction{▻}\AgdaSpace{}%
\AgdaSymbol{(}\AgdaInductiveConstructor{GStream}\AgdaSpace{}%
\AgdaBound{A}\AgdaSymbol{)}\<%
\end{code}
}
\newcommand{\tmGConsTy}{%
  \begin{code}[hide]%
\>[0]\AgdaFunction{\AgdaUnderscore{}}\AgdaSpace{}%
\AgdaSymbol{:}\AgdaSpace{}%
\AgdaDatatype{TyExpr}\AgdaSpace{}%
\AgdaInductiveConstructor{★}\AgdaSpace{}%
\AgdaSymbol{→}\AgdaSpace{}%
\AgdaDatatype{TyExpr}\AgdaSpace{}%
\AgdaInductiveConstructor{ω}\<%
\\
\>[0]\AgdaSymbol{\AgdaUnderscore{}}\AgdaSpace{}%
\AgdaSymbol{=}\AgdaSpace{}%
\AgdaSymbol{λ}%
\>[124I]\AgdaBound{A}\AgdaSpace{}%
\AgdaSymbol{→}\<%
\end{code}%
  \begin{code}[inline]%
\>[.][@{}l@{}]\<[124I]%
\>[6]\AgdaOperator{\AgdaInductiveConstructor{⟨}}\AgdaSpace{}%
\AgdaInductiveConstructor{constantly}\AgdaSpace{}%
\AgdaOperator{\AgdaInductiveConstructor{∣}}\AgdaSpace{}%
\AgdaBound{A}\AgdaSpace{}%
\AgdaOperator{\AgdaInductiveConstructor{⟩}}\AgdaSpace{}%
\AgdaOperator{\AgdaInductiveConstructor{⇛}}\AgdaSpace{}%
\AgdaFunction{▻}\AgdaSpace{}%
\AgdaSymbol{(}\AgdaInductiveConstructor{GStream}\AgdaSpace{}%
\AgdaBound{A}\AgdaSymbol{)}\AgdaSpace{}%
\AgdaOperator{\AgdaInductiveConstructor{⇛}}\AgdaSpace{}%
\AgdaInductiveConstructor{GStream}\AgdaSpace{}%
\AgdaBound{A}\<%
\end{code}
}

\newcommand{\GMTForeverLater}{%
  \begin{code}[hide]%
\>[0]\AgdaFunction{\AgdaUnderscore{}}\AgdaSpace{}%
\AgdaSymbol{:}\AgdaSpace{}%
\AgdaDatatype{ModalityExpr}\AgdaSpace{}%
\AgdaInductiveConstructor{ω}\AgdaSpace{}%
\AgdaInductiveConstructor{★}\<%
\\
\>[0]\AgdaSymbol{\AgdaUnderscore{}}%
\>[141I]\AgdaSymbol{=}\<%
\end{code}%
  \begin{code}[inline]%
\>[.][@{}l@{}]\<[141I]%
\>[2]\AgdaInductiveConstructor{forever}\AgdaSpace{}%
\AgdaOperator{\AgdaInductiveConstructor{ⓜ}}\AgdaSpace{}%
\AgdaInductiveConstructor{later}\<%
\end{code}%
}
\newcommand{\GMTForeverConstantly}{%
  \begin{code}[hide]%
\>[0]\AgdaFunction{\AgdaUnderscore{}}\AgdaSpace{}%
\AgdaSymbol{:}\AgdaSpace{}%
\AgdaDatatype{ModalityExpr}\AgdaSpace{}%
\AgdaInductiveConstructor{★}\AgdaSpace{}%
\AgdaInductiveConstructor{★}\<%
\\
\>[0]\AgdaSymbol{\AgdaUnderscore{}}%
\>[148I]\AgdaSymbol{=}\<%
\end{code}%
  \begin{code}[inline]%
\>[.][@{}l@{}]\<[148I]%
\>[2]\AgdaInductiveConstructor{forever}\AgdaSpace{}%
\AgdaOperator{\AgdaInductiveConstructor{ⓜ}}\AgdaSpace{}%
\AgdaInductiveConstructor{constantly}\<%
\end{code}%
}
\newcommand{\GMTUnitLaterTwoCell}{%
  \begin{code}[hide]%
\>[0]\AgdaFunction{\AgdaUnderscore{}}%
\>[151I]\AgdaSymbol{=}\<%
\end{code}%
  \begin{code}[inline]%
\>[.][@{}l@{}]\<[151I]%
\>[2]\AgdaInductiveConstructor{𝟙-to-later}\<%
\end{code}%
}
\newcommand{\GMTConstantlyForever}{%
  \begin{code}[hide]%
\>[0]\AgdaFunction{\AgdaUnderscore{}}\AgdaSpace{}%
\AgdaSymbol{:}\AgdaSpace{}%
\AgdaDatatype{ModalityExpr}\AgdaSpace{}%
\AgdaInductiveConstructor{ω}\AgdaSpace{}%
\AgdaInductiveConstructor{ω}\<%
\\
\>[0]\AgdaSymbol{\AgdaUnderscore{}}%
\>[156I]\AgdaSymbol{=}\<%
\end{code}%
  \begin{code}[inline]%
\>[.][@{}l@{}]\<[156I]%
\>[2]\AgdaSymbol{(}\AgdaInductiveConstructor{constantly}\AgdaSpace{}%
\AgdaOperator{\AgdaInductiveConstructor{ⓜ}}\AgdaSpace{}%
\AgdaInductiveConstructor{forever}\AgdaSymbol{)}\<%
\end{code}%
}
\newcommand{\GMTConstForevUnitTwoCell}{%
  \begin{code}[hide]%
\>[0]\AgdaFunction{\AgdaUnderscore{}}\AgdaSpace{}%
\AgdaSymbol{:}\AgdaSpace{}%
\AgdaDatatype{TwoCellExpr}\<%
\\
\>[0]\AgdaSymbol{\AgdaUnderscore{}}%
\>[161I]\AgdaSymbol{=}\<%
\end{code}%
  \begin{code}[inline]%
\>[.][@{}l@{}]\<[161I]%
\>[2]\AgdaInductiveConstructor{const∘forev-to-𝟙}\<%
\end{code}%
}

\begin{figure*}[tb]
\footnotesize
\centering
\begin{minipage}{.42\textwidth}
\begin{subfigure}{\linewidth}
\begin{AgdaMultiCode}
\begin{code}%
\>[0]\AgdaKeyword{record}\AgdaSpace{}%
\AgdaRecord{Stream}\AgdaSpace{}%
\AgdaSymbol{(}\AgdaBound{A}\AgdaSpace{}%
\AgdaSymbol{:}\AgdaSpace{}%
\AgdaPrimitive{Set}\AgdaSymbol{)}\AgdaSpace{}%
\AgdaSymbol{:}\AgdaSpace{}%
\AgdaPrimitive{Set}\AgdaSpace{}%
\AgdaKeyword{where}\<%
\\
\>[0][@{}l@{\AgdaIndent{0}}]%
\>[2]\AgdaKeyword{coinductive}\<%
\\
\>[2]\AgdaKeyword{field}%
\>[169I]\AgdaField{head}\AgdaSpace{}%
\AgdaSymbol{:}\AgdaSpace{}%
\AgdaBound{A}\<%
\\
\>[.][@{}l@{}]\<[169I]%
\>[8]\AgdaField{tail}\AgdaSpace{}%
\AgdaSymbol{:}\AgdaSpace{}%
\AgdaRecord{Stream}\AgdaSpace{}%
\AgdaBound{A}\<%
\end{code}%
\begin{code}[hide]%
\>[0]\AgdaKeyword{open}\AgdaSpace{}%
\AgdaModule{Stream}\AgdaSpace{}%
\AgdaKeyword{public}\<%
\end{code}%
\begin{code}[hide]%
\>[0]\AgdaKeyword{open}\AgdaSpace{}%
\AgdaKeyword{import}\AgdaSpace{}%
\AgdaModule{Data.Nat}\<%
\end{code}%
\begin{code}%
\>[0]\<%
\\
\>[0]\AgdaFunction{zeros}\AgdaSpace{}%
\AgdaSymbol{:}\AgdaSpace{}%
\AgdaRecord{Stream}\AgdaSpace{}%
\AgdaDatatype{ℕ}\<%
\\
\>[0]\AgdaField{head}\AgdaSpace{}%
\AgdaFunction{zeros}\AgdaSpace{}%
\AgdaSymbol{=}\AgdaSpace{}%
\AgdaNumber{0}\<%
\\
\>[0]\AgdaField{tail}\AgdaSpace{}%
\AgdaFunction{zeros}\AgdaSpace{}%
\AgdaSymbol{=}\AgdaSpace{}%
\AgdaFunction{zeros}\<%
\end{code}%
\begin{code}%
\>[0]\<%
\\
\>[0]\AgdaFunction{map}\AgdaSpace{}%
\AgdaSymbol{:}\AgdaSpace{}%
\AgdaSymbol{(}\AgdaBound{A}\AgdaSpace{}%
\AgdaSymbol{→}\AgdaSpace{}%
\AgdaBound{B}\AgdaSymbol{)}\AgdaSpace{}%
\AgdaSymbol{→}\AgdaSpace{}%
\AgdaRecord{Stream}\AgdaSpace{}%
\AgdaBound{A}\AgdaSpace{}%
\AgdaSymbol{→}\AgdaSpace{}%
\AgdaRecord{Stream}\AgdaSpace{}%
\AgdaBound{B}\<%
\\
\>[0]\AgdaField{head}\AgdaSpace{}%
\AgdaSymbol{(}\AgdaFunction{map}\AgdaSpace{}%
\AgdaBound{f}\AgdaSpace{}%
\AgdaBound{as}\AgdaSymbol{)}\AgdaSpace{}%
\AgdaSymbol{=}\AgdaSpace{}%
\AgdaBound{f}\AgdaSpace{}%
\AgdaSymbol{(}\AgdaField{head}\AgdaSpace{}%
\AgdaBound{as}\AgdaSymbol{)}\<%
\\
\>[0]\AgdaField{tail}\AgdaSpace{}%
\AgdaSymbol{(}\AgdaFunction{map}\AgdaSpace{}%
\AgdaBound{f}\AgdaSpace{}%
\AgdaBound{as}\AgdaSymbol{)}\AgdaSpace{}%
\AgdaSymbol{=}\AgdaSpace{}%
\AgdaFunction{map}\AgdaSpace{}%
\AgdaBound{f}\AgdaSpace{}%
\AgdaSymbol{(}\AgdaField{tail}\AgdaSpace{}%
\AgdaBound{as}\AgdaSymbol{)}\<%
\\
\>[0]\<%
\end{code}
    \input{content-lagda/invalid-agda/nats-code}%
  \end{AgdaMultiCode}
  \caption{Defining and working with coinductive streams in Agda.\textsuperscript{\ref{fn:salmon}}}
  \label{fig:streams-coinduction}
\end{subfigure}

\bigskip

\noindent%
\begin{subfigure}{\linewidth}
  \begin{center}
    \begin{tikzcd}
      \AIC{ω} \arrow[rr, "\AIC{forever}", bend left] \arrow["\AIC{later}", loop, distance=2em, in=145, out=215] & & \AIC{★} \arrow[ll, "\AIC{constantly}", bend left]
    \end{tikzcd}
  \end{center}
    \GMTForeverLater{} $\modeq$ \AIC{forever} \\
    \GMTForeverConstantly{} $\modeq$ \AIC{𝟙} \\
    \GMTUnitLaterTwoCell{} $\in$ \AIC{𝟙} $\Rightarrow$ \AIC{later} \\
    \GMTConstForevUnitTwoCell{} $\in$ \GMTConstantlyForever{} $\Rightarrow$ \AIC{𝟙}
  \caption{Mode theory for guarded recursion.}
  \label{fig:guarded-mode-theory}
\end{subfigure}%
\end{minipage}%
\hspace{\fill}
\begin{minipage}{.55\textwidth}
\begin{subfigure}{\linewidth}
\begin{AgdaMultiCode}

\begin{code}%
\>[0]\AgdaFunction{g-map}\AgdaSpace{}%
\AgdaSymbol{:}\AgdaSpace{}%
\AgdaDatatype{TyExpr}\AgdaSpace{}%
\AgdaInductiveConstructor{★}\AgdaSpace{}%
\AgdaSymbol{→}\AgdaSpace{}%
\AgdaDatatype{TyExpr}\AgdaSpace{}%
\AgdaInductiveConstructor{★}\AgdaSpace{}%
\AgdaSymbol{→}\AgdaSpace{}%
\AgdaDatatype{TmExpr}\AgdaSpace{}%
\AgdaInductiveConstructor{ω}\<%
\\
\>[0]\AgdaFunction{g-map}\AgdaSpace{}%
\AgdaBound{A}\AgdaSpace{}%
\AgdaBound{B}\AgdaSpace{}%
\AgdaSymbol{=}\AgdaSpace{}%
\AgdaOperator{\AgdaInductiveConstructor{ann}}\AgdaSpace{}%
\AgdaSymbol{(}\<%
\\
\>[0][@{}l@{\AgdaIndent{0}}]%
\>[2]\AgdaOperator{\AgdaFunction{lam[}}\AgdaSpace{}%
\AgdaInductiveConstructor{constantly}\AgdaSpace{}%
\AgdaOperator{\AgdaFunction{∣}}\AgdaSpace{}%
\AgdaString{"f"}\AgdaSpace{}%
\AgdaOperator{\AgdaFunction{∈}}\AgdaSpace{}%
\AgdaBound{A}\AgdaSpace{}%
\AgdaOperator{\AgdaInductiveConstructor{⇛}}\AgdaSpace{}%
\AgdaBound{B}\AgdaSpace{}%
\AgdaOperator{\AgdaFunction{]}}\<%
\\
\>[2][@{}l@{\AgdaIndent{0}}]%
\>[4]\AgdaOperator{\AgdaInductiveConstructor{löb[later∣}}\AgdaSpace{}%
\AgdaString{"m"}\AgdaSpace{}%
\AgdaOperator{\AgdaInductiveConstructor{∈}}\AgdaSpace{}%
\AgdaInductiveConstructor{GStream}\AgdaSpace{}%
\AgdaBound{A}\AgdaSpace{}%
\AgdaOperator{\AgdaInductiveConstructor{⇛}}\AgdaSpace{}%
\AgdaInductiveConstructor{GStream}\AgdaSpace{}%
\AgdaBound{B}\AgdaSpace{}%
\AgdaOperator{\AgdaInductiveConstructor{]}}\<%
\\
\>[4][@{}l@{\AgdaIndent{0}}]%
\>[6]\AgdaOperator{\AgdaInductiveConstructor{lam[}}\AgdaSpace{}%
\AgdaString{"s"}\AgdaSpace{}%
\AgdaOperator{\AgdaInductiveConstructor{∈}}\AgdaSpace{}%
\AgdaInductiveConstructor{GStream}\AgdaSpace{}%
\AgdaBound{A}\AgdaSpace{}%
\AgdaOperator{\AgdaInductiveConstructor{]}}\<%
\\
\>[6][@{}l@{\AgdaIndent{0}}]%
\>[8]\AgdaFunction{let'}\AgdaSpace{}%
\AgdaFunction{mod⟨}\AgdaSpace{}%
\AgdaInductiveConstructor{constantly}\AgdaSpace{}%
\AgdaFunction{⟩}\AgdaSpace{}%
\AgdaString{"head-s"}\AgdaSpace{}%
\AgdaFunction{←}\AgdaSpace{}%
\AgdaInductiveConstructor{g-head}\AgdaSpace{}%
\AgdaOperator{\AgdaInductiveConstructor{∙}}\AgdaSpace{}%
\AgdaFunction{svar}\AgdaSpace{}%
\AgdaString{"s"}\AgdaSpace{}%
\AgdaFunction{in'}\<%
\\
\>[8]\AgdaFunction{let'}\AgdaSpace{}%
\AgdaFunction{mod⟨}\AgdaSpace{}%
\AgdaInductiveConstructor{later}\AgdaSpace{}%
\AgdaFunction{⟩}\AgdaSpace{}%
\AgdaString{"tail-s"}\AgdaSpace{}%
\AgdaFunction{←}\AgdaSpace{}%
\AgdaInductiveConstructor{g-tail}\AgdaSpace{}%
\AgdaOperator{\AgdaInductiveConstructor{∙}}\AgdaSpace{}%
\AgdaFunction{svar}\AgdaSpace{}%
\AgdaString{"s"}\AgdaSpace{}%
\AgdaFunction{in'}\<%
\\
\>[8]\AgdaInductiveConstructor{g-cons}%
\>[276I]\AgdaOperator{\AgdaFunction{∙⟨}}\AgdaSpace{}%
\AgdaInductiveConstructor{constantly}\AgdaSpace{}%
\AgdaOperator{\AgdaFunction{⟩}}\AgdaSpace{}%
\AgdaSymbol{(}\AgdaFunction{svar}\AgdaSpace{}%
\AgdaString{"f"}\AgdaSpace{}%
\AgdaOperator{\AgdaInductiveConstructor{∙}}\AgdaSpace{}%
\AgdaFunction{svar}\AgdaSpace{}%
\AgdaString{"head-s"}\AgdaSymbol{)}\<%
\\
\>[.][@{}l@{}]\<[276I]%
\>[27]\AgdaOperator{\AgdaFunction{∙⟨}}\AgdaSpace{}%
\AgdaInductiveConstructor{later}\AgdaSpace{}%
\AgdaOperator{\AgdaFunction{⟩}}\AgdaSpace{}%
\AgdaSymbol{(}\AgdaFunction{svar}\AgdaSpace{}%
\AgdaString{"m"}\AgdaSpace{}%
\AgdaOperator{\AgdaInductiveConstructor{∙}}\AgdaSpace{}%
\AgdaFunction{svar}\AgdaSpace{}%
\AgdaString{"tail-s"}\AgdaSymbol{)}\<%
\\
\>[2]\AgdaSymbol{)}\AgdaSpace{}%
\AgdaOperator{\AgdaInductiveConstructor{∈}}\AgdaSpace{}%
\AgdaSymbol{(}\AgdaOperator{\AgdaInductiveConstructor{⟨}}\AgdaSpace{}%
\AgdaInductiveConstructor{constantly}\AgdaSpace{}%
\AgdaOperator{\AgdaInductiveConstructor{∣}}\AgdaSpace{}%
\AgdaBound{A}\AgdaSpace{}%
\AgdaOperator{\AgdaInductiveConstructor{⇛}}\AgdaSpace{}%
\AgdaBound{B}\AgdaSpace{}%
\AgdaOperator{\AgdaInductiveConstructor{⟩}}\AgdaSpace{}%
\AgdaOperator{\AgdaInductiveConstructor{⇛}}\AgdaSpace{}%
\AgdaInductiveConstructor{GStream}\AgdaSpace{}%
\AgdaBound{A}\AgdaSpace{}%
\AgdaOperator{\AgdaInductiveConstructor{⇛}}\AgdaSpace{}%
\AgdaInductiveConstructor{GStream}\AgdaSpace{}%
\AgdaBound{B}\AgdaSymbol{)}\<%
\\
\>[0]\<%
\end{code}
\begin{code} \label{code:g-nats}%
\>[0]\AgdaFunction{g-nats}\AgdaSpace{}%
\AgdaSymbol{:}\AgdaSpace{}%
\AgdaDatatype{TmExpr}\AgdaSpace{}%
\AgdaInductiveConstructor{ω}\<%
\\
\>[0]\AgdaFunction{g-nats}\AgdaSpace{}%
\AgdaSymbol{=}\AgdaSpace{}%
\AgdaOperator{\AgdaInductiveConstructor{löb[later∣}}\AgdaSpace{}%
\AgdaString{"s"}\AgdaSpace{}%
\AgdaOperator{\AgdaInductiveConstructor{∈}}\AgdaSpace{}%
\AgdaInductiveConstructor{GStream}\AgdaSpace{}%
\AgdaInductiveConstructor{Nat}\AgdaSpace{}%
\AgdaOperator{\AgdaInductiveConstructor{]}}\<%
\\
\>[0][@{}l@{\AgdaIndent{0}}]%
\>[2]\AgdaInductiveConstructor{g-cons}%
\>[316I]\AgdaOperator{\AgdaFunction{∙⟨}}\AgdaSpace{}%
\AgdaInductiveConstructor{constantly}\AgdaSpace{}%
\AgdaOperator{\AgdaFunction{⟩}}\AgdaSpace{}%
\AgdaInductiveConstructor{lit}\AgdaSpace{}%
\AgdaNumber{0}\<%
\\
\>[.][@{}l@{}]\<[316I]%
\>[23]\AgdaOperator{\AgdaFunction{∙⟨}}\AgdaSpace{}%
\AgdaInductiveConstructor{later}\AgdaSpace{}%
\AgdaOperator{\AgdaFunction{⟩}}\AgdaSpace{}%
\AgdaSymbol{(}\AgdaFunction{g-map}\AgdaSpace{}%
\AgdaOperator{\AgdaFunction{∙⟨}}\AgdaSpace{}%
\AgdaInductiveConstructor{constantly}\AgdaSpace{}%
\AgdaOperator{\AgdaFunction{⟩}}\AgdaSpace{}%
\AgdaInductiveConstructor{suc}\AgdaSpace{}%
\AgdaOperator{\AgdaInductiveConstructor{∙}}\AgdaSpace{}%
\AgdaFunction{svar}\AgdaSpace{}%
\AgdaString{"s"}\AgdaSymbol{)}\<%
\\
\>[0]\<%
\end{code}
\begin{code}%
\>[0]\AgdaFunction{Stream'}\AgdaSpace{}%
\AgdaSymbol{:}\AgdaSpace{}%
\AgdaDatatype{TyExpr}\AgdaSpace{}%
\AgdaInductiveConstructor{★}\AgdaSpace{}%
\AgdaSymbol{→}\AgdaSpace{}%
\AgdaDatatype{TyExpr}\AgdaSpace{}%
\AgdaInductiveConstructor{★}\<%
\\
\>[0]\AgdaFunction{Stream'}\AgdaSpace{}%
\AgdaBound{A}\AgdaSpace{}%
\AgdaSymbol{=}\AgdaSpace{}%
\AgdaOperator{\AgdaInductiveConstructor{⟨}}\AgdaSpace{}%
\AgdaInductiveConstructor{forever}\AgdaSpace{}%
\AgdaOperator{\AgdaInductiveConstructor{∣}}\AgdaSpace{}%
\AgdaInductiveConstructor{GStream}\AgdaSpace{}%
\AgdaBound{A}\AgdaSpace{}%
\AgdaOperator{\AgdaInductiveConstructor{⟩}}\<%
\\
\>[0]\<%
\end{code}
  
\begin{code}%
\>[0]\AgdaFunction{cons'}\AgdaSpace{}%
\AgdaSymbol{:}\AgdaSpace{}%
\AgdaDatatype{TyExpr}\AgdaSpace{}%
\AgdaInductiveConstructor{★}\AgdaSpace{}%
\AgdaSymbol{→}\AgdaSpace{}%
\AgdaDatatype{TmExpr}\AgdaSpace{}%
\AgdaInductiveConstructor{★}\<%
\\
\>[0]\AgdaFunction{cons'}\AgdaSpace{}%
\AgdaBound{A}\AgdaSpace{}%
\AgdaSymbol{=}\AgdaSpace{}%
\AgdaOperator{\AgdaInductiveConstructor{lam[}}\AgdaSpace{}%
\AgdaString{"a"}\AgdaSpace{}%
\AgdaOperator{\AgdaInductiveConstructor{∈}}\AgdaSpace{}%
\AgdaBound{A}\AgdaSpace{}%
\AgdaOperator{\AgdaInductiveConstructor{]}}\AgdaSpace{}%
\AgdaOperator{\AgdaInductiveConstructor{lam[}}\AgdaSpace{}%
\AgdaString{"as"}\AgdaSpace{}%
\AgdaOperator{\AgdaInductiveConstructor{∈}}\AgdaSpace{}%
\AgdaFunction{Stream'}\AgdaSpace{}%
\AgdaBound{A}\AgdaSpace{}%
\AgdaOperator{\AgdaInductiveConstructor{]}}\<%
\\
\>[0][@{}l@{\AgdaIndent{0}}]%
\>[2]\AgdaFunction{let'}\AgdaSpace{}%
\AgdaFunction{mod⟨}\AgdaSpace{}%
\AgdaInductiveConstructor{forever}\AgdaSpace{}%
\AgdaFunction{⟩}\AgdaSpace{}%
\AgdaString{"g-as"}\AgdaSpace{}%
\AgdaFunction{←}\AgdaSpace{}%
\AgdaFunction{svar}\AgdaSpace{}%
\AgdaString{"as"}\AgdaSpace{}%
\AgdaFunction{in'}\<%
\\
\>[2]\AgdaSymbol{(}\AgdaOperator{\AgdaInductiveConstructor{mod⟨}}\AgdaSpace{}%
\AgdaInductiveConstructor{forever}\AgdaSpace{}%
\AgdaOperator{\AgdaInductiveConstructor{⟩}}\AgdaSpace{}%
\AgdaSymbol{(}\AgdaInductiveConstructor{g-cons}\AgdaSpace{}%
\AgdaBound{A}%
\>[378I]\AgdaOperator{\AgdaFunction{∙⟨}}\AgdaSpace{}%
\AgdaInductiveConstructor{constantly}\AgdaSpace{}%
\AgdaOperator{\AgdaFunction{⟩}}\AgdaSpace{}%
\AgdaFunction{svar}\AgdaSpace{}%
\AgdaString{"a"}\<%
\\
\>[.][@{}l@{}]\<[378I]%
\>[28]\AgdaOperator{\AgdaFunction{∙⟨}}\AgdaSpace{}%
\AgdaInductiveConstructor{later}\AgdaSpace{}%
\AgdaOperator{\AgdaFunction{⟩}}\AgdaSpace{}%
\AgdaFunction{svar}\AgdaSpace{}%
\AgdaString{"g-as"}\AgdaSymbol{))}\<%
\end{code}
\begin{code}%
\>[0]\AgdaFunction{nats}\AgdaSpace{}%
\AgdaSymbol{:}\AgdaSpace{}%
\AgdaDatatype{TmExpr}\AgdaSpace{}%
\AgdaInductiveConstructor{★}\<%
\\
\>[0]\AgdaFunction{nats}\AgdaSpace{}%
\AgdaSymbol{=}\AgdaSpace{}%
\AgdaOperator{\AgdaInductiveConstructor{mod⟨}}\AgdaSpace{}%
\AgdaInductiveConstructor{forever}\AgdaSpace{}%
\AgdaOperator{\AgdaInductiveConstructor{⟩}}\AgdaSpace{}%
\AgdaFunction{g-nats}\<%
\end{code}
\end{AgdaMultiCode}
  \caption{Sikkel code for guarded type theory.}
  \label{fig:code-guarded}
\end{subfigure}%
\end{minipage}%
\caption{Streams in Agda and guarded recursive streams in Sikkel (Section~\ref{sec:appl-guard-recurs}).}
\end{figure*}


To illustrate the use of Sikkel in practice, we demonstrate the use of guarded recursion.
To motivate guarded recursion, consider the definitions in Figure~\ref{fig:streams-coinduction} which illustrate Agda's support for infinite structures using coinduction and copatterns.
The code defines infinite streams as a coinductive record, essentially by specifying how they can be destructed: by taking their \AFi{head} of type \AB{A} and their \AFi{tail} which is again of type \AR{Stream}\AS{}\AB{A}.
Values of such a coinductive type can be created using copattern matching and corecursion.
This is used to define a stream of zeros by specifying its \AFi{head} and its \AFi{tail}.
Note how the \AFi{tail} is defined corecursively by referring to the stream we are defining.
Slightly more complicated is the definition of \AF{map}, which consumes one stream while producing another one.

For Agda to be sound as a logic and for type checking to be decidable, corecursive definitions must be productive, meaning that any element of the stream must be computable in a finite amount of time.
To enforce productivity, Agda's productivity checker only allows a corecursive call if the operations applied to its result form a strict suffix of the coinductive destructors applied to the left-hand side of the clause.
This check is conservative and sometimes quite restrictive, as illustrated by the \AF{nats} example.%
\footnote{\label{fn:salmon}The salmon-colored background is applied by Agda to inform that there are termination or productivity issues.}
This example is rejected by Agda's productivity checker, because the right-hand side of the last clause corecursively refers to \AF{nats} in an application of \AF{map}\AS\AIC{suc}, which is not a strict suffix of \AFi{tail}.

\newcommand{\mapSuc}{
  \begin{code}[hide]%
\>[0]\AgdaFunction{\AgdaUnderscore{}}\AgdaSpace{}%
\AgdaSymbol{:}\AgdaSpace{}%
\AgdaRecord{Stream}\AgdaSpace{}%
\AgdaDatatype{ℕ}\AgdaSpace{}%
\AgdaSymbol{→}\AgdaSpace{}%
\AgdaRecord{Stream}\AgdaSpace{}%
\AgdaDatatype{ℕ}\<%
\\
\>[0]\AgdaSymbol{\AgdaUnderscore{}}%
\>[401I]\AgdaSymbol{=}\<%
\end{code}%
  \begin{code}[inline]%
\>[.][@{}l@{}]\<[401I]%
\>[2]\AgdaFunction{map}\AgdaSpace{}%
\AgdaInductiveConstructor{suc}\<%
\end{code}%
}%
\newcommand{\headTailNats}{\AFi{head}\AS\ASy{(}\AFi{tail}\AS\AF{nats}\ASy{)}}
The definition of \AF{nats} is actually productive: computing the $(n+1)$th element only requires the $n$-th element of the stream for any $n$.
Agda does not see this because it depends on a property of the function\mapSuc{}: the fact that it only needs the $n$-th element of the input stream to produce the $n$-th element of its output stream.
If we replaced\mapSuc{} with a function \AF{flipFst} (of the same type
\begin{code}[hide]%
\>[0]\AgdaFunction{\AgdaUnderscore{}}%
\>[403I]\AgdaSymbol{:}\<%
\end{code}
\begin{code}[inline]%
\>[.][@{}l@{}]\<[403I]%
\>[2]\AgdaRecord{Stream}\AgdaSpace{}%
\AgdaDatatype{ℕ}\AgdaSpace{}%
\AgdaSymbol{→}\AgdaSpace{}%
\AgdaRecord{Stream}\AgdaSpace{}%
\AgdaDatatype{ℕ}\<%
\end{code}
\begin{code}[hide]%
\>[0]\AgdaSymbol{\AgdaUnderscore{}}\AgdaSpace{}%
\AgdaSymbol{=}\AgdaSpace{}%
\AgdaFunction{map}\AgdaSpace{}%
\AgdaInductiveConstructor{suc}\<%
\end{code})
that flips the first two elements of a stream, we would be defining \headTailNats{} as \headTailNats{} itself, which would not be productive.


Guarded recursive type theory, originally proposed by Nakano~\cite{nakano00-modality}, allows to express the difference between\mapSuc{} and \AF{flipFst} in the type system and can therefore accept more productive definitions.
In order to work with guarded recursion in Sikkel, we use the same mode theory as Gratzer et al.~\cite{gratzer20-multimodal} which is specified in Figure~\ref{fig:guarded-mode-theory}.
There is a time-independent or trivial mode \stmode{} and a time-dependent mode \omode{}.
Intuitively, one can imagine terms at mode \omode{} to unfold over time, revealing at every time step a new piece of information as specified by their type, whereas terms at mode \stmode{} can best be thought of as ordinary Agda values.
There is a modality \AIC{later} from \omode{} to \omode{}, whose action on types we will denote by \AF{▻}\AS{}\AB{T} instead of
\begin{code}[hide]%
\>[0]\AgdaFunction{\AgdaUnderscore{}}\AgdaSpace{}%
\AgdaSymbol{:}\AgdaSpace{}%
\AgdaDatatype{TyExpr}\AgdaSpace{}%
\AgdaInductiveConstructor{ω}\AgdaSpace{}%
\AgdaSymbol{→}\AgdaSpace{}%
\AgdaDatatype{TyExpr}\AgdaSpace{}%
\AgdaInductiveConstructor{ω}\<%
\\
\>[0]\AgdaSymbol{\AgdaUnderscore{}}%
\>[417I]\AgdaSymbol{=}\AgdaSpace{}%
\AgdaSymbol{λ}\AgdaSpace{}%
\AgdaBound{T}\AgdaSpace{}%
\AgdaSymbol{→}\<%
\end{code}
\begin{code}[inline]%
\>[.][@{}l@{}]\<[417I]%
\>[2]\AgdaOperator{\AgdaInductiveConstructor{⟨}}\AgdaSpace{}%
\AgdaInductiveConstructor{later}\AgdaSpace{}%
\AgdaOperator{\AgdaInductiveConstructor{∣}}\AgdaSpace{}%
\AgdaBound{T}\AgdaSpace{}%
\AgdaOperator{\AgdaInductiveConstructor{⟩}}\<%
\end{code}.
Intuitively, a term of type \AF{▻}\AS{}\AB{T} is just a term of type \AB{T} whose unfolding process is delayed by one step: the information that would be available today is actually only available tomorrow.
The intuition behind the \AIC{constantly} modality is that it takes a static, time-independent value and embeds it as a term in mode \omode{} which does not unfold: it is constantly the same and all information is present from the beginning.
The \AIC{forever} modality, on the other hand, converts a time-dependent value to a time-independent one by applying the unfolding \textit{forever}, obtaining the fully unfolded value with all information present.



Following Veltri and van der Weide~\cite{veltri19-guarded} and Gratzer et al.~\cite{gratzer20-multimodal}, we then add a new type constructor \AIC{GStream} of \emph{guarded} streams to our language.
It takes a type in mode \stmode{} and produces a type in mode \omode{}, whose values can be thought of as streams that unfold over time by making available one additional element at every time step.
As a result, the tail of such a guarded stream will have type
\begin{code}[hide]%
\>[0]\AgdaFunction{\AgdaUnderscore{}}\AgdaSpace{}%
\AgdaSymbol{:}\AgdaSpace{}%
\AgdaDatatype{TyExpr}\AgdaSpace{}%
\AgdaInductiveConstructor{★}\AgdaSpace{}%
\AgdaSymbol{→}\AgdaSpace{}%
\AgdaDatatype{TyExpr}\AgdaSpace{}%
\AgdaInductiveConstructor{ω}\<%
\\
\>[0]\AgdaSymbol{\AgdaUnderscore{}}%
\>[431I]\AgdaSymbol{=}\AgdaSpace{}%
\AgdaSymbol{λ}\AgdaSpace{}%
\AgdaBound{A}\AgdaSpace{}%
\AgdaSymbol{→}\<%
\end{code}
\begin{code}[inline*]%
\>[.][@{}l@{}]\<[431I]%
\>[2]\AgdaFunction{▻}\AgdaSpace{}%
\AgdaSymbol{(}\AgdaInductiveConstructor{GStream}\AgdaSpace{}%
\AgdaBound{A}\AgdaSymbol{)}\<%
\end{code}
as its first element will only be available tomorrow since it is the second element of the original stream.
The other built-in operations for guarded streams can be found at the top of Figure~\ref{fig:typing-rules-guarded}.
Note that the types of \AIC{g-head} and \AIC{g-cons} use the \AIC{constantly} modality to embed a time-independent type into mode \omode{}.

\begin{figure}[htb]
    \begin{center}
    \begin{tabular}{ll}
      \AB{Γ} $\vdash$ \text{\AIC{g-head}} $:$ \text{\tmGHeadTy{}} & \quad \rulename{(Tm-GHead)} \\
      \AB{Γ} $\vdash$ \text{\AIC{g-tail}} $:$ \text{\tmGTailTy{}} & \quad \rulename{(Tm-GTail)} \\
      \AB{Γ} $\vdash$ \text{\AIC{g-cons}} $:$ \text{\tmGConsTy{}} & \quad \rulename{(Tm-GCons)}
    \end{tabular}
    \end{center}
    \begin{equation*}
      \prftree[r]{\rulename{Tm-L\"ob}}{
        \prfassumption{\text{\tmLobCtx{}} \vdash \AB{t} : \AB{S}}}{
        \prfassumption{\AB{T} \tyeq \AB{S}}}{
        \AB{Γ} \vdash \text{\tmLobTm{}} : \AB{T}}
    \end{equation*}
  \caption{Additional typing rules for the Sikkel implementation of guarded recursive type theory.}
  \label{fig:typing-rules-guarded}
\end{figure}

Guarded recursive type theory offers a built-in induction primitive called L\"ob induction (\rulename{Tm-L\"ob}, see Figure~\ref{fig:typing-rules-guarded}).
When constructing a value of any type \AB{T} at mode \omode{}, it allows us to use a variable of type \AB{T} that appears under the later modality.
This variable will represent a (co)recursive call, and the fact that it appears under the later modality prevents us from writing unproductive definitions.

We can now implement a map operation \AF{g-map}
for guarded streams in Sikkel (Fig.~\ref{fig:code-guarded}).
We use modal elimination twice to make sure that there are variables representing the head and the tail of \ASt{"s"} and that they appear in the context under the right modalities to be used in the application of the modal function \AIC{g-cons}.
The variable \ASt{"m"} bound by L\"ob induction 
is used to corecursively map \ASt{"f"} over the tail of \ASt{"s"}.
It is now also possible to write a valid definition of the guarded stream of natural numbers \AF{g-nats} (Fig.~\ref{fig:code-guarded}).
Here \AIC{lit} introduces natural number literals and \AIC{suc} is the successor function.

Note that it is possible to implement a function \AF{g-flipFst} that flips the first two elements of a guarded stream.
However, it will have type
\begin{code}[hide]%
\>[0]\AgdaFunction{\AgdaUnderscore{}}\AgdaSpace{}%
\AgdaSymbol{:}\AgdaSpace{}%
\AgdaDatatype{TyExpr}\AgdaSpace{}%
\AgdaInductiveConstructor{★}\AgdaSpace{}%
\AgdaSymbol{→}\AgdaSpace{}%
\AgdaDatatype{TyExpr}\AgdaSpace{}%
\AgdaInductiveConstructor{ω}\<%
\\
\>[0]\AgdaSymbol{\AgdaUnderscore{}}%
\>[443I]\AgdaSymbol{=}\AgdaSpace{}%
\AgdaSymbol{λ}\AgdaSpace{}%
\AgdaBound{A}\AgdaSpace{}%
\AgdaSymbol{→}\<%
\end{code}
\begin{code}[inline*]%
\>[.][@{}l@{}]\<[443I]%
\>[2]\AgdaInductiveConstructor{GStream}\AgdaSpace{}%
\AgdaBound{A}\AgdaSpace{}%
\AgdaOperator{\AgdaInductiveConstructor{⇛}}\AgdaSpace{}%
\AgdaFunction{▻}\AgdaSpace{}%
\AgdaSymbol{(}\AgdaInductiveConstructor{GStream}\AgdaSpace{}%
\AgdaBound{A}\AgdaSymbol{)}\<%
\end{code}
because the head of the resulting stream is the second element of the argument stream, which is only available tomorrow.
As a result, replacing
\begin{code}[hide]%
\>[0]\AgdaFunction{\AgdaUnderscore{}}\AgdaSpace{}%
\AgdaSymbol{:}\AgdaSpace{}%
\AgdaDatatype{TmExpr}\AgdaSpace{}%
\AgdaInductiveConstructor{ω}\<%
\\
\>[0]\AgdaSymbol{\AgdaUnderscore{}}%
\>[455I]\AgdaSymbol{=}\<%
\end{code}
\begin{code}[inline*]%
\>[.][@{}l@{}]\<[455I]%
\>[2]\AgdaFunction{g-map}\AgdaSpace{}%
\AgdaOperator{\AgdaFunction{∙⟨}}\AgdaSpace{}%
\AgdaInductiveConstructor{constantly}\AgdaSpace{}%
\AgdaOperator{\AgdaFunction{⟩}}\AgdaSpace{}%
\AgdaInductiveConstructor{suc}\<%
\end{code}
with \AF{g-flipFst} in the definition of \AF{g-nats} results in an ill-typed term.

Guarded streams do not behave in the same way as Agda's coinductive streams.
An important difference is the occurrence of the modalities in the type of \AIC{g-cons}.
We can however define a type constructor \AF{Stream'} of standard streams (as fully unfolded guarded streams) at mode \stmode{} by applying the \AIC{forever} modality to the type of guarded streams \cite{veltri19-guarded,gratzer20-multimodal}
(Fig.~\ref{fig:code-guarded}).
The \AF{cons'} function then has the expected type
\begin{code}[hide]%
\>[0]\AgdaFunction{\AgdaUnderscore{}}\AgdaSpace{}%
\AgdaSymbol{:}\AgdaSpace{}%
\AgdaDatatype{TyExpr}\AgdaSpace{}%
\AgdaInductiveConstructor{★}\AgdaSpace{}%
\AgdaSymbol{→}\AgdaSpace{}%
\AgdaDatatype{TyExpr}\AgdaSpace{}%
\AgdaInductiveConstructor{★}\<%
\\
\>[0]\AgdaSymbol{\AgdaUnderscore{}}%
\>[468I]\AgdaSymbol{=}\AgdaSpace{}%
\AgdaSymbol{λ}\AgdaSpace{}%
\AgdaBound{A}\AgdaSpace{}%
\AgdaSymbol{→}\<%
\end{code}
\begin{code}[inline]%
\>[.][@{}l@{}]\<[468I]%
\>[2]\AgdaBound{A}\AgdaSpace{}%
\AgdaOperator{\AgdaInductiveConstructor{⇛}}\AgdaSpace{}%
\AgdaFunction{Stream'}\AgdaSpace{}%
\AgdaBound{A}\AgdaSpace{}%
\AgdaOperator{\AgdaInductiveConstructor{⇛}}\AgdaSpace{}%
\AgdaFunction{Stream'}\AgdaSpace{}%
\AgdaBound{A}\<%
\end{code}.
It is well-typed because of the two modality equivalences from Figure~\ref{fig:guarded-mode-theory}, which let us use the variables \ASt{"a"} and \ASt{"g-as"} via the trivial 2-cell.
We can construct the standard stream \AF{nats} of natural numbers using \AF{g-nats} (Fig.~\ref{fig:code-guarded}).
Many more examples, including all the examples involving streams by Clouston et al.~\cite[pp.\ 8--12]{clouston17-guarded}, can be found in the Agda implementation.


%% file: content-lagda/invalid-agda/nats-code.tex
\begin{code}%
\>[0]\AgdaTerminationProblem{\AgdaFunction{nats}}\AgdaSpace{}%
\AgdaSymbol{:}\AgdaSpace{}%
\AgdaRecord{Stream}\AgdaSpace{}%
\AgdaDatatype{ℕ}\<%
\\
\>[0]\AgdaField{head}\AgdaSpace{}%
\AgdaFunction{nats}\AgdaSpace{}%
\AgdaSymbol{=}\AgdaSpace{}%
\AgdaNumber{0}\<%
\\
\>[0]\AgdaField{tail}\AgdaSpace{}%
\AgdaFunction{nats}\AgdaSpace{}%
\AgdaSymbol{=}\AgdaSpace{}%
\AgdaFunction{map}\AgdaSpace{}%
\AgdaInductiveConstructor{suc}\AgdaSpace{}%
\AgdaTerminationProblem{\AgdaFunction{nats}}\<%
\end{code}%

%% file: content-lagda/presheaves.tex
\section{Presheaf Models}
\label{sec:presheaves}

\begin{code}[hide]%
\>[0]\AgdaKeyword{module}\AgdaSpace{}%
\AgdaModule{presheaves}\AgdaSpace{}%
\AgdaKeyword{where}\<%
\\
\\[\AgdaEmptyExtraSkip]%
\>[0]\AgdaKeyword{open}\AgdaSpace{}%
\AgdaKeyword{import}\AgdaSpace{}%
\AgdaModule{Data.Unit}\AgdaSpace{}%
\AgdaKeyword{using}\AgdaSpace{}%
\AgdaSymbol{(}\AgdaRecord{⊤}\AgdaSymbol{;}\AgdaSpace{}%
\AgdaInductiveConstructor{tt}\AgdaSymbol{)}\<%
\\
\>[0]\AgdaKeyword{open}\AgdaSpace{}%
\AgdaKeyword{import}\AgdaSpace{}%
\AgdaModule{Relation.Binary.PropositionalEquality}\<%
\end{code}

For now, the definition of \AF{nats} in Fig.~\ref{fig:code-guarded} gives us just the syntax of an MSTT term.
However, Sikkel is intended to make modal primitives available for defining regular Agda values, not just terms in a deeply embedded syntax.
A standard approach to do this would be to write an interpreter that translates MSTT types and terms to Agda types and terms.
However, modal types or term formers like \AIC{l\"ob} would be hard to translate in an off-the-shelf version of Agda.
Instead, we will first interpret MSTT types and terms in Sikkel's second, semantic layer: a presheaf model that supports the new primitives of the extended type theory.
All of the intended applications of Sikkel listed in the introduction have such a presheaf model.

The intuition behind presheaf models is covered in Section~\ref{sec:presheaf-motivation} and we continue with details about the actual formalization of presheaf models in Section~\ref{sec:presheaf-models-agda}. We refer to Appendix~\ref{sec:psh-sem-guard} for details about the implementation of the semantics of guarded recursion.

\subsection{Some Intuition}
\label{sec:presheaf-motivation}

We can explain the intuition behind presheaf models using the example of guarded recursion from Section~\ref{sec:appl-guard-recurs}.
Recall that a value at mode \omode{} can be seen as if it unfolds over time, making available some new information at every time step.
Consequently, we can model a type at mode \omode{} as a sequence of Agda types, the $n$-th type representing the unfolding after $n$ steps.
For example, the type \AIC{GStream}\AS{}\AB{A} will be interpreted as the following diagram of Agda types and functions.
\begin{equation}
  \label{eq:diagram}
  \begin{tikzcd}
    \ADT{Vec}_1\, A & \ADT{Vec}_2\, A \arrow[l, "\AF{init}"'] & \ldots \arrow[l, "\AF{init}"'] & \ADT{Vec}_n\, A \arrow[l, "\AF{init}"'] & \ldots \arrow[l, "\AF{init}"']
  \end{tikzcd}
\end{equation}
Here $\ADT{Vec}_n$ is the Agda type constructor for lists of length $n$ and \AF{init} drops the last element of such a list.
A term of type \AIC{GStream}\AS{}\AB{A} will then be modeled as a sequence of Agda values of these types compatible with the \AF{init} functions, i.e.\ effectively as a vector gaining more elements over time where every element, once present, must remain unchanged.%
\footnote{\adapted{It might be surprising that a guarded stream is represented as a sequence of vectors (lists), rather than as a sequence of values of type \AB{A}.
  However, in the $n$-th type of the sequence, we want to keep track of \emph{all} information present after $n$ steps, not just the newly added data.
  The functions in diagram \eqref{eq:diagram} will then tell how to forget the new information when going from the $(n+1)$th to the $n$-th type.
  This is the standard way to interpret guarded streams in a presheaf model \cite{birkedal12-first}.
  Moreover, by changing the types and functions in diagram \eqref{eq:diagram}, one can model structures different from streams (e.g.\ replacing $\ADT{Vec}_n$ with the type constructor for binary trees of height $n$ allows to model infinite binary trees).}}

\adapted{In general, a presheaf model is parametrized by a base category.
Types at a mode $m$ are modeled as diagrams such as the one above, called \emph{presheaves}, whose shape is determined by the base category that corresponds to $m$.
Depending on the shapes of these diagrams, one can implement operations on semantic types which will serve as the interpretation of modalities or other primitives added to an instance of MSTT.
For example, since every type at mode \omode{} is represented as a diagram with a shape as in \eqref{eq:diagram}, we can implement the later modality by shifting the sequence one step to the right and adding Agda's unit type \ADT{⊤}, containing no information, to the front (i.e.\ delaying the unfolding process by one step).}

\subsection{Presheaf Models in Agda}
\label{sec:presheaf-models-agda}

Our formalization of presheaf models in Agda follows the general construction by Hofmann~\cite{Hofmann97-presheaf-chapter} and is structured as an internal Category with Families (CwF) \cite{dybjer96-internal}.

A presheaf model is parametrized by a base category \AB{C} : \AR{BaseCategory}, which can be any small category.
Its object and morphism types will be denoted \AF{Ob} and \AF{Hom} \AB{x} \AB{y}.
\begin{code}[hide]%
\>[0]\AgdaKeyword{record}\AgdaSpace{}%
\AgdaRecord{BaseCategory}\AgdaSpace{}%
\AgdaSymbol{:}\AgdaSpace{}%
\AgdaPrimitive{Set₁}\AgdaSpace{}%
\AgdaKeyword{where}\<%
\\
\>[0][@{}l@{\AgdaIndent{0}}]%
\>[2]\AgdaKeyword{field}\<%
\\
\>[2][@{}l@{\AgdaIndent{0}}]%
\>[4]\AgdaField{Ob}\AgdaSpace{}%
\AgdaSymbol{:}\AgdaSpace{}%
\AgdaPrimitive{Set}\<%
\\
\>[4]\AgdaField{Hom}\AgdaSpace{}%
\AgdaSymbol{:}\AgdaSpace{}%
\AgdaField{Ob}\AgdaSpace{}%
\AgdaSymbol{→}\AgdaSpace{}%
\AgdaField{Ob}\AgdaSpace{}%
\AgdaSymbol{→}\AgdaSpace{}%
\AgdaPrimitive{Set}\<%
\\
\>[4]\AgdaField{hom-id}\AgdaSpace{}%
\AgdaSymbol{:}\AgdaSpace{}%
\AgdaField{Hom}\AgdaSpace{}%
\AgdaBound{x}\AgdaSpace{}%
\AgdaBound{x}\<%
\\
\>[4]\AgdaOperator{\AgdaField{\AgdaUnderscore{}∙\AgdaUnderscore{}}}\AgdaSpace{}%
\AgdaSymbol{:}\AgdaSpace{}%
\AgdaField{Hom}\AgdaSpace{}%
\AgdaBound{y}\AgdaSpace{}%
\AgdaBound{z}\AgdaSpace{}%
\AgdaSymbol{→}\AgdaSpace{}%
\AgdaField{Hom}\AgdaSpace{}%
\AgdaBound{x}\AgdaSpace{}%
\AgdaBound{y}\AgdaSpace{}%
\AgdaSymbol{→}\AgdaSpace{}%
\AgdaField{Hom}\AgdaSpace{}%
\AgdaBound{x}\AgdaSpace{}%
\AgdaBound{z}\<%
\end{code}
\begin{code}[hide]%
\>[0]\AgdaKeyword{private}\AgdaSpace{}%
\AgdaKeyword{variable}\<%
\\
\>[0][@{}l@{\AgdaIndent{0}}]%
\>[2]\AgdaGeneralizable{C}\AgdaSpace{}%
\AgdaSymbol{:}\AgdaSpace{}%
\AgdaRecord{BaseCategory}\<%
\end{code}
A semantic context \AB{Γ} : \AR{Ctx} \AB{C} is now defined as a presheaf (i.e.\ a \AF{Set}-valued contravariant functor) over the base category. In the following record type, as well as in the rest of this section, we omit fields expressing equality laws.%
\begin{code}%
\>[0]\AgdaKeyword{record}\AgdaSpace{}%
\AgdaRecord{Ctx}\AgdaSpace{}%
\AgdaSymbol{(}\AgdaBound{C}\AgdaSpace{}%
\AgdaSymbol{:}\AgdaSpace{}%
\AgdaRecord{BaseCategory}\AgdaSymbol{)}\AgdaSpace{}%
\AgdaSymbol{:}\AgdaSpace{}%
\AgdaPrimitive{Set₁}\AgdaSpace{}%
\AgdaKeyword{where}\<%
\\
\>[0][@{}l@{\AgdaIndent{0}}]%
\>[2]\AgdaKeyword{field}%
\>[58I]\AgdaField{ctx-cell}\AgdaSpace{}%
\AgdaSymbol{:}\AgdaSpace{}%
\AgdaFunction{Ob}\AgdaSpace{}%
\AgdaSymbol{→}\AgdaSpace{}%
\AgdaPrimitive{Set}\<%
\\
\>[.][@{}l@{}]\<[58I]%
\>[13]\AgdaField{ctx-hom}\AgdaSpace{}%
\AgdaSymbol{:}\AgdaSpace{}%
\AgdaFunction{Hom}\AgdaSpace{}%
\AgdaBound{x}\AgdaSpace{}%
\AgdaBound{y}\AgdaSpace{}%
\AgdaSymbol{→}\AgdaSpace{}%
\AgdaField{ctx-cell}\AgdaSpace{}%
\AgdaBound{y}\AgdaSpace{}%
\AgdaSymbol{→}\AgdaSpace{}%
\AgdaField{ctx-cell}\AgdaSpace{}%
\AgdaBound{x}\<%
\end{code}
\begin{code}[hide]%
\>[0]\AgdaKeyword{open}\AgdaSpace{}%
\AgdaModule{Ctx}\AgdaSpace{}%
\AgdaKeyword{renaming}\AgdaSpace{}%
\AgdaSymbol{(}\AgdaField{ctx-cell}\AgdaSpace{}%
\AgdaSymbol{to}\AgdaSpace{}%
\AgdaField{\AgdaUnderscore{}⟨\AgdaUnderscore{}⟩}\AgdaSymbol{;}\AgdaSpace{}%
\AgdaField{ctx-hom}\AgdaSpace{}%
\AgdaSymbol{to}\AgdaSpace{}%
\AgdaField{\AgdaUnderscore{}⟪\AgdaUnderscore{}⟫\AgdaUnderscore{}}\AgdaSymbol{)}\<%
\\
\\[\AgdaEmptyExtraSkip]%
\>[0]\AgdaKeyword{private}\AgdaSpace{}%
\AgdaKeyword{variable}\<%
\\
\>[0][@{}l@{\AgdaIndent{0}}]%
\>[2]\AgdaGeneralizable{Γ}\AgdaSpace{}%
\AgdaSymbol{:}\AgdaSpace{}%
\AgdaRecord{Ctx}\AgdaSpace{}%
\AgdaGeneralizable{C}\<%
\end{code}
We use the notation
\begin{code}[hide]%
\>[0]\AgdaKeyword{module}\AgdaSpace{}%
\AgdaModule{\AgdaUnderscore{}}\AgdaSpace{}%
\AgdaSymbol{(}\AgdaBound{C}\AgdaSpace{}%
\AgdaSymbol{:}\AgdaSpace{}%
\AgdaRecord{BaseCategory}\AgdaSymbol{)}\AgdaSpace{}%
\AgdaSymbol{(}\AgdaBound{Γ}\AgdaSpace{}%
\AgdaSymbol{:}\AgdaSpace{}%
\AgdaRecord{Ctx}\AgdaSpace{}%
\AgdaBound{C}\AgdaSymbol{)}\AgdaSpace{}%
\AgdaKeyword{where}\<%
\\
\>[0][@{}l@{\AgdaIndent{0}}]%
\>[2]\AgdaKeyword{open}\AgdaSpace{}%
\AgdaModule{BaseCategory}\AgdaSpace{}%
\AgdaBound{C}\<%
\\
\\[\AgdaEmptyExtraSkip]%
\>[2]\AgdaFunction{\AgdaUnderscore{}}\AgdaSpace{}%
\AgdaSymbol{:}\AgdaSpace{}%
\AgdaFunction{Ob}\AgdaSpace{}%
\AgdaSymbol{→}\AgdaSpace{}%
\AgdaPrimitive{Set}\<%
\\
\>[2]\AgdaSymbol{\AgdaUnderscore{}}%
\>[104I]\AgdaSymbol{=}\AgdaSpace{}%
\AgdaSymbol{λ}\AgdaSpace{}%
\AgdaBound{x}\AgdaSpace{}%
\AgdaSymbol{→}\<%
\end{code}
\begin{code}[inline*]%
\>[.][@{}l@{}]\<[104I]%
\>[4]\AgdaBound{Γ}\AgdaSpace{}%
\AgdaOperator{\AgdaField{⟨}}\AgdaSpace{}%
\AgdaBound{x}\AgdaSpace{}%
\AgdaOperator{\AgdaField{⟩}}\<%
\end{code}
for the type of cells over \AB{x} : \AF{Ob} and
\begin{code}[hide]%
\>[2]\AgdaFunction{\AgdaUnderscore{}}\AgdaSpace{}%
\AgdaSymbol{:}\AgdaSpace{}%
\AgdaSymbol{(}\AgdaBound{x}\AgdaSpace{}%
\AgdaBound{y}\AgdaSpace{}%
\AgdaSymbol{:}\AgdaSpace{}%
\AgdaFunction{Ob}\AgdaSymbol{)}\AgdaSpace{}%
\AgdaSymbol{(}\AgdaBound{f}\AgdaSpace{}%
\AgdaSymbol{:}\AgdaSpace{}%
\AgdaFunction{Hom}\AgdaSpace{}%
\AgdaBound{x}\AgdaSpace{}%
\AgdaBound{y}\AgdaSymbol{)}\AgdaSpace{}%
\AgdaSymbol{→}\AgdaSpace{}%
\AgdaBound{Γ}\AgdaSpace{}%
\AgdaOperator{\AgdaField{⟨}}\AgdaSpace{}%
\AgdaBound{y}\AgdaSpace{}%
\AgdaOperator{\AgdaField{⟩}}\AgdaSpace{}%
\AgdaSymbol{→}\AgdaSpace{}%
\AgdaBound{Γ}\AgdaSpace{}%
\AgdaOperator{\AgdaField{⟨}}\AgdaSpace{}%
\AgdaBound{x}\AgdaSpace{}%
\AgdaOperator{\AgdaField{⟩}}\<%
\\
\>[2]\AgdaSymbol{\AgdaUnderscore{}}%
\>[131I]\AgdaSymbol{=}\AgdaSpace{}%
\AgdaSymbol{λ}\AgdaSpace{}%
\AgdaBound{x}\AgdaSpace{}%
\AgdaBound{y}\AgdaSpace{}%
\AgdaBound{f}\AgdaSpace{}%
\AgdaBound{γ}\AgdaSpace{}%
\AgdaSymbol{→}\<%
\end{code}
\begin{code}[inline*]%
\>[.][@{}l@{}]\<[131I]%
\>[4]\AgdaBound{Γ}\AgdaSpace{}%
\AgdaOperator{\AgdaField{⟪}}\AgdaSpace{}%
\AgdaBound{f}\AgdaSpace{}%
\AgdaOperator{\AgdaField{⟫}}\AgdaSpace{}%
\AgdaBound{γ}\<%
\end{code}
for the restriction of \AB{γ} : \AB{Γ} \AFi{⟨} \AB{y} \AFi{⟩} by \AB{f} : \AF{Hom} \AB{x} \AB{y}.
A semantic context
\begin{code}[hide]%
\>[0]\AgdaKeyword{module}\AgdaSpace{}%
\AgdaModule{\AgdaUnderscore{}}\AgdaSpace{}%
\AgdaSymbol{(}\AgdaBound{C}\AgdaSpace{}%
\AgdaSymbol{:}\AgdaSpace{}%
\AgdaRecord{BaseCategory}\AgdaSymbol{)}\AgdaSpace{}%
\AgdaSymbol{(}\<%
\end{code}
\begin{code}[inline*]%
\>[0][@{}l@{\AgdaIndent{1}}]%
\>[2]\AgdaBound{Γ}\AgdaSpace{}%
\AgdaSymbol{:}\AgdaSpace{}%
\AgdaRecord{Ctx}\AgdaSpace{}%
\AgdaBound{C}\<%
\end{code}
\begin{code}[hide]%
\>[2]\AgdaSymbol{)}\AgdaSpace{}%
\AgdaKeyword{where}\<%
\end{code},
is a diagram as in \eqref{eq:diagram}, with a shape determined by the base category $C$.
\adapted{Indeed, for every object $x$ of $C$ there is a node in the diagram containing the Agda type
\begin{code}[hide]%
\>[2]\AgdaKeyword{open}\AgdaSpace{}%
\AgdaModule{BaseCategory}\AgdaSpace{}%
\AgdaBound{C}\<%
\\
\>[2]\AgdaFunction{\AgdaUnderscore{}}\AgdaSpace{}%
\AgdaSymbol{:}\AgdaSpace{}%
\AgdaFunction{Ob}\AgdaSpace{}%
\AgdaSymbol{→}\AgdaSpace{}%
\AgdaPrimitive{Set}\<%
\\
\>[2]\AgdaSymbol{\AgdaUnderscore{}}%
\>[157I]\AgdaSymbol{=}\AgdaSpace{}%
\AgdaSymbol{λ}\AgdaSpace{}%
\AgdaBound{x}\AgdaSpace{}%
\AgdaSymbol{→}\<%
\end{code}
\begin{code}[inline*]%
\>[.][@{}l@{}]\<[157I]%
\>[4]\AgdaBound{Γ}\AgdaSpace{}%
\AgdaOperator{\AgdaField{⟨}}\AgdaSpace{}%
\AgdaBound{x}\AgdaSpace{}%
\AgdaOperator{\AgdaField{⟩}}\<%
\end{code}
 and for every morphism $f$ from $x$ to $y$ in $C$ there is an arrow in the diagram representing an Agda function
\begin{code}[hide]%
\>[2]\AgdaFunction{\AgdaUnderscore{}}\AgdaSpace{}%
\AgdaSymbol{:}\AgdaSpace{}%
\AgdaSymbol{(}\AgdaBound{x}\AgdaSpace{}%
\AgdaBound{y}\AgdaSpace{}%
\AgdaSymbol{:}\AgdaSpace{}%
\AgdaFunction{Ob}\AgdaSymbol{)}\AgdaSpace{}%
\AgdaSymbol{→}\AgdaSpace{}%
\AgdaFunction{Hom}\AgdaSpace{}%
\AgdaBound{x}\AgdaSpace{}%
\AgdaBound{y}\AgdaSpace{}%
\AgdaSymbol{→}\AgdaSpace{}%
\AgdaBound{Γ}\AgdaSpace{}%
\AgdaOperator{\AgdaField{⟨}}\AgdaSpace{}%
\AgdaBound{y}\AgdaSpace{}%
\AgdaOperator{\AgdaField{⟩}}\AgdaSpace{}%
\AgdaSymbol{→}\AgdaSpace{}%
\AgdaBound{Γ}\AgdaSpace{}%
\AgdaOperator{\AgdaField{⟨}}\AgdaSpace{}%
\AgdaBound{x}\AgdaSpace{}%
\AgdaOperator{\AgdaField{⟩}}\<%
\\
\>[2]\AgdaSymbol{\AgdaUnderscore{}}%
\>[183I]\AgdaSymbol{=}\AgdaSpace{}%
\AgdaSymbol{λ}\AgdaSpace{}%
\AgdaBound{x}\AgdaSpace{}%
\AgdaBound{y}\AgdaSpace{}%
\AgdaBound{f}\AgdaSpace{}%
\AgdaSymbol{→}\<%
\end{code}
\begin{code}[inline*]%
\>[.][@{}l@{}]\<[183I]%
\>[4]\AgdaBound{Γ}\AgdaSpace{}%
\AgdaOperator{\AgdaField{⟪}}\AgdaSpace{}%
\AgdaBound{f}\AgdaSpace{}%
\AgdaOperator{\AgdaField{⟫\AgdaUnderscore{}}}\<%
\end{code}
\begin{code}[hide]%
\>[2]\AgdaKeyword{module}\AgdaSpace{}%
\AgdaModule{\AgdaUnderscore{}}\AgdaSpace{}%
\AgdaSymbol{(}\AgdaBound{x}\AgdaSpace{}%
\AgdaBound{y}\AgdaSpace{}%
\AgdaSymbol{:}\AgdaSpace{}%
\AgdaFunction{Ob}\AgdaSymbol{)}\AgdaSpace{}%
\AgdaSymbol{(}\AgdaBound{foo}\<%
\end{code}
\begin{code}[inline]%
\>[2][@{}l@{\AgdaIndent{1}}]%
\>[4]\AgdaSymbol{:}\AgdaSpace{}%
\AgdaBound{Γ}\AgdaSpace{}%
\AgdaOperator{\AgdaField{⟨}}\AgdaSpace{}%
\AgdaBound{y}\AgdaSpace{}%
\AgdaOperator{\AgdaField{⟩}}\AgdaSpace{}%
\AgdaSymbol{→}\AgdaSpace{}%
\AgdaBound{Γ}\AgdaSpace{}%
\AgdaOperator{\AgdaField{⟨}}\AgdaSpace{}%
\AgdaBound{x}\AgdaSpace{}%
\AgdaOperator{\AgdaField{⟩}}\<%
\end{code}
\begin{code}[hide]%
\>[4]\AgdaSymbol{)}\AgdaSpace{}%
\AgdaKeyword{where}\<%
\end{code}
.
The base category that determines the shape of diagram \eqref{eq:diagram} has the natural numbers as objects, giving rise to a sequence of Agda types.
Furthermore, it has exactly one morphism from $m$ to $n$ if and only if $m \leq n$.
This means that diagram  \eqref{eq:diagram} actually has more arrows than shown (e.g.\ from $\ADT{Vec}_{n+2}\, A$ to $\ADT{Vec}_n\, A$), but the omitted equality laws in the definition of \AR{Ctx} above make sure that all of these can be obtained by composing the \AF{init} functions.}

There is a semantic empty context \AF{◇} that is defined as
\begin{code}[hide]%
\>[0]\AgdaFunction{◇}\AgdaSpace{}%
\AgdaSymbol{:}\AgdaSpace{}%
\AgdaRecord{Ctx}\AgdaSpace{}%
\AgdaGeneralizable{C}\<%
\end{code}
\begin{code}[inline*]%
\>[0]\AgdaFunction{◇}\AgdaSpace{}%
\AgdaOperator{\AgdaField{⟨}}\AgdaSpace{}%
\AgdaBound{x}\AgdaSpace{}%
\AgdaOperator{\AgdaField{⟩}}\AgdaSpace{}%
\AgdaSymbol{=}\AgdaSpace{}%
\AgdaRecord{⊤}\<%
\end{code}
\begin{code}[hide]%
\>[0]\AgdaFunction{◇}\AgdaSpace{}%
\AgdaOperator{\AgdaField{⟪}}\AgdaSpace{}%
\AgdaBound{f}\AgdaSpace{}%
\AgdaOperator{\AgdaField{⟫}}\AgdaSpace{}%
\AgdaSymbol{\AgdaUnderscore{}}\AgdaSpace{}%
\AgdaSymbol{=}\AgdaSpace{}%
\AgdaInductiveConstructor{tt}\<%
\end{code}
for every $x$.
Furthermore, a CwF has the notion of semantic substitutions between two contexts.
These are implemented in Sikkel, but not needed in the rest of the paper.

\adapted{Although MSTT is currently not dependently typed, in the future we do plan to support theories where types may depend on variables.}
Our formalization of presheaf models already anticipates this and every semantic type lives in a certain context.
As a result, the representation of semantic types is more complicated than the diagram from \eqref{eq:diagram}.
We define the type of semantic types in a given context as the following record type (omitting equality laws).%
\footnote{Here \ADT{≡} is Agda's identity type with reflexivity constructor \AIC{refl}.
  Arguments between curly brackets are implicit and will be inferred by Agda.}%
\begin{code}%
\>[0]\AgdaKeyword{record}\AgdaSpace{}%
\AgdaRecord{Ty}\AgdaSpace{}%
\AgdaSymbol{(}\AgdaBound{Γ}\AgdaSpace{}%
\AgdaSymbol{:}\AgdaSpace{}%
\AgdaRecord{Ctx}\AgdaSpace{}%
\AgdaGeneralizable{C}\AgdaSymbol{)}\AgdaSpace{}%
\AgdaSymbol{:}\AgdaSpace{}%
\AgdaPrimitive{Set₁}\AgdaSpace{}%
\AgdaKeyword{where}\<%
\\
\>[0][@{}l@{\AgdaIndent{0}}]%
\>[2]\AgdaKeyword{field}%
\>[232I]\AgdaField{ty-cell}\AgdaSpace{}%
\AgdaSymbol{:}\AgdaSpace{}%
\AgdaSymbol{(}\AgdaBound{x}\AgdaSpace{}%
\AgdaSymbol{:}\AgdaSpace{}%
\AgdaFunction{Ob}\AgdaSymbol{)}\AgdaSpace{}%
\AgdaSymbol{(}\AgdaBound{γ}\AgdaSpace{}%
\AgdaSymbol{:}\AgdaSpace{}%
\AgdaBound{Γ}\AgdaSpace{}%
\AgdaOperator{\AgdaField{⟨}}\AgdaSpace{}%
\AgdaBound{x}\AgdaSpace{}%
\AgdaOperator{\AgdaField{⟩}}\AgdaSymbol{)}\AgdaSpace{}%
\AgdaSymbol{→}\AgdaSpace{}%
\AgdaPrimitive{Set}\<%
\\
\>[.][@{}l@{}]\<[232I]%
\>[13]\AgdaField{ty-hom}\AgdaSpace{}%
\AgdaSymbol{:}\AgdaSpace{}%
\AgdaSymbol{(}\AgdaBound{f}\AgdaSpace{}%
\AgdaSymbol{:}\AgdaSpace{}%
\AgdaFunction{Hom}\AgdaSpace{}%
\AgdaBound{x}\AgdaSpace{}%
\AgdaBound{y}\AgdaSymbol{)}\AgdaSpace{}%
\AgdaSymbol{\{}\AgdaBound{γy}\AgdaSpace{}%
\AgdaSymbol{:}\AgdaSpace{}%
\AgdaBound{Γ}\AgdaSpace{}%
\AgdaOperator{\AgdaField{⟨}}\AgdaSpace{}%
\AgdaBound{y}\AgdaSpace{}%
\AgdaOperator{\AgdaField{⟩}}\AgdaSymbol{\}}\AgdaSpace{}%
\AgdaSymbol{\{}\AgdaBound{γx}\AgdaSpace{}%
\AgdaSymbol{:}\AgdaSpace{}%
\AgdaBound{Γ}\AgdaSpace{}%
\AgdaOperator{\AgdaField{⟨}}\AgdaSpace{}%
\AgdaBound{x}\AgdaSpace{}%
\AgdaOperator{\AgdaField{⟩}}\AgdaSymbol{\}}\AgdaSpace{}%
\AgdaSymbol{→}\AgdaSpace{}%
\AgdaBound{Γ}\AgdaSpace{}%
\AgdaOperator{\AgdaField{⟪}}\AgdaSpace{}%
\AgdaBound{f}\AgdaSpace{}%
\AgdaOperator{\AgdaField{⟫}}\AgdaSpace{}%
\AgdaBound{γy}\AgdaSpace{}%
\AgdaOperator{\AgdaDatatype{≡}}\AgdaSpace{}%
\AgdaBound{γx}\AgdaSpace{}%
\AgdaSymbol{→}\AgdaSpace{}%
\AgdaField{ty-cell}\AgdaSpace{}%
\AgdaBound{y}\AgdaSpace{}%
\AgdaBound{γy}\AgdaSpace{}%
\AgdaSymbol{→}\AgdaSpace{}%
\AgdaField{ty-cell}\AgdaSpace{}%
\AgdaBound{x}\AgdaSpace{}%
\AgdaBound{γx}\<%
\end{code}
\begin{code}[hide]%
\>[13]\AgdaField{ty-hom'}\AgdaSpace{}%
\AgdaSymbol{:}\AgdaSpace{}%
\AgdaSymbol{∀}\AgdaSpace{}%
\AgdaSymbol{\{}\AgdaBound{x}\AgdaSpace{}%
\AgdaBound{y}\AgdaSymbol{\}}\AgdaSpace{}%
\AgdaSymbol{\{}\AgdaBound{γy}\AgdaSpace{}%
\AgdaSymbol{:}\AgdaSpace{}%
\AgdaBound{Γ}\AgdaSpace{}%
\AgdaOperator{\AgdaField{⟨}}\AgdaSpace{}%
\AgdaBound{y}\AgdaSpace{}%
\AgdaOperator{\AgdaField{⟩}}\AgdaSymbol{\}}\<%
\end{code}
\newcommand{\alternativeTyHom}{%
\begin{code}[inline]
\>[13][@{}l@{\AgdaIndent{1}}]%
\>[15]\AgdaSymbol{(}\AgdaBound{f}\AgdaSpace{}%
\AgdaSymbol{:}\AgdaSpace{}%
\AgdaFunction{Hom}\AgdaSpace{}%
\AgdaBound{x}\AgdaSpace{}%
\AgdaBound{y}\AgdaSymbol{)}\AgdaSpace{}%
\AgdaSymbol{→}\AgdaSpace{}%
\AgdaField{ty-cell}\AgdaSpace{}%
\AgdaBound{y}\AgdaSpace{}%
\AgdaBound{γy}\AgdaSpace{}%
\AgdaSymbol{→}\AgdaSpace{}%
\AgdaField{ty-cell}\AgdaSpace{}%
\AgdaBound{x}\AgdaSpace{}%
\AgdaSymbol{(}\AgdaBound{Γ}\AgdaSpace{}%
\AgdaOperator{\AgdaField{⟪}}\AgdaSpace{}%
\AgdaBound{f}\AgdaSpace{}%
\AgdaOperator{\AgdaField{⟫}}\AgdaSpace{}%
\AgdaBound{γy}\AgdaSymbol{)}\<%
\end{code}}%
\begin{code}[hide]%
\>[0]\AgdaKeyword{open}\AgdaSpace{}%
\AgdaModule{Ty}\AgdaSpace{}%
\AgdaKeyword{renaming}\AgdaSpace{}%
\AgdaSymbol{(}\AgdaField{ty-cell}\AgdaSpace{}%
\AgdaSymbol{to}\AgdaSpace{}%
\AgdaField{\AgdaUnderscore{}⟨\AgdaUnderscore{},\AgdaUnderscore{}⟩}\AgdaSymbol{;}\AgdaSpace{}%
\AgdaField{ty-hom}\AgdaSpace{}%
\AgdaSymbol{to}\AgdaSpace{}%
\AgdaField{\AgdaUnderscore{}⟪\AgdaUnderscore{},\AgdaUnderscore{}⟫\AgdaUnderscore{}}\AgdaSymbol{)}\<%
\end{code}
Again, we introduce the notation
\begin{code}[hide]%
\>[0]\AgdaKeyword{module}\AgdaSpace{}%
\AgdaModule{\AgdaUnderscore{}}\AgdaSpace{}%
\AgdaSymbol{(}\AgdaBound{C}\AgdaSpace{}%
\AgdaSymbol{:}\AgdaSpace{}%
\AgdaRecord{BaseCategory}\AgdaSymbol{)}\AgdaSpace{}%
\AgdaSymbol{(}\AgdaBound{Γ}\AgdaSpace{}%
\AgdaSymbol{:}\AgdaSpace{}%
\AgdaRecord{Ctx}\AgdaSpace{}%
\AgdaBound{C}\AgdaSymbol{)}\AgdaSpace{}%
\AgdaSymbol{(}\AgdaBound{T}\AgdaSpace{}%
\AgdaSymbol{:}\AgdaSpace{}%
\AgdaRecord{Ty}\AgdaSpace{}%
\AgdaBound{Γ}\AgdaSymbol{)}\AgdaSpace{}%
\AgdaKeyword{where}\<%
\\
\>[0][@{}l@{\AgdaIndent{0}}]%
\>[2]\AgdaKeyword{open}\AgdaSpace{}%
\AgdaModule{BaseCategory}\AgdaSpace{}%
\AgdaBound{C}\<%
\\
\\[\AgdaEmptyExtraSkip]%
\>[2]\AgdaFunction{\AgdaUnderscore{}}\AgdaSpace{}%
\AgdaSymbol{:}\AgdaSpace{}%
\AgdaSymbol{(}\AgdaBound{x}\AgdaSpace{}%
\AgdaSymbol{:}\AgdaSpace{}%
\AgdaFunction{Ob}\AgdaSymbol{)}\AgdaSpace{}%
\AgdaSymbol{→}\AgdaSpace{}%
\AgdaBound{Γ}\AgdaSpace{}%
\AgdaOperator{\AgdaField{⟨}}\AgdaSpace{}%
\AgdaBound{x}\AgdaSpace{}%
\AgdaOperator{\AgdaField{⟩}}\AgdaSpace{}%
\AgdaSymbol{→}\AgdaSpace{}%
\AgdaPrimitive{Set}\<%
\\
\>[2]\AgdaSymbol{\AgdaUnderscore{}}%
\>[342I]\AgdaSymbol{=}\AgdaSpace{}%
\AgdaSymbol{λ}\AgdaSpace{}%
\AgdaBound{x}\AgdaSpace{}%
\AgdaBound{γ}\AgdaSpace{}%
\AgdaSymbol{→}\<%
\end{code}
\begin{code}[inline*]%
\>[342I][@{}l@{\AgdaIndent{1}}]%
\>[5]\AgdaBound{T}\AgdaSpace{}%
\AgdaOperator{\AgdaField{⟨}}\AgdaSpace{}%
\AgdaBound{x}\AgdaSpace{}%
\AgdaOperator{\AgdaField{,}}\AgdaSpace{}%
\AgdaBound{γ}\AgdaSpace{}%
\AgdaOperator{\AgdaField{⟩}}\<%
\end{code}
$=$ \AFi{ty-cell} $T$ $x$ $\gamma$ and
\begin{code}[hide]%
\>[2]\AgdaFunction{\AgdaUnderscore{}}\AgdaSpace{}%
\AgdaSymbol{:}\AgdaSpace{}%
\AgdaSymbol{\{}\AgdaBound{x}\AgdaSpace{}%
\AgdaBound{y}\AgdaSpace{}%
\AgdaSymbol{:}\AgdaSpace{}%
\AgdaFunction{Ob}\AgdaSymbol{\}}\AgdaSpace{}%
\AgdaSymbol{(}\AgdaBound{f}\AgdaSpace{}%
\AgdaSymbol{:}\AgdaSpace{}%
\AgdaFunction{Hom}\AgdaSpace{}%
\AgdaBound{x}\AgdaSpace{}%
\AgdaBound{y}\AgdaSymbol{)}\AgdaSpace{}%
\AgdaSymbol{\{}\AgdaBound{γx}\AgdaSpace{}%
\AgdaSymbol{:}\AgdaSpace{}%
\AgdaBound{Γ}\AgdaSpace{}%
\AgdaOperator{\AgdaField{⟨}}\AgdaSpace{}%
\AgdaBound{x}\AgdaSpace{}%
\AgdaOperator{\AgdaField{⟩}}\AgdaSymbol{\}}\AgdaSpace{}%
\AgdaSymbol{\{}\AgdaBound{γy}\AgdaSpace{}%
\AgdaSymbol{:}\AgdaSpace{}%
\AgdaBound{Γ}\AgdaSpace{}%
\AgdaOperator{\AgdaField{⟨}}\AgdaSpace{}%
\AgdaBound{y}\AgdaSpace{}%
\AgdaOperator{\AgdaField{⟩}}\AgdaSymbol{\}}\AgdaSpace{}%
\AgdaSymbol{(}\AgdaBound{e}\AgdaSpace{}%
\AgdaSymbol{:}\AgdaSpace{}%
\AgdaBound{Γ}\AgdaSpace{}%
\AgdaOperator{\AgdaField{⟪}}\AgdaSpace{}%
\AgdaBound{f}\AgdaSpace{}%
\AgdaOperator{\AgdaField{⟫}}\AgdaSpace{}%
\AgdaBound{γy}\AgdaSpace{}%
\AgdaOperator{\AgdaDatatype{≡}}\AgdaSpace{}%
\AgdaBound{γx}\AgdaSymbol{)}\AgdaSpace{}%
\AgdaSymbol{→}\AgdaSpace{}%
\AgdaBound{T}\AgdaSpace{}%
\AgdaOperator{\AgdaField{⟨}}\AgdaSpace{}%
\AgdaBound{y}\AgdaSpace{}%
\AgdaOperator{\AgdaField{,}}\AgdaSpace{}%
\AgdaBound{γy}\AgdaSpace{}%
\AgdaOperator{\AgdaField{⟩}}\AgdaSpace{}%
\AgdaSymbol{→}\AgdaSpace{}%
\AgdaBound{T}\AgdaSpace{}%
\AgdaOperator{\AgdaField{⟨}}\AgdaSpace{}%
\AgdaBound{x}\AgdaSpace{}%
\AgdaOperator{\AgdaField{,}}\AgdaSpace{}%
\AgdaBound{γx}\AgdaSpace{}%
\AgdaOperator{\AgdaField{⟩}}\<%
\\
\>[2]\AgdaSymbol{\AgdaUnderscore{}}\AgdaSpace{}%
\AgdaSymbol{=}%
\>[398I]\AgdaSymbol{λ}\AgdaSpace{}%
\AgdaBound{f}\AgdaSpace{}%
\AgdaBound{e}\AgdaSpace{}%
\AgdaBound{t}\AgdaSpace{}%
\AgdaSymbol{→}\<%
\end{code}
\begin{code}[inline*]%
\>[.][@{}l@{}]\<[398I]%
\>[6]\AgdaBound{T}\AgdaSpace{}%
\AgdaOperator{\AgdaField{⟪}}\AgdaSpace{}%
\AgdaBound{f}\AgdaSpace{}%
\AgdaOperator{\AgdaField{,}}\AgdaSpace{}%
\AgdaBound{e}\AgdaSpace{}%
\AgdaOperator{\AgdaField{⟫}}\AgdaSpace{}%
\AgdaBound{t}\<%
\end{code}
$=$ \AFi{ty-hom} $T$ $f$ $e$ $t$.
If $T$ is a type in context \AB{Γ}, we get an Agda type
\begin{code}[hide]%
\>[0]\AgdaKeyword{module}\AgdaSpace{}%
\AgdaModule{UseTy}\<%
\\
\>[0][@{}l@{\AgdaIndent{0}}]%
\>[2]\AgdaSymbol{(}\AgdaBound{C}\AgdaSpace{}%
\AgdaSymbol{:}\AgdaSpace{}%
\AgdaRecord{BaseCategory}\AgdaSymbol{)}\AgdaSpace{}%
\AgdaSymbol{(}\AgdaBound{x}\AgdaSpace{}%
\AgdaBound{y}\AgdaSpace{}%
\AgdaSymbol{:}\AgdaSpace{}%
\AgdaField{BaseCategory.Ob}\AgdaSpace{}%
\AgdaBound{C}\AgdaSymbol{)}\AgdaSpace{}%
\AgdaSymbol{(}\AgdaBound{f}\AgdaSpace{}%
\AgdaSymbol{:}\AgdaSpace{}%
\AgdaField{BaseCategory.Hom}\AgdaSpace{}%
\AgdaBound{C}\AgdaSpace{}%
\AgdaBound{x}\AgdaSpace{}%
\AgdaBound{y}\AgdaSymbol{)}\<%
\\
\>[2]\AgdaSymbol{(}\AgdaBound{Γ}\AgdaSpace{}%
\AgdaSymbol{:}\AgdaSpace{}%
\AgdaRecord{Ctx}\AgdaSpace{}%
\AgdaBound{C}\AgdaSymbol{)}\AgdaSpace{}%
\AgdaSymbol{(}\AgdaBound{γ}\AgdaSpace{}%
\AgdaBound{γx}\AgdaSpace{}%
\AgdaSymbol{:}\AgdaSpace{}%
\AgdaBound{Γ}\AgdaSpace{}%
\AgdaOperator{\AgdaField{⟨}}\AgdaSpace{}%
\AgdaBound{x}\AgdaSpace{}%
\AgdaOperator{\AgdaField{⟩}}\AgdaSymbol{)}\AgdaSpace{}%
\AgdaSymbol{(}\AgdaBound{γy}\AgdaSpace{}%
\AgdaSymbol{:}\AgdaSpace{}%
\AgdaBound{Γ}\AgdaSpace{}%
\AgdaOperator{\AgdaField{⟨}}\AgdaSpace{}%
\AgdaBound{y}\AgdaSpace{}%
\AgdaOperator{\AgdaField{⟩}}\AgdaSymbol{)}\AgdaSpace{}%
\AgdaSymbol{(}\AgdaBound{T}\AgdaSpace{}%
\AgdaSymbol{:}\AgdaSpace{}%
\AgdaRecord{Ty}\AgdaSpace{}%
\AgdaBound{Γ}\AgdaSymbol{)}\AgdaSpace{}%
\AgdaSymbol{(}\AgdaBound{e}\AgdaSpace{}%
\AgdaSymbol{:}\AgdaSpace{}%
\AgdaBound{Γ}\AgdaSpace{}%
\AgdaOperator{\AgdaField{⟪}}\AgdaSpace{}%
\AgdaBound{f}\AgdaSpace{}%
\AgdaOperator{\AgdaField{⟫}}\AgdaSpace{}%
\AgdaBound{γy}\AgdaSpace{}%
\AgdaOperator{\AgdaDatatype{≡}}\AgdaSpace{}%
\AgdaBound{γx}\AgdaSymbol{)}\<%
\\
\>[2]\AgdaKeyword{where}\<%
\\
\>[2]\AgdaFunction{\AgdaUnderscore{}}\AgdaSpace{}%
\AgdaSymbol{:}\AgdaSpace{}%
\AgdaPrimitive{Set}\<%
\\
\>[2]\AgdaSymbol{\AgdaUnderscore{}}%
\>[454I]\AgdaSymbol{=}\<%
\end{code}
\begin{code}[inline*]%
\>[.][@{}l@{}]\<[454I]%
\>[4]\AgdaBound{T}\AgdaSpace{}%
\AgdaOperator{\AgdaField{⟨}}\AgdaSpace{}%
\AgdaBound{x}\AgdaSpace{}%
\AgdaOperator{\AgdaField{,}}\AgdaSpace{}%
\AgdaBound{γ}\AgdaSpace{}%
\AgdaOperator{\AgdaField{⟩}}\<%
\end{code}
for every object $x$ in the base category $C$ and every cell
\AB{γ}\AS\ASy{:}\AS\AB{Γ}\AS\AFi{⟨}\AS\AB{x}\AS\AFi{⟩}.
(In practice, all types in this paper will be closed and will hence not depend on the cell \AB{γ}.)
Furthermore, there is a restriction map
\begin{code}[hide]%
\>[2]\AgdaFunction{\AgdaUnderscore{}}%
\>[460I]\AgdaSymbol{:}\<%
\end{code}%
\newcommand{\TytoTx}{%
  \begin{code}[inline]%
\>[.][@{}l@{}]\<[460I]%
\>[4]\AgdaBound{T}\AgdaSpace{}%
\AgdaOperator{\AgdaField{⟨}}\AgdaSpace{}%
\AgdaBound{y}\AgdaSpace{}%
\AgdaOperator{\AgdaField{,}}\AgdaSpace{}%
\AgdaBound{γy}\AgdaSpace{}%
\AgdaOperator{\AgdaField{⟩}}\AgdaSpace{}%
\AgdaSymbol{→}\AgdaSpace{}%
\AgdaBound{T}\AgdaSpace{}%
\AgdaOperator{\AgdaField{⟨}}\AgdaSpace{}%
\AgdaBound{x}\AgdaSpace{}%
\AgdaOperator{\AgdaField{,}}\AgdaSpace{}%
\AgdaBound{γx}\AgdaSpace{}%
\AgdaOperator{\AgdaField{⟩}}\<%
\end{code}
}%
\begin{code}[hide]%
\>[2]\AgdaSymbol{\AgdaUnderscore{}}%
\>[473I]\AgdaSymbol{=}\<%
\end{code}%
\newcommand{\Trestr}{%
\begin{code}[inline*]%
\>[.][@{}l@{}]\<[473I]%
\>[4]\AgdaBound{T}\AgdaSpace{}%
\AgdaOperator{\AgdaField{⟪}}\AgdaSpace{}%
\AgdaBound{f}\AgdaSpace{}%
\AgdaOperator{\AgdaField{,}}\AgdaSpace{}%
\AgdaBound{e}\AgdaSpace{}%
\AgdaOperator{\AgdaField{⟫\AgdaUnderscore{}}}\<%
\end{code}
}%
\Trestr{}\ASy{:}\AS\TytoTx{}
for every morphism $f$ from $x$ to $y$ in $C$ and for all cells \AB{γx} and \AB{γy} that satisfy
\begin{code}[hide]%
\>[0]\AgdaKeyword{module}\AgdaSpace{}%
\AgdaModule{\AgdaUnderscore{}}%
\>[480I]\AgdaSymbol{(}\AgdaBound{C}\AgdaSpace{}%
\AgdaSymbol{:}\AgdaSpace{}%
\AgdaRecord{BaseCategory}\AgdaSymbol{)}\AgdaSpace{}%
\AgdaSymbol{(}\AgdaBound{x}\AgdaSpace{}%
\AgdaBound{y}\AgdaSpace{}%
\AgdaSymbol{:}\AgdaSpace{}%
\AgdaField{BaseCategory.Ob}\AgdaSpace{}%
\AgdaBound{C}\AgdaSymbol{)}\AgdaSpace{}%
\AgdaSymbol{(}\AgdaBound{f}\AgdaSpace{}%
\AgdaSymbol{:}\AgdaSpace{}%
\AgdaField{BaseCategory.Hom}\AgdaSpace{}%
\AgdaBound{C}\AgdaSpace{}%
\AgdaBound{x}\AgdaSpace{}%
\AgdaBound{y}\AgdaSymbol{)}\<%
\\
\>[.][@{}l@{}]\<[480I]%
\>[9]\AgdaSymbol{(}\AgdaBound{Γ}\AgdaSpace{}%
\AgdaSymbol{:}\AgdaSpace{}%
\AgdaRecord{Ctx}\AgdaSpace{}%
\AgdaBound{C}\AgdaSymbol{)}\AgdaSpace{}%
\AgdaSymbol{(}\AgdaBound{γx}\AgdaSpace{}%
\AgdaSymbol{:}\AgdaSpace{}%
\AgdaBound{Γ}\AgdaSpace{}%
\AgdaOperator{\AgdaField{⟨}}\AgdaSpace{}%
\AgdaBound{x}\AgdaSpace{}%
\AgdaOperator{\AgdaField{⟩}}\AgdaSymbol{)}\AgdaSpace{}%
\AgdaSymbol{(}\AgdaBound{γy}\AgdaSpace{}%
\AgdaSymbol{:}\AgdaSpace{}%
\AgdaBound{Γ}\AgdaSpace{}%
\AgdaOperator{\AgdaField{⟨}}\AgdaSpace{}%
\AgdaBound{y}\AgdaSpace{}%
\AgdaOperator{\AgdaField{⟩}}\AgdaSymbol{)}\AgdaSpace{}%
\AgdaSymbol{(}\AgdaBound{T}\AgdaSpace{}%
\AgdaSymbol{:}\AgdaSpace{}%
\AgdaRecord{Ty}\AgdaSpace{}%
\AgdaBound{Γ}\AgdaSymbol{)}\AgdaSpace{}%
\AgdaKeyword{where}\<%
\\
\>[0][@{}l@{\AgdaIndent{0}}]%
\>[2]\AgdaFunction{\AgdaUnderscore{}}\AgdaSpace{}%
\AgdaSymbol{:}\AgdaSpace{}%
\AgdaPrimitive{Set}\<%
\\
\>[2]\AgdaSymbol{\AgdaUnderscore{}}%
\>[516I]\AgdaSymbol{=}\<%
\end{code}
\begin{code}[inline]%
\>[.][@{}l@{}]\<[516I]%
\>[4]\AgdaBound{Γ}\AgdaSpace{}%
\AgdaOperator{\AgdaField{⟪}}\AgdaSpace{}%
\AgdaBound{f}\AgdaSpace{}%
\AgdaOperator{\AgdaField{⟫}}\AgdaSpace{}%
\AgdaBound{γy}\AgdaSpace{}%
\AgdaOperator{\AgdaDatatype{≡}}\AgdaSpace{}%
\AgdaBound{γx}\<%
\end{code}.
This makes $T$ a presheaf over the category of elements of \AB{Γ}.
\adapted{In standard mathematical presentations of presheaf models, \AFi{ty-hom} has the type \alternativeTyHom{}.
  However, in our formalization this turns out to make the implementation of type substitution considerably more complicated.
  In the context of indexed inductive type families, the technique to introduce a variable \AB{γx} together with a propositional equality constraint is known as fording \cite{mcbride-phd} (named after a quote by Henry Ford: ``Any customer can have a car painted any color that he wants so long as it is black.'').
}

A semantic term of type $T$ in context \AB{Γ} then specifies for every object $x$ in the base category and every cell
\begin{code}[hide]%
\>[0]\AgdaKeyword{module}\AgdaSpace{}%
\AgdaModule{\AgdaUnderscore{}}\AgdaSpace{}%
\AgdaSymbol{(}\AgdaBound{Γ}\AgdaSpace{}%
\AgdaSymbol{:}\AgdaSpace{}%
\AgdaRecord{Ctx}\AgdaSpace{}%
\AgdaGeneralizable{C}\AgdaSymbol{)}\AgdaSpace{}%
\AgdaSymbol{(}\AgdaBound{x}\AgdaSpace{}%
\AgdaSymbol{:}\AgdaSpace{}%
\AgdaField{BaseCategory.Ob}\AgdaSpace{}%
\AgdaGeneralizable{C}\AgdaSymbol{)}\AgdaSpace{}%
\AgdaSymbol{(}\<%
\end{code}
\begin{code}[inline*]%
\>[0][@{}l@{\AgdaIndent{1}}]%
\>[2]\AgdaBound{γ}\AgdaSpace{}%
\AgdaSymbol{:}\AgdaSpace{}%
\AgdaBound{Γ}\AgdaSpace{}%
\AgdaOperator{\AgdaField{⟨}}\AgdaSpace{}%
\AgdaBound{x}\AgdaSpace{}%
\AgdaOperator{\AgdaField{⟩}}\<%
\end{code}
\begin{code}[hide]%
\>[2]\AgdaSymbol{)}\AgdaSpace{}%
\AgdaSymbol{(}\AgdaBound{T}\AgdaSpace{}%
\AgdaSymbol{:}\AgdaSpace{}%
\AgdaRecord{Ty}\AgdaSpace{}%
\AgdaBound{Γ}\AgdaSymbol{)}\AgdaSpace{}%
\AgdaKeyword{where}\<%
\end{code}
an Agda value of type
\begin{code}[hide]%
\>[2]\AgdaFunction{\AgdaUnderscore{}}\AgdaSpace{}%
\AgdaSymbol{:}\AgdaSpace{}%
\AgdaPrimitive{Set}\<%
\\
\>[2]\AgdaSymbol{\AgdaUnderscore{}}%
\>[545I]\AgdaSymbol{=}\<%
\end{code}
\begin{code}[inline]%
\>[.][@{}l@{}]\<[545I]%
\>[4]\AgdaBound{T}\AgdaSpace{}%
\AgdaOperator{\AgdaField{⟨}}\AgdaSpace{}%
\AgdaBound{x}\AgdaSpace{}%
\AgdaOperator{\AgdaField{,}}\AgdaSpace{}%
\AgdaBound{γ}\AgdaSpace{}%
\AgdaOperator{\AgdaField{⟩}}\<%
\end{code}.
\begin{code}%
\>[0]\AgdaKeyword{record}%
\>[551I]\AgdaRecord{Tm}\AgdaSpace{}%
\AgdaSymbol{(}\AgdaBound{Γ}\AgdaSpace{}%
\AgdaSymbol{:}\AgdaSpace{}%
\AgdaRecord{Ctx}\AgdaSpace{}%
\AgdaGeneralizable{C}\AgdaSymbol{)}\AgdaSpace{}%
\AgdaSymbol{(}\AgdaBound{T}\AgdaSpace{}%
\AgdaSymbol{:}\AgdaSpace{}%
\AgdaRecord{Ty}\AgdaSpace{}%
\AgdaBound{Γ}\AgdaSymbol{)}\AgdaSpace{}%
\AgdaSymbol{:}\AgdaSpace{}%
\AgdaPrimitive{Set}\AgdaSpace{}%
\AgdaKeyword{where}\AgdaSpace{}%
\AgdaKeyword{field}\AgdaSpace{}%
\AgdaField{term}\AgdaSpace{}%
\AgdaSymbol{:}\AgdaSpace{}%
\AgdaSymbol{(}\AgdaBound{x}\AgdaSpace{}%
\AgdaSymbol{:}\AgdaSpace{}%
\AgdaFunction{Ob}\AgdaSymbol{)}\AgdaSpace{}%
\AgdaSymbol{(}\AgdaBound{γ}\AgdaSpace{}%
\AgdaSymbol{:}\AgdaSpace{}%
\AgdaBound{Γ}\AgdaSpace{}%
\AgdaOperator{\AgdaField{⟨}}\AgdaSpace{}%
\AgdaBound{x}\AgdaSpace{}%
\AgdaOperator{\AgdaField{⟩}}\AgdaSymbol{)}\AgdaSpace{}%
\AgdaSymbol{→}\AgdaSpace{}%
\AgdaBound{T}\AgdaSpace{}%
\AgdaOperator{\AgdaField{⟨}}\AgdaSpace{}%
\AgdaBound{x}\AgdaSpace{}%
\AgdaOperator{\AgdaField{,}}\AgdaSpace{}%
\AgdaBound{γ}\AgdaSpace{}%
\AgdaOperator{\AgdaField{⟩}}\<%
\end{code}
\begin{code}[hide]%
\>[0]\AgdaKeyword{open}\AgdaSpace{}%
\AgdaModule{Tm}\AgdaSpace{}%
\AgdaKeyword{renaming}\AgdaSpace{}%
\AgdaSymbol{(}\AgdaField{term}\AgdaSpace{}%
\AgdaSymbol{to}\AgdaSpace{}%
\AgdaField{\AgdaUnderscore{}⟨\AgdaUnderscore{},\AgdaUnderscore{}⟩'}\AgdaSymbol{)}\<%
\end{code}
We omit a naturality condition assuring that these Agda values are stable under the restriction maps of $T$. We use the notation
\begin{code}[hide]%
\>[0]\AgdaFunction{\AgdaUnderscore{}}\AgdaSpace{}%
\AgdaSymbol{:}\AgdaSpace{}%
\AgdaSymbol{\{}\AgdaBound{Γ}\AgdaSpace{}%
\AgdaSymbol{:}\AgdaSpace{}%
\AgdaRecord{Ctx}\AgdaSpace{}%
\AgdaGeneralizable{C}\AgdaSymbol{\}}\AgdaSpace{}%
\AgdaSymbol{\{}\AgdaBound{T}\AgdaSpace{}%
\AgdaSymbol{:}\AgdaSpace{}%
\AgdaRecord{Ty}\AgdaSpace{}%
\AgdaBound{Γ}\AgdaSymbol{\}}\AgdaSpace{}%
\AgdaSymbol{(}\AgdaBound{t}\AgdaSpace{}%
\AgdaSymbol{:}\AgdaSpace{}%
\AgdaRecord{Tm}\AgdaSpace{}%
\AgdaBound{Γ}\AgdaSpace{}%
\AgdaBound{T}\AgdaSymbol{)}\AgdaSpace{}%
\AgdaSymbol{(}\AgdaBound{x}\AgdaSpace{}%
\AgdaSymbol{:}\AgdaSpace{}%
\AgdaField{BaseCategory.Ob}\AgdaSpace{}%
\AgdaGeneralizable{C}\AgdaSymbol{)}\AgdaSpace{}%
\AgdaSymbol{(}\AgdaBound{γ}\AgdaSpace{}%
\AgdaSymbol{:}\AgdaSpace{}%
\AgdaBound{Γ}\AgdaSpace{}%
\AgdaOperator{\AgdaField{⟨}}\AgdaSpace{}%
\AgdaBound{x}\AgdaSpace{}%
\AgdaOperator{\AgdaField{⟩}}\AgdaSymbol{)}\AgdaSpace{}%
\AgdaSymbol{→}\AgdaSpace{}%
\AgdaBound{T}\AgdaSpace{}%
\AgdaOperator{\AgdaField{⟨}}\AgdaSpace{}%
\AgdaBound{x}\AgdaSpace{}%
\AgdaOperator{\AgdaField{,}}\AgdaSpace{}%
\AgdaBound{γ}\AgdaSpace{}%
\AgdaOperator{\AgdaField{⟩}}\<%
\\
\>[0]\AgdaSymbol{\AgdaUnderscore{}}%
\>[620I]\AgdaSymbol{=}\AgdaSpace{}%
\AgdaSymbol{λ}\AgdaSpace{}%
\AgdaBound{t}\AgdaSpace{}%
\AgdaBound{x}\AgdaSpace{}%
\AgdaBound{γ}\AgdaSpace{}%
\AgdaSymbol{→}\<%
\end{code}
\begin{code}[inline*]%
\>[.][@{}l@{}]\<[620I]%
\>[2]\AgdaBound{t}\AgdaSpace{}%
\AgdaOperator{\AgdaField{⟨}}\AgdaSpace{}%
\AgdaBound{x}\AgdaSpace{}%
\AgdaOperator{\AgdaField{,}}\AgdaSpace{}%
\AgdaBound{γ}\AgdaSpace{}%
\AgdaOperator{\AgdaField{⟩'}}\<%
\end{code}
$=$ \AFi{term} $t$ $x$ $\gamma$.

Types in MSTT do not depend on variables in the context.
As a result, they can be interpreted as closed types which can live in any semantic context.
\begin{code}%
\>[0]\AgdaFunction{ClosedTy}\AgdaSpace{}%
\AgdaSymbol{:}\AgdaSpace{}%
\AgdaRecord{BaseCategory}\AgdaSpace{}%
\AgdaSymbol{→}\AgdaSpace{}%
\AgdaPrimitive{Set₁}\<%
\\
\>[0]\AgdaFunction{ClosedTy}\AgdaSpace{}%
\AgdaBound{C}\AgdaSpace{}%
\AgdaSymbol{=}\AgdaSpace{}%
\AgdaSymbol{\{}\AgdaBound{Γ}\AgdaSpace{}%
\AgdaSymbol{:}\AgdaSpace{}%
\AgdaRecord{Ctx}\AgdaSpace{}%
\AgdaBound{C}\AgdaSymbol{\}}\AgdaSpace{}%
\AgdaSymbol{→}\AgdaSpace{}%
\AgdaRecord{Ty}\AgdaSpace{}%
\AgdaBound{Γ}\<%
\end{code}
\adapted{(We omit a condition that expresses naturality w.r.t.\ substitutions; this condition implies that a closed type is entirely
determined by its manifestation in the empty context.)}

Sikkel provides an equivalence relation \AR{≅ᵗʸ} on semantic types in the same context.
It is defined as natural isomorphism, so a proof of
\begin{code}[hide]%
\>[0]\AgdaKeyword{record}\AgdaSpace{}%
\AgdaOperator{\AgdaRecord{\AgdaUnderscore{}≅ᵗʸ\AgdaUnderscore{}}}\AgdaSpace{}%
\AgdaSymbol{\{}\AgdaBound{Γ}\AgdaSpace{}%
\AgdaSymbol{:}\AgdaSpace{}%
\AgdaRecord{Ctx}\AgdaSpace{}%
\AgdaGeneralizable{C}\AgdaSymbol{\}}\AgdaSpace{}%
\AgdaSymbol{(}\AgdaBound{T}\AgdaSpace{}%
\AgdaBound{S}\AgdaSpace{}%
\AgdaSymbol{:}\AgdaSpace{}%
\AgdaRecord{Ty}\AgdaSpace{}%
\AgdaBound{Γ}\AgdaSymbol{)}\AgdaSpace{}%
\AgdaSymbol{:}\AgdaSpace{}%
\AgdaPrimitive{Set}\AgdaSpace{}%
\AgdaKeyword{where}\<%
\\
\>[0][@{}l@{\AgdaIndent{0}}]%
\>[2]\AgdaKeyword{field}\<%
\\
\>[2][@{}l@{\AgdaIndent{0}}]%
\>[4]\AgdaField{tm-conv}\AgdaSpace{}%
\AgdaSymbol{:}\AgdaSpace{}%
\AgdaRecord{Tm}\AgdaSpace{}%
\AgdaBound{Γ}\AgdaSpace{}%
\AgdaBound{S}\AgdaSpace{}%
\AgdaSymbol{→}\AgdaSpace{}%
\AgdaRecord{Tm}\AgdaSpace{}%
\AgdaBound{Γ}\AgdaSpace{}%
\AgdaBound{T}\<%
\\
\\[\AgdaEmptyExtraSkip]%
\>[0]\AgdaKeyword{module}\AgdaSpace{}%
\AgdaModule{\AgdaUnderscore{}}\AgdaSpace{}%
\AgdaSymbol{(}\AgdaBound{Γ}\AgdaSpace{}%
\AgdaSymbol{:}\AgdaSpace{}%
\AgdaRecord{Ctx}\AgdaSpace{}%
\AgdaGeneralizable{C}\AgdaSymbol{)}\AgdaSpace{}%
\AgdaSymbol{(}\AgdaBound{T}\AgdaSpace{}%
\AgdaBound{S}\AgdaSpace{}%
\AgdaSymbol{:}\AgdaSpace{}%
\AgdaRecord{Ty}\AgdaSpace{}%
\AgdaBound{Γ}\AgdaSymbol{)}\AgdaSpace{}%
\AgdaSymbol{(}\AgdaBound{x}\AgdaSpace{}%
\AgdaSymbol{:}\AgdaSpace{}%
\AgdaField{BaseCategory.Ob}\AgdaSpace{}%
\AgdaGeneralizable{C}\AgdaSymbol{)}\AgdaSpace{}%
\AgdaSymbol{(}\AgdaBound{γ}\AgdaSpace{}%
\AgdaSymbol{:}\AgdaSpace{}%
\AgdaBound{Γ}\AgdaSpace{}%
\AgdaOperator{\AgdaField{⟨}}\AgdaSpace{}%
\AgdaBound{x}\AgdaSpace{}%
\AgdaOperator{\AgdaField{⟩}}\AgdaSymbol{)}\AgdaSpace{}%
\AgdaKeyword{where}\<%
\\
\>[0][@{}l@{\AgdaIndent{0}}]%
\>[2]\AgdaFunction{\AgdaUnderscore{}}\AgdaSpace{}%
\AgdaSymbol{:}\AgdaSpace{}%
\AgdaPrimitive{Set}\<%
\\
\>[2]\AgdaSymbol{\AgdaUnderscore{}}%
\>[688I]\AgdaSymbol{=}\<%
\end{code}
\begin{code}[inline*]%
\>[.][@{}l@{}]\<[688I]%
\>[4]\AgdaBound{T}\AgdaSpace{}%
\AgdaOperator{\AgdaRecord{≅ᵗʸ}}\AgdaSpace{}%
\AgdaBound{S}\<%
\end{code}
amounts to a collection of Agda isomorphisms between
\begin{code}[hide]%
\>[2]\AgdaFunction{\AgdaUnderscore{}}\AgdaSpace{}%
\AgdaSymbol{:}\AgdaSpace{}%
\AgdaPrimitive{Set}\<%
\\
\>[2]\AgdaSymbol{\AgdaUnderscore{}}%
\>[693I]\AgdaSymbol{=}\<%
\end{code}
\begin{code}[inline*]%
\>[.][@{}l@{}]\<[693I]%
\>[4]\AgdaBound{T}\AgdaSpace{}%
\AgdaOperator{\AgdaField{⟨}}\AgdaSpace{}%
\AgdaBound{x}\AgdaSpace{}%
\AgdaOperator{\AgdaField{,}}\AgdaSpace{}%
\AgdaBound{γ}\AgdaSpace{}%
\AgdaOperator{\AgdaField{⟩}}\<%
\end{code}
and
\begin{code}[hide]%
\>[2]\AgdaFunction{\AgdaUnderscore{}}\AgdaSpace{}%
\AgdaSymbol{:}\AgdaSpace{}%
\AgdaPrimitive{Set}\<%
\\
\>[2]\AgdaSymbol{\AgdaUnderscore{}}%
\>[701I]\AgdaSymbol{=}\<%
\end{code}
\begin{code}[inline*]%
\>[.][@{}l@{}]\<[701I]%
\>[4]\AgdaBound{S}\AgdaSpace{}%
\AgdaOperator{\AgdaField{⟨}}\AgdaSpace{}%
\AgdaBound{x}\AgdaSpace{}%
\AgdaOperator{\AgdaField{,}}\AgdaSpace{}%
\AgdaBound{γ}\AgdaSpace{}%
\AgdaOperator{\AgdaField{⟩}}\<%
\end{code}
that is compatible with the restriction maps.
Such a proof allows to convert terms of type $S$ to type $T$ via the operation
\begin{code}[inline]%
\>[0]\AgdaOperator{\AgdaFunction{ι[\AgdaUnderscore{}]\AgdaUnderscore{}}}\AgdaSpace{}%
\AgdaSymbol{:}\AgdaSpace{}%
\AgdaBound{T}\AgdaSpace{}%
\AgdaOperator{\AgdaRecord{≅ᵗʸ}}\AgdaSpace{}%
\AgdaBound{S}\AgdaSpace{}%
\AgdaSymbol{→}\AgdaSpace{}%
\AgdaRecord{Tm}\AgdaSpace{}%
\AgdaBound{Γ}\AgdaSpace{}%
\AgdaBound{S}\AgdaSpace{}%
\AgdaSymbol{→}\AgdaSpace{}%
\AgdaRecord{Tm}\AgdaSpace{}%
\AgdaBound{Γ}\AgdaSpace{}%
\AgdaBound{T}\<%
\end{code}
\begin{code}[hide]%
\>[0]\AgdaOperator{\AgdaFunction{ι[\AgdaUnderscore{}]\AgdaUnderscore{}}}\AgdaSpace{}%
\AgdaSymbol{=}\AgdaSpace{}%
\AgdaField{\AgdaUnderscore{}≅ᵗʸ\AgdaUnderscore{}.tm-conv}\<%
\end{code}
\ whose implementation we omit.

Regardless of the base category, every presheaf model supports the standard type and term formers for booleans, natural numbers, function types, product types, etc.
We will discuss some details about the implementation of function types and refer to the Agda code for the other type formers.
If $T$ and $S$ are seen as diagrams, a term of type
\begin{code}[hide]%
\>[0]\AgdaKeyword{module}\AgdaSpace{}%
\AgdaModule{\AgdaUnderscore{}}\AgdaSpace{}%
\AgdaSymbol{\{}\AgdaBound{C}\AgdaSpace{}%
\AgdaSymbol{:}\AgdaSpace{}%
\AgdaRecord{BaseCategory}\AgdaSymbol{\}}\AgdaSpace{}%
\AgdaKeyword{where}\<%
\\
\>[0][@{}l@{\AgdaIndent{0}}]%
\>[2]\AgdaKeyword{open}\AgdaSpace{}%
\AgdaModule{BaseCategory}\AgdaSpace{}%
\AgdaBound{C}\<%
\end{code}
\newcommand{\pshfun}{
  \begin{code}%
\>[2]\AgdaKeyword{record}\AgdaSpace{}%
\AgdaRecord{PshFun}\AgdaSpace{}%
\AgdaSymbol{(}\AgdaBound{T}\AgdaSpace{}%
\AgdaBound{S}\AgdaSpace{}%
\AgdaSymbol{:}\AgdaSpace{}%
\AgdaRecord{Ty}\AgdaSpace{}%
\AgdaBound{Γ}\AgdaSymbol{)}\AgdaSpace{}%
\AgdaSymbol{(}\AgdaBound{x}\AgdaSpace{}%
\AgdaSymbol{:}\AgdaSpace{}%
\AgdaField{Ob}\AgdaSymbol{)}\AgdaSpace{}%
\AgdaSymbol{(}\AgdaBound{γx}\AgdaSpace{}%
\AgdaSymbol{:}\AgdaSpace{}%
\AgdaBound{Γ}\AgdaSpace{}%
\AgdaOperator{\AgdaField{⟨}}\AgdaSpace{}%
\AgdaBound{x}\AgdaSpace{}%
\AgdaOperator{\AgdaField{⟩}}\AgdaSymbol{)}\AgdaSpace{}%
\AgdaSymbol{:}\AgdaSpace{}%
\AgdaPrimitive{Set}\AgdaSpace{}%
\AgdaKeyword{where}\<%
\\
\>[2][@{}l@{\AgdaIndent{0}}]%
\>[4]\AgdaKeyword{field}\AgdaSpace{}%
\AgdaField{fun}\AgdaSpace{}%
\AgdaSymbol{:}\AgdaSpace{}%
\AgdaSymbol{∀}\AgdaSpace{}%
\AgdaSymbol{\{}\AgdaBound{y}\AgdaSymbol{\}}\AgdaSpace{}%
\AgdaSymbol{(}\AgdaBound{ρ}\AgdaSpace{}%
\AgdaSymbol{:}\AgdaSpace{}%
\AgdaField{Hom}\AgdaSpace{}%
\AgdaBound{y}\AgdaSpace{}%
\AgdaBound{x}\AgdaSymbol{)}\AgdaSpace{}%
\AgdaSymbol{\{}\AgdaBound{γy}\AgdaSpace{}%
\AgdaSymbol{:}\AgdaSpace{}%
\AgdaBound{Γ}\AgdaSpace{}%
\AgdaOperator{\AgdaField{⟨}}\AgdaSpace{}%
\AgdaBound{y}\AgdaSpace{}%
\AgdaOperator{\AgdaField{⟩}}\AgdaSymbol{\}}\AgdaSpace{}%
\AgdaSymbol{→}\AgdaSpace{}%
\AgdaBound{Γ}\AgdaSpace{}%
\AgdaOperator{\AgdaField{⟪}}\AgdaSpace{}%
\AgdaBound{ρ}\AgdaSpace{}%
\AgdaOperator{\AgdaField{⟫}}\AgdaSpace{}%
\AgdaBound{γx}\AgdaSpace{}%
\AgdaOperator{\AgdaDatatype{≡}}\AgdaSpace{}%
\AgdaBound{γy}\AgdaSpace{}%
\AgdaSymbol{→}\AgdaSpace{}%
\AgdaBound{T}\AgdaSpace{}%
\AgdaOperator{\AgdaField{⟨}}\AgdaSpace{}%
\AgdaBound{y}\AgdaSpace{}%
\AgdaOperator{\AgdaField{,}}\AgdaSpace{}%
\AgdaBound{γy}\AgdaSpace{}%
\AgdaOperator{\AgdaField{⟩}}\AgdaSpace{}%
\AgdaSymbol{→}\AgdaSpace{}%
\AgdaBound{S}\AgdaSpace{}%
\AgdaOperator{\AgdaField{⟨}}\AgdaSpace{}%
\AgdaBound{y}\AgdaSpace{}%
\AgdaOperator{\AgdaField{,}}\AgdaSpace{}%
\AgdaBound{γy}\AgdaSpace{}%
\AgdaOperator{\AgdaField{⟩}}\<%
\end{code}
}%
\begin{code}[hide]%
\>[2]\AgdaKeyword{open}\AgdaSpace{}%
\AgdaModule{PshFun}\AgdaSpace{}%
\AgdaKeyword{public}\AgdaSpace{}%
\AgdaKeyword{renaming}\AgdaSpace{}%
\AgdaSymbol{(}\AgdaField{fun}\AgdaSpace{}%
\AgdaSymbol{to}\AgdaSpace{}%
\AgdaField{\AgdaUnderscore{}\$⟨\AgdaUnderscore{},\AgdaUnderscore{}⟩\AgdaUnderscore{}}\AgdaSymbol{)}\<%
\\
\\[\AgdaEmptyExtraSkip]%
\>[2]\AgdaSymbol{\{-\#}\AgdaSpace{}%
\AgdaKeyword{NON\AgdaUnderscore{}COVERING}\AgdaSpace{}%
\AgdaSymbol{\#-\}}\<%
\\
\>[2]\AgdaOperator{\AgdaFunction{\AgdaUnderscore{}⇛\AgdaUnderscore{}}}\AgdaSpace{}%
\AgdaSymbol{:}\AgdaSpace{}%
\AgdaSymbol{\{}\AgdaBound{Γ}\AgdaSpace{}%
\AgdaSymbol{:}\AgdaSpace{}%
\AgdaRecord{Ctx}\AgdaSpace{}%
\AgdaBound{C}\AgdaSymbol{\}}\AgdaSpace{}%
\AgdaSymbol{→}\AgdaSpace{}%
\AgdaRecord{Ty}\AgdaSpace{}%
\AgdaBound{Γ}\AgdaSpace{}%
\AgdaSymbol{→}\AgdaSpace{}%
\AgdaRecord{Ty}\AgdaSpace{}%
\AgdaBound{Γ}\AgdaSpace{}%
\AgdaSymbol{→}\AgdaSpace{}%
\AgdaRecord{Ty}\AgdaSpace{}%
\AgdaBound{Γ}\<%
\\
\>[2]\AgdaSymbol{(}\AgdaBound{T}\AgdaSpace{}%
\AgdaOperator{\AgdaFunction{⇛}}\AgdaSpace{}%
\AgdaBound{S}\AgdaSymbol{)}\AgdaSpace{}%
\AgdaOperator{\AgdaField{⟨}}\AgdaSpace{}%
\AgdaBound{x}\AgdaSpace{}%
\AgdaOperator{\AgdaField{,}}\AgdaSpace{}%
\AgdaBound{γx}\AgdaSpace{}%
\AgdaOperator{\AgdaField{⟩}}\AgdaSpace{}%
\AgdaSymbol{=}\AgdaSpace{}%
\AgdaRecord{PshFun}\AgdaSpace{}%
\AgdaBound{T}\AgdaSpace{}%
\AgdaBound{S}\AgdaSpace{}%
\AgdaBound{x}\AgdaSpace{}%
\AgdaBound{γx}\<%
\\
\>[0]\<%
\\
\>[0]\AgdaFunction{\AgdaUnderscore{}}\AgdaSpace{}%
\AgdaSymbol{:}\AgdaSpace{}%
\AgdaSymbol{\{}\AgdaBound{Γ}\AgdaSpace{}%
\AgdaSymbol{:}\AgdaSpace{}%
\AgdaRecord{Ctx}\AgdaSpace{}%
\AgdaGeneralizable{C}\AgdaSymbol{\}}\AgdaSpace{}%
\AgdaSymbol{→}\AgdaSpace{}%
\AgdaRecord{Ty}\AgdaSpace{}%
\AgdaBound{Γ}\AgdaSpace{}%
\AgdaSymbol{→}\AgdaSpace{}%
\AgdaRecord{Ty}\AgdaSpace{}%
\AgdaBound{Γ}\AgdaSpace{}%
\AgdaSymbol{→}\AgdaSpace{}%
\AgdaRecord{Ty}\AgdaSpace{}%
\AgdaBound{Γ}\<%
\\
\>[0]\AgdaSymbol{\AgdaUnderscore{}}%
\>[845I]\AgdaSymbol{=}\AgdaSpace{}%
\AgdaSymbol{λ}\AgdaSpace{}%
\AgdaBound{T}\AgdaSpace{}%
\AgdaBound{S}\AgdaSpace{}%
\AgdaSymbol{→}\<%
\end{code}
\begin{code}[inline*]%
\>[.][@{}l@{}]\<[845I]%
\>[2]\AgdaBound{T}\AgdaSpace{}%
\AgdaOperator{\AgdaFunction{⇛}}\AgdaSpace{}%
\AgdaBound{S}\<%
\end{code}
consists of a function from every Agda type of $T$ to the corresponding type of $S$ such that all squares arising in this way commute.
Hence, one could naively try to define the Agda type%
\begin{code}[hide]%
\>[0]\AgdaKeyword{module}\AgdaSpace{}%
\AgdaModule{\AgdaUnderscore{}}\AgdaSpace{}%
\AgdaSymbol{\{}\AgdaBound{Γ}\AgdaSpace{}%
\AgdaSymbol{:}\AgdaSpace{}%
\AgdaRecord{Ctx}\AgdaSpace{}%
\AgdaGeneralizable{C}\AgdaSymbol{\}}\AgdaSpace{}%
\AgdaSymbol{(}\AgdaBound{T}\AgdaSpace{}%
\AgdaBound{S}\AgdaSpace{}%
\AgdaSymbol{:}\AgdaSpace{}%
\AgdaRecord{Ty}\AgdaSpace{}%
\AgdaBound{Γ}\AgdaSymbol{)}\AgdaSpace{}%
\AgdaSymbol{(}\AgdaBound{x}\AgdaSpace{}%
\AgdaSymbol{:}\AgdaSpace{}%
\AgdaField{BaseCategory.Ob}\AgdaSpace{}%
\AgdaGeneralizable{C}\AgdaSymbol{)}\AgdaSpace{}%
\AgdaSymbol{(}\AgdaBound{γ}\AgdaSpace{}%
\AgdaSymbol{:}\AgdaSpace{}%
\AgdaBound{Γ}\AgdaSpace{}%
\AgdaOperator{\AgdaField{⟨}}\AgdaSpace{}%
\AgdaBound{x}\AgdaSpace{}%
\AgdaOperator{\AgdaField{⟩}}\AgdaSymbol{)}\AgdaSpace{}%
\AgdaKeyword{where}\<%
\\
\>[0][@{}l@{\AgdaIndent{0}}]%
\>[2]\AgdaFunction{\AgdaUnderscore{}}\AgdaSpace{}%
\AgdaSymbol{:}\AgdaSpace{}%
\AgdaPrimitive{Set}\<%
\\
\>[2]\AgdaSymbol{\AgdaUnderscore{}}%
\>[875I]\AgdaSymbol{=}\<%
\end{code}%
\newcommand{\functionCell}{
\begin{code}[inline*]%
\>[.][@{}l@{}]\<[875I]%
\>[4]\AgdaSymbol{(}\AgdaBound{T}\AgdaSpace{}%
\AgdaOperator{\AgdaFunction{⇛}}\AgdaSpace{}%
\AgdaBound{S}\AgdaSymbol{)}\AgdaSpace{}%
\AgdaOperator{\AgdaField{⟨}}\AgdaSpace{}%
\AgdaBound{x}\AgdaSpace{}%
\AgdaOperator{\AgdaField{,}}\AgdaSpace{}%
\AgdaBound{γ}\AgdaSpace{}%
\AgdaOperator{\AgdaField{⟩}}\<%
\end{code}%
}%
\functionCell{}%
as
\begin{code}[hide]%
\>[2]\AgdaFunction{\AgdaUnderscore{}}\AgdaSpace{}%
\AgdaSymbol{:}\AgdaSpace{}%
\AgdaPrimitive{Set}\<%
\\
\>[2]\AgdaSymbol{\AgdaUnderscore{}}%
\>[885I]\AgdaSymbol{=}\<%
\end{code}
\begin{code}[inline]%
\>[.][@{}l@{}]\<[885I]%
\>[4]\AgdaBound{T}\AgdaSpace{}%
\AgdaOperator{\AgdaField{⟨}}\AgdaSpace{}%
\AgdaBound{x}\AgdaSpace{}%
\AgdaOperator{\AgdaField{,}}\AgdaSpace{}%
\AgdaBound{γ}\AgdaSpace{}%
\AgdaOperator{\AgdaField{⟩}}\AgdaSpace{}%
\AgdaSymbol{→}\AgdaSpace{}%
\AgdaBound{S}\AgdaSpace{}%
\AgdaOperator{\AgdaField{⟨}}\AgdaSpace{}%
\AgdaBound{x}\AgdaSpace{}%
\AgdaOperator{\AgdaField{,}}\AgdaSpace{}%
\AgdaBound{γ}\AgdaSpace{}%
\AgdaOperator{\AgdaField{⟩}}\<%
\end{code},
but that definition does not allow to implement the restriction maps for
\begin{code}[hide]%
\>[2]\AgdaFunction{\AgdaUnderscore{}}\AgdaSpace{}%
\AgdaSymbol{:}\AgdaSpace{}%
\AgdaRecord{Ty}\AgdaSpace{}%
\AgdaBound{Γ}\<%
\\
\>[2]\AgdaSymbol{\AgdaUnderscore{}}%
\>[901I]\AgdaSymbol{=}\<%
\end{code}
\begin{code}[inline]%
\>[.][@{}l@{}]\<[901I]%
\>[4]\AgdaBound{T}\AgdaSpace{}%
\AgdaOperator{\AgdaFunction{⇛}}\AgdaSpace{}%
\AgdaBound{S}\<%
\end{code}.
The solution is to require such an Agda function not only for $x$, but for every object $y$ and every morphism from $y$ to $x$.
Consequently,\functionCell{}%
is defined as
\begin{code}[hide]%
\>[2]\AgdaFunction{\AgdaUnderscore{}}\AgdaSpace{}%
\AgdaSymbol{:}\AgdaSpace{}%
\AgdaPrimitive{Set}\<%
\\
\>[2]\AgdaSymbol{\AgdaUnderscore{}}%
\>[906I]\AgdaSymbol{=}\<%
\end{code}
\begin{code}[inline]%
\>[.][@{}l@{}]\<[906I]%
\>[4]\AgdaRecord{PshFun}\AgdaSpace{}%
\AgdaBound{T}\AgdaSpace{}%
\AgdaBound{S}\AgdaSpace{}%
\AgdaBound{x}\AgdaSpace{}%
\AgdaBound{γ}\<%
\end{code},
where \AR{PshFun} is the following record type:
\pshfun{}%
Again a naturality condition has been omitted. We use the notation
\begin{code}[hide]%
\>[2]\AgdaFunction{\AgdaUnderscore{}}\AgdaSpace{}%
\AgdaSymbol{:}\AgdaSpace{}%
\AgdaSymbol{∀}\AgdaSpace{}%
\AgdaSymbol{(}\AgdaBound{f}\AgdaSpace{}%
\AgdaSymbol{:}\AgdaSpace{}%
\AgdaRecord{PshFun}\AgdaSpace{}%
\AgdaBound{T}\AgdaSpace{}%
\AgdaBound{S}\AgdaSpace{}%
\AgdaBound{x}\AgdaSpace{}%
\AgdaBound{γ}\AgdaSymbol{)}\AgdaSpace{}%
\AgdaSymbol{\{}\AgdaBound{y}\AgdaSymbol{\}}\AgdaSpace{}%
\AgdaSymbol{(}\AgdaBound{ρ}\AgdaSpace{}%
\AgdaSymbol{:}\AgdaSpace{}%
\AgdaField{BaseCategory.Hom}\AgdaSpace{}%
\AgdaBound{C}\AgdaSpace{}%
\AgdaBound{y}\AgdaSpace{}%
\AgdaBound{x}\AgdaSymbol{)}\AgdaSpace{}%
\AgdaSymbol{\{}\AgdaBound{γy}\AgdaSpace{}%
\AgdaSymbol{:}\AgdaSpace{}%
\AgdaBound{Γ}\AgdaSpace{}%
\AgdaOperator{\AgdaField{⟨}}\AgdaSpace{}%
\AgdaBound{y}\AgdaSpace{}%
\AgdaOperator{\AgdaField{⟩}}\AgdaSymbol{\}}\AgdaSpace{}%
\AgdaSymbol{→}\AgdaSpace{}%
\AgdaBound{Γ}\AgdaSpace{}%
\AgdaOperator{\AgdaField{⟪}}\AgdaSpace{}%
\AgdaBound{ρ}\AgdaSpace{}%
\AgdaOperator{\AgdaField{⟫}}\AgdaSpace{}%
\AgdaBound{γ}\AgdaSpace{}%
\AgdaOperator{\AgdaDatatype{≡}}\AgdaSpace{}%
\AgdaBound{γy}\AgdaSpace{}%
\AgdaSymbol{→}\AgdaSpace{}%
\AgdaBound{T}\AgdaSpace{}%
\AgdaOperator{\AgdaField{⟨}}\AgdaSpace{}%
\AgdaBound{y}\AgdaSpace{}%
\AgdaOperator{\AgdaField{,}}\AgdaSpace{}%
\AgdaBound{γy}\AgdaSpace{}%
\AgdaOperator{\AgdaField{⟩}}\AgdaSpace{}%
\AgdaSymbol{→}\AgdaSpace{}%
\AgdaBound{S}\AgdaSpace{}%
\AgdaOperator{\AgdaField{⟨}}\AgdaSpace{}%
\AgdaBound{y}\AgdaSpace{}%
\AgdaOperator{\AgdaField{,}}\AgdaSpace{}%
\AgdaBound{γy}\AgdaSpace{}%
\AgdaOperator{\AgdaField{⟩}}\<%
\\
\>[2]\AgdaSymbol{\AgdaUnderscore{}}%
\>[955I]\AgdaSymbol{=}\AgdaSpace{}%
\AgdaSymbol{λ}\AgdaSpace{}%
\AgdaBound{f}\AgdaSpace{}%
\AgdaBound{ρ}\AgdaSpace{}%
\AgdaBound{e}\AgdaSpace{}%
\AgdaBound{t}\AgdaSpace{}%
\AgdaSymbol{→}\<%
\end{code}
\begin{code}[inline*]%
\>[.][@{}l@{}]\<[955I]%
\>[4]\AgdaBound{f}\AgdaSpace{}%
\AgdaOperator{\AgdaField{\$⟨}}\AgdaSpace{}%
\AgdaBound{ρ}\AgdaSpace{}%
\AgdaOperator{\AgdaField{,}}\AgdaSpace{}%
\AgdaBound{e}\AgdaSpace{}%
\AgdaOperator{\AgdaField{⟩}}\AgdaSpace{}%
\AgdaBound{t}\<%
\end{code}
$=$ \AFi{fun} $f$ $\rho$ $e$ $t$.
The restriction maps can then be implemented by using composition in the base category. 
\adapted{Just like for \AFi{ty-hom} in the definition of \AR{Ty}, we use fording for the field \AFi{fun} to make the implementation of function types considerably easier.}

A modality will be interpreted as a dependent right adjunction (DRA) defined by Birkedal et al.~\cite{birkedal20-modal}.%
\footnote{In fact a DRA requires the \AFi{lock} operation to be a functor between context categories that also acts on semantic substitutions.
  We follow this in the Sikkel implementation, but simplified the presentation in the paper.}
\begin{code}%
\>[0]\AgdaKeyword{record}\AgdaSpace{}%
\AgdaRecord{Modality}\AgdaSpace{}%
\AgdaSymbol{(}\AgdaBound{C}\AgdaSpace{}%
\AgdaBound{D}\AgdaSpace{}%
\AgdaSymbol{:}\AgdaSpace{}%
\AgdaRecord{BaseCategory}\AgdaSymbol{)}\AgdaSpace{}%
\AgdaSymbol{:}\AgdaSpace{}%
\AgdaPrimitive{Set₁}\AgdaSpace{}%
\AgdaKeyword{where}\<%
\\
\>[0][@{}l@{\AgdaIndent{0}}]%
\>[2]\AgdaKeyword{field}%
\>[976I]\AgdaField{lock}\AgdaSpace{}%
\AgdaSymbol{:}\AgdaSpace{}%
\AgdaRecord{Ctx}\AgdaSpace{}%
\AgdaBound{D}\AgdaSpace{}%
\AgdaSymbol{→}\AgdaSpace{}%
\AgdaRecord{Ctx}\AgdaSpace{}%
\AgdaBound{C}\<%
\\
\>[.][@{}l@{}]\<[976I]%
\>[8]\AgdaField{mod}\AgdaSpace{}%
\AgdaSymbol{:}\AgdaSpace{}%
\AgdaRecord{Ty}\AgdaSpace{}%
\AgdaSymbol{(}\AgdaField{lock}\AgdaSpace{}%
\AgdaBound{Γ}\AgdaSymbol{)}\AgdaSpace{}%
\AgdaSymbol{→}\AgdaSpace{}%
\AgdaRecord{Ty}\AgdaSpace{}%
\AgdaBound{Γ}\<%
\\
\>[8]\AgdaField{mod-intro}\AgdaSpace{}%
\AgdaSymbol{:}\AgdaSpace{}%
\AgdaRecord{Tm}\AgdaSpace{}%
\AgdaSymbol{(}\AgdaField{lock}\AgdaSpace{}%
\AgdaBound{Γ}\AgdaSymbol{)}\AgdaSpace{}%
\AgdaBound{T}\AgdaSpace{}%
\AgdaSymbol{→}\AgdaSpace{}%
\AgdaRecord{Tm}\AgdaSpace{}%
\AgdaBound{Γ}\AgdaSpace{}%
\AgdaSymbol{(}\AgdaField{mod}\AgdaSpace{}%
\AgdaBound{T}\AgdaSymbol{)}\<%
\\
\>[8]\AgdaField{mod-elim}\AgdaSpace{}%
\AgdaSymbol{:}\AgdaSpace{}%
\AgdaRecord{Tm}\AgdaSpace{}%
\AgdaBound{Γ}\AgdaSpace{}%
\AgdaSymbol{(}\AgdaField{mod}\AgdaSpace{}%
\AgdaBound{T}\AgdaSymbol{)}\AgdaSpace{}%
\AgdaSymbol{→}\AgdaSpace{}%
\AgdaRecord{Tm}\AgdaSpace{}%
\AgdaSymbol{(}\AgdaField{lock}\AgdaSpace{}%
\AgdaBound{Γ}\AgdaSymbol{)}\AgdaSpace{}%
\AgdaBound{T}\<%
\end{code}
\begin{code}[hide]%
\>[0]\AgdaKeyword{open}\AgdaSpace{}%
\AgdaModule{Modality}\AgdaSpace{}%
\AgdaKeyword{renaming}\AgdaSpace{}%
\AgdaSymbol{(}\AgdaField{mod}\AgdaSpace{}%
\AgdaSymbol{to}\AgdaSpace{}%
\AgdaField{⟨\AgdaUnderscore{}∣\AgdaUnderscore{}⟩}\AgdaSymbol{)}\<%
\end{code}
For readability we use the notation
\begin{code}[hide]%
\>[0]\AgdaFunction{\AgdaUnderscore{}}\AgdaSpace{}%
\AgdaSymbol{:}\AgdaSpace{}%
\AgdaSymbol{∀}\AgdaSpace{}%
\AgdaSymbol{\{}\AgdaBound{C}\AgdaSpace{}%
\AgdaBound{D}\AgdaSymbol{\}}\AgdaSpace{}%
\AgdaSymbol{(}\AgdaBound{μ}\AgdaSpace{}%
\AgdaSymbol{:}\AgdaSpace{}%
\AgdaRecord{Modality}\AgdaSpace{}%
\AgdaBound{C}\AgdaSpace{}%
\AgdaBound{D}\AgdaSymbol{)}\AgdaSpace{}%
\AgdaSymbol{→}\AgdaSpace{}%
\AgdaSymbol{\{}\AgdaBound{Γ}\AgdaSpace{}%
\AgdaSymbol{:}\AgdaSpace{}%
\AgdaRecord{Ctx}\AgdaSpace{}%
\AgdaBound{D}\AgdaSymbol{\}}\AgdaSpace{}%
\AgdaSymbol{→}\AgdaSpace{}%
\AgdaRecord{Ty}\AgdaSpace{}%
\AgdaSymbol{(}\AgdaField{lock}\AgdaSpace{}%
\AgdaBound{μ}\AgdaSpace{}%
\AgdaBound{Γ}\AgdaSymbol{)}\AgdaSpace{}%
\AgdaSymbol{→}\AgdaSpace{}%
\AgdaRecord{Ty}\AgdaSpace{}%
\AgdaBound{Γ}\<%
\\
\>[0]\AgdaSymbol{\AgdaUnderscore{}}%
\>[1062I]\AgdaSymbol{=}\AgdaSpace{}%
\AgdaSymbol{λ}\AgdaSpace{}%
\AgdaBound{μ}\AgdaSpace{}%
\AgdaBound{T}\AgdaSpace{}%
\AgdaSymbol{→}\<%
\end{code}
\begin{code}[inline*]%
\>[.][@{}l@{}]\<[1062I]%
\>[2]\AgdaOperator{\AgdaField{⟨}}\AgdaSpace{}%
\AgdaBound{μ}\AgdaSpace{}%
\AgdaOperator{\AgdaField{∣}}\AgdaSpace{}%
\AgdaBound{T}\AgdaSpace{}%
\AgdaOperator{\AgdaField{⟩}}\<%
\end{code}
$=$ \AFi{mod} $\mu$ $T$.
As we can see, a DRA specifies the semantic counterparts of the context and type formers associated with modalities in MSTT, \adapted{as well as the interpretation of the modal term constructor}.
The eliminator \AFi{mod-elim} is the inverse of \AFi{mod-intro} and is hence different from the rule \inlinerulename{Tm-ModElim} in Figure~\ref{fig:typing-rules}.
This difference will be handled by the type checker when interpreting MSTT syntax in the model.
There is a unit DRA \AF{𝟙} for every base category (whose constituent operations are just the identity) as well as composition of DRAs (by composing the constituent operations).
Furthermore, Sikkel has an equivalence relation \AR{≅ᵐ} for DRAs according to which composition is associative and the unit DRA is a unit.
We will not discuss the definition of \AR{≅ᵐ}, but it is sufficient to know that if
\begin{code}[hide]%
\>[0]\AgdaKeyword{record}\AgdaSpace{}%
\AgdaOperator{\AgdaRecord{\AgdaUnderscore{}≅ᵐ\AgdaUnderscore{}}}\AgdaSpace{}%
\AgdaSymbol{\{}\AgdaBound{C}\AgdaSpace{}%
\AgdaBound{D}\AgdaSymbol{\}}\AgdaSpace{}%
\AgdaSymbol{(}\AgdaBound{μ}\AgdaSpace{}%
\AgdaBound{ρ}\AgdaSpace{}%
\AgdaSymbol{:}\AgdaSpace{}%
\AgdaRecord{Modality}\AgdaSpace{}%
\AgdaBound{C}\AgdaSpace{}%
\AgdaBound{D}\AgdaSymbol{)}\AgdaSpace{}%
\AgdaSymbol{:}\AgdaSpace{}%
\AgdaPrimitive{Set₁}\AgdaSpace{}%
\AgdaKeyword{where}\<%
\\
\>[0][@{}l@{\AgdaIndent{0}}]%
\>[2]\AgdaKeyword{field}\<%
\\
\>[2][@{}l@{\AgdaIndent{0}}]%
\>[4]\AgdaField{eq-mod-closed}\AgdaSpace{}%
\AgdaSymbol{:}\AgdaSpace{}%
\AgdaSymbol{(}\AgdaBound{A}\AgdaSpace{}%
\AgdaSymbol{:}\AgdaSpace{}%
\AgdaFunction{ClosedTy}\AgdaSpace{}%
\AgdaBound{C}\AgdaSymbol{)}\AgdaSpace{}%
\AgdaSymbol{→}\AgdaSpace{}%
\AgdaSymbol{∀}\AgdaSpace{}%
\AgdaSymbol{\{}\AgdaBound{Γ}\AgdaSymbol{\}}\AgdaSpace{}%
\AgdaSymbol{→}\AgdaSpace{}%
\AgdaOperator{\AgdaField{⟨}}\AgdaSpace{}%
\AgdaBound{μ}\AgdaSpace{}%
\AgdaOperator{\AgdaField{∣}}\AgdaSpace{}%
\AgdaBound{A}\AgdaSpace{}%
\AgdaSymbol{\{}\AgdaField{lock}\AgdaSpace{}%
\AgdaBound{μ}\AgdaSpace{}%
\AgdaBound{Γ}\AgdaSymbol{\}}\AgdaSpace{}%
\AgdaOperator{\AgdaField{⟩}}\AgdaSpace{}%
\AgdaOperator{\AgdaRecord{≅ᵗʸ}}\AgdaSpace{}%
\AgdaOperator{\AgdaField{⟨}}\AgdaSpace{}%
\AgdaBound{ρ}\AgdaSpace{}%
\AgdaOperator{\AgdaField{∣}}\AgdaSpace{}%
\AgdaBound{A}\AgdaSpace{}%
\AgdaOperator{\AgdaField{⟩}}\<%
\\
\\[\AgdaEmptyExtraSkip]%
\>[0]\AgdaKeyword{module}\AgdaSpace{}%
\AgdaModule{\AgdaUnderscore{}}\AgdaSpace{}%
\AgdaSymbol{\{}\AgdaBound{C}\AgdaSpace{}%
\AgdaBound{D}\AgdaSymbol{\}}\AgdaSpace{}%
\AgdaSymbol{(}\AgdaBound{μ}\AgdaSpace{}%
\AgdaBound{ρ}\AgdaSpace{}%
\AgdaSymbol{:}\AgdaSpace{}%
\AgdaRecord{Modality}\AgdaSpace{}%
\AgdaBound{C}\AgdaSpace{}%
\AgdaBound{D}\AgdaSymbol{)}\AgdaSpace{}%
\AgdaSymbol{(}\AgdaBound{A}\AgdaSpace{}%
\AgdaSymbol{:}\AgdaSpace{}%
\AgdaFunction{ClosedTy}\AgdaSpace{}%
\AgdaBound{C}\AgdaSymbol{)}\AgdaSpace{}%
\AgdaKeyword{where}\<%
\\
\>[0][@{}l@{\AgdaIndent{0}}]%
\>[2]\AgdaFunction{\AgdaUnderscore{}}\AgdaSpace{}%
\AgdaSymbol{:}\AgdaSpace{}%
\AgdaPrimitive{Set₁}\<%
\\
\>[2]\AgdaSymbol{\AgdaUnderscore{}}%
\>[1122I]\AgdaSymbol{=}\<%
\end{code}
\begin{code}[inline]%
\>[.][@{}l@{}]\<[1122I]%
\>[4]\AgdaBound{μ}\AgdaSpace{}%
\AgdaOperator{\AgdaRecord{≅ᵐ}}\AgdaSpace{}%
\AgdaBound{ρ}\<%
\end{code},
then it follows that
\begin{code}[hide]%
\>[2]\AgdaFunction{\AgdaUnderscore{}}\AgdaSpace{}%
\AgdaSymbol{:}\AgdaSpace{}%
\AgdaRecord{Ctx}\AgdaSpace{}%
\AgdaBound{D}\AgdaSpace{}%
\AgdaSymbol{→}\AgdaSpace{}%
\AgdaPrimitive{Set}\<%
\\
\>[2]\AgdaSymbol{\AgdaUnderscore{}}%
\>[1130I]\AgdaSymbol{=}\AgdaSpace{}%
\AgdaSymbol{λ}\AgdaSpace{}%
\AgdaBound{Γ}\AgdaSpace{}%
\AgdaSymbol{→}\<%
\end{code}
\begin{code}[inline*]%
\>[.][@{}l@{}]\<[1130I]%
\>[4]\AgdaOperator{\AgdaField{⟨}}\AgdaSpace{}%
\AgdaBound{μ}\AgdaSpace{}%
\AgdaOperator{\AgdaField{∣}}\AgdaSpace{}%
\AgdaBound{A}\AgdaSpace{}%
\AgdaOperator{\AgdaField{⟩}}\AgdaSpace{}%
\AgdaOperator{\AgdaRecord{≅ᵗʸ}}\AgdaSpace{}%
\AgdaOperator{\AgdaField{⟨}}\AgdaSpace{}%
\AgdaBound{ρ}\AgdaSpace{}%
\AgdaOperator{\AgdaField{∣}}\AgdaSpace{}%
\AgdaBound{A}\AgdaSpace{}%
\AgdaOperator{\AgdaField{⟩}}\<%
\end{code}
for every
\begin{code}[hide]%
\>[0]\AgdaKeyword{module}\AgdaSpace{}%
\AgdaModule{\AgdaUnderscore{}}\AgdaSpace{}%
\AgdaSymbol{(}\<%
\end{code}
\begin{code}[inline]%
\>[0][@{}l@{\AgdaIndent{1}}]%
\>[2]\AgdaBound{A}\AgdaSpace{}%
\AgdaSymbol{:}\AgdaSpace{}%
\AgdaFunction{ClosedTy}\AgdaSpace{}%
\AgdaGeneralizable{C}\<%
\end{code}
\begin{code}[hide]%
\>[2]\AgdaSymbol{)}\AgdaSpace{}%
\AgdaKeyword{where}\<%
\end{code}.
Finally, there is a type family \AR{TwoCell}, indexed by two DRAs, whose values are essentially natural transformations between the \AFi{lock} context functors of the DRAs.
\adapted{We refer to the Agda code for details.}


%% file: content-lagda/sound-typechecker.tex
\section{A Sound Type Checker}
\label{sec:sound-typechecker}

\begin{code}[hide]%
\>[0]\AgdaKeyword{module}\AgdaSpace{}%
\AgdaModule{sound-typechecker}\AgdaSpace{}%
\AgdaKeyword{where}\<%
\\
\\[\AgdaEmptyExtraSkip]%
\>[0]\AgdaKeyword{open}\AgdaSpace{}%
\AgdaKeyword{import}\AgdaSpace{}%
\AgdaModule{Data.Empty}\<%
\\
\>[0]\AgdaKeyword{open}\AgdaSpace{}%
\AgdaKeyword{import}\AgdaSpace{}%
\AgdaModule{Data.Nat}\<%
\\
\>[0]\AgdaKeyword{open}\AgdaSpace{}%
\AgdaKeyword{import}\AgdaSpace{}%
\AgdaModule{Data.String}\AgdaSpace{}%
\AgdaKeyword{using}\AgdaSpace{}%
\AgdaSymbol{(}\AgdaPostulate{String}\AgdaSymbol{)}\<%
\\
\>[0]\AgdaKeyword{open}\AgdaSpace{}%
\AgdaKeyword{import}\AgdaSpace{}%
\AgdaModule{Data.Unit}\<%
\\
\>[0]\AgdaKeyword{open}\AgdaSpace{}%
\AgdaKeyword{import}\AgdaSpace{}%
\AgdaModule{Level}\AgdaSpace{}%
\AgdaKeyword{using}\AgdaSpace{}%
\AgdaSymbol{(}\AgdaPostulate{Level}\AgdaSymbol{)}\<%
\\
\>[0]\AgdaKeyword{open}\AgdaSpace{}%
\AgdaKeyword{import}\AgdaSpace{}%
\AgdaModule{Relation.Binary.PropositionalEquality}\<%
\\
\>[0]\AgdaKeyword{open}\AgdaSpace{}%
\AgdaKeyword{import}\AgdaSpace{}%
\AgdaModule{Function}\AgdaSpace{}%
\AgdaKeyword{using}\AgdaSpace{}%
\AgdaSymbol{(}\AgdaOperator{\AgdaFunction{\AgdaUnderscore{}∋\AgdaUnderscore{}}}\AgdaSymbol{)}\<%
\\
\\[\AgdaEmptyExtraSkip]%
\>[0]\AgdaKeyword{open}\AgdaSpace{}%
\AgdaKeyword{import}\AgdaSpace{}%
\AgdaModule{Model.BaseCategory}\AgdaSpace{}%
\AgdaSymbol{as}\AgdaSpace{}%
\AgdaModule{M}\<%
\\
\>[0]\AgdaKeyword{open}\AgdaSpace{}%
\AgdaKeyword{import}\AgdaSpace{}%
\AgdaModule{Model.Modality}\AgdaSpace{}%
\AgdaSymbol{as}\AgdaSpace{}%
\AgdaModule{M}\AgdaSpace{}%
\AgdaKeyword{using}\AgdaSpace{}%
\AgdaSymbol{(}\AgdaRecord{Modality}\AgdaSymbol{;}\AgdaSpace{}%
\AgdaOperator{\AgdaRecord{\AgdaUnderscore{}≅ᵐ\AgdaUnderscore{}}}\AgdaSymbol{;}\AgdaSpace{}%
\AgdaFunction{lock}\AgdaSymbol{;}\AgdaSpace{}%
\AgdaRecord{TwoCell}\AgdaSymbol{;}\AgdaSpace{}%
\AgdaField{mod-intro}\AgdaSymbol{)}\<%
\\
\>[0]\AgdaKeyword{open}\AgdaSpace{}%
\AgdaKeyword{import}\AgdaSpace{}%
\AgdaModule{Model.CwF-Structure}\AgdaSpace{}%
\AgdaSymbol{as}\AgdaSpace{}%
\AgdaModule{M}\AgdaSpace{}%
\AgdaKeyword{using}\AgdaSpace{}%
\AgdaSymbol{(}\AgdaRecord{Ctx}\AgdaSymbol{;}\AgdaSpace{}%
\AgdaRecord{Ty}\AgdaSymbol{;}\AgdaSpace{}%
\AgdaRecord{Tm}\AgdaSymbol{;}\AgdaSpace{}%
\AgdaFunction{ClosedTy}\AgdaSymbol{;}\AgdaSpace{}%
\AgdaOperator{\AgdaFunction{\AgdaUnderscore{},,\AgdaUnderscore{}}}\AgdaSymbol{;}\AgdaSpace{}%
\AgdaOperator{\AgdaRecord{\AgdaUnderscore{}≅ᵗʸ\AgdaUnderscore{}}}\AgdaSymbol{;}\AgdaSpace{}%
\AgdaOperator{\AgdaFunction{ι[\AgdaUnderscore{}]\AgdaUnderscore{}}}\AgdaSymbol{;}\AgdaSpace{}%
\AgdaField{\AgdaUnderscore{}⟨\AgdaUnderscore{},\AgdaUnderscore{}⟩'}\AgdaSymbol{)}\<%
\\
\>[0]\AgdaKeyword{import}\AgdaSpace{}%
\AgdaModule{Model.Type.Function}\AgdaSpace{}%
\AgdaSymbol{as}\AgdaSpace{}%
\AgdaModule{M}\<%
\\
\>[0]\AgdaKeyword{import}\AgdaSpace{}%
\AgdaModule{Model.Type.Discrete}\AgdaSpace{}%
\AgdaSymbol{as}\AgdaSpace{}%
\AgdaModule{M}\AgdaSpace{}%
\AgdaKeyword{renaming}\AgdaSpace{}%
\AgdaSymbol{(}\AgdaFunction{Nat'}\AgdaSpace{}%
\AgdaSymbol{to}\AgdaSpace{}%
\AgdaFunction{Nat}\AgdaSymbol{;}\AgdaSpace{}%
\AgdaFunction{Bool'}\AgdaSpace{}%
\AgdaSymbol{to}\AgdaSpace{}%
\AgdaFunction{Bool}\AgdaSymbol{)}\<%
\\
\>[0]\AgdaKeyword{import}\AgdaSpace{}%
\AgdaModule{Applications.GuardedRecursion.Model.Modalities}\AgdaSpace{}%
\AgdaSymbol{as}\AgdaSpace{}%
\AgdaModule{M}\<%
\\
\>[0]\AgdaKeyword{import}\AgdaSpace{}%
\AgdaModule{Applications.GuardedRecursion.Model.Streams.Guarded}\AgdaSpace{}%
\AgdaSymbol{as}\AgdaSpace{}%
\AgdaModule{M}\<%
\\
\\[\AgdaEmptyExtraSkip]%
\>[0]\AgdaKeyword{import}\AgdaSpace{}%
\AgdaModule{MSTT.TCMonad}\AgdaSpace{}%
\AgdaSymbol{as}\AgdaSpace{}%
\AgdaModule{TCMonad}\<%
\\
\>[0]\AgdaKeyword{open}\AgdaSpace{}%
\AgdaKeyword{import}\AgdaSpace{}%
\AgdaModule{MSTT.Parameter.ModeTheory}\<%
\\
\>[0]\AgdaKeyword{open}\AgdaSpace{}%
\AgdaKeyword{import}\AgdaSpace{}%
\AgdaModule{MSTT.Parameter.TypeExtension}\<%
\\
\>[0]\AgdaKeyword{open}\AgdaSpace{}%
\AgdaKeyword{import}\AgdaSpace{}%
\AgdaModule{MSTT.Parameter.TermExtension}\<%
\\
\\[\AgdaEmptyExtraSkip]%
\>[0]\AgdaKeyword{open}\AgdaSpace{}%
\AgdaKeyword{import}\AgdaSpace{}%
\AgdaModule{guarded-recursion}\AgdaSpace{}%
\AgdaKeyword{using}\AgdaSpace{}%
\AgdaSymbol{(}\AgdaFunction{g-nats}\AgdaSymbol{;}\AgdaSpace{}%
\AgdaFunction{nats}\AgdaSymbol{)}\<%
\\
\\[\AgdaEmptyExtraSkip]%
\>[0]\AgdaKeyword{private}\AgdaSpace{}%
\AgdaKeyword{variable}\<%
\\
\>[0][@{}l@{\AgdaIndent{0}}]%
\>[2]\AgdaGeneralizable{ℓ}\AgdaSpace{}%
\AgdaSymbol{:}\AgdaSpace{}%
\AgdaPostulate{Level}\<%
\\
\\[\AgdaEmptyExtraSkip]%
\>[0]\AgdaKeyword{module}\AgdaSpace{}%
\AgdaModule{DefineTypeChecker}\AgdaSpace{}%
\AgdaSymbol{(}\AgdaBound{mt}\AgdaSpace{}%
\AgdaSymbol{:}\AgdaSpace{}%
\AgdaRecord{ModeTheory}\AgdaSymbol{)}\AgdaSpace{}%
\AgdaKeyword{where}\<%
\\
\>[0][@{}l@{\AgdaIndent{0}}]%
\>[2]\AgdaFunction{ty-ext}\AgdaSpace{}%
\AgdaSymbol{:}\AgdaSpace{}%
\AgdaRecord{TyExt}\AgdaSpace{}%
\AgdaBound{mt}\<%
\\
\>[2]\AgdaField{TyExt.TyExtCode}\AgdaSpace{}%
\AgdaFunction{ty-ext}\AgdaSpace{}%
\AgdaSymbol{\AgdaUnderscore{}}\AgdaSpace{}%
\AgdaSymbol{\AgdaUnderscore{}}\AgdaSpace{}%
\AgdaSymbol{=}\AgdaSpace{}%
\AgdaDatatype{⊥}\<%
\\
\>[2]\AgdaOperator{\AgdaField{TyExt.\AgdaUnderscore{}≟code\AgdaUnderscore{}}}\AgdaSpace{}%
\AgdaFunction{ty-ext}\AgdaSpace{}%
\AgdaSymbol{()}\AgdaSpace{}%
\AgdaSymbol{()}\<%
\\
\>[2]\AgdaField{TyExt.show-code}\AgdaSpace{}%
\AgdaFunction{ty-ext}\AgdaSpace{}%
\AgdaSymbol{()}\<%
\\
\>[2]\AgdaField{TyExt.interpret-code}\AgdaSpace{}%
\AgdaFunction{ty-ext}\AgdaSpace{}%
\AgdaSymbol{()}\<%
\\
\>[2]\AgdaField{TyExt.interpret-code-natural}\AgdaSpace{}%
\AgdaFunction{ty-ext}\AgdaSpace{}%
\AgdaSymbol{()}\<%
\\
\>[2]\AgdaField{TyExt.interpret-code-cong}\AgdaSpace{}%
\AgdaFunction{ty-ext}\AgdaSpace{}%
\AgdaSymbol{()}\<%
\\
\\[\AgdaEmptyExtraSkip]%
\>[2]\AgdaFunction{tm-ext}\AgdaSpace{}%
\AgdaSymbol{:}\AgdaSpace{}%
\AgdaRecord{TmExt}\AgdaSpace{}%
\AgdaBound{mt}\AgdaSpace{}%
\AgdaFunction{ty-ext}\<%
\\
\>[2]\AgdaField{TmExt.TmExtCode}\AgdaSpace{}%
\AgdaFunction{tm-ext}\AgdaSpace{}%
\AgdaSymbol{\AgdaUnderscore{}}\AgdaSpace{}%
\AgdaSymbol{\AgdaUnderscore{}}\AgdaSpace{}%
\AgdaSymbol{=}\AgdaSpace{}%
\AgdaDatatype{⊥}\<%
\\
\>[2]\AgdaField{TmExt.infer-interpret-code}\AgdaSpace{}%
\AgdaFunction{tm-ext}\AgdaSpace{}%
\AgdaSymbol{()}\<%
\\
\\[\AgdaEmptyExtraSkip]%
\>[2]\AgdaKeyword{open}\AgdaSpace{}%
\AgdaModule{ModeTheory}\AgdaSpace{}%
\AgdaBound{mt}\AgdaSpace{}%
\AgdaKeyword{hiding}\AgdaSpace{}%
\AgdaSymbol{(}\AgdaOperator{\AgdaField{⟦\AgdaUnderscore{}⟧mode}}\AgdaSymbol{;}\AgdaSpace{}%
\AgdaOperator{\AgdaField{⟦\AgdaUnderscore{}⟧modality}}\AgdaSymbol{;}\AgdaSpace{}%
\AgdaOperator{\AgdaField{\AgdaUnderscore{}≟mode\AgdaUnderscore{}}}\AgdaSymbol{;}\AgdaSpace{}%
\AgdaOperator{\AgdaField{\AgdaUnderscore{}≃ᵐ?\AgdaUnderscore{}}}\AgdaSymbol{)}\<%
\\
\>[2]\AgdaKeyword{open}\AgdaSpace{}%
\AgdaKeyword{import}\AgdaSpace{}%
\AgdaModule{MSTT.Syntax.Type}\AgdaSpace{}%
\AgdaBound{mt}\AgdaSpace{}%
\AgdaFunction{ty-ext}\AgdaSpace{}%
\AgdaKeyword{hiding}\AgdaSpace{}%
\AgdaSymbol{(}\AgdaFunction{is-modal-ty}\AgdaSymbol{;}\AgdaSpace{}%
\AgdaDatatype{IsModalTyExpr}\AgdaSymbol{)}\AgdaSpace{}%
\AgdaKeyword{renaming}\AgdaSpace{}%
\AgdaSymbol{(}\AgdaInductiveConstructor{Nat'}\AgdaSpace{}%
\AgdaSymbol{to}\AgdaSpace{}%
\AgdaInductiveConstructor{Nat}\AgdaSymbol{;}\AgdaSpace{}%
\AgdaInductiveConstructor{Bool'}\AgdaSpace{}%
\AgdaSymbol{to}\AgdaSpace{}%
\AgdaInductiveConstructor{Bool}\AgdaSymbol{)}\<%
\\
\>[2]\AgdaKeyword{open}\AgdaSpace{}%
\AgdaKeyword{import}\AgdaSpace{}%
\AgdaModule{MSTT.Syntax.Context}\AgdaSpace{}%
\AgdaBound{mt}\AgdaSpace{}%
\AgdaFunction{ty-ext}\<%
\\
\>[2]\AgdaKeyword{open}\AgdaSpace{}%
\AgdaKeyword{import}\AgdaSpace{}%
\AgdaModule{MSTT.Syntax.Term}\AgdaSpace{}%
\AgdaBound{mt}\AgdaSpace{}%
\AgdaFunction{ty-ext}\AgdaSpace{}%
\AgdaFunction{tm-ext}\<%
\\
\>[2]\AgdaKeyword{private}\AgdaSpace{}%
\AgdaKeyword{variable}\<%
\\
\>[2][@{}l@{\AgdaIndent{0}}]%
\>[4]\AgdaGeneralizable{m}\AgdaSpace{}%
\AgdaGeneralizable{n}\AgdaSpace{}%
\AgdaSymbol{:}\AgdaSpace{}%
\AgdaField{ModeExpr}\<%
\end{code}

The typing relation in Figure~\ref{fig:typing-rules} is not formalized as a relation in Agda,
but as an algorithm that checks whether a term is well-typed.
In fact, as pointed out before, the terms of MSTT provide enough information for types to be inferred rather than checked so there is a function \AF{infer-type} that returns a term's type, provided that it is well-typed.

Moreover, Sikkel's type checker is sound by construction, so on top of returning a well-typed term's type, it also returns its denotation in the presheaf model.
As such, the type checker bridges Sikkel's syntactic and semantic layers.
\adapted{We will} use the prefix ``\ASy{M.}'' (for model) to disambiguate names that are defined in both the semantic and syntactic layer.

\begin{figure}[htb]
\begin{minipage}{.48\textwidth}
\begin{subfigure}{\linewidth}
\begin{AgdaMultiCode}
\begin{code}[hide]%
\>[2]\AgdaKeyword{module}\AgdaSpace{}%
\AgdaModule{InterpretModeTheory}\AgdaSpace{}%
\AgdaKeyword{where}\<%
\\
\>[2][@{}l@{\AgdaIndent{0}}]%
\>[4]\AgdaKeyword{record}\AgdaSpace{}%
\AgdaRecord{NewFields}\AgdaSpace{}%
\AgdaSymbol{:}\AgdaSpace{}%
\AgdaPrimitive{Set₁}\AgdaSpace{}%
\AgdaKeyword{where}\<%
\\
\>[4][@{}l@{\AgdaIndent{0}}]%
\>[6]\AgdaKeyword{field}\<%
\end{code}
\begin{code}%
\>[6][@{}l@{\AgdaIndent{1}}]%
\>[8]\AgdaOperator{\AgdaField{⟦\AgdaUnderscore{}⟧mode}}\AgdaSpace{}%
\AgdaSymbol{:}\AgdaSpace{}%
\AgdaField{ModeExpr}\AgdaSpace{}%
\AgdaSymbol{→}\AgdaSpace{}%
\AgdaRecord{BaseCategory}\<%
\\
\>[8]\AgdaOperator{\AgdaField{⟦\AgdaUnderscore{}⟧modality}}\AgdaSpace{}%
\AgdaSymbol{:}%
\>[165I]\AgdaField{ModalityExpr}\AgdaSpace{}%
\AgdaGeneralizable{m}\AgdaSpace{}%
\AgdaGeneralizable{n}\AgdaSpace{}%
\AgdaSymbol{→}\<%
\\
\>[.][@{}l@{}]\<[165I]%
\>[22]\AgdaRecord{Modality}\AgdaSpace{}%
\AgdaOperator{\AgdaField{⟦}}\AgdaSpace{}%
\AgdaGeneralizable{m}\AgdaSpace{}%
\AgdaOperator{\AgdaField{⟧mode}}\AgdaSpace{}%
\AgdaOperator{\AgdaField{⟦}}\AgdaSpace{}%
\AgdaGeneralizable{n}\AgdaSpace{}%
\AgdaOperator{\AgdaField{⟧mode}}\<%
\end{code}
\begin{code}[hide]%
\>[2]\AgdaKeyword{open}\AgdaSpace{}%
\AgdaModule{ModeTheory}\AgdaSpace{}%
\AgdaBound{mt}\AgdaSpace{}%
\AgdaKeyword{using}\AgdaSpace{}%
\AgdaSymbol{(}\AgdaOperator{\AgdaField{⟦\AgdaUnderscore{}⟧mode}}\AgdaSymbol{;}\AgdaSpace{}%
\AgdaOperator{\AgdaField{⟦\AgdaUnderscore{}⟧modality}}\AgdaSymbol{)}\<%
\end{code}
\end{AgdaMultiCode}
\caption{Modes and modalities (these are additional fields to \AR{ModeTheory}, Fig.~\ref{fig:modetheorycomplete}).}
\label{fig:interp-mode-theory}
\end{subfigure} \\
\begin{subfigure}{\linewidth}
\begin{AgdaMultiCode}
\begin{code}[hide]%
\>[2]\AgdaSymbol{\{-\#}\AgdaSpace{}%
\AgdaKeyword{NON\AgdaUnderscore{}COVERING}\AgdaSpace{}%
\AgdaSymbol{\#-\}}\<%
\end{code}
\begin{code}%
\>[2]\AgdaOperator{\AgdaFunction{⟦\AgdaUnderscore{}⟧ty}}\AgdaSpace{}%
\AgdaSymbol{:}\AgdaSpace{}%
\AgdaDatatype{TyExpr}\AgdaSpace{}%
\AgdaGeneralizable{m}\AgdaSpace{}%
\AgdaSymbol{→}\AgdaSpace{}%
\AgdaFunction{ClosedTy}\AgdaSpace{}%
\AgdaOperator{\AgdaField{⟦}}\AgdaSpace{}%
\AgdaGeneralizable{m}\AgdaSpace{}%
\AgdaOperator{\AgdaField{⟧mode}}\<%
\\
\>[2]\AgdaOperator{\AgdaFunction{⟦}}\AgdaSpace{}%
\AgdaBound{T}\AgdaSpace{}%
\AgdaOperator{\AgdaInductiveConstructor{⇛}}\AgdaSpace{}%
\AgdaBound{S}\AgdaSpace{}%
\AgdaOperator{\AgdaFunction{⟧ty}}\AgdaSpace{}%
\AgdaSymbol{=}\AgdaSpace{}%
\AgdaOperator{\AgdaFunction{⟦}}\AgdaSpace{}%
\AgdaBound{T}\AgdaSpace{}%
\AgdaOperator{\AgdaFunction{⟧ty}}\AgdaSpace{}%
\AgdaOperator{\AgdaFunction{M.⇛}}\AgdaSpace{}%
\AgdaOperator{\AgdaFunction{⟦}}\AgdaSpace{}%
\AgdaBound{S}\AgdaSpace{}%
\AgdaOperator{\AgdaFunction{⟧ty}}\<%
\\
\>[2]\AgdaOperator{\AgdaFunction{⟦}}\AgdaSpace{}%
\AgdaOperator{\AgdaInductiveConstructor{⟨}}\AgdaSpace{}%
\AgdaBound{μ}\AgdaSpace{}%
\AgdaOperator{\AgdaInductiveConstructor{∣}}\AgdaSpace{}%
\AgdaBound{T}\AgdaSpace{}%
\AgdaOperator{\AgdaInductiveConstructor{⟩}}\AgdaSpace{}%
\AgdaOperator{\AgdaFunction{⟧ty}}\AgdaSpace{}%
\AgdaSymbol{=}\AgdaSpace{}%
\AgdaOperator{\AgdaField{M.⟨}}\AgdaSpace{}%
\AgdaOperator{\AgdaField{⟦}}\AgdaSpace{}%
\AgdaBound{μ}\AgdaSpace{}%
\AgdaOperator{\AgdaField{⟧modality}}\AgdaSpace{}%
\AgdaOperator{\AgdaField{∣}}\AgdaSpace{}%
\AgdaOperator{\AgdaFunction{⟦}}\AgdaSpace{}%
\AgdaBound{T}\AgdaSpace{}%
\AgdaOperator{\AgdaFunction{⟧ty}}\AgdaSpace{}%
\AgdaOperator{\AgdaField{⟩}}\<%
\end{code}
\end{AgdaMultiCode}
\caption{Types (similar clauses for other types are omitted).}
\label{fig:interp-ty}
\end{subfigure}
\end{minipage}
\hspace{\fill}
\begin{minipage}{.48\textwidth}
\begin{subfigure}{\linewidth}
\begin{AgdaMultiCode}
\begin{code}%
\>[2]\AgdaOperator{\AgdaFunction{⟦\AgdaUnderscore{}⟧ctx}}\AgdaSpace{}%
\AgdaSymbol{:}\AgdaSpace{}%
\AgdaDatatype{CtxExpr}\AgdaSpace{}%
\AgdaGeneralizable{m}\AgdaSpace{}%
\AgdaSymbol{→}\AgdaSpace{}%
\AgdaRecord{Ctx}\AgdaSpace{}%
\AgdaOperator{\AgdaField{⟦}}\AgdaSpace{}%
\AgdaGeneralizable{m}\AgdaSpace{}%
\AgdaOperator{\AgdaField{⟧mode}}\<%
\\
\>[2]\AgdaOperator{\AgdaFunction{⟦}}\AgdaSpace{}%
\AgdaInductiveConstructor{◇}\AgdaSpace{}%
\AgdaOperator{\AgdaFunction{⟧ctx}}\AgdaSpace{}%
\AgdaSymbol{=}\AgdaSpace{}%
\AgdaFunction{M.◇}\<%
\\
\>[2]\AgdaOperator{\AgdaFunction{⟦}}%
\>[230I]\AgdaBound{Γ}\AgdaSpace{}%
\AgdaOperator{\AgdaInductiveConstructor{,}}\AgdaSpace{}%
\AgdaBound{μ}\AgdaSpace{}%
\AgdaOperator{\AgdaInductiveConstructor{∣}}\AgdaSpace{}%
\AgdaBound{x}\AgdaSpace{}%
\AgdaOperator{\AgdaInductiveConstructor{∈}}\AgdaSpace{}%
\AgdaBound{T}\AgdaSpace{}%
\AgdaOperator{\AgdaFunction{⟧ctx}}\AgdaSpace{}%
\AgdaSymbol{=}\<%
\\
\>[.][@{}l@{}]\<[230I]%
\>[4]\AgdaOperator{\AgdaFunction{⟦}}\AgdaSpace{}%
\AgdaBound{Γ}\AgdaSpace{}%
\AgdaOperator{\AgdaFunction{⟧ctx}}\AgdaSpace{}%
\AgdaOperator{\AgdaFunction{,,}}\AgdaSpace{}%
\AgdaOperator{\AgdaField{M.⟨}}\AgdaSpace{}%
\AgdaOperator{\AgdaField{⟦}}\AgdaSpace{}%
\AgdaBound{μ}\AgdaSpace{}%
\AgdaOperator{\AgdaField{⟧modality}}\AgdaSpace{}%
\AgdaOperator{\AgdaField{∣}}\AgdaSpace{}%
\AgdaOperator{\AgdaFunction{⟦}}\AgdaSpace{}%
\AgdaBound{T}\AgdaSpace{}%
\AgdaOperator{\AgdaFunction{⟧ty}}\AgdaSpace{}%
\AgdaOperator{\AgdaField{⟩}}\<%
\\
\>[2]\AgdaOperator{\AgdaFunction{⟦}}\AgdaSpace{}%
\AgdaBound{Γ}\AgdaSpace{}%
\AgdaOperator{\AgdaInductiveConstructor{,lock⟨}}\AgdaSpace{}%
\AgdaBound{μ}\AgdaSpace{}%
\AgdaOperator{\AgdaInductiveConstructor{⟩}}\AgdaSpace{}%
\AgdaOperator{\AgdaFunction{⟧ctx}}\AgdaSpace{}%
\AgdaSymbol{=}\AgdaSpace{}%
\AgdaFunction{lock}\AgdaSpace{}%
\AgdaOperator{\AgdaField{⟦}}\AgdaSpace{}%
\AgdaBound{μ}\AgdaSpace{}%
\AgdaOperator{\AgdaField{⟧modality}}\AgdaSpace{}%
\AgdaOperator{\AgdaFunction{⟦}}\AgdaSpace{}%
\AgdaBound{Γ}\AgdaSpace{}%
\AgdaOperator{\AgdaFunction{⟧ctx}}\<%
\end{code}
\end{AgdaMultiCode}
\caption{Contexts. Here, 
\begin{code}[hide]%
\>[2]\AgdaFunction{\AgdaUnderscore{}}\AgdaSpace{}%
\AgdaSymbol{:}\AgdaSpace{}%
\AgdaSymbol{∀}\AgdaSpace{}%
\AgdaSymbol{\{}\AgdaBound{C}\AgdaSymbol{\}}\AgdaSpace{}%
\AgdaSymbol{(}\AgdaBound{Γ}\AgdaSpace{}%
\AgdaSymbol{:}\AgdaSpace{}%
\AgdaRecord{Ctx}\AgdaSpace{}%
\AgdaBound{C}\AgdaSymbol{)}\AgdaSpace{}%
\AgdaSymbol{→}\AgdaSpace{}%
\AgdaRecord{Ty}\AgdaSpace{}%
\AgdaBound{Γ}\AgdaSpace{}%
\AgdaSymbol{→}\AgdaSpace{}%
\AgdaRecord{Ctx}\AgdaSpace{}%
\AgdaBound{C}\<%
\\
\>[2]\AgdaSymbol{\AgdaUnderscore{}}%
\>[277I]\AgdaSymbol{=}\<%
\end{code}
\begin{code}[inline*]%
\>[.][@{}l@{}]\<[277I]%
\>[4]\AgdaOperator{\AgdaFunction{\AgdaUnderscore{},,\AgdaUnderscore{}}}\<%
\end{code}
is semantic context extension (the semantic layer uses a de Bruijn representation of variables).}
\label{fig:interp-ctx}
\end{subfigure}
\end{minipage}
\caption{Interpretation functions for necessarily well-typed expressions.}
\end{figure}

Before we can interpret terms, we must know how to interpret types, modes and modalities.
As already mentioned in Section~\ref{sec:presheaves}, modes in MSTT will be interpreted as base categories and modalities as DRAs.
Correspondingly, in Fig.~\ref{fig:interp-mode-theory} we add two fields to the definition of \AR{ModeTheory} from Fig.~\ref{fig:modetheorycomplete}.
Furthermore, \AR{ModeTheory} additionally requires the interpretation of the modality \AFi{𝟙} to be equivalent to the DRA \AF{M.𝟙} according to \AR{≅ᵐ}, and similarly for composition of modalities resp.\ DRAs.
MSTT types are interpreted as closed semantic types and MSTT contexts as semantic contexts (Figs.~\ref{fig:interp-ty} and \ref{fig:interp-ctx}).

In order to handle type errors, the type checker makes use of a type checking monad (Fig.~\ref{fig:tcm}).%
\footnote{The $\ell$ in \APt{Set}\AS$\ell$ is a universe level.
  This means that \AR{TCM} can act on any Agda type, not just types in one particular universe.}
A value of type \ADT{TCM}\AS{}\AB{A} represents either a type error (with a string as message) or a value of type \AB{A}.
\ADT{TCM} has the structure of a monad, and in fact it is just a reformulation of Haskell's error monad.

An additional field \AFi{⟦_∈_⇒_⟧two-cell} is included in the definition of \AR{ModeTheory} (Fig.~\ref{fig:interpret-2-cell}).
It will result in a type error if the given modalities cannot be the domain and codomain of the given 2-cell, and otherwise it will return the interpretation of the 2-cell in the presheaf model.

The type checker will also need to test whether two modes or two modalities are equivalent.
Therefore, we add two more fields to the definition of \AR{ModeTheory} (Fig.~\ref{fig:mode-theory-eq}).
The first function takes two syntactic modes and either provides a proof that these modes are syntactically equal or results in a type error if they are not.%
\footnote{We check syntactic equality, as the semantic layer does not account for equivalence of modes, i.e.\ base categories.}
The second function is the long awaited specification of the modality equivalence relation $\modeq$, not as a predicate but as a sound decision procedure.
As such, it tests whether two syntactic modalities of the same domain and codomain are equivalent, and if so, it produces a proof that the modalities' interpretations as DRAs are equivalent.
As mentioned in \ref{sec:mode-theory}, it is expected that associativity and unit laws of \AFi{ⓜ} hold up to $\modeq$, but it is not necessary to prove this explicitly.

\newcommand{\TCM}{%
\begin{code}[hide]%
\>[2]\AgdaKeyword{module}\AgdaSpace{}%
\AgdaModule{TCMDef}\AgdaSpace{}%
\AgdaKeyword{where}\<%
\end{code}
\begin{code}%
\>[2][@{}l@{\AgdaIndent{1}}]%
\>[4]\AgdaKeyword{data}\AgdaSpace{}%
\AgdaDatatype{TCM}\AgdaSpace{}%
\AgdaSymbol{(}\AgdaBound{A}\AgdaSpace{}%
\AgdaSymbol{:}\AgdaSpace{}%
\AgdaPrimitive{Set}\AgdaSpace{}%
\AgdaGeneralizable{ℓ}\AgdaSymbol{)}\AgdaSpace{}%
\AgdaSymbol{:}\AgdaSpace{}%
\AgdaPrimitive{Set}\AgdaSpace{}%
\AgdaBound{ℓ}\AgdaSpace{}%
\AgdaKeyword{where}\<%
\\
\>[4][@{}l@{\AgdaIndent{0}}]%
\>[6]\AgdaInductiveConstructor{type-error}\AgdaSpace{}%
\AgdaSymbol{:}\AgdaSpace{}%
\AgdaPostulate{String}\AgdaSpace{}%
\AgdaSymbol{→}\AgdaSpace{}%
\AgdaDatatype{TCM}\AgdaSpace{}%
\AgdaBound{A}\<%
\\
\>[6]\AgdaInductiveConstructor{ok}\AgdaSpace{}%
\AgdaSymbol{:}\AgdaSpace{}%
\AgdaBound{A}\AgdaSpace{}%
\AgdaSymbol{→}\AgdaSpace{}%
\AgdaDatatype{TCM}\AgdaSpace{}%
\AgdaBound{A}\<%
\end{code}
\begin{code}[hide]%
\>[2]\AgdaKeyword{open}\AgdaSpace{}%
\AgdaModule{TCMonad}\<%
\end{code}
}
\newcommand{\CheckTyEq}{%
\begin{code}[inline]%
\>[2]\AgdaOperator{\AgdaFunction{\AgdaUnderscore{}≃ᵗʸ?\AgdaUnderscore{}}}\AgdaSpace{}%
\AgdaSymbol{:}\AgdaSpace{}%
\AgdaSymbol{(}\AgdaBound{T}\AgdaSpace{}%
\AgdaBound{S}\AgdaSpace{}%
\AgdaSymbol{:}\AgdaSpace{}%
\AgdaDatatype{TyExpr}\AgdaSpace{}%
\AgdaGeneralizable{m}\AgdaSymbol{)}\AgdaSpace{}%
\AgdaSymbol{→}\AgdaSpace{}%
\AgdaDatatype{TCM}\AgdaSpace{}%
\AgdaSymbol{(}\AgdaOperator{\AgdaFunction{⟦}}\AgdaSpace{}%
\AgdaBound{T}\AgdaSpace{}%
\AgdaOperator{\AgdaFunction{⟧ty}}\AgdaSpace{}%
\AgdaOperator{\AgdaRecord{≅ᵗʸ}}\AgdaSpace{}%
\AgdaOperator{\AgdaFunction{⟦}}\AgdaSpace{}%
\AgdaBound{S}\AgdaSpace{}%
\AgdaOperator{\AgdaFunction{⟧ty}}\AgdaSymbol{)}\<%
\end{code}
\begin{code}[hide]%
\>[2]\AgdaBound{T}\AgdaSpace{}%
\AgdaOperator{\AgdaFunction{≃ᵗʸ?}}\AgdaSpace{}%
\AgdaBound{S}\AgdaSpace{}%
\AgdaSymbol{=}\AgdaSpace{}%
\AgdaInductiveConstructor{type-error}\AgdaSpace{}%
\AgdaString{""}\<%
\end{code}
}

\begin{figure}[htb]
\begin{minipage}{.4\textwidth}
\begin{subfigure}{\linewidth}
\begin{AgdaMultiCode}
\TCM
\end{AgdaMultiCode}
\caption{Type-checking monad.}
\label{fig:tcm}
\end{subfigure} \\
\begin{subfigure}{\linewidth}
\begin{code}[hide]%
\>[2]\AgdaKeyword{module}\AgdaSpace{}%
\AgdaModule{ModeTheoryTwoCellInterpr}\AgdaSpace{}%
\AgdaKeyword{where}\<%
\\
\>[2][@{}l@{\AgdaIndent{0}}]%
\>[4]\AgdaKeyword{record}\AgdaSpace{}%
\AgdaRecord{ExtraFields}\AgdaSpace{}%
\AgdaSymbol{:}\AgdaSpace{}%
\AgdaPrimitive{Set₁}\AgdaSpace{}%
\AgdaKeyword{where}\<%
\\
\>[4][@{}l@{\AgdaIndent{0}}]%
\>[6]\AgdaKeyword{field}\<%
\end{code}
\begin{AgdaSuppressSpace}
\begin{code}%
\>[6][@{}l@{\AgdaIndent{1}}]%
\>[8]\AgdaOperator{\AgdaField{⟦\AgdaUnderscore{}∈\AgdaUnderscore{}⇒\AgdaUnderscore{}⟧two-cell}}\AgdaSpace{}%
\AgdaSymbol{:}\AgdaSpace{}%
\AgdaField{TwoCellExpr}\AgdaSpace{}%
\AgdaSymbol{→}\<%
\\
\>[8][@{}l@{\AgdaIndent{0}}]%
\>[11]\AgdaSymbol{(}\AgdaBound{μ}\AgdaSpace{}%
\AgdaBound{ρ}\AgdaSpace{}%
\AgdaSymbol{:}\AgdaSpace{}%
\AgdaField{ModalityExpr}\AgdaSpace{}%
\AgdaGeneralizable{m}\AgdaSpace{}%
\AgdaGeneralizable{n}\AgdaSymbol{)}\AgdaSpace{}%
\AgdaSymbol{→}\<%
\\
\>[11]\AgdaDatatype{TCM}\AgdaSpace{}%
\AgdaSymbol{(}\AgdaRecord{TwoCell}\AgdaSpace{}%
\AgdaOperator{\AgdaField{⟦}}\AgdaSpace{}%
\AgdaBound{μ}\AgdaSpace{}%
\AgdaOperator{\AgdaField{⟧modality}}\AgdaSpace{}%
\AgdaOperator{\AgdaField{⟦}}\AgdaSpace{}%
\AgdaBound{ρ}\AgdaSpace{}%
\AgdaOperator{\AgdaField{⟧modality}}\AgdaSymbol{)}\<%
\end{code}
\end{AgdaSuppressSpace}
\caption{Checking and interpretation of 2-cells (an additional field to \AR{ModeTheory}, Fig.~\ref{fig:modetheorycomplete}).}
\label{fig:interpret-2-cell}
\end{subfigure}
\begin{subfigure}{\linewidth}
\begin{AgdaMultiCode}
\begin{code}[hide]%
\>[2]\AgdaKeyword{infix}\AgdaSpace{}%
\AgdaNumber{1}\AgdaSpace{}%
\AgdaOperator{\AgdaInductiveConstructor{\AgdaUnderscore{},\AgdaUnderscore{}}}\<%
\end{code}
\begin{code}%
\>[2]\AgdaKeyword{record}%
\>[348I]\AgdaRecord{InferInterpretResult}\<%
\\
\>[348I][@{}l@{\AgdaIndent{0}}]%
\>[11]\AgdaSymbol{(}\AgdaBound{Γ}\AgdaSpace{}%
\AgdaSymbol{:}\AgdaSpace{}%
\AgdaDatatype{CtxExpr}\AgdaSpace{}%
\AgdaGeneralizable{m}\AgdaSymbol{)}\AgdaSpace{}%
\AgdaSymbol{:}\AgdaSpace{}%
\AgdaPrimitive{Set}\AgdaSpace{}%
\AgdaKeyword{where}\<%
\\
\>[2][@{}l@{\AgdaIndent{0}}]%
\>[4]\AgdaKeyword{constructor}\AgdaSpace{}%
\AgdaOperator{\AgdaInductiveConstructor{\AgdaUnderscore{},\AgdaUnderscore{}}}\<%
\\
\>[4]\AgdaKeyword{field}%
\>[356I]\AgdaField{type}\AgdaSpace{}%
\AgdaSymbol{:}\AgdaSpace{}%
\AgdaDatatype{TyExpr}\AgdaSpace{}%
\AgdaBound{m}\<%
\\
\>[.][@{}l@{}]\<[356I]%
\>[10]\AgdaField{interpretation}\AgdaSpace{}%
\AgdaSymbol{:}\AgdaSpace{}%
\AgdaRecord{Tm}\AgdaSpace{}%
\AgdaOperator{\AgdaFunction{⟦}}\AgdaSpace{}%
\AgdaBound{Γ}\AgdaSpace{}%
\AgdaOperator{\AgdaFunction{⟧ctx}}\AgdaSpace{}%
\AgdaOperator{\AgdaFunction{⟦}}\AgdaSpace{}%
\AgdaField{type}\AgdaSpace{}%
\AgdaOperator{\AgdaFunction{⟧ty}}\<%
\end{code}
\end{AgdaMultiCode}
\caption{Output type of successful type inference.}
\label{fig:infer-interpret-result}
\end{subfigure} \\
\end{minipage}
\hspace{\fill}
\begin{minipage}{.5\textwidth}
\begin{subfigure}{\linewidth}
\begin{AgdaMultiCode}
\begin{code}[hide]%
\>[2]\AgdaKeyword{module}\AgdaSpace{}%
\AgdaModule{ModeTheoryDecisions}\AgdaSpace{}%
\AgdaKeyword{where}\<%
\\
\>[2][@{}l@{\AgdaIndent{0}}]%
\>[4]\AgdaKeyword{record}\AgdaSpace{}%
\AgdaRecord{ExtraFields}\AgdaSpace{}%
\AgdaSymbol{:}\AgdaSpace{}%
\AgdaPrimitive{Set₁}\AgdaSpace{}%
\AgdaKeyword{where}\<%
\\
\>[4][@{}l@{\AgdaIndent{0}}]%
\>[6]\AgdaKeyword{field}\<%
\end{code}
\begin{code}%
\>[6][@{}l@{\AgdaIndent{1}}]%
\>[8]\AgdaOperator{\AgdaField{\AgdaUnderscore{}=mode?\AgdaUnderscore{}}}\AgdaSpace{}%
\AgdaSymbol{:}\AgdaSpace{}%
\AgdaSymbol{(}\AgdaBound{m}\AgdaSpace{}%
\AgdaBound{n}\AgdaSpace{}%
\AgdaSymbol{:}\AgdaSpace{}%
\AgdaField{ModeExpr}\AgdaSymbol{)}\AgdaSpace{}%
\AgdaSymbol{→}\AgdaSpace{}%
\AgdaDatatype{TCM}\AgdaSpace{}%
\AgdaSymbol{(}\AgdaBound{m}\AgdaSpace{}%
\AgdaOperator{\AgdaDatatype{≡}}\AgdaSpace{}%
\AgdaBound{n}\AgdaSymbol{)}\<%
\\
\>[8]\AgdaOperator{\AgdaField{\AgdaUnderscore{}≃ᵐ?\AgdaUnderscore{}}}\AgdaSpace{}%
\AgdaSymbol{:}%
\>[385I]\AgdaSymbol{(}\AgdaBound{μ}\AgdaSpace{}%
\AgdaBound{ρ}\AgdaSpace{}%
\AgdaSymbol{:}\AgdaSpace{}%
\AgdaField{ModalityExpr}\AgdaSpace{}%
\AgdaBound{m}\AgdaSpace{}%
\AgdaBound{n}\AgdaSymbol{)}\AgdaSpace{}%
\AgdaSymbol{→}\<%
\\
\>[.][@{}l@{}]\<[385I]%
\>[16]\AgdaDatatype{TCM}\AgdaSpace{}%
\AgdaSymbol{(}\AgdaOperator{\AgdaField{⟦}}\AgdaSpace{}%
\AgdaBound{μ}\AgdaSpace{}%
\AgdaOperator{\AgdaField{⟧modality}}\AgdaSpace{}%
\AgdaOperator{\AgdaRecord{≅ᵐ}}\AgdaSpace{}%
\AgdaOperator{\AgdaField{⟦}}\AgdaSpace{}%
\AgdaBound{ρ}\AgdaSpace{}%
\AgdaOperator{\AgdaField{⟧modality}}\AgdaSymbol{)}\<%
\end{code}
\begin{code}[hide]%
\>[2]\AgdaKeyword{open}\AgdaSpace{}%
\AgdaModule{ModeTheory}\AgdaSpace{}%
\AgdaBound{mt}\AgdaSpace{}%
\AgdaKeyword{renaming}\AgdaSpace{}%
\AgdaSymbol{(}\AgdaOperator{\AgdaField{\AgdaUnderscore{}≟mode\AgdaUnderscore{}}}\AgdaSpace{}%
\AgdaSymbol{to}\AgdaSpace{}%
\AgdaOperator{\AgdaField{\AgdaUnderscore{}=mode?\AgdaUnderscore{}}}\AgdaSymbol{)}\AgdaSpace{}%
\AgdaKeyword{using}\AgdaSpace{}%
\AgdaSymbol{(}\AgdaOperator{\AgdaField{\AgdaUnderscore{}≃ᵐ?\AgdaUnderscore{}}}\AgdaSymbol{)}\<%
\end{code}
\end{AgdaMultiCode}
\caption{Equivalence checking of modes and modalities (these are additional fields to \AR{ModeTheory}, Fig.~\ref{fig:modetheorycomplete}).}
\label{fig:mode-theory-eq}
\end{subfigure} \\
\begin{subfigure}{\linewidth}
\begin{AgdaMultiCode}
\newcommand{\IsModalTy}{
  \begin{code}%
\>[2]\AgdaKeyword{data}\AgdaSpace{}%
\AgdaDatatype{IsModalTyExpr}\AgdaSpace{}%
\AgdaSymbol{:}\AgdaSpace{}%
\AgdaDatatype{TyExpr}\AgdaSpace{}%
\AgdaGeneralizable{n}\AgdaSpace{}%
\AgdaSymbol{→}\AgdaSpace{}%
\AgdaPrimitive{Set}\AgdaSpace{}%
\AgdaKeyword{where}\<%
\\
\>[2][@{}l@{\AgdaIndent{0}}]%
\>[4]\AgdaInductiveConstructor{modal-ty}\AgdaSpace{}%
\AgdaSymbol{:}%
\>[419I]\AgdaSymbol{(}\AgdaBound{m}\AgdaSpace{}%
\AgdaSymbol{:}\AgdaSpace{}%
\AgdaField{ModeExpr}\AgdaSymbol{)}\AgdaSpace{}%
\AgdaSymbol{(}\AgdaBound{μ}\AgdaSpace{}%
\AgdaSymbol{:}\AgdaSpace{}%
\AgdaField{ModalityExpr}\AgdaSpace{}%
\AgdaBound{m}\AgdaSpace{}%
\AgdaGeneralizable{n}\AgdaSymbol{)}\<%
\\
\>[.][@{}l@{}]\<[419I]%
\>[15]\AgdaSymbol{(}\AgdaBound{T}\AgdaSpace{}%
\AgdaSymbol{:}\AgdaSpace{}%
\AgdaDatatype{TyExpr}\AgdaSpace{}%
\AgdaBound{m}\AgdaSymbol{)}\AgdaSpace{}%
\AgdaSymbol{→}\AgdaSpace{}%
\AgdaDatatype{IsModalTyExpr}\AgdaSpace{}%
\AgdaOperator{\AgdaInductiveConstructor{⟨}}\AgdaSpace{}%
\AgdaBound{μ}\AgdaSpace{}%
\AgdaOperator{\AgdaInductiveConstructor{∣}}\AgdaSpace{}%
\AgdaBound{T}\AgdaSpace{}%
\AgdaOperator{\AgdaInductiveConstructor{⟩}}\<%
\end{code}%
}%
\begin{code}[hide]%
\>[2]\AgdaFunction{is-modal-ty}\AgdaSpace{}%
\AgdaSymbol{:}\AgdaSpace{}%
\AgdaSymbol{(}\AgdaBound{T}\AgdaSpace{}%
\AgdaSymbol{:}\AgdaSpace{}%
\AgdaDatatype{TyExpr}\AgdaSpace{}%
\AgdaGeneralizable{n}\AgdaSymbol{)}\AgdaSpace{}%
\AgdaSymbol{→}\AgdaSpace{}%
\AgdaDatatype{TCM}\AgdaSpace{}%
\AgdaSymbol{(}\AgdaDatatype{IsModalTyExpr}\AgdaSpace{}%
\AgdaBound{T}\AgdaSymbol{)}\<%
\\
\>[2]\AgdaFunction{is-modal-ty}\AgdaSpace{}%
\AgdaOperator{\AgdaInductiveConstructor{⟨}}\AgdaSpace{}%
\AgdaBound{μ}\AgdaSpace{}%
\AgdaOperator{\AgdaInductiveConstructor{∣}}\AgdaSpace{}%
\AgdaBound{T}\AgdaSpace{}%
\AgdaOperator{\AgdaInductiveConstructor{⟩}}\AgdaSpace{}%
\AgdaSymbol{=}\AgdaSpace{}%
\AgdaFunction{return}\AgdaSpace{}%
\AgdaSymbol{(}\AgdaInductiveConstructor{modal-ty}\AgdaSpace{}%
\AgdaSymbol{\AgdaUnderscore{}}\AgdaSpace{}%
\AgdaBound{μ}\AgdaSpace{}%
\AgdaBound{T}\AgdaSymbol{)}\<%
\\
\>[2]\AgdaCatchallClause{\AgdaFunction{is-modal-ty}}\AgdaSpace{}%
\AgdaCatchallClause{\AgdaSymbol{\AgdaUnderscore{}}}\AgdaSpace{}%
\AgdaSymbol{=}\AgdaSpace{}%
\AgdaInductiveConstructor{type-error}\AgdaSpace{}%
\AgdaString{""}\<%
\end{code}%
\begin{code}%
\>[2]\AgdaFunction{infer-interpret}\AgdaSpace{}%
\AgdaSymbol{:}%
\>[462I]\AgdaDatatype{TmExpr}\AgdaSpace{}%
\AgdaGeneralizable{m}\AgdaSpace{}%
\AgdaSymbol{→}\AgdaSpace{}%
\AgdaSymbol{(}\AgdaBound{Γ}\AgdaSpace{}%
\AgdaSymbol{:}\AgdaSpace{}%
\AgdaDatatype{CtxExpr}\AgdaSpace{}%
\AgdaGeneralizable{m}\AgdaSymbol{)}\AgdaSpace{}%
\AgdaSymbol{→}\<%
\\
\>[.][@{}l@{}]\<[462I]%
\>[20]\AgdaDatatype{TCM}\AgdaSpace{}%
\AgdaSymbol{(}\AgdaRecord{InferInterpretResult}\AgdaSpace{}%
\AgdaBound{Γ}\AgdaSymbol{)}\<%
\\
\>[2]\AgdaFunction{infer-interpret}\AgdaSpace{}%
\AgdaSymbol{(}\AgdaOperator{\AgdaInductiveConstructor{mod⟨}}\AgdaSpace{}%
\AgdaBound{μ}\AgdaSpace{}%
\AgdaOperator{\AgdaInductiveConstructor{⟩}}\AgdaSpace{}%
\AgdaBound{t}\AgdaSymbol{)}\AgdaSpace{}%
\AgdaBound{Γ}\AgdaSpace{}%
\AgdaSymbol{=}\AgdaSpace{}%
\AgdaKeyword{do}\<%
\\
\>[2][@{}l@{\AgdaIndent{0}}]%
\>[4]\AgdaBound{T}\AgdaSpace{}%
\AgdaOperator{\AgdaInductiveConstructor{,}}\AgdaSpace{}%
\AgdaBound{sem-t}\AgdaSpace{}%
\AgdaOperator{\AgdaFunction{←}}\AgdaSpace{}%
\AgdaFunction{infer-interpret}\AgdaSpace{}%
\AgdaBound{t}\AgdaSpace{}%
\AgdaSymbol{(}\AgdaBound{Γ}\AgdaSpace{}%
\AgdaOperator{\AgdaInductiveConstructor{,lock⟨}}\AgdaSpace{}%
\AgdaBound{μ}\AgdaSpace{}%
\AgdaOperator{\AgdaInductiveConstructor{⟩}}\AgdaSymbol{)}\<%
\\
\>[4]\AgdaFunction{return}\AgdaSpace{}%
\AgdaSymbol{(}\AgdaOperator{\AgdaInductiveConstructor{⟨}}\AgdaSpace{}%
\AgdaBound{μ}\AgdaSpace{}%
\AgdaOperator{\AgdaInductiveConstructor{∣}}\AgdaSpace{}%
\AgdaBound{T}\AgdaSpace{}%
\AgdaOperator{\AgdaInductiveConstructor{⟩}}\AgdaSpace{}%
\AgdaOperator{\AgdaInductiveConstructor{,}}\AgdaSpace{}%
\AgdaField{mod-intro}\AgdaSpace{}%
\AgdaOperator{\AgdaField{⟦}}\AgdaSpace{}%
\AgdaBound{μ}\AgdaSpace{}%
\AgdaOperator{\AgdaField{⟧modality}}\AgdaSpace{}%
\AgdaBound{sem-t}\AgdaSymbol{)}\<%
\\
\>[2]\AgdaFunction{infer-interpret}\AgdaSpace{}%
\AgdaSymbol{(}\AgdaOperator{\AgdaInductiveConstructor{ann}}\AgdaSpace{}%
\AgdaBound{t}\AgdaSpace{}%
\AgdaOperator{\AgdaInductiveConstructor{∈}}\AgdaSpace{}%
\AgdaBound{S}\AgdaSymbol{)}\AgdaSpace{}%
\AgdaBound{Γ}\AgdaSpace{}%
\AgdaSymbol{=}\AgdaSpace{}%
\AgdaKeyword{do}\<%
\\
\>[2][@{}l@{\AgdaIndent{0}}]%
\>[4]\AgdaBound{T}\AgdaSpace{}%
\AgdaOperator{\AgdaInductiveConstructor{,}}\AgdaSpace{}%
\AgdaBound{sem-t}\AgdaSpace{}%
\AgdaOperator{\AgdaFunction{←}}\AgdaSpace{}%
\AgdaFunction{infer-interpret}\AgdaSpace{}%
\AgdaBound{t}\AgdaSpace{}%
\AgdaBound{Γ}\<%
\\
\>[4]\AgdaBound{e}\AgdaSpace{}%
\AgdaOperator{\AgdaFunction{←}}\AgdaSpace{}%
\AgdaBound{S}\AgdaSpace{}%
\AgdaOperator{\AgdaFunction{≃ᵗʸ?}}\AgdaSpace{}%
\AgdaBound{T}\<%
\\
\>[4]\AgdaFunction{return}\AgdaSpace{}%
\AgdaSymbol{(}\AgdaBound{S}\AgdaSpace{}%
\AgdaOperator{\AgdaInductiveConstructor{,}}\AgdaSpace{}%
\AgdaOperator{\AgdaFunction{ι[}}\AgdaSpace{}%
\AgdaBound{e}\AgdaSpace{}%
\AgdaOperator{\AgdaFunction{]}}\AgdaSpace{}%
\AgdaBound{sem-t}\AgdaSymbol{)}\<%
\end{code}
\begin{code}[hide]%
\>[2]\AgdaComment{\{-infer-interpret\ (coe\ \{mμ\}\ μ\ ρ\ α\ t)\ Γ\ =\ do
\ \ \ \ T\ ,\ ⟦t⟧\ ←\ infer-interpret\ t\ Γ
\ \ \ \ modal-ty\ mκ\ κ\ A\ ←\ is-modal-ty\ T
\ \ \ \ refl\ ←\ mμ\ =mode?\ mκ
\ \ \ \ μ=κ\ ←\ μ\ ≃ᵐ?\ κ
\ \ \ \ type-error\ ""
\ \ \ \ --\ return\ (⟨\ ρ\ ∣\ A\ ⟩\ ,\ M.coe-closed\ ⟦\ α\ ⟧two-cell\ \{\{⟦⟧ty-natural\ A\}\}\ (ι[\ eq-mod-closed\ μ=κ\ ⟦\ A\ ⟧ty\ \{\{⟦⟧ty-natural\ A\}\}\ ]\ ⟦t⟧))-\}}\<%
\\
\>[2]\AgdaCatchallClause{\AgdaFunction{infer-interpret}}\AgdaSpace{}%
\AgdaCatchallClause{\AgdaSymbol{\AgdaUnderscore{}}}\AgdaSpace{}%
\AgdaCatchallClause{\AgdaBound{Γ}}\AgdaSpace{}%
\AgdaSymbol{=}\AgdaSpace{}%
\AgdaInductiveConstructor{type-error}\AgdaSpace{}%
\AgdaString{""}\<%
\end{code}
\end{AgdaMultiCode}
\caption{Type inference (most clauses are omitted).}
\label{fig:infer-interpret}
\end{subfigure}
\end{minipage}
\caption{Equivalence checking and type inference.}
\end{figure}

Analogously, the Sikkel type checker also provides a function
\CheckTyEq{}
to test whether two types are equivalent according to the relation $\tyeq$.

If an MSTT term is well-typed, the type checker will return both its type and its denotation in the presheaf model.
We pack this result in the record type \AR{InferInterpretResult} (Fig.~\ref{fig:infer-interpret-result}).
In Fig.~\ref{fig:infer-interpret}, we look at some of the cases in the implementation of the type checker.
We make use of Agda's do-notation for the \ADT{TCM} monad.
When inferring the type of
\begin{code}[hide]%
\>[2]\AgdaKeyword{module}\AgdaSpace{}%
\AgdaModule{\AgdaUnderscore{}}\AgdaSpace{}%
\AgdaSymbol{(}\AgdaBound{μ}\AgdaSpace{}%
\AgdaSymbol{:}\AgdaSpace{}%
\AgdaField{ModalityExpr}\AgdaSpace{}%
\AgdaGeneralizable{m}\AgdaSpace{}%
\AgdaGeneralizable{n}\AgdaSymbol{)}\AgdaSpace{}%
\AgdaSymbol{(}\AgdaBound{t}\AgdaSpace{}%
\AgdaSymbol{:}\AgdaSpace{}%
\AgdaDatatype{TmExpr}\AgdaSpace{}%
\AgdaGeneralizable{m}\AgdaSymbol{)}\AgdaSpace{}%
\AgdaSymbol{(}\AgdaBound{Γ}\AgdaSpace{}%
\AgdaSymbol{:}\AgdaSpace{}%
\AgdaDatatype{CtxExpr}\AgdaSpace{}%
\AgdaGeneralizable{n}\AgdaSymbol{)}\AgdaSpace{}%
\AgdaKeyword{where}\<%
\\
\>[2][@{}l@{\AgdaIndent{0}}]%
\>[4]\AgdaFunction{\AgdaUnderscore{}}\AgdaSpace{}%
\AgdaSymbol{:}\AgdaSpace{}%
\AgdaDatatype{TmExpr}\AgdaSpace{}%
\AgdaBound{n}\<%
\\
\>[4]\AgdaSymbol{\AgdaUnderscore{}}%
\>[545I]\AgdaSymbol{=}\<%
\end{code}
\begin{code}[inline*]%
\>[.][@{}l@{}]\<[545I]%
\>[6]\AgdaOperator{\AgdaInductiveConstructor{mod⟨}}\AgdaSpace{}%
\AgdaBound{μ}\AgdaSpace{}%
\AgdaOperator{\AgdaInductiveConstructor{⟩}}\AgdaSpace{}%
\AgdaBound{t}\<%
\end{code}
in a context \AB{Γ}, we first infer the type of $t$ in the locked context
\begin{code}[hide]%
\>[4]\AgdaFunction{\AgdaUnderscore{}}\AgdaSpace{}%
\AgdaSymbol{:}\AgdaSpace{}%
\AgdaDatatype{CtxExpr}\AgdaSpace{}%
\AgdaBound{m}\<%
\\
\>[4]\AgdaSymbol{\AgdaUnderscore{}}%
\>[552I]\AgdaSymbol{=}\<%
\end{code}
\begin{code}[inline*]%
\>[.][@{}l@{}]\<[552I]%
\>[6]\AgdaBound{Γ}\AgdaSpace{}%
\AgdaOperator{\AgdaInductiveConstructor{,lock⟨}}\AgdaSpace{}%
\AgdaBound{μ}\AgdaSpace{}%
\AgdaOperator{\AgdaInductiveConstructor{⟩}}\<%
\end{code}
and bind the result to $T$.
Simultaneously, the denotation of $t$ in the presheaf model is computed and bound to the variable \AB{sem-t}, which will have the Agda type
\begin{code}[hide]%
\>[4]\AgdaFunction{\AgdaUnderscore{}}\AgdaSpace{}%
\AgdaSymbol{:}\AgdaSpace{}%
\AgdaDatatype{TyExpr}\AgdaSpace{}%
\AgdaBound{m}\AgdaSpace{}%
\AgdaSymbol{→}\AgdaSpace{}%
\AgdaPrimitive{Set}\<%
\\
\>[4]\AgdaSymbol{\AgdaUnderscore{}}%
\>[561I]\AgdaSymbol{=}\AgdaSpace{}%
\AgdaSymbol{λ}\AgdaSpace{}%
\AgdaBound{T}\AgdaSpace{}%
\AgdaSymbol{→}\<%
\end{code}
\begin{code}[inline]%
\>[.][@{}l@{}]\<[561I]%
\>[6]\AgdaRecord{Tm}\AgdaSpace{}%
\AgdaSymbol{(}\AgdaFunction{lock}\AgdaSpace{}%
\AgdaOperator{\AgdaField{⟦}}\AgdaSpace{}%
\AgdaBound{μ}\AgdaSpace{}%
\AgdaOperator{\AgdaField{⟧modality}}\AgdaSpace{}%
\AgdaOperator{\AgdaFunction{⟦}}\AgdaSpace{}%
\AgdaBound{Γ}\AgdaSpace{}%
\AgdaOperator{\AgdaFunction{⟧ctx}}\AgdaSymbol{)}\AgdaSpace{}%
\AgdaOperator{\AgdaFunction{⟦}}\AgdaSpace{}%
\AgdaBound{T}\AgdaSpace{}%
\AgdaOperator{\AgdaFunction{⟧ty}}\<%
\end{code}.
Then we infer that the type of the original term is
\begin{code}[hide]%
\>[4]\AgdaFunction{\AgdaUnderscore{}}\AgdaSpace{}%
\AgdaSymbol{:}\AgdaSpace{}%
\AgdaDatatype{TyExpr}\AgdaSpace{}%
\AgdaBound{m}\AgdaSpace{}%
\AgdaSymbol{→}\AgdaSpace{}%
\AgdaDatatype{TyExpr}\AgdaSpace{}%
\AgdaBound{n}\<%
\\
\>[4]\AgdaSymbol{\AgdaUnderscore{}}%
\>[581I]\AgdaSymbol{=}\AgdaSpace{}%
\AgdaSymbol{λ}\AgdaSpace{}%
\AgdaBound{T}\AgdaSpace{}%
\AgdaSymbol{→}\<%
\end{code}
\begin{code}[inline*]%
\>[.][@{}l@{}]\<[581I]%
\>[6]\AgdaOperator{\AgdaInductiveConstructor{⟨}}\AgdaSpace{}%
\AgdaBound{μ}\AgdaSpace{}%
\AgdaOperator{\AgdaInductiveConstructor{∣}}\AgdaSpace{}%
\AgdaBound{T}\AgdaSpace{}%
\AgdaOperator{\AgdaInductiveConstructor{⟩}}\<%
\end{code}
and we construct the required semantic term of semantic type
\begin{code}[hide]%
\>[4]\AgdaFunction{\AgdaUnderscore{}}\AgdaSpace{}%
\AgdaSymbol{:}\AgdaSpace{}%
\AgdaDatatype{TyExpr}\AgdaSpace{}%
\AgdaBound{m}\AgdaSpace{}%
\AgdaSymbol{→}\AgdaSpace{}%
\AgdaRecord{Ty}\AgdaSpace{}%
\AgdaOperator{\AgdaFunction{⟦}}\AgdaSpace{}%
\AgdaBound{Γ}\AgdaSpace{}%
\AgdaOperator{\AgdaFunction{⟧ctx}}\<%
\\
\>[4]\AgdaSymbol{\AgdaUnderscore{}}%
\>[597I]\AgdaSymbol{=}\AgdaSpace{}%
\AgdaSymbol{λ}\AgdaSpace{}%
\AgdaBound{T}\AgdaSpace{}%
\AgdaSymbol{→}\<%
\end{code}
\begin{code}[inline*]%
\>[.][@{}l@{}]\<[597I]%
\>[6]\AgdaOperator{\AgdaField{M.⟨}}\AgdaSpace{}%
\AgdaOperator{\AgdaField{⟦}}\AgdaSpace{}%
\AgdaBound{μ}\AgdaSpace{}%
\AgdaOperator{\AgdaField{⟧modality}}\AgdaSpace{}%
\AgdaOperator{\AgdaField{∣}}\AgdaSpace{}%
\AgdaOperator{\AgdaFunction{⟦}}\AgdaSpace{}%
\AgdaBound{T}\AgdaSpace{}%
\AgdaOperator{\AgdaFunction{⟧ty}}\AgdaSpace{}%
\AgdaOperator{\AgdaField{⟩}}\<%
\end{code}
by applying the introduction operation for DRAs to \AB{sem-t}.
In the case of an annotated term
\begin{code}[hide]%
\>[2]\AgdaKeyword{module}\AgdaSpace{}%
\AgdaModule{\AgdaUnderscore{}}\AgdaSpace{}%
\AgdaSymbol{(}\AgdaBound{t}\AgdaSpace{}%
\AgdaSymbol{:}\AgdaSpace{}%
\AgdaDatatype{TmExpr}\AgdaSpace{}%
\AgdaGeneralizable{m}\AgdaSymbol{)}\AgdaSpace{}%
\AgdaSymbol{(}\AgdaBound{Γ}\AgdaSpace{}%
\AgdaSymbol{:}\AgdaSpace{}%
\AgdaDatatype{CtxExpr}\AgdaSpace{}%
\AgdaGeneralizable{m}\AgdaSymbol{)}\AgdaSpace{}%
\AgdaSymbol{(}\AgdaBound{T}\AgdaSpace{}%
\AgdaBound{S}\AgdaSpace{}%
\AgdaSymbol{:}\AgdaSpace{}%
\AgdaDatatype{TyExpr}\AgdaSpace{}%
\AgdaGeneralizable{m}\AgdaSymbol{)}\AgdaSpace{}%
\AgdaKeyword{where}\<%
\\
\>[2][@{}l@{\AgdaIndent{0}}]%
\>[4]\AgdaFunction{\AgdaUnderscore{}}\AgdaSpace{}%
\AgdaSymbol{:}\AgdaSpace{}%
\AgdaDatatype{TmExpr}\AgdaSpace{}%
\AgdaBound{m}\<%
\\
\>[4]\AgdaSymbol{\AgdaUnderscore{}}%
\>[627I]\AgdaSymbol{=}\<%
\end{code}
\begin{code}[inline]%
\>[.][@{}l@{}]\<[627I]%
\>[6]\AgdaOperator{\AgdaInductiveConstructor{ann}}\AgdaSpace{}%
\AgdaBound{t}\AgdaSpace{}%
\AgdaOperator{\AgdaInductiveConstructor{∈}}\AgdaSpace{}%
\AgdaBound{S}\<%
\end{code},
we first infer the type and denotation of $t$.
The variable \AB{sem-t} will now have the Agda type
\begin{code}[hide]%
\>[4]\AgdaFunction{\AgdaUnderscore{}}\AgdaSpace{}%
\AgdaSymbol{:}\AgdaSpace{}%
\AgdaPrimitive{Set}\<%
\\
\>[4]\AgdaSymbol{\AgdaUnderscore{}}%
\>[633I]\AgdaSymbol{=}\<%
\end{code}
\begin{code}[inline]%
\>[.][@{}l@{}]\<[633I]%
\>[6]\AgdaRecord{Tm}\AgdaSpace{}%
\AgdaOperator{\AgdaFunction{⟦}}\AgdaSpace{}%
\AgdaBound{Γ}\AgdaSpace{}%
\AgdaOperator{\AgdaFunction{⟧ctx}}\AgdaSpace{}%
\AgdaOperator{\AgdaFunction{⟦}}\AgdaSpace{}%
\AgdaBound{T}\AgdaSpace{}%
\AgdaOperator{\AgdaFunction{⟧ty}}\<%
\end{code}.
Next, we verify that $S$ and $T$ are equivalent types and as a result we get a proof
\begin{code}[hide]%
\>[4]\AgdaKeyword{module}\AgdaSpace{}%
\AgdaModule{\AgdaUnderscore{}}\AgdaSpace{}%
\AgdaSymbol{(}\<%
\end{code}
\begin{code}[inline*]%
\>[4][@{}l@{\AgdaIndent{1}}]%
\>[6]\AgdaBound{e}\AgdaSpace{}%
\AgdaSymbol{:}\AgdaSpace{}%
\AgdaOperator{\AgdaFunction{⟦}}\AgdaSpace{}%
\AgdaBound{S}\AgdaSpace{}%
\AgdaOperator{\AgdaFunction{⟧ty}}\AgdaSpace{}%
\AgdaOperator{\AgdaRecord{≅ᵗʸ}}\AgdaSpace{}%
\AgdaOperator{\AgdaFunction{⟦}}\AgdaSpace{}%
\AgdaBound{T}\AgdaSpace{}%
\AgdaOperator{\AgdaFunction{⟧ty}}\<%
\end{code}
\begin{code}[hide]%
\>[6]\AgdaSymbol{)}\AgdaSpace{}%
\AgdaKeyword{where}\<%
\end{code}
that the denotations of $S$ and $T$ in the presheaf model are isomorphic.
Finally, we say that the annotated term has type $S$ and we produce a semantic term of type
\begin{code}[hide]%
\>[4]\AgdaFunction{\AgdaUnderscore{}}\AgdaSpace{}%
\AgdaSymbol{:}\AgdaSpace{}%
\AgdaFunction{ClosedTy}\AgdaSpace{}%
\AgdaOperator{\AgdaField{⟦}}\AgdaSpace{}%
\AgdaBound{m}\AgdaSpace{}%
\AgdaOperator{\AgdaField{⟧mode}}\<%
\\
\>[4]\AgdaSymbol{\AgdaUnderscore{}}%
\>[658I]\AgdaSymbol{=}\<%
\end{code}
\begin{code}[inline*]%
\>[.][@{}l@{}]\<[658I]%
\>[6]\AgdaOperator{\AgdaFunction{⟦}}\AgdaSpace{}%
\AgdaBound{S}\AgdaSpace{}%
\AgdaOperator{\AgdaFunction{⟧ty}}\<%
\end{code}
by transporting \AB{sem-t} over the isomorphism $e$.
We refer to the Agda implementation for all other clauses of \AF{infer-interpret}.%

\begin{code}[hide]%
\>[2]\AgdaFunction{infer-type}\AgdaSpace{}%
\AgdaSymbol{:}\AgdaSpace{}%
\AgdaDatatype{TmExpr}\AgdaSpace{}%
\AgdaGeneralizable{m}\AgdaSpace{}%
\AgdaSymbol{→}\AgdaSpace{}%
\AgdaDatatype{CtxExpr}\AgdaSpace{}%
\AgdaGeneralizable{m}\AgdaSpace{}%
\AgdaSymbol{→}\AgdaSpace{}%
\AgdaDatatype{TCM}\AgdaSpace{}%
\AgdaSymbol{(}\AgdaDatatype{TyExpr}\AgdaSpace{}%
\AgdaGeneralizable{m}\AgdaSymbol{)}\<%
\\
\>[2]\AgdaFunction{infer-type}\AgdaSpace{}%
\AgdaBound{t}\AgdaSpace{}%
\AgdaBound{Γ}\AgdaSpace{}%
\AgdaSymbol{=}\AgdaSpace{}%
\AgdaField{InferInterpretResult.type}\AgdaSpace{}%
\AgdaOperator{\AgdaFunction{<\$>}}\AgdaSpace{}%
\AgdaFunction{infer-interpret}\AgdaSpace{}%
\AgdaBound{t}\AgdaSpace{}%
\AgdaBound{Γ}\<%
\\
\\[\AgdaEmptyExtraSkip]%
\>[2]\AgdaOperator{\AgdaFunction{⟦\AgdaUnderscore{}⟧tm-in\AgdaUnderscore{}}}\AgdaSpace{}%
\AgdaSymbol{:}\AgdaSpace{}%
\AgdaSymbol{(}\AgdaBound{t}\AgdaSpace{}%
\AgdaSymbol{:}\AgdaSpace{}%
\AgdaDatatype{TmExpr}\AgdaSpace{}%
\AgdaGeneralizable{m}\AgdaSymbol{)}\AgdaSpace{}%
\AgdaSymbol{(}\AgdaBound{Γ}\AgdaSpace{}%
\AgdaSymbol{:}\AgdaSpace{}%
\AgdaDatatype{CtxExpr}\AgdaSpace{}%
\AgdaGeneralizable{m}\AgdaSymbol{)}\AgdaSpace{}%
\AgdaSymbol{→}\AgdaSpace{}%
\AgdaFunction{tcm-elim}\AgdaSpace{}%
\AgdaSymbol{(λ}\AgdaSpace{}%
\AgdaSymbol{\AgdaUnderscore{}}\AgdaSpace{}%
\AgdaSymbol{→}\AgdaSpace{}%
\AgdaRecord{⊤}\AgdaSymbol{)}\AgdaSpace{}%
\AgdaSymbol{(λ}\AgdaSpace{}%
\AgdaBound{T}\AgdaSpace{}%
\AgdaSymbol{→}\AgdaSpace{}%
\AgdaRecord{Tm}\AgdaSpace{}%
\AgdaOperator{\AgdaFunction{⟦}}\AgdaSpace{}%
\AgdaBound{Γ}\AgdaSpace{}%
\AgdaOperator{\AgdaFunction{⟧ctx}}\AgdaSpace{}%
\AgdaOperator{\AgdaFunction{⟦}}\AgdaSpace{}%
\AgdaBound{T}\AgdaSpace{}%
\AgdaOperator{\AgdaFunction{⟧ty}}\AgdaSymbol{)}\AgdaSpace{}%
\AgdaSymbol{(}\AgdaFunction{infer-type}\AgdaSpace{}%
\AgdaBound{t}\AgdaSpace{}%
\AgdaBound{Γ}\AgdaSymbol{)}\<%
\\
\>[2]\AgdaOperator{\AgdaFunction{⟦}}\AgdaSpace{}%
\AgdaBound{t}\AgdaSpace{}%
\AgdaOperator{\AgdaFunction{⟧tm-in}}\AgdaSpace{}%
\AgdaBound{Γ}\AgdaSpace{}%
\AgdaKeyword{with}\AgdaSpace{}%
\AgdaFunction{infer-interpret}\AgdaSpace{}%
\AgdaBound{t}\AgdaSpace{}%
\AgdaBound{Γ}\<%
\\
\>[2]\AgdaOperator{\AgdaFunction{⟦}}\AgdaSpace{}%
\AgdaBound{t}\AgdaSpace{}%
\AgdaOperator{\AgdaFunction{⟧tm-in}}\AgdaSpace{}%
\AgdaBound{Γ}\AgdaSpace{}%
\AgdaSymbol{|}\AgdaSpace{}%
\AgdaInductiveConstructor{type-error}\AgdaSpace{}%
\AgdaSymbol{\AgdaUnderscore{}}\AgdaSpace{}%
\AgdaSymbol{=}\AgdaSpace{}%
\AgdaInductiveConstructor{tt}\<%
\\
\>[2]\AgdaOperator{\AgdaFunction{⟦}}\AgdaSpace{}%
\AgdaBound{t}\AgdaSpace{}%
\AgdaOperator{\AgdaFunction{⟧tm-in}}\AgdaSpace{}%
\AgdaBound{Γ}\AgdaSpace{}%
\AgdaSymbol{|}\AgdaSpace{}%
\AgdaInductiveConstructor{ok}\AgdaSpace{}%
\AgdaSymbol{(}\AgdaBound{T}\AgdaSpace{}%
\AgdaOperator{\AgdaInductiveConstructor{,}}\AgdaSpace{}%
\AgdaBound{⟦t⟧}\AgdaSymbol{)}\AgdaSpace{}%
\AgdaSymbol{=}\AgdaSpace{}%
\AgdaBound{⟦t⟧}\<%
\end{code}
Using \AF{infer-interpret}, we can implement an operation
\begin{code}[hide]%
\>[2]\AgdaFunction{\AgdaUnderscore{}}\AgdaSpace{}%
\AgdaSymbol{:}\AgdaSpace{}%
\AgdaSymbol{(}\AgdaBound{t}\AgdaSpace{}%
\AgdaSymbol{:}\AgdaSpace{}%
\AgdaDatatype{TmExpr}\AgdaSpace{}%
\AgdaGeneralizable{m}\AgdaSymbol{)}\AgdaSpace{}%
\AgdaSymbol{(}\AgdaBound{Γ}\AgdaSpace{}%
\AgdaSymbol{:}\AgdaSpace{}%
\AgdaDatatype{CtxExpr}\AgdaSpace{}%
\AgdaGeneralizable{m}\AgdaSymbol{)}\AgdaSpace{}%
\AgdaSymbol{→}\AgdaSpace{}%
\AgdaFunction{tcm-elim}\AgdaSpace{}%
\AgdaSymbol{(λ}\AgdaSpace{}%
\AgdaSymbol{\AgdaUnderscore{}}\AgdaSpace{}%
\AgdaSymbol{→}\AgdaSpace{}%
\AgdaRecord{⊤}\AgdaSymbol{)}\AgdaSpace{}%
\AgdaSymbol{(λ}\AgdaSpace{}%
\AgdaBound{T}\AgdaSpace{}%
\AgdaSymbol{→}\AgdaSpace{}%
\AgdaRecord{Tm}\AgdaSpace{}%
\AgdaOperator{\AgdaFunction{⟦}}\AgdaSpace{}%
\AgdaBound{Γ}\AgdaSpace{}%
\AgdaOperator{\AgdaFunction{⟧ctx}}\AgdaSpace{}%
\AgdaOperator{\AgdaFunction{⟦}}\AgdaSpace{}%
\AgdaBound{T}\AgdaSpace{}%
\AgdaOperator{\AgdaFunction{⟧ty}}\AgdaSymbol{)}\AgdaSpace{}%
\AgdaSymbol{(}\AgdaFunction{infer-type}\AgdaSpace{}%
\AgdaBound{t}\AgdaSpace{}%
\AgdaBound{Γ}\AgdaSymbol{)}\<%
\\
\>[2]\AgdaSymbol{\AgdaUnderscore{}}%
\>[760I]\AgdaSymbol{=}\<%
\end{code}
\begin{code}[inline*]%
\>[.][@{}l@{}]\<[760I]%
\>[4]\AgdaOperator{\AgdaFunction{⟦\AgdaUnderscore{}⟧tm-in\AgdaUnderscore{}}}\<%
\end{code}
that accepts an MSTT term and context and returns the term's denotation if it is well-typed or a value of the unit type \AR{⊤} if it is not (hence this operation is a dependent function whose result type depends on the success of the Sikkel type checker).
After extending Sikkel's generic type checker with cases for L\"ob induction and the operations for guarded streams, we can then interpret the examples from Section~\ref{sec:appl-guard-recurs} in the presheaf model for guarded recursion.
For example, the denotations of the streams \AF{g-nats} and \AF{nats} from Fig.~\ref{fig:code-guarded} in the empty context are computed as follows.

\noindent
\begin{minipage}{.48\textwidth}
\begin{code}[hide]%
\>[0]\AgdaKeyword{open}\AgdaSpace{}%
\AgdaKeyword{import}\AgdaSpace{}%
\AgdaModule{Applications.GuardedRecursion.MSTT}\<%
\end{code}
\begin{code}%
\>[0]\AgdaFunction{g-nats-sem}\AgdaSpace{}%
\AgdaSymbol{:}\AgdaSpace{}%
\AgdaRecord{Tm}\AgdaSpace{}%
\AgdaSymbol{\{}\AgdaFunction{M.ω}\AgdaSymbol{\}}\AgdaSpace{}%
\AgdaFunction{M.◇}\<%
\\
\>[0][@{}l@{\AgdaIndent{0}}]%
\>[2]\AgdaSymbol{(}\AgdaFunction{M.GStream}\AgdaSpace{}%
\AgdaFunction{M.Nat}\AgdaSymbol{)}\<%
\\
\>[0]\AgdaFunction{g-nats-sem}\AgdaSpace{}%
\AgdaSymbol{=}\AgdaSpace{}%
\AgdaOperator{\AgdaFunction{⟦}}\AgdaSpace{}%
\AgdaFunction{g-nats}\AgdaSpace{}%
\AgdaOperator{\AgdaFunction{⟧tm-in}}\AgdaSpace{}%
\AgdaInductiveConstructor{◇}\<%
\end{code}
\end{minipage}
\hspace{\fill}
\noindent
\begin{minipage}{.48\textwidth}
\begin{code}%
\>[0]\AgdaFunction{nats-sem}\AgdaSpace{}%
\AgdaSymbol{:}\AgdaSpace{}%
\AgdaRecord{Tm}\AgdaSpace{}%
\AgdaSymbol{\{}\AgdaFunction{M.★}\AgdaSymbol{\}}\AgdaSpace{}%
\AgdaFunction{M.◇}\<%
\\
\>[0][@{}l@{\AgdaIndent{0}}]%
\>[2]\AgdaOperator{\AgdaField{M.⟨}}\AgdaSpace{}%
\AgdaFunction{M.forever}\AgdaSpace{}%
\AgdaOperator{\AgdaField{∣}}\AgdaSpace{}%
\AgdaFunction{M.GStream}\AgdaSpace{}%
\AgdaFunction{M.Nat}\AgdaSpace{}%
\AgdaOperator{\AgdaField{⟩}}\<%
\\
\>[0]\AgdaFunction{nats-sem}\AgdaSpace{}%
\AgdaSymbol{=}\AgdaSpace{}%
\AgdaOperator{\AgdaFunction{⟦}}\AgdaSpace{}%
\AgdaFunction{nats}\AgdaSpace{}%
\AgdaOperator{\AgdaFunction{⟧tm-in}}\AgdaSpace{}%
\AgdaInductiveConstructor{◇}\<%
\end{code}
\end{minipage}

\noindent
Note that these have computational content in Agda, e.g.\ 
\begin{code}[hide]%
\>[0]\AgdaKeyword{open}\AgdaSpace{}%
\AgdaKeyword{import}\AgdaSpace{}%
\AgdaModule{Data.Vec}\<%
\\
\>[0]\AgdaFunction{\AgdaUnderscore{}}\AgdaSpace{}%
\AgdaSymbol{:}\AgdaSpace{}%
\AgdaSymbol{(}\AgdaBound{n}\AgdaSpace{}%
\AgdaSymbol{:}\AgdaSpace{}%
\AgdaDatatype{ℕ}\AgdaSymbol{)}\AgdaSpace{}%
\AgdaSymbol{→}\AgdaSpace{}%
\AgdaDatatype{Vec}\AgdaSpace{}%
\AgdaDatatype{ℕ}\AgdaSpace{}%
\AgdaSymbol{(}\AgdaInductiveConstructor{suc}\AgdaSpace{}%
\AgdaBound{n}\AgdaSymbol{)}\<%
\\
\>[0]\AgdaSymbol{\AgdaUnderscore{}}%
\>[798I]\AgdaSymbol{=}\AgdaSpace{}%
\AgdaSymbol{λ}\AgdaSpace{}%
\AgdaBound{n}\AgdaSpace{}%
\AgdaSymbol{→}\<%
\end{code}
\begin{code}[inline*]%
\>[.][@{}l@{}]\<[798I]%
\>[2]\AgdaFunction{g-nats-sem}\AgdaSpace{}%
\AgdaOperator{\AgdaField{⟨}}\AgdaSpace{}%
\AgdaBound{n}\AgdaSpace{}%
\AgdaOperator{\AgdaField{,}}\AgdaSpace{}%
\AgdaInductiveConstructor{tt}\AgdaSpace{}%
\AgdaOperator{\AgdaField{⟩'}}\<%
\end{code}
is the Agda list consisting of the first $n + 1$ natural numbers for any $n$.


%% file: content-lagda/extraction.tex
\section{Extraction to the Meta-level}
\label{sec:extraction}

\begin{code}[hide]%
\>[0]\AgdaKeyword{module}\AgdaSpace{}%
\AgdaModule{extraction}\AgdaSpace{}%
\AgdaKeyword{where}\<%
\\
\\[\AgdaEmptyExtraSkip]%
\>[0]\AgdaKeyword{open}\AgdaSpace{}%
\AgdaKeyword{import}\AgdaSpace{}%
\AgdaModule{Data.Nat}\<%
\\
\>[0]\AgdaKeyword{open}\AgdaSpace{}%
\AgdaKeyword{import}\AgdaSpace{}%
\AgdaModule{Data.Unit}\<%
\\
\>[0]\AgdaKeyword{open}\AgdaSpace{}%
\AgdaKeyword{import}\AgdaSpace{}%
\AgdaModule{Data.Vec}\<%
\\
\\[\AgdaEmptyExtraSkip]%
\>[0]\AgdaKeyword{open}\AgdaSpace{}%
\AgdaKeyword{import}\AgdaSpace{}%
\AgdaModule{Model.BaseCategory}\<%
\\
\>[0]\AgdaKeyword{open}\AgdaSpace{}%
\AgdaKeyword{import}\AgdaSpace{}%
\AgdaModule{Model.CwF-Structure}\<%
\\
\>[0]\AgdaKeyword{open}\AgdaSpace{}%
\AgdaKeyword{import}\AgdaSpace{}%
\AgdaModule{Model.Modality}\<%
\\
\>[0]\AgdaKeyword{open}\AgdaSpace{}%
\AgdaKeyword{import}\AgdaSpace{}%
\AgdaModule{Model.Type.Discrete}\AgdaSpace{}%
\AgdaKeyword{renaming}\AgdaSpace{}%
\AgdaSymbol{(}\AgdaFunction{Nat'}\AgdaSpace{}%
\AgdaSymbol{to}\AgdaSpace{}%
\AgdaFunction{Nat}\AgdaSymbol{)}\<%
\\
\>[0]\AgdaKeyword{open}\AgdaSpace{}%
\AgdaKeyword{import}\AgdaSpace{}%
\AgdaModule{Model.Type.Function}\<%
\\
\>[0]\AgdaKeyword{open}\AgdaSpace{}%
\AgdaKeyword{import}\AgdaSpace{}%
\AgdaModule{Applications.GuardedRecursion.Model.Modalities}\<%
\\
\>[0]\AgdaKeyword{open}\AgdaSpace{}%
\AgdaKeyword{import}\AgdaSpace{}%
\AgdaModule{Applications.GuardedRecursion.Model.Streams.Guarded}\<%
\\
\\[\AgdaEmptyExtraSkip]%
\>[0]\AgdaKeyword{open}\AgdaSpace{}%
\AgdaKeyword{import}\AgdaSpace{}%
\AgdaModule{guarded-recursion}\<%
\\
\>[0]\AgdaKeyword{open}\AgdaSpace{}%
\AgdaKeyword{import}\AgdaSpace{}%
\AgdaModule{sound-typechecker}\<%
\end{code}

The trivial base category \AF{★} corresponding to the trivial mode for guarded recursion has one object (its type \AFi{Ob} of objects is Agda's unit type \AR{⊤}) and one trivial morphism.
In the presheaf model over \AF{★}, a type $T$ and a term $t$ in the empty context are nothing more than respectively an Agda type
\begin{code}[hide]%
\>[0]\AgdaFunction{\AgdaUnderscore{}}\AgdaSpace{}%
\AgdaSymbol{:}\AgdaSpace{}%
\AgdaRecord{Ty}\AgdaSpace{}%
\AgdaSymbol{\{}\AgdaArgument{C}\AgdaSpace{}%
\AgdaSymbol{=}\AgdaSpace{}%
\AgdaFunction{★}\AgdaSymbol{\}}\AgdaSpace{}%
\AgdaFunction{◇}\AgdaSpace{}%
\AgdaSymbol{→}\AgdaSpace{}%
\AgdaPrimitive{Set}\<%
\\
\>[0]\AgdaSymbol{\AgdaUnderscore{}}%
\>[38I]\AgdaSymbol{=}\AgdaSpace{}%
\AgdaSymbol{λ}\AgdaSpace{}%
\AgdaBound{T}\AgdaSpace{}%
\AgdaSymbol{→}\<%
\end{code}
\begin{code}[inline*]%
\>[.][@{}l@{}]\<[38I]%
\>[2]\AgdaBound{T}\AgdaSpace{}%
\AgdaOperator{\AgdaField{⟨}}\AgdaSpace{}%
\AgdaInductiveConstructor{tt}\AgdaSpace{}%
\AgdaOperator{\AgdaField{,}}\AgdaSpace{}%
\AgdaInductiveConstructor{tt}\AgdaSpace{}%
\AgdaOperator{\AgdaField{⟩}}\<%
\end{code}
and an Agda term
\begin{code}[hide]%
\>[0]\AgdaFunction{\AgdaUnderscore{}}\AgdaSpace{}%
\AgdaSymbol{:}\AgdaSpace{}%
\AgdaSymbol{(}\AgdaBound{T}\AgdaSpace{}%
\AgdaSymbol{:}\AgdaSpace{}%
\AgdaRecord{Ty}\AgdaSpace{}%
\AgdaSymbol{\{}\AgdaArgument{C}\AgdaSpace{}%
\AgdaSymbol{=}\AgdaSpace{}%
\AgdaFunction{★}\AgdaSymbol{\}}\AgdaSpace{}%
\AgdaFunction{◇}\AgdaSymbol{)}\AgdaSpace{}%
\AgdaSymbol{→}\AgdaSpace{}%
\AgdaRecord{Tm}\AgdaSpace{}%
\AgdaFunction{◇}\AgdaSpace{}%
\AgdaBound{T}\AgdaSpace{}%
\AgdaSymbol{→}\AgdaSpace{}%
\AgdaBound{T}\AgdaSpace{}%
\AgdaOperator{\AgdaField{⟨}}\AgdaSpace{}%
\AgdaInductiveConstructor{tt}\AgdaSpace{}%
\AgdaOperator{\AgdaField{,}}\AgdaSpace{}%
\AgdaInductiveConstructor{tt}\AgdaSpace{}%
\AgdaOperator{\AgdaField{⟩}}\<%
\\
\>[0]\AgdaSymbol{\AgdaUnderscore{}}%
\>[66I]\AgdaSymbol{=}\AgdaSpace{}%
\AgdaSymbol{λ}\AgdaSpace{}%
\AgdaBound{T}\AgdaSpace{}%
\AgdaBound{t}\AgdaSpace{}%
\AgdaSymbol{→}\<%
\end{code}
\begin{code}[inline]%
\>[.][@{}l@{}]\<[66I]%
\>[2]\AgdaBound{t}\AgdaSpace{}%
\AgdaOperator{\AgdaField{⟨}}\AgdaSpace{}%
\AgdaInductiveConstructor{tt}\AgdaSpace{}%
\AgdaOperator{\AgdaField{,}}\AgdaSpace{}%
\AgdaInductiveConstructor{tt}\AgdaSpace{}%
\AgdaOperator{\AgdaField{⟩'}}\<%
\end{code}.
In fact, the presheaf model over \AF{★} is the standard set model of type theory.

However, we cannot directly use this to extract a value out of a semantic term $t$ (in the empty context with base category \AF{★}) to get to the third layer in Figure~\ref{fig:sikkel-architecture}.
The reason is that this value does not always have the desired type on the nose.
For example, if $t$ has type
\begin{code}[hide]%
\>[0]\AgdaFunction{\AgdaUnderscore{}}\AgdaSpace{}%
\AgdaSymbol{:}\AgdaSpace{}%
\AgdaSymbol{(}\AgdaBound{T}\AgdaSpace{}%
\AgdaBound{S}\AgdaSpace{}%
\AgdaSymbol{:}\AgdaSpace{}%
\AgdaRecord{Ty}\AgdaSpace{}%
\AgdaSymbol{\{}\AgdaArgument{C}\AgdaSpace{}%
\AgdaSymbol{=}\AgdaSpace{}%
\AgdaFunction{★}\AgdaSymbol{\}}\AgdaSpace{}%
\AgdaFunction{◇}\AgdaSymbol{)}\AgdaSpace{}%
\AgdaSymbol{→}\AgdaSpace{}%
\AgdaRecord{Ty}\AgdaSpace{}%
\AgdaFunction{◇}\<%
\\
\>[0]\AgdaSymbol{\AgdaUnderscore{}}%
\>[88I]\AgdaSymbol{=}\AgdaSpace{}%
\AgdaSymbol{λ}\AgdaSpace{}%
\AgdaBound{T}\AgdaSpace{}%
\AgdaBound{S}\AgdaSpace{}%
\AgdaSymbol{→}\<%
\end{code}
\begin{code}[inline]%
\>[.][@{}l@{}]\<[88I]%
\>[2]\AgdaBound{T}\AgdaSpace{}%
\AgdaOperator{\AgdaFunction{⇛}}\AgdaSpace{}%
\AgdaBound{S}\<%
\end{code},
then
\begin{code}[hide]%
\>[0]\AgdaFunction{\AgdaUnderscore{}}\AgdaSpace{}%
\AgdaSymbol{:}\AgdaSpace{}%
\AgdaSymbol{(}\AgdaBound{T}\AgdaSpace{}%
\AgdaSymbol{:}\AgdaSpace{}%
\AgdaRecord{Ty}\AgdaSpace{}%
\AgdaSymbol{\{}\AgdaArgument{C}\AgdaSpace{}%
\AgdaSymbol{=}\AgdaSpace{}%
\AgdaFunction{★}\AgdaSymbol{\}}\AgdaSpace{}%
\AgdaFunction{◇}\AgdaSymbol{)}\AgdaSpace{}%
\AgdaSymbol{→}\AgdaSpace{}%
\AgdaRecord{Tm}\AgdaSpace{}%
\AgdaFunction{◇}\AgdaSpace{}%
\AgdaBound{T}\AgdaSpace{}%
\AgdaSymbol{→}\AgdaSpace{}%
\AgdaBound{T}\AgdaSpace{}%
\AgdaOperator{\AgdaField{⟨}}\AgdaSpace{}%
\AgdaInductiveConstructor{tt}\AgdaSpace{}%
\AgdaOperator{\AgdaField{,}}\AgdaSpace{}%
\AgdaInductiveConstructor{tt}\AgdaSpace{}%
\AgdaOperator{\AgdaField{⟩}}\<%
\\
\>[0]\AgdaSymbol{\AgdaUnderscore{}}%
\>[114I]\AgdaSymbol{=}\AgdaSpace{}%
\AgdaSymbol{λ}\AgdaSpace{}%
\AgdaBound{T}\AgdaSpace{}%
\AgdaBound{t}\AgdaSpace{}%
\AgdaSymbol{→}\<%
\end{code}
\begin{code}[inline*]%
\>[.][@{}l@{}]\<[114I]%
\>[2]\AgdaBound{t}\AgdaSpace{}%
\AgdaOperator{\AgdaField{⟨}}\AgdaSpace{}%
\AgdaInductiveConstructor{tt}\AgdaSpace{}%
\AgdaOperator{\AgdaField{,}}\AgdaSpace{}%
\AgdaInductiveConstructor{tt}\AgdaSpace{}%
\AgdaOperator{\AgdaField{⟩'}}\<%
\end{code}
will have the Agda type
\begin{code}[hide]%
\>[0]\AgdaFunction{\AgdaUnderscore{}}\AgdaSpace{}%
\AgdaSymbol{:}\AgdaSpace{}%
\AgdaSymbol{(}\AgdaBound{T}\AgdaSpace{}%
\AgdaBound{S}\AgdaSpace{}%
\AgdaSymbol{:}\AgdaSpace{}%
\AgdaRecord{Ty}\AgdaSpace{}%
\AgdaSymbol{\{}\AgdaArgument{C}\AgdaSpace{}%
\AgdaSymbol{=}\AgdaSpace{}%
\AgdaFunction{★}\AgdaSymbol{\}}\AgdaSpace{}%
\AgdaFunction{◇}\AgdaSymbol{)}\AgdaSpace{}%
\AgdaSymbol{→}\AgdaSpace{}%
\AgdaPrimitive{Set}\<%
\\
\>[0]\AgdaSymbol{\AgdaUnderscore{}}%
\>[135I]\AgdaSymbol{=}\AgdaSpace{}%
\AgdaSymbol{λ}\AgdaSpace{}%
\AgdaBound{T}\AgdaSpace{}%
\AgdaBound{S}\AgdaSpace{}%
\AgdaSymbol{→}\<%
\end{code}
\begin{code}[inline*]%
\>[.][@{}l@{}]\<[135I]%
\>[2]\AgdaRecord{PshFun}\AgdaSpace{}%
\AgdaBound{T}\AgdaSpace{}%
\AgdaBound{S}\AgdaSpace{}%
\AgdaInductiveConstructor{tt}\AgdaSpace{}%
\AgdaInductiveConstructor{tt}\<%
\end{code}
and not the function type
\begin{code}[hide]%
\>[0]\AgdaFunction{\AgdaUnderscore{}}\AgdaSpace{}%
\AgdaSymbol{:}\AgdaSpace{}%
\AgdaSymbol{(}\AgdaBound{T}\AgdaSpace{}%
\AgdaBound{S}\AgdaSpace{}%
\AgdaSymbol{:}\AgdaSpace{}%
\AgdaRecord{Ty}\AgdaSpace{}%
\AgdaSymbol{\{}\AgdaArgument{C}\AgdaSpace{}%
\AgdaSymbol{=}\AgdaSpace{}%
\AgdaFunction{★}\AgdaSymbol{\}}\AgdaSpace{}%
\AgdaFunction{◇}\AgdaSymbol{)}\AgdaSpace{}%
\AgdaSymbol{→}\AgdaSpace{}%
\AgdaPrimitive{Set}\<%
\\
\>[0]\AgdaSymbol{\AgdaUnderscore{}}%
\>[155I]\AgdaSymbol{=}\AgdaSpace{}%
\AgdaSymbol{λ}\AgdaSpace{}%
\AgdaBound{T}\AgdaSpace{}%
\AgdaBound{S}\AgdaSpace{}%
\AgdaSymbol{→}\<%
\end{code}
\begin{code}[inline]%
\>[.][@{}l@{}]\<[155I]%
\>[2]\AgdaBound{T}\AgdaSpace{}%
\AgdaOperator{\AgdaField{⟨}}\AgdaSpace{}%
\AgdaInductiveConstructor{tt}\AgdaSpace{}%
\AgdaOperator{\AgdaField{,}}\AgdaSpace{}%
\AgdaInductiveConstructor{tt}\AgdaSpace{}%
\AgdaOperator{\AgdaField{⟩}}\AgdaSpace{}%
\AgdaSymbol{→}\AgdaSpace{}%
\AgdaBound{S}\AgdaSpace{}%
\AgdaOperator{\AgdaField{⟨}}\AgdaSpace{}%
\AgdaInductiveConstructor{tt}\AgdaSpace{}%
\AgdaOperator{\AgdaField{,}}\AgdaSpace{}%
\AgdaInductiveConstructor{tt}\AgdaSpace{}%
\AgdaOperator{\AgdaField{⟩}}\<%
\end{code},
although these two types are isomorphic.
Similarly, the Agda value obtained from a semantic term of type
\begin{code}[hide]%
\>[0]\AgdaFunction{\AgdaUnderscore{}}\AgdaSpace{}%
\AgdaSymbol{:}\AgdaSpace{}%
\AgdaRecord{Ty}\AgdaSpace{}%
\AgdaSymbol{\{}\AgdaArgument{C}\AgdaSpace{}%
\AgdaSymbol{=}\AgdaSpace{}%
\AgdaFunction{★}\AgdaSymbol{\}}\AgdaSpace{}%
\AgdaFunction{◇}\<%
\\
\>[0]\AgdaSymbol{\AgdaUnderscore{}}%
\>[178I]\AgdaSymbol{=}\<%
\end{code}
\begin{code}[inline*]%
\>[.][@{}l@{}]\<[178I]%
\>[2]\AgdaOperator{\AgdaField{⟨}}\AgdaSpace{}%
\AgdaFunction{forever}\AgdaSpace{}%
\AgdaOperator{\AgdaField{∣}}\AgdaSpace{}%
\AgdaFunction{GStream}\AgdaSpace{}%
\AgdaFunction{Nat}\AgdaSpace{}%
\AgdaOperator{\AgdaField{⟩}}\<%
\end{code}
would have type
\begin{code}[hide]%
\>[0]\AgdaFunction{\AgdaUnderscore{}}\AgdaSpace{}%
\AgdaSymbol{:}\AgdaSpace{}%
\AgdaPrimitive{Set}\<%
\\
\>[0]\AgdaSymbol{\AgdaUnderscore{}}%
\>[186I]\AgdaSymbol{=}\<%
\end{code}
\begin{code}[inline*]%
\>[.][@{}l@{}]\<[186I]%
\>[2]\AgdaSymbol{∀}\AgdaSpace{}%
\AgdaBound{n}\AgdaSpace{}%
\AgdaSymbol{→}\AgdaSpace{}%
\AgdaDatatype{Vec}\AgdaSpace{}%
\AgdaDatatype{ℕ}\AgdaSpace{}%
\AgdaSymbol{(}\AgdaInductiveConstructor{suc}\AgdaSpace{}%
\AgdaBound{n}\AgdaSymbol{)}\<%
\end{code}
(together with a naturality condition) and not
\begin{code}[hide]%
\>[0]\AgdaFunction{\AgdaUnderscore{}}\AgdaSpace{}%
\AgdaSymbol{:}\AgdaSpace{}%
\AgdaPrimitive{Set}\<%
\\
\>[0]\AgdaSymbol{\AgdaUnderscore{}}%
\>[195I]\AgdaSymbol{=}\<%
\end{code}
\begin{code}[inline]%
\>[.][@{}l@{}]\<[195I]%
\>[2]\AgdaRecord{Stream}\AgdaSpace{}%
\AgdaDatatype{ℕ}\<%
\end{code}.
Again, these two Agda types are isomorphic.

To bridge this gap, Sikkel has a type class \AR{Extractable} for closed semantic types over \AF{★}.
\begin{code}%
\>[0]\AgdaKeyword{record}\AgdaSpace{}%
\AgdaRecord{Extractable}\AgdaSpace{}%
\AgdaSymbol{(}\AgdaBound{T}\AgdaSpace{}%
\AgdaSymbol{:}\AgdaSpace{}%
\AgdaFunction{ClosedTy}\AgdaSpace{}%
\AgdaFunction{★}\AgdaSymbol{)}\AgdaSpace{}%
\AgdaSymbol{:}\AgdaSpace{}%
\AgdaPrimitive{Set₁}\AgdaSpace{}%
\AgdaKeyword{where}\<%
\\
\>[0][@{}l@{\AgdaIndent{0}}]%
\>[2]\AgdaKeyword{field}%
\>[205I]\AgdaField{translated-type}\AgdaSpace{}%
\AgdaSymbol{:}\AgdaSpace{}%
\AgdaPrimitive{Set}\<%
\\
\>[.][@{}l@{}]\<[205I]%
\>[8]\AgdaField{extract-term}\AgdaSpace{}%
\AgdaSymbol{:}\AgdaSpace{}%
\AgdaRecord{Tm}\AgdaSpace{}%
\AgdaFunction{◇}\AgdaSpace{}%
\AgdaBound{T}\AgdaSpace{}%
\AgdaSymbol{→}\AgdaSpace{}%
\AgdaField{translated-type}\<%
\\
\>[8]\AgdaField{embed-term}\AgdaSpace{}%
\AgdaSymbol{:}\AgdaSpace{}%
\AgdaField{translated-type}\AgdaSpace{}%
\AgdaSymbol{→}\AgdaSpace{}%
\AgdaRecord{Tm}\AgdaSpace{}%
\AgdaFunction{◇}\AgdaSpace{}%
\AgdaBound{T}\<%
\end{code}
\begin{code}[hide]%
\>[0]\AgdaKeyword{open}\AgdaSpace{}%
\AgdaModule{Extractable}\AgdaSpace{}%
\AgdaSymbol{\{\{...\}\}}\<%
\end{code}
In order for a semantic type to be an instance of this type class, we must specify its intended translation as an Agda type.
Furthermore, we must be able to produce an Agda value of this type given a semantic term in the empty context and vice versa.
\adapted{It is expected that these functions constitute an isomorphism, but this is not formally required and there are no corresponding proof obligations (for some types this will only be provable when assuming certain axioms, e.g.\ for function types, one should assume function extensionality).}
The field \AFi{embed-term} is essentially needed for the translation of function types.

Sikkel provides \AFi{Extractable} instances for natural numbers, booleans, functions and products.
Moreover, we can also provide an instance for
\begin{code}[hide]%
\>[0]\AgdaFunction{\AgdaUnderscore{}}\AgdaSpace{}%
\AgdaSymbol{:}\AgdaSpace{}%
\AgdaFunction{ClosedTy}\AgdaSpace{}%
\AgdaFunction{★}\AgdaSpace{}%
\AgdaSymbol{→}\AgdaSpace{}%
\AgdaRecord{Ty}\AgdaSpace{}%
\AgdaSymbol{\{}\AgdaArgument{C}\AgdaSpace{}%
\AgdaSymbol{=}\AgdaSpace{}%
\AgdaFunction{★}\AgdaSymbol{\}}\AgdaSpace{}%
\AgdaFunction{◇}\<%
\\
\>[0]\AgdaSymbol{\AgdaUnderscore{}}%
\>[231I]\AgdaSymbol{=}\AgdaSpace{}%
\AgdaSymbol{λ}\AgdaSpace{}%
\AgdaBound{A}\AgdaSpace{}%
\AgdaSymbol{→}\<%
\end{code}
\begin{code}[inline*]%
\>[.][@{}l@{}]\<[231I]%
\>[2]\AgdaOperator{\AgdaField{⟨}}\AgdaSpace{}%
\AgdaFunction{forever}\AgdaSpace{}%
\AgdaOperator{\AgdaField{∣}}\AgdaSpace{}%
\AgdaFunction{GStream}\AgdaSpace{}%
\AgdaBound{A}\AgdaSpace{}%
\AgdaOperator{\AgdaField{⟩}}\<%
\end{code}
whenever $A$ is extractable, the extraction resulting in an Agda stream.
In this way, we can finally construct the Agda stream of all natural numbers.
\begin{code}[hide]%
\>[0]\AgdaKeyword{instance}\<%
\\
\>[0][@{}l@{\AgdaIndent{0}}]%
\>[2]\AgdaSymbol{\{-\#}\AgdaSpace{}%
\AgdaKeyword{NON\AgdaUnderscore{}COVERING}\AgdaSpace{}%
\AgdaSymbol{\#-\}}\<%
\\
\>[2]\AgdaFunction{nat-extractable}\AgdaSpace{}%
\AgdaSymbol{:}\AgdaSpace{}%
\AgdaRecord{Extractable}\AgdaSpace{}%
\AgdaFunction{Nat}\<%
\\
\>[2]\AgdaField{Extractable.translated-type}\AgdaSpace{}%
\AgdaFunction{nat-extractable}\AgdaSpace{}%
\AgdaSymbol{=}\AgdaSpace{}%
\AgdaDatatype{ℕ}\<%
\\
\\[\AgdaEmptyExtraSkip]%
\>[2]\AgdaSymbol{\{-\#}\AgdaSpace{}%
\AgdaKeyword{NON\AgdaUnderscore{}COVERING}\AgdaSpace{}%
\AgdaSymbol{\#-\}}\<%
\\
\>[2]\AgdaFunction{stream-extractable}\AgdaSpace{}%
\AgdaSymbol{:}\AgdaSpace{}%
\AgdaSymbol{\{}\AgdaBound{A}\AgdaSpace{}%
\AgdaSymbol{:}\AgdaSpace{}%
\AgdaFunction{ClosedTy}\AgdaSpace{}%
\AgdaFunction{★}\AgdaSymbol{\}}\AgdaSpace{}%
\AgdaSymbol{→}\AgdaSpace{}%
\AgdaSymbol{\{\{}\AgdaRecord{Extractable}\AgdaSpace{}%
\AgdaBound{A}\AgdaSymbol{\}\}}\AgdaSpace{}%
\AgdaSymbol{→}\AgdaSpace{}%
\AgdaRecord{Extractable}\AgdaSpace{}%
\AgdaOperator{\AgdaField{⟨}}\AgdaSpace{}%
\AgdaFunction{forever}\AgdaSpace{}%
\AgdaOperator{\AgdaField{∣}}\AgdaSpace{}%
\AgdaFunction{GStream}\AgdaSpace{}%
\AgdaBound{A}\AgdaSpace{}%
\AgdaOperator{\AgdaField{⟩}}\<%
\\
\>[2]\AgdaField{Extractable.translated-type}\AgdaSpace{}%
\AgdaSymbol{(}\AgdaFunction{stream-extractable}\AgdaSpace{}%
\AgdaSymbol{\{}\AgdaBound{A}\AgdaSymbol{\})}\AgdaSpace{}%
\AgdaSymbol{=}\AgdaSpace{}%
\AgdaRecord{Stream}\AgdaSpace{}%
\AgdaSymbol{(}\AgdaField{translated-type}\AgdaSpace{}%
\AgdaSymbol{\{}\AgdaBound{A}\AgdaSymbol{\})}\<%
\end{code}
\begin{code}%
\>[0]\AgdaFunction{nats-agda}\AgdaSpace{}%
\AgdaSymbol{:}\AgdaSpace{}%
\AgdaRecord{Stream}\AgdaSpace{}%
\AgdaDatatype{ℕ}\<%
\\
\>[0]\AgdaFunction{nats-agda}\AgdaSpace{}%
\AgdaSymbol{=}\AgdaSpace{}%
\AgdaField{extract-term}\AgdaSpace{}%
\AgdaFunction{nats-sem}\<%
\end{code}


%% file: content-lagda/parametricity.tex
\section{Application 2: Representation Independence through Parametricity}
\label{sec:parametricity}

\begin{code}[hide]%
\>[0]\AgdaKeyword{module}\AgdaSpace{}%
\AgdaModule{parametricity}\AgdaSpace{}%
\AgdaKeyword{where}\<%
\\
\\[\AgdaEmptyExtraSkip]%
\>[0]\AgdaKeyword{open}\AgdaSpace{}%
\AgdaKeyword{import}\AgdaSpace{}%
\AgdaModule{Data.Product}\AgdaSpace{}%
\AgdaKeyword{using}\AgdaSpace{}%
\AgdaSymbol{(}\AgdaField{proj₂}\AgdaSymbol{;}\AgdaSpace{}%
\AgdaFunction{Σ-syntax}\AgdaSymbol{;}\AgdaSpace{}%
\AgdaOperator{\AgdaInductiveConstructor{\AgdaUnderscore{},\AgdaUnderscore{}}}\AgdaSymbol{)}\AgdaSpace{}%
\AgdaComment{--\ renaming\ (\AgdaUnderscore{},\AgdaUnderscore{}\ to\ [\AgdaUnderscore{},\AgdaUnderscore{}])}\<%
\\
\>[0]\AgdaKeyword{open}\AgdaSpace{}%
\AgdaKeyword{import}\AgdaSpace{}%
\AgdaModule{Data.Unit}\<%
\\
\>[0]\AgdaKeyword{open}\AgdaSpace{}%
\AgdaKeyword{import}\AgdaSpace{}%
\AgdaModule{Relation.Binary.PropositionalEquality}\AgdaSpace{}%
\AgdaKeyword{hiding}\AgdaSpace{}%
\AgdaSymbol{(}\AgdaOperator{\AgdaInductiveConstructor{[\AgdaUnderscore{}]}}\AgdaSymbol{)}\<%
\\
\\[\AgdaEmptyExtraSkip]%
\>[0]\AgdaKeyword{open}\AgdaSpace{}%
\AgdaKeyword{import}\AgdaSpace{}%
\AgdaModule{Model.BaseCategory}\AgdaSpace{}%
\AgdaSymbol{as}\AgdaSpace{}%
\AgdaModule{M}\AgdaSpace{}%
\AgdaKeyword{hiding}\AgdaSpace{}%
\AgdaSymbol{(}\AgdaFunction{⋀}\AgdaSymbol{;}\AgdaSpace{}%
\AgdaFunction{★}\AgdaSymbol{)}\<%
\\
\>[0]\AgdaKeyword{open}\AgdaSpace{}%
\AgdaKeyword{import}\AgdaSpace{}%
\AgdaModule{Model.CwF-Structure}\AgdaSpace{}%
\AgdaSymbol{as}\AgdaSpace{}%
\AgdaModule{M}\AgdaSpace{}%
\AgdaKeyword{hiding}\AgdaSpace{}%
\AgdaSymbol{(}\AgdaFunction{◇}\AgdaSymbol{;}\AgdaSpace{}%
\AgdaField{\AgdaUnderscore{}⟨\AgdaUnderscore{}⟩}\AgdaSymbol{;}\AgdaSpace{}%
\AgdaOperator{\AgdaFunction{\AgdaUnderscore{}[\AgdaUnderscore{}]}}\AgdaSymbol{)}\<%
\\
\>[0]\AgdaKeyword{import}\AgdaSpace{}%
\AgdaModule{Applications.Parametricity.Model}\AgdaSpace{}%
\AgdaSymbol{as}\AgdaSpace{}%
\AgdaModule{PM}\<%
\\
\>[0]\AgdaKeyword{open}\AgdaSpace{}%
\AgdaKeyword{import}\AgdaSpace{}%
\AgdaModule{Applications.Parametricity.IntegerExample}\<%
\\
\>[0][@{}l@{\AgdaIndent{0}}]%
\>[2]\AgdaKeyword{hiding}\AgdaSpace{}%
\AgdaSymbol{(}\AgdaRecord{IntStructure}\AgdaSymbol{;}\AgdaSpace{}%
\AgdaFunction{subtract}\AgdaSymbol{;}\AgdaSpace{}%
\AgdaFunction{subtract★-left}\AgdaSymbol{;}\AgdaSpace{}%
\AgdaFunction{subtract★-right}\AgdaSymbol{;}\AgdaSpace{}%
\AgdaFunction{subtract-∼}\AgdaSymbol{)}\<%
\\
\>[0]\AgdaKeyword{open}\AgdaSpace{}%
\AgdaKeyword{import}\AgdaSpace{}%
\AgdaModule{Applications.Parametricity.MSTT}\AgdaSpace{}%
\AgdaFunction{z-rel-ext}\AgdaSpace{}%
\AgdaKeyword{hiding}\AgdaSpace{}%
\AgdaSymbol{(}\AgdaOperator{\AgdaInductiveConstructor{\AgdaUnderscore{},\AgdaUnderscore{}}}\AgdaSymbol{)}\<%
\\
\>[0]\AgdaKeyword{open}\AgdaSpace{}%
\AgdaKeyword{import}\AgdaSpace{}%
\AgdaModule{MSTT.TCMonad}\<%
\\
\>[0]\AgdaKeyword{open}\AgdaSpace{}%
\AgdaKeyword{import}\AgdaSpace{}%
\AgdaModule{Extraction}\<%
\\
\>[0]\AgdaKeyword{open}\AgdaSpace{}%
\AgdaKeyword{import}\AgdaSpace{}%
\AgdaModule{Model.Type.Function}\AgdaSpace{}%
\AgdaKeyword{using}\AgdaSpace{}%
\AgdaSymbol{(}\AgdaOperator{\AgdaField{\AgdaUnderscore{}\$⟨\AgdaUnderscore{},\AgdaUnderscore{}⟩\AgdaUnderscore{}}}\AgdaSymbol{)}\<%
\\
\\[\AgdaEmptyExtraSkip]%
\>[0]\AgdaKeyword{private}\AgdaSpace{}%
\AgdaKeyword{variable}\<%
\\
\>[0][@{}l@{\AgdaIndent{0}}]%
\>[2]\AgdaGeneralizable{m}\AgdaSpace{}%
\AgdaSymbol{:}\AgdaSpace{}%
\AgdaDatatype{ModeExpr}\<%
\end{code}

\newcommand{\nxn}{%
\begin{code}[hide]%
\>[0]\AgdaKeyword{open}\AgdaSpace{}%
\AgdaKeyword{import}\AgdaSpace{}%
\AgdaModule{Data.Nat}\<%
\\
\>[0]\AgdaKeyword{open}\AgdaSpace{}%
\AgdaKeyword{import}\AgdaSpace{}%
\AgdaModule{Data.Product}\AgdaSpace{}%
\AgdaKeyword{using}\AgdaSpace{}%
\AgdaSymbol{(}\AgdaOperator{\AgdaFunction{\AgdaUnderscore{}×\AgdaUnderscore{}}}\AgdaSymbol{)}\<%
\\
\>[0]\AgdaFunction{\AgdaUnderscore{}}\AgdaSpace{}%
\AgdaSymbol{:}\AgdaSpace{}%
\AgdaPrimitive{Set}\<%
\\
\>[0]\AgdaSymbol{\AgdaUnderscore{}}%
\>[64I]\AgdaSymbol{=}\<%
\end{code}
\begin{code}[inline]%
\>[.][@{}l@{}]\<[64I]%
\>[2]\AgdaDatatype{ℕ}\AgdaSpace{}%
\AgdaOperator{\AgdaFunction{×}}\AgdaSpace{}%
\AgdaDatatype{ℕ}\<%
\end{code}
}
\newcommand{\signxn}{%
\begin{code}[hide]%
\>[0]\AgdaFunction{\AgdaUnderscore{}}\AgdaSpace{}%
\AgdaSymbol{:}\AgdaSpace{}%
\AgdaPrimitive{Set}\<%
\\
\>[0]\AgdaSymbol{\AgdaUnderscore{}}%
\>[69I]\AgdaSymbol{=}\<%
\end{code}
\begin{code}[inline]%
\>[.][@{}l@{}]\<[69I]%
\>[2]\AgdaDatatype{Sign}\AgdaSpace{}%
\AgdaOperator{\AgdaFunction{×}}\AgdaSpace{}%
\AgdaDatatype{ℕ}\<%
\end{code}
}

An important characteristic of Sikkel is that its different layers are parametrized by the mode theory and base category.
As a result, Sikkel can be applied to extend type theory with other features than guarded recursion.
In this section, we demonstrate a basic parametric type theory using a toy example where we define subtraction of integers in terms of addition and negation.

Consider the minimalistic interface for integers in Figure~\ref{fig:integer-agda}.
This interface can for instance be implemented by the types \nxn{} (representing differences of natural numbers), or \signxn{} (representing natural numbers with a sign, with two representations for zero).
We can use the operations in \AF{IntStructure} to implement new functions such as \AF{subtract} in Figure~\ref{fig:integer-agda}.

\begin{figure}[htb]
\begin{minipage}{0.4\textwidth}
\begin{subfigure}{\linewidth}
\begin{AgdaMultiCode}
\begin{code}[hide]%
\>[0]\AgdaKeyword{module}\AgdaSpace{}%
\AgdaModule{Introduction}\AgdaSpace{}%
\AgdaKeyword{where}\<%
\end{code}
\begin{code}%
\>[0][@{}l@{\AgdaIndent{1}}]%
\>[2]\AgdaKeyword{record}\AgdaSpace{}%
\AgdaRecord{IntStructure}\AgdaSpace{}%
\AgdaSymbol{(}\AgdaBound{A}\AgdaSpace{}%
\AgdaSymbol{:}\AgdaSpace{}%
\AgdaPrimitive{Set}\AgdaSymbol{)}\AgdaSpace{}%
\AgdaSymbol{:}\AgdaSpace{}%
\AgdaPrimitive{Set}\AgdaSpace{}%
\AgdaKeyword{where}\<%
\\
\>[2][@{}l@{\AgdaIndent{0}}]%
\>[4]\AgdaKeyword{field}%
\>[81I]\AgdaField{add}\AgdaSpace{}%
\AgdaSymbol{:}\AgdaSpace{}%
\AgdaBound{A}\AgdaSpace{}%
\AgdaSymbol{→}\AgdaSpace{}%
\AgdaBound{A}\AgdaSpace{}%
\AgdaSymbol{→}\AgdaSpace{}%
\AgdaBound{A}\<%
\\
\>[.][@{}l@{}]\<[81I]%
\>[10]\AgdaField{negate}\AgdaSpace{}%
\AgdaSymbol{:}\AgdaSpace{}%
\AgdaBound{A}\AgdaSpace{}%
\AgdaSymbol{→}\AgdaSpace{}%
\AgdaBound{A}\<%
\end{code}
\begin{code}[hide]%
\>[2]\AgdaKeyword{open}\AgdaSpace{}%
\AgdaModule{IntStructure}\<%
\end{code}
\begin{code}%
\>[2]\AgdaFunction{subtract}\AgdaSpace{}%
\AgdaSymbol{:}\AgdaSpace{}%
\AgdaRecord{IntStructure}\AgdaSpace{}%
\AgdaBound{A}\AgdaSpace{}%
\AgdaSymbol{→}\AgdaSpace{}%
\AgdaBound{A}\AgdaSpace{}%
\AgdaSymbol{→}\AgdaSpace{}%
\AgdaBound{A}\AgdaSpace{}%
\AgdaSymbol{→}\AgdaSpace{}%
\AgdaBound{A}\<%
\\
\>[2]\AgdaFunction{subtract}\AgdaSpace{}%
\AgdaBound{s}\AgdaSpace{}%
\AgdaBound{x}\AgdaSpace{}%
\AgdaBound{y}\AgdaSpace{}%
\AgdaSymbol{=}\AgdaSpace{}%
\AgdaField{add}\AgdaSpace{}%
\AgdaBound{s}\AgdaSpace{}%
\AgdaBound{x}\AgdaSpace{}%
\AgdaSymbol{(}\AgdaField{negate}\AgdaSpace{}%
\AgdaBound{s}\AgdaSpace{}%
\AgdaBound{y}\AgdaSymbol{)}\<%
\end{code}        
\end{AgdaMultiCode}
\caption{For Agda types.}
\label{fig:integer-agda}
\end{subfigure}
\end{minipage}
\hspace{\fill}
\begin{minipage}{0.5\textwidth}
\begin{subfigure}{\linewidth}
\begin{AgdaMultiCode}
\begin{code}%
\>[0]\AgdaKeyword{record}\AgdaSpace{}%
\AgdaRecord{IntStructure}\AgdaSpace{}%
\AgdaSymbol{(}\AgdaBound{A}\AgdaSpace{}%
\AgdaSymbol{:}\AgdaSpace{}%
\AgdaDatatype{TyExpr}\AgdaSpace{}%
\AgdaGeneralizable{m}\AgdaSymbol{)}\AgdaSpace{}%
\AgdaSymbol{:}\AgdaSpace{}%
\AgdaPrimitive{Set}\AgdaSpace{}%
\AgdaKeyword{where}\<%
\\
\>[0][@{}l@{\AgdaIndent{0}}]%
\>[2]\AgdaKeyword{field}\AgdaSpace{}%
\AgdaField{add}\AgdaSpace{}%
\AgdaField{negate}\AgdaSpace{}%
\AgdaSymbol{:}\AgdaSpace{}%
\AgdaDatatype{TmExpr}\AgdaSpace{}%
\AgdaBound{m}\<%
\end{code}
\begin{code}[hide]%
\>[0]\AgdaKeyword{open}\AgdaSpace{}%
\AgdaModule{IntStructure}\<%
\end{code}
\begin{code}%
\>[0]\AgdaFunction{subtract}\AgdaSpace{}%
\AgdaSymbol{:}\AgdaSpace{}%
\AgdaSymbol{(}\AgdaRecord{IntStructure}\AgdaSpace{}%
\AgdaBound{A}\AgdaSymbol{)}\AgdaSpace{}%
\AgdaSymbol{→}\AgdaSpace{}%
\AgdaDatatype{TmExpr}\AgdaSpace{}%
\AgdaGeneralizable{m}\<%
\\
\>[0]\AgdaFunction{subtract}\AgdaSpace{}%
\AgdaBound{s}\AgdaSpace{}%
\AgdaSymbol{=}%
\>[142I]\AgdaOperator{\AgdaInductiveConstructor{lam[}}\AgdaSpace{}%
\AgdaString{"a"}\AgdaSpace{}%
\AgdaOperator{\AgdaInductiveConstructor{∈}}%
\>[41]\AgdaBound{A}\AgdaSpace{}%
\AgdaOperator{\AgdaInductiveConstructor{]}}\AgdaSpace{}%
\AgdaOperator{\AgdaInductiveConstructor{lam[}}\AgdaSpace{}%
\AgdaString{"b"}\AgdaSpace{}%
\AgdaOperator{\AgdaInductiveConstructor{∈}}\AgdaSpace{}%
\AgdaBound{A}\AgdaSpace{}%
\AgdaOperator{\AgdaInductiveConstructor{]}}\<%
\\
\>[.][@{}l@{}]\<[142I]%
\>[29]\AgdaField{add}\AgdaSpace{}%
\AgdaBound{s}\AgdaSpace{}%
\AgdaOperator{\AgdaInductiveConstructor{∙}}\AgdaSpace{}%
\AgdaFunction{svar}\AgdaSpace{}%
\AgdaString{"a"}\AgdaSpace{}%
\AgdaOperator{\AgdaInductiveConstructor{∙}}\AgdaSpace{}%
\AgdaSymbol{(}\AgdaField{negate}\AgdaSpace{}%
\AgdaBound{s}\AgdaSpace{}%
\AgdaOperator{\AgdaInductiveConstructor{∙}}\AgdaSpace{}%
\AgdaFunction{svar}\AgdaSpace{}%
\AgdaString{"b"}\AgdaSymbol{)}\<%
\end{code}
\end{AgdaMultiCode}
\caption{For Sikkel types.
}
\label{fig:integer-sikkel}
\end{subfigure}
\end{minipage}
\caption{Minimalistic interface for integers and implementation of subtraction.}
\end{figure}

The function \AF{subtract} operates on an abstract type \AB{A}, with only the operations exposed in the \AR{IntStructure} interface.
Because of this, it will satisfy a form of \emph{representation independence}: its behavior will not depend on the underlying representation of the abstract type $A$, but only on the behavior exposed through the interface.
This is traditionally formalized as a form of parametricity: for any relation $R$ between two types implementing \AR{IntStructure}, $R$ will be preserved by \AF{subtract} if it is preserved by the operations \AFi{add} and \AFi{negate} \cite{reynolds83-types}.

\begin{code}[hide]%
\>[0]\AgdaKeyword{module}\AgdaSpace{}%
\AgdaModule{⋀-Diagram}\AgdaSpace{}%
\AgdaSymbol{(}\AgdaBound{T}\AgdaSpace{}%
\AgdaSymbol{:}\AgdaSpace{}%
\AgdaRecord{Ty}\AgdaSpace{}%
\AgdaSymbol{\{}\AgdaArgument{C}\AgdaSpace{}%
\AgdaSymbol{=}\AgdaSpace{}%
\AgdaFunction{M.⋀}\AgdaSymbol{\}}\AgdaSpace{}%
\AgdaFunction{M.◇}\AgdaSymbol{)}\AgdaSpace{}%
\AgdaKeyword{where}\<%
\end{code}
\newcommand{\paramTyRel}{%
  \begin{code}[hide]%
\>[0][@{}l@{\AgdaIndent{1}}]%
\>[2]\AgdaFunction{\AgdaUnderscore{}}\AgdaSpace{}%
\AgdaSymbol{:}\AgdaSpace{}%
\AgdaPrimitive{Set}\<%
\\
\>[2]\AgdaSymbol{\AgdaUnderscore{}}%
\>[172I]\AgdaSymbol{=}\<%
\end{code}%
  \begin{code}[inline*]%
\>[.][@{}l@{}]\<[172I]%
\>[4]\AgdaBound{T}\AgdaSpace{}%
\AgdaOperator{\AgdaField{⟨}}\AgdaSpace{}%
\AgdaInductiveConstructor{relation}\AgdaSpace{}%
\AgdaOperator{\AgdaField{,}}\AgdaSpace{}%
\AgdaInductiveConstructor{tt}\AgdaSpace{}%
\AgdaOperator{\AgdaField{⟩}}\<%
\end{code}%
}%
\newcommand{\paramTyLeft}{%
  \begin{code}[hide]%
\>[2]\AgdaFunction{\AgdaUnderscore{}}\AgdaSpace{}%
\AgdaSymbol{:}\AgdaSpace{}%
\AgdaPrimitive{Set}\<%
\\
\>[2]\AgdaSymbol{\AgdaUnderscore{}}%
\>[180I]\AgdaSymbol{=}\<%
\end{code}%
  \begin{code}[inline*]%
\>[.][@{}l@{}]\<[180I]%
\>[4]\AgdaBound{T}\AgdaSpace{}%
\AgdaOperator{\AgdaField{⟨}}\AgdaSpace{}%
\AgdaInductiveConstructor{left}\AgdaSpace{}%
\AgdaOperator{\AgdaField{,}}\AgdaSpace{}%
\AgdaInductiveConstructor{tt}\AgdaSpace{}%
\AgdaOperator{\AgdaField{⟩}}\<%
\end{code}%
}%
\newcommand{\paramTyRight}{%
  \begin{code}[hide]%
\>[2]\AgdaFunction{\AgdaUnderscore{}}\AgdaSpace{}%
\AgdaSymbol{:}\AgdaSpace{}%
\AgdaPrimitive{Set}\<%
\\
\>[2]\AgdaSymbol{\AgdaUnderscore{}}%
\>[188I]\AgdaSymbol{=}\<%
\end{code}%
  \begin{code}[inline*]%
\>[.][@{}l@{}]\<[188I]%
\>[4]\AgdaBound{T}\AgdaSpace{}%
\AgdaOperator{\AgdaField{⟨}}\AgdaSpace{}%
\AgdaInductiveConstructor{right}\AgdaSpace{}%
\AgdaOperator{\AgdaField{,}}\AgdaSpace{}%
\AgdaInductiveConstructor{tt}\AgdaSpace{}%
\AgdaOperator{\AgdaField{⟩}}\<%
\end{code}%
}%
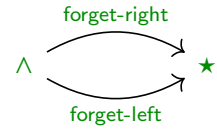
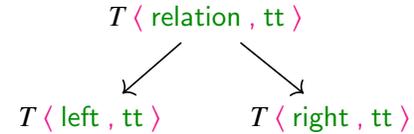
\begin{wrapfigure}[16]{R}{0.35\textwidth}
  \vspace{-4pt}
  \begin{subfigure}{\linewidth}
    \centering
    \begin{tikzcd}
      \wsmode{} \arrow[rr, "\AIC{forget-right}", bend left] \arrow[rr, "\AIC{forget-left}"', bend right] &  & \stmode{}
    \end{tikzcd}
    \caption{Mode theory.\\ \phantom{a}}
    \label{fig:param-mode-theory}
  \end{subfigure}
  \begin{subfigure}{\linewidth}
    \centering
    \begin{tikzcd}[column sep=-3em]
      & \text{\paramTyRel{}} \arrow[ld] \arrow[rd] & \\
      \text{\paramTyLeft{}} & & \text{\paramTyRight{}}
    \end{tikzcd}
    \caption{A semantic type $T$ in the empty context over base category \AF{M.⋀}.}
    \label{fig:param-diagram}
  \end{subfigure}
  \caption{Mode theory for parametricity.}
  \label{fig:parametricity}
\end{wrapfigure}

In Sikkel, we support such results in a parametric type theory with mode theory given in Figure~\ref{fig:param-mode-theory}.
Mode \stmode{} is again the trivial mode, interpreted as the trivial base category.
There is also a parametric mode, denoted by a wedge (\wsmode{}) and interpreted as the walking cospan category \AF{M.⋀}.
This category has 3 objects \AIC{left}, \AIC{right} and \AIC{relation} and 2 non-trivial morphisms: one from both \AIC{left} and \AIC{right} to \AIC{relation}.
Consequently, a presheaf over \AF{M.⋀} is a diagram as depicted in Figure~\ref{fig:param-diagram}, i.e.\ a span of types.
We can think of such a presheaf as consisting of two Agda types, a left one and a right one, together with a relation between them which is represented as the type of related pairs.
The 2 restriction maps will map every related pair to its left and right components.

We can reformulate the interface for MSTT types and implement subtraction as shown in Figure~\ref{fig:integer-sikkel}.
Since MSTT is extrinsically typed, we cannot directly require \AFi{add} and \AFi{negate} to have a specific type.
The actual definition of \AR{IntStructure} therefore contains two more fields requiring that the types inferred for \AFi{add} and \AFi{negate} should be equal to respectively
\begin{code}[hide]%
\>[0]\AgdaFunction{\AgdaUnderscore{}}\AgdaSpace{}%
\AgdaSymbol{:}\AgdaSpace{}%
\AgdaDatatype{TyExpr}\AgdaSpace{}%
\AgdaGeneralizable{m}\AgdaSpace{}%
\AgdaSymbol{→}\AgdaSpace{}%
\AgdaDatatype{TyExpr}\AgdaSpace{}%
\AgdaGeneralizable{m}\<%
\\
\>[0]\AgdaSymbol{\AgdaUnderscore{}}%
\>[200I]\AgdaSymbol{=}\AgdaSpace{}%
\AgdaSymbol{λ}\AgdaSpace{}%
\AgdaBound{A}\AgdaSpace{}%
\AgdaSymbol{→}\<%
\end{code}
\begin{code}[inline*]%
\>[.][@{}l@{}]\<[200I]%
\>[2]\AgdaBound{A}\AgdaSpace{}%
\AgdaOperator{\AgdaInductiveConstructor{⇛}}\AgdaSpace{}%
\AgdaBound{A}\AgdaSpace{}%
\AgdaOperator{\AgdaInductiveConstructor{⇛}}\AgdaSpace{}%
\AgdaBound{A}\<%
\end{code}
and
\begin{code}[hide]%
\>[0]\AgdaFunction{\AgdaUnderscore{}}\AgdaSpace{}%
\AgdaSymbol{:}\AgdaSpace{}%
\AgdaDatatype{TyExpr}\AgdaSpace{}%
\AgdaGeneralizable{m}\AgdaSpace{}%
\AgdaSymbol{→}\AgdaSpace{}%
\AgdaDatatype{TyExpr}\AgdaSpace{}%
\AgdaGeneralizable{m}\<%
\\
\>[0]\AgdaSymbol{\AgdaUnderscore{}}%
\>[214I]\AgdaSymbol{=}\AgdaSpace{}%
\AgdaSymbol{λ}\AgdaSpace{}%
\AgdaBound{A}\AgdaSpace{}%
\AgdaSymbol{→}\<%
\end{code}
\begin{code}[inline]%
\>[.][@{}l@{}]\<[214I]%
\>[2]\AgdaBound{A}\AgdaSpace{}%
\AgdaOperator{\AgdaInductiveConstructor{⇛}}\AgdaSpace{}%
\AgdaBound{A}\<%
\end{code}.

This application adds a new type former \AIC{FromRel} to MSTT, which allows a user to create new built-in types in mode \wsmode{} that represent two user-specified Agda types with a relation between them.
In our case, we add the built-in type \AF{ℤ} that is interpreted as \nxn{} and \signxn{} with the relation \AF{∼} expressing that a pair of natural numbers and a signed natural number represent the same integer.
When interpreting this type \AF{ℤ} in the presheaf model, the Agda type at the top of the diagram in Figure~\ref{fig:param-diagram} will be
\begin{code}[hide]%
\>[0]\AgdaFunction{\AgdaUnderscore{}}\AgdaSpace{}%
\AgdaSymbol{:}\AgdaSpace{}%
\AgdaPrimitive{Set}\<%
\\
\>[0]\AgdaSymbol{\AgdaUnderscore{}}%
\>[222I]\AgdaSymbol{=}\<%
\end{code}
\begin{code}[inline]%
\>[.][@{}l@{}]\<[222I]%
\>[2]\AgdaFunction{Σ[}\AgdaSpace{}%
\AgdaBound{(}\AgdaBound{i}\AgdaSpace{}%
\AgdaOperator{\AgdaInductiveConstructor{,}}\AgdaSpace{}%
\AgdaBound{j}\AgdaBound{)}\AgdaSpace{}%
\AgdaFunction{∈}\AgdaSpace{}%
\AgdaSymbol{(}\AgdaDatatype{ℕ}\AgdaSpace{}%
\AgdaOperator{\AgdaFunction{×}}\AgdaSpace{}%
\AgdaDatatype{ℕ}\AgdaSymbol{)}\AgdaSpace{}%
\AgdaOperator{\AgdaFunction{×}}\AgdaSpace{}%
\AgdaSymbol{(}\AgdaDatatype{Sign}\AgdaSpace{}%
\AgdaOperator{\AgdaFunction{×}}\AgdaSpace{}%
\AgdaDatatype{ℕ}\AgdaSymbol{)}\AgdaSpace{}%
\AgdaFunction{]}\AgdaSpace{}%
\AgdaBound{i}\AgdaSpace{}%
\AgdaOperator{\AgdaDatatype{∼}}\AgdaSpace{}%
\AgdaBound{j}\<%
\end{code},
i.e.\ the type of related pairs (using Agda's \AR{Σ} types).
Furthermore, new term formers allow to construct MSTT functions between these built-in types given the corresponding Agda functions for the left and right types and a proof that these functions preserve the built-in relations.
This means that we can create an implementation \AF{ℤ-int} of \AR{IntStructure} for \AF{ℤ} from implementations of addition and negation for \nxn{} and \signxn{} and proofs that they respect the relation \AF{∼}.
\begin{code}[hide]%
\>[0]\AgdaFunction{ℤ-int}\AgdaSpace{}%
\AgdaSymbol{:}\AgdaSpace{}%
\AgdaRecord{IntStructure}\AgdaSpace{}%
\AgdaFunction{ℤ}\<%
\\
\>[0]\AgdaField{IntStructure.add}\AgdaSpace{}%
\AgdaFunction{ℤ-int}\AgdaSpace{}%
\AgdaSymbol{=}\AgdaSpace{}%
\AgdaInductiveConstructor{from-rel2}\AgdaSpace{}%
\AgdaInductiveConstructor{ℤ-code}\AgdaSpace{}%
\AgdaInductiveConstructor{ℤ-code}\AgdaSpace{}%
\AgdaInductiveConstructor{ℤ-code}\AgdaSpace{}%
\AgdaOperator{\AgdaFunction{\AgdaUnderscore{}+D\AgdaUnderscore{}}}\AgdaSpace{}%
\AgdaOperator{\AgdaFunction{\AgdaUnderscore{}+S\AgdaUnderscore{}}}\AgdaSpace{}%
\AgdaFunction{+-preserves-∼}\<%
\\
\>[0]\AgdaField{IntStructure.negate}\AgdaSpace{}%
\AgdaFunction{ℤ-int}\AgdaSpace{}%
\AgdaSymbol{=}\AgdaSpace{}%
\AgdaInductiveConstructor{from-rel1}\AgdaSpace{}%
\AgdaInductiveConstructor{ℤ-code}\AgdaSpace{}%
\AgdaInductiveConstructor{ℤ-code}\AgdaSpace{}%
\AgdaFunction{negateD}\AgdaSpace{}%
\AgdaFunction{negateS}\AgdaSpace{}%
\AgdaFunction{negate-preserves-∼}\<%
\end{code}

Recalling the implementation of function types in a presheaf model, the interpretation of
\begin{code}[hide]%
\>[0]\AgdaFunction{\AgdaUnderscore{}}\AgdaSpace{}%
\AgdaSymbol{:}\AgdaSpace{}%
\AgdaDatatype{TmExpr}\AgdaSpace{}%
\AgdaInductiveConstructor{⋀}\<%
\\
\>[0]\AgdaSymbol{\AgdaUnderscore{}}%
\>[261I]\AgdaSymbol{=}\<%
\end{code}
\begin{code}[inline*]%
\>[.][@{}l@{}]\<[261I]%
\>[2]\AgdaFunction{subtract}\AgdaSpace{}%
\AgdaFunction{ℤ-int}\<%
\end{code}
in the semantic layer consists of the implementation of subtraction for \nxn{} and \signxn{}, together with a proof that they respect the relation \AF{∼}.
We can obtain the subtraction implementations at the Agda level by using the modalities from the mode theory.
The \AIC{forget-right} modality turns a \wsmode{}-type into a \stmode{}-type by forgetting about the right type and the relation.
Hence we can write an MSTT function
\begin{code}%
\>[0]\AgdaFunction{subtract★-left}\AgdaSpace{}%
\AgdaSymbol{:}\AgdaSpace{}%
\AgdaDatatype{TmExpr}\AgdaSpace{}%
\AgdaInductiveConstructor{★}\<%
\\
\>[0]\AgdaFunction{subtract★-left}\AgdaSpace{}%
\AgdaSymbol{=}\AgdaSpace{}%
\AgdaOperator{\AgdaFunction{lam[}}\AgdaSpace{}%
\AgdaInductiveConstructor{forget-right}\AgdaSpace{}%
\AgdaOperator{\AgdaFunction{∣}}\AgdaSpace{}%
\AgdaString{"x"}\AgdaSpace{}%
\AgdaOperator{\AgdaFunction{∈}}\AgdaSpace{}%
\AgdaFunction{ℤ}\AgdaSpace{}%
\AgdaOperator{\AgdaFunction{]}}\AgdaSpace{}%
\AgdaOperator{\AgdaFunction{lam[}}\AgdaSpace{}%
\AgdaInductiveConstructor{forget-right}\AgdaSpace{}%
\AgdaOperator{\AgdaFunction{∣}}\AgdaSpace{}%
\AgdaString{"y"}\AgdaSpace{}%
\AgdaOperator{\AgdaFunction{∈}}\AgdaSpace{}%
\AgdaFunction{ℤ}\AgdaSpace{}%
\AgdaOperator{\AgdaFunction{]}}\<%
\\
\>[0][@{}l@{\AgdaIndent{0}}]%
\>[2]\AgdaOperator{\AgdaInductiveConstructor{mod⟨}}\AgdaSpace{}%
\AgdaInductiveConstructor{forget-right}\AgdaSpace{}%
\AgdaOperator{\AgdaInductiveConstructor{⟩}}\AgdaSpace{}%
\AgdaSymbol{(}\AgdaFunction{subtract}\AgdaSpace{}%
\AgdaFunction{ℤ-int}\AgdaSpace{}%
\AgdaOperator{\AgdaInductiveConstructor{∙}}\AgdaSpace{}%
\AgdaFunction{svar}\AgdaSpace{}%
\AgdaString{"x"}\AgdaSpace{}%
\AgdaOperator{\AgdaInductiveConstructor{∙}}\AgdaSpace{}%
\AgdaFunction{svar}\AgdaSpace{}%
\AgdaString{"y"}\AgdaSymbol{)}\<%
\end{code}
of type
\begin{code}[hide]%
\>[0]\AgdaFunction{\AgdaUnderscore{}}\AgdaSpace{}%
\AgdaSymbol{:}\AgdaSpace{}%
\AgdaDatatype{TyExpr}\AgdaSpace{}%
\AgdaInductiveConstructor{★}\<%
\\
\>[0]\AgdaSymbol{\AgdaUnderscore{}}%
\>[294I]\AgdaSymbol{=}\<%
\end{code}
\begin{code}[inline*]%
\>[.][@{}l@{}]\<[294I]%
\>[2]\AgdaOperator{\AgdaInductiveConstructor{⟨}}\AgdaSpace{}%
\AgdaInductiveConstructor{forget-right}\AgdaSpace{}%
\AgdaOperator{\AgdaInductiveConstructor{∣}}\AgdaSpace{}%
\AgdaFunction{ℤ}\AgdaSpace{}%
\AgdaOperator{\AgdaInductiveConstructor{⟩}}\AgdaSpace{}%
\AgdaOperator{\AgdaInductiveConstructor{⇛}}\AgdaSpace{}%
\AgdaOperator{\AgdaInductiveConstructor{⟨}}\AgdaSpace{}%
\AgdaInductiveConstructor{forget-right}\AgdaSpace{}%
\AgdaOperator{\AgdaInductiveConstructor{∣}}\AgdaSpace{}%
\AgdaFunction{ℤ}\AgdaSpace{}%
\AgdaOperator{\AgdaInductiveConstructor{⟩}}\AgdaSpace{}%
\AgdaOperator{\AgdaInductiveConstructor{⇛}}\AgdaSpace{}%
\AgdaOperator{\AgdaInductiveConstructor{⟨}}\AgdaSpace{}%
\AgdaInductiveConstructor{forget-right}\AgdaSpace{}%
\AgdaOperator{\AgdaInductiveConstructor{∣}}\AgdaSpace{}%
\AgdaFunction{ℤ}\AgdaSpace{}%
\AgdaOperator{\AgdaInductiveConstructor{⟩}}\<%
\end{code}
and use the sound type-checker and extraction to obtain a function \AF{subtract-ℕ×ℕ} for \nxn{}, and similarly \AF{subtract-Sign×ℕ}.%
\footnote{It may seem laborious to define two analogues of \AF{subtract★-left} for every parametric function we define. Instead we can define an applicative operator for arbitrary modalities \AB{μ} and arbitrary functions \AB{f} with $k$ arguments. The instance for \AB{μ} $=$ \AIC{forget-right} and \AB{f} $=$ \AF{subtract} \AF{ℤ-int} and $k = 2$ yields \AF{subtract★-left}.}
Finally, we have a proof that these functions respect the relation \AF{∼}.
\begin{code}[hide]%
\>[0]\AgdaFunction{subtract★-right}\AgdaSpace{}%
\AgdaSymbol{:}\AgdaSpace{}%
\AgdaDatatype{TmExpr}\AgdaSpace{}%
\AgdaInductiveConstructor{★}\<%
\\
\>[0]\AgdaFunction{subtract★-right}\AgdaSpace{}%
\AgdaSymbol{=}\AgdaSpace{}%
\AgdaFunction{liftA2}\AgdaSpace{}%
\AgdaInductiveConstructor{forget-left}\AgdaSpace{}%
\AgdaOperator{\AgdaInductiveConstructor{∙}}\AgdaSpace{}%
\AgdaSymbol{(}\AgdaOperator{\AgdaInductiveConstructor{mod⟨}}\AgdaSpace{}%
\AgdaInductiveConstructor{forget-left}\AgdaSpace{}%
\AgdaOperator{\AgdaInductiveConstructor{⟩}}\AgdaSpace{}%
\AgdaFunction{subtract}\AgdaSpace{}%
\AgdaFunction{ℤ-int}\AgdaSymbol{)}\<%
\\
\\[\AgdaEmptyExtraSkip]%
\>[0]\AgdaFunction{subtract-ℕ×ℕ}\AgdaSpace{}%
\AgdaSymbol{:}\AgdaSpace{}%
\AgdaFunction{DiffNat}\AgdaSpace{}%
\AgdaSymbol{→}\AgdaSpace{}%
\AgdaFunction{DiffNat}\AgdaSpace{}%
\AgdaSymbol{→}\AgdaSpace{}%
\AgdaFunction{DiffNat}\<%
\\
\>[0]\AgdaFunction{subtract-ℕ×ℕ}\AgdaSpace{}%
\AgdaSymbol{=}\AgdaSpace{}%
\AgdaField{extract-term}\AgdaSpace{}%
\AgdaSymbol{(}\AgdaOperator{\AgdaFunction{⟦}}\AgdaSpace{}%
\AgdaFunction{subtract★-left}\AgdaSpace{}%
\AgdaOperator{\AgdaFunction{⟧tm-in}}\AgdaSpace{}%
\AgdaInductiveConstructor{◇}\AgdaSymbol{)}\<%
\\
\\[\AgdaEmptyExtraSkip]%
\>[0]\AgdaFunction{subtract-Sign×ℕ}\AgdaSpace{}%
\AgdaSymbol{:}\AgdaSpace{}%
\AgdaFunction{SignNat}\AgdaSpace{}%
\AgdaSymbol{→}\AgdaSpace{}%
\AgdaFunction{SignNat}\AgdaSpace{}%
\AgdaSymbol{→}\AgdaSpace{}%
\AgdaFunction{SignNat}\<%
\\
\>[0]\AgdaFunction{subtract-Sign×ℕ}\AgdaSpace{}%
\AgdaSymbol{=}\AgdaSpace{}%
\AgdaField{extract-term}\AgdaSpace{}%
\AgdaSymbol{(}\AgdaOperator{\AgdaFunction{⟦}}\AgdaSpace{}%
\AgdaFunction{subtract★-right}\AgdaSpace{}%
\AgdaOperator{\AgdaFunction{⟧tm-in}}\AgdaSpace{}%
\AgdaInductiveConstructor{◇}\AgdaSymbol{)}\<%
\end{code}
\begin{code}%
\>[0]\AgdaFunction{subtract-∼}\AgdaSpace{}%
\AgdaSymbol{:}\AgdaSpace{}%
\AgdaSymbol{∀}\AgdaSpace{}%
\AgdaBound{i1}\AgdaSpace{}%
\AgdaBound{j1}\AgdaSpace{}%
\AgdaBound{i2}\AgdaSpace{}%
\AgdaBound{j2}\AgdaSpace{}%
\AgdaSymbol{→}\AgdaSpace{}%
\AgdaBound{i1}\AgdaSpace{}%
\AgdaOperator{\AgdaDatatype{∼}}\AgdaSpace{}%
\AgdaBound{i2}\AgdaSpace{}%
\AgdaSymbol{→}\AgdaSpace{}%
\AgdaBound{j1}\AgdaSpace{}%
\AgdaOperator{\AgdaDatatype{∼}}\AgdaSpace{}%
\AgdaBound{j2}\AgdaSpace{}%
\AgdaSymbol{→}\AgdaSpace{}%
\AgdaFunction{subtract-ℕ×ℕ}\AgdaSpace{}%
\AgdaBound{i1}\AgdaSpace{}%
\AgdaBound{j1}\AgdaSpace{}%
\AgdaOperator{\AgdaDatatype{∼}}\AgdaSpace{}%
\AgdaFunction{subtract-Sign×ℕ}\AgdaSpace{}%
\AgdaBound{i2}\AgdaSpace{}%
\AgdaBound{j2}\<%
\\
\>[0]\AgdaFunction{subtract-∼}\AgdaSpace{}%
\AgdaBound{i1}\AgdaSpace{}%
\AgdaBound{j1}\AgdaSpace{}%
\AgdaBound{i2}\AgdaSpace{}%
\AgdaBound{j2}\AgdaSpace{}%
\AgdaBound{ri}\AgdaSpace{}%
\AgdaBound{rj}\AgdaSpace{}%
\AgdaSymbol{=}\AgdaSpace{}%
\AgdaField{proj₂}\AgdaSpace{}%
\AgdaSymbol{(((}\AgdaOperator{\AgdaFunction{⟦}}\AgdaSpace{}%
\AgdaFunction{subtract}\AgdaSpace{}%
\AgdaFunction{ℤ-int}\AgdaSpace{}%
\AgdaOperator{\AgdaFunction{⟧tm-in}}\AgdaSpace{}%
\AgdaInductiveConstructor{◇}\AgdaSymbol{)}\AgdaSpace{}%
\AgdaOperator{\AgdaField{⟨}}\AgdaSpace{}%
\AgdaInductiveConstructor{relation}\AgdaSpace{}%
\AgdaOperator{\AgdaField{,}}\AgdaSpace{}%
\AgdaInductiveConstructor{tt}\AgdaSpace{}%
\AgdaOperator{\AgdaField{⟩'}}\<%
\\
\>[0][@{}l@{\AgdaIndent{0}}]%
\>[4]\AgdaOperator{\AgdaField{\$⟨}}\AgdaSpace{}%
\AgdaInductiveConstructor{relation-id}\AgdaSpace{}%
\AgdaOperator{\AgdaField{,}}\AgdaSpace{}%
\AgdaInductiveConstructor{refl}\AgdaSpace{}%
\AgdaOperator{\AgdaField{⟩}}\AgdaSpace{}%
\AgdaSymbol{((}\AgdaBound{i1}\AgdaSpace{}%
\AgdaOperator{\AgdaInductiveConstructor{,}}\AgdaSpace{}%
\AgdaBound{i2}\AgdaSymbol{)}\AgdaSpace{}%
\AgdaOperator{\AgdaInductiveConstructor{,}}\AgdaSpace{}%
\AgdaBound{ri}\AgdaSymbol{))}\AgdaSpace{}%
\AgdaOperator{\AgdaField{\$⟨}}\AgdaSpace{}%
\AgdaInductiveConstructor{relation-id}\AgdaSpace{}%
\AgdaOperator{\AgdaField{,}}\AgdaSpace{}%
\AgdaInductiveConstructor{refl}\AgdaSpace{}%
\AgdaOperator{\AgdaField{⟩}}\AgdaSpace{}%
\AgdaSymbol{((}\AgdaBound{j1}\AgdaSpace{}%
\AgdaOperator{\AgdaInductiveConstructor{,}}\AgdaSpace{}%
\AgdaBound{j2}\AgdaSymbol{)}\AgdaSpace{}%
\AgdaOperator{\AgdaInductiveConstructor{,}}\AgdaSpace{}%
\AgdaBound{rj}\AgdaSymbol{))}\<%
\end{code}


%% file: content-tex/discussion-relatedwork.tex
\section{Discussion, Related and Future Work}
\label{sec:discussion}

\adapted{One of Sikkel's design choices is to make its syntactic layer extrinsically typed, i.e.\ the Agda type \ADT{TmExpr} is not indexed by a context or type of the object theory.
  A downside of this approach is that a user of the library cannot rely on Agda to indicate the expected types of subterms when writing programs in Sikkel.
  However, an intrinsically typed syntax would require explicit casting by the programmer whenever an equivalence of modalities is used, something which is now handled by Sikkel's type checker.
  Furthermore, an extrinsically typed syntax allows for easier implementation of named variables.
}

Sikkel's goal of extending type theories with primitives like guarded recursion or parametricity can be approached in many other ways as well.

One can simply modify an existing implementation of type theory, as was done in Agda-parametric \cite{nuyts17-parametric}, cubical Agda \cite{vezzosi-cubical-2021} or guarded cubical Agda \cite{veltri20-pi-calculus}.
Trading type-checking performance for flexibility, one can alternatively rely on postulates and rewrite rules \cite{cockx-taming-2021}.
Compared to Sikkel, such approaches have benefits and downsides: the result can be very user-friendly (since both syntax and typing rules can be controlled) but implementation errors may compromise soundness of the theory, it is not possible to restrict the effect of the modifications to a part of a larger codebase and it can be hard to deviate strongly from the structural rules of the system, as required for some type theory extensions, e.g. \cite{param-app,abstract-atomicity}.

Some other proposals are more closely related to Sikkel.
Veltri and van der Weide~\cite{veltri19-guarded} present an implementation of guarded recursion, which is closely related but uses a less general model and relies on Agda's experimental support for sized types.
Jaber et al.~\cite{jaber12-extending} translate an extension of the calculus of constructions (CoC) to presheaves over an arbitrary preorder in regular CoC.
Lacking modes (and the trivial mode), they do not support extracting code in the extension to regular CoC.
Bach Poulsen et al.\ have implemented guarded type theory as ordered families of equivalences (OFEs) \cite{bachpoulsen17-intrinsically}.\footnote{Available at \url{https://github.com/metaborg/mj.agda/tree/develop/src/Categorical}.}
These are also used in the model of the Iris logic \cite{jung-iris-2018} which comes with support for interactive proofs in the Iris Proof Mode \cite{krebbers-interactive-2017}. 
Finally, presheaves have been formalized for different purposes than Sikkel \cite{bickford18-formalizing,hu21-agdacategories,timany16-category}, but these formalizations do not provide all of Sikkel's features.

In future work, we plan to extend Sikkel with support for dependent types.
\adapted{As mentioned in Section~\ref{sec:presheaves}, our formalization of presheaf models already anticipates this.
  However, with a dependently-typed syntax the interpretation functions for contexts, types and terms become mutually recursive and it is a significant challenge to satisfy Agda's termination checker.
  Furthermore, the addition of a Hofmann-Streicher universe to our presheaf model turns out to be non-trivial.}
Apart from the applications worked out in this paper, we intend to use Sikkel for other type theory extensions like nominal \cite{freshmltt}, directed \cite{weaver-licata-dua} or univalent type theory \cite{cubical}, as well as for the implementation of additional primitives like the transpension type \cite{nuyts-transpension-2020}

\adapted{
\paragraph{Acknowledgements}
Joris Ceulemans and Andreas Nuyts hold a PhD Fellowship and a Postdoctoral Fellowship, respectively, from the Research Foundation -- Flanders (FWO).
This work was partially supported by the research project G0G0519N of the Research Foundation -- Flanders (FWO).
}

%% file: content-lagda/appendix-presheaves.tex
\section{Presheaf Semantics of Guarded Recursion}
\label{sec:psh-sem-guard}

\begin{code}[hide]%
\>[0]\AgdaKeyword{module}\AgdaSpace{}%
\AgdaModule{appendix-presheaves}\AgdaSpace{}%
\AgdaKeyword{where}\<%
\\
\\[\AgdaEmptyExtraSkip]%
\>[0]\AgdaKeyword{open}\AgdaSpace{}%
\AgdaKeyword{import}\AgdaSpace{}%
\AgdaModule{Data.Nat}\<%
\\
\>[0]\AgdaKeyword{open}\AgdaSpace{}%
\AgdaKeyword{import}\AgdaSpace{}%
\AgdaModule{Data.Nat.Properties}\AgdaSpace{}%
\AgdaKeyword{using}\AgdaSpace{}%
\AgdaSymbol{(}\AgdaFunction{≤-refl}\AgdaSymbol{;}\AgdaSpace{}%
\AgdaFunction{≤-trans}\AgdaSymbol{)}\<%
\\
\>[0]\AgdaKeyword{open}\AgdaSpace{}%
\AgdaKeyword{import}\AgdaSpace{}%
\AgdaModule{Data.Unit}\AgdaSpace{}%
\AgdaKeyword{using}\AgdaSpace{}%
\AgdaSymbol{(}\AgdaRecord{⊤}\AgdaSymbol{;}\AgdaSpace{}%
\AgdaInductiveConstructor{tt}\AgdaSymbol{)}\<%
\\
\>[0]\AgdaKeyword{open}\AgdaSpace{}%
\AgdaKeyword{import}\AgdaSpace{}%
\AgdaModule{Data.Vec}\<%
\\
\>[0]\AgdaKeyword{open}\AgdaSpace{}%
\AgdaKeyword{import}\AgdaSpace{}%
\AgdaModule{Relation.Binary.PropositionalEquality}\<%
\\
\\[\AgdaEmptyExtraSkip]%
\>[0]\AgdaKeyword{open}\AgdaSpace{}%
\AgdaKeyword{import}\AgdaSpace{}%
\AgdaModule{Model.BaseCategory}\AgdaSpace{}%
\AgdaKeyword{hiding}\AgdaSpace{}%
\AgdaSymbol{(}\AgdaFunction{ω}\AgdaSymbol{;}\AgdaSpace{}%
\AgdaFunction{★}\AgdaSymbol{)}\<%
\\
\>[0]\AgdaKeyword{open}\AgdaSpace{}%
\AgdaKeyword{import}\AgdaSpace{}%
\AgdaModule{Model.CwF-Structure}\<%
\\
\>[0]\AgdaKeyword{open}\AgdaSpace{}%
\AgdaKeyword{import}\AgdaSpace{}%
\AgdaModule{Model.Modality}\<%
\\
\\[\AgdaEmptyExtraSkip]%
\>[0]\AgdaKeyword{private}\AgdaSpace{}%
\AgdaKeyword{variable}\<%
\\
\>[0][@{}l@{\AgdaIndent{0}}]%
\>[2]\AgdaGeneralizable{C}\AgdaSpace{}%
\AgdaSymbol{:}\AgdaSpace{}%
\AgdaRecord{BaseCategory}\<%
\\
\>[2]\AgdaGeneralizable{Γ}\AgdaSpace{}%
\AgdaSymbol{:}\AgdaSpace{}%
\AgdaRecord{Ctx}\AgdaSpace{}%
\AgdaGeneralizable{C}\<%
\end{code}

In this appendix we take a closer look at the semantics of some of the primitives from Section~\ref{sec:appl-guard-recurs}.

The trivial mode corresponds to the trivial base category \AF{★} with one object and one (identity) morphism from that object to itself.
Presheaves over \AF{★} are essentially sets.
On the other hand, the base category \AF{ω} corresponding to the time-dependent mode has the natural numbers as objects and the type of morphisms from numbers $m$ to $n$ is Agda's inequality type
\begin{code}[hide]%
\>[0]\AgdaKeyword{module}\AgdaSpace{}%
\AgdaModule{\AgdaUnderscore{}}\AgdaSpace{}%
\AgdaSymbol{(}\AgdaBound{m}\AgdaSpace{}%
\AgdaBound{n}\AgdaSpace{}%
\AgdaSymbol{:}\AgdaSpace{}%
\AgdaDatatype{ℕ}\AgdaSymbol{)}\AgdaSpace{}%
\AgdaSymbol{(}\AgdaBound{ineq}\AgdaSpace{}%
\AgdaSymbol{:}\<%
\end{code}
\begin{code}[inline]%
\>[0][@{}l@{\AgdaIndent{1}}]%
\>[2]\AgdaBound{m}\AgdaSpace{}%
\AgdaOperator{\AgdaDatatype{≤}}\AgdaSpace{}%
\AgdaBound{n}\<%
\end{code}
\begin{code}[hide]%
\>[0][@{}l@{\AgdaIndent{2}}]%
\>[1]\AgdaSymbol{)}\AgdaSpace{}%
\AgdaSymbol{(}\AgdaBound{zn}\AgdaSpace{}%
\AgdaSymbol{:}\AgdaSpace{}%
\AgdaSymbol{∀}\AgdaSpace{}%
\AgdaSymbol{\{}\AgdaBound{n}\AgdaSymbol{\}}\AgdaSpace{}%
\AgdaSymbol{→}\<%
\end{code}.
This inequality type has two constructors \AIC{z≤n}\AS\ASy{:}
\begin{code}[inline*]%
\>[1][@{}l@{\AgdaIndent{1}}]%
\>[2]\AgdaNumber{0}\AgdaSpace{}%
\AgdaOperator{\AgdaDatatype{≤}}\AgdaSpace{}%
\AgdaBound{n}\<%
\end{code}
\begin{code}[hide]%
\>[2]\AgdaSymbol{)}\AgdaSpace{}%
\AgdaSymbol{(}\AgdaBound{ss}\AgdaSpace{}%
\AgdaSymbol{:}\AgdaSpace{}%
\AgdaSymbol{∀}\AgdaSpace{}%
\AgdaSymbol{\{}\AgdaBound{m}\AgdaSpace{}%
\AgdaBound{n}\AgdaSymbol{\}}\AgdaSpace{}%
\AgdaSymbol{→}\<%
\end{code}
and \AIC{s≤s}\AS\ASy{:}
\begin{code}[inline]%
\>[2]\AgdaBound{m}\AgdaSpace{}%
\AgdaOperator{\AgdaDatatype{≤}}\AgdaSpace{}%
\AgdaBound{n}\AgdaSpace{}%
\AgdaSymbol{→}\AgdaSpace{}%
\AgdaInductiveConstructor{suc}\AgdaSpace{}%
\AgdaBound{m}\AgdaSpace{}%
\AgdaOperator{\AgdaDatatype{≤}}\AgdaSpace{}%
\AgdaInductiveConstructor{suc}\AgdaSpace{}%
\AgdaBound{n}\<%
\end{code}
\begin{code}[hide]%
\>[2]\AgdaSymbol{)}\AgdaSpace{}%
\AgdaKeyword{where}\<%
\end{code}.
As such, a presheaf over \AF{ω} is a diagram with a shape as in \eqref{eq:diagram}.
The presheaf category over \AF{ω} is called the topos of trees and is a well-known model for guarded recursion \cite{birkedal12-first}.

As mentioned in Section~\ref{sec:presheaf-motivation}, we can implement the the later modality by shifting a diagram to the right and adding Agda's unit type to the front.
The corresponding lock operation is called \AF{◄} (earlier) and does the opposite: it shifts a diagram to the left.%
\begin{code}[hide]%
\>[0]\AgdaSymbol{\{-\#}\AgdaSpace{}%
\AgdaKeyword{NON\AgdaUnderscore{}COVERING}\AgdaSpace{}%
\AgdaSymbol{\#-\}}\<%
\\
\>[0]\AgdaFunction{ω}\AgdaSpace{}%
\AgdaSymbol{:}\AgdaSpace{}%
\AgdaRecord{BaseCategory}\<%
\\
\>[0]\AgdaField{BaseCategory.Ob}\AgdaSpace{}%
\AgdaFunction{ω}\AgdaSpace{}%
\AgdaSymbol{=}\AgdaSpace{}%
\AgdaDatatype{ℕ}\<%
\\
\>[0]\AgdaField{BaseCategory.Hom}\AgdaSpace{}%
\AgdaFunction{ω}\AgdaSpace{}%
\AgdaSymbol{=}\AgdaSpace{}%
\AgdaSymbol{λ}\AgdaSpace{}%
\AgdaBound{m}\AgdaSpace{}%
\AgdaBound{n}\AgdaSpace{}%
\AgdaSymbol{→}\AgdaSpace{}%
\AgdaBound{m}\AgdaSpace{}%
\AgdaOperator{\AgdaDatatype{≤}}\AgdaSpace{}%
\AgdaBound{n}\<%
\\
\>[0]\AgdaField{BaseCategory.hom-id}\AgdaSpace{}%
\AgdaFunction{ω}\AgdaSpace{}%
\AgdaSymbol{=}\AgdaSpace{}%
\AgdaFunction{≤-refl}\<%
\\
\>[0]\AgdaOperator{\AgdaField{BaseCategory.\AgdaUnderscore{}∙\AgdaUnderscore{}}}\AgdaSpace{}%
\AgdaFunction{ω}\AgdaSpace{}%
\AgdaSymbol{=}\AgdaSpace{}%
\AgdaSymbol{λ}\AgdaSpace{}%
\AgdaBound{m≤n}\AgdaSpace{}%
\AgdaBound{k≤m}\AgdaSpace{}%
\AgdaSymbol{→}\AgdaSpace{}%
\AgdaFunction{≤-trans}\AgdaSpace{}%
\AgdaBound{k≤m}\AgdaSpace{}%
\AgdaBound{m≤n}\<%
\end{code}
\begin{code}[hide]%
\>[0]\AgdaSymbol{\{-\#}\AgdaSpace{}%
\AgdaKeyword{NON\AgdaUnderscore{}COVERING}\AgdaSpace{}%
\AgdaSymbol{\#-\}}\<%
\end{code}
\begin{code}%
\>[0]\AgdaFunction{◄}\AgdaSpace{}%
\AgdaSymbol{:}\AgdaSpace{}%
\AgdaRecord{Ctx}\AgdaSpace{}%
\AgdaFunction{ω}\AgdaSpace{}%
\AgdaSymbol{→}\AgdaSpace{}%
\AgdaRecord{Ctx}\AgdaSpace{}%
\AgdaFunction{ω}\<%
\\
\>[0]\AgdaFunction{◄}\AgdaSpace{}%
\AgdaBound{Γ}\AgdaSpace{}%
\AgdaOperator{\AgdaField{⟨}}\AgdaSpace{}%
\AgdaBound{n}\AgdaSpace{}%
\AgdaOperator{\AgdaField{⟩}}\AgdaSpace{}%
\AgdaSymbol{=}\AgdaSpace{}%
\AgdaBound{Γ}\AgdaSpace{}%
\AgdaOperator{\AgdaField{⟨}}\AgdaSpace{}%
\AgdaInductiveConstructor{suc}\AgdaSpace{}%
\AgdaBound{n}\AgdaSpace{}%
\AgdaOperator{\AgdaField{⟩}}\<%
\\
\>[0]\AgdaFunction{◄}\AgdaSpace{}%
\AgdaBound{Γ}\AgdaSpace{}%
\AgdaOperator{\AgdaField{⟪}}\AgdaSpace{}%
\AgdaBound{m≤n}\AgdaSpace{}%
\AgdaOperator{\AgdaField{⟫}}\AgdaSpace{}%
\AgdaBound{γ}\AgdaSpace{}%
\AgdaSymbol{=}\AgdaSpace{}%
\AgdaBound{Γ}\AgdaSpace{}%
\AgdaOperator{\AgdaField{⟪}}\AgdaSpace{}%
\AgdaInductiveConstructor{s≤s}\AgdaSpace{}%
\AgdaBound{m≤n}\AgdaSpace{}%
\AgdaOperator{\AgdaField{⟫}}\AgdaSpace{}%
\AgdaBound{γ}\<%
\\
\\[\AgdaEmptyExtraSkip]%
\>[0]\AgdaFunction{▻}\AgdaSpace{}%
\AgdaSymbol{:}\AgdaSpace{}%
\AgdaRecord{Ty}\AgdaSpace{}%
\AgdaSymbol{(}\AgdaFunction{◄}\AgdaSpace{}%
\AgdaGeneralizable{Γ}\AgdaSymbol{)}\AgdaSpace{}%
\AgdaSymbol{→}\AgdaSpace{}%
\AgdaRecord{Ty}\AgdaSpace{}%
\AgdaGeneralizable{Γ}\<%
\\
\>[0]\AgdaFunction{▻}\AgdaSpace{}%
\AgdaBound{T}\AgdaSpace{}%
\AgdaOperator{\AgdaField{⟨}}\AgdaSpace{}%
\AgdaInductiveConstructor{zero}\AgdaSpace{}%
\AgdaOperator{\AgdaField{,}}%
\>[14]\AgdaBound{γ}\AgdaSpace{}%
\AgdaOperator{\AgdaField{⟩}}\AgdaSpace{}%
\AgdaSymbol{=}\AgdaSpace{}%
\AgdaRecord{⊤}\<%
\\
\>[0]\AgdaFunction{▻}\AgdaSpace{}%
\AgdaBound{T}\AgdaSpace{}%
\AgdaOperator{\AgdaField{⟨}}\AgdaSpace{}%
\AgdaInductiveConstructor{suc}\AgdaSpace{}%
\AgdaBound{n}\AgdaSpace{}%
\AgdaOperator{\AgdaField{,}}\AgdaSpace{}%
\AgdaBound{γ}\AgdaSpace{}%
\AgdaOperator{\AgdaField{⟩}}\AgdaSpace{}%
\AgdaSymbol{=}\AgdaSpace{}%
\AgdaBound{T}\AgdaSpace{}%
\AgdaOperator{\AgdaField{⟨}}\AgdaSpace{}%
\AgdaBound{n}\AgdaSpace{}%
\AgdaOperator{\AgdaField{,}}\AgdaSpace{}%
\AgdaBound{γ}\AgdaSpace{}%
\AgdaOperator{\AgdaField{⟩}}\<%
\\
\>[0]\AgdaFunction{▻}\AgdaSpace{}%
\AgdaBound{T}\AgdaSpace{}%
\AgdaOperator{\AgdaField{⟪}}\AgdaSpace{}%
\AgdaInductiveConstructor{z≤n}\AgdaSpace{}%
\AgdaOperator{\AgdaField{,}}\AgdaSpace{}%
\AgdaBound{e}\AgdaSpace{}%
\AgdaOperator{\AgdaField{⟫}}\AgdaSpace{}%
\AgdaBound{t}\AgdaSpace{}%
\AgdaSymbol{=}\AgdaSpace{}%
\AgdaInductiveConstructor{tt}\<%
\\
\>[0]\AgdaFunction{▻}\AgdaSpace{}%
\AgdaBound{T}\AgdaSpace{}%
\AgdaOperator{\AgdaField{⟪}}\AgdaSpace{}%
\AgdaInductiveConstructor{s≤s}\AgdaSpace{}%
\AgdaBound{m≤n}\AgdaSpace{}%
\AgdaOperator{\AgdaField{,}}\AgdaSpace{}%
\AgdaBound{e}\AgdaSpace{}%
\AgdaOperator{\AgdaField{⟫}}\AgdaSpace{}%
\AgdaBound{t}\AgdaSpace{}%
\AgdaSymbol{=}\AgdaSpace{}%
\AgdaBound{T}\AgdaSpace{}%
\AgdaOperator{\AgdaField{⟪}}\AgdaSpace{}%
\AgdaBound{m≤n}\AgdaSpace{}%
\AgdaOperator{\AgdaField{,}}\AgdaSpace{}%
\AgdaBound{e}\AgdaSpace{}%
\AgdaOperator{\AgdaField{⟫}}\AgdaSpace{}%
\AgdaBound{t}\<%
\end{code}
\begin{code}[hide]%
\>[0]\AgdaField{ty-cong}\AgdaSpace{}%
\AgdaSymbol{(}\AgdaFunction{▻}\AgdaSpace{}%
\AgdaBound{T}\AgdaSymbol{)}\AgdaSpace{}%
\AgdaSymbol{\{}\AgdaArgument{f}\AgdaSpace{}%
\AgdaSymbol{=}\AgdaSpace{}%
\AgdaInductiveConstructor{z≤n}\AgdaSymbol{\}}%
\>[28]\AgdaInductiveConstructor{refl}\AgdaSpace{}%
\AgdaSymbol{=}\AgdaSpace{}%
\AgdaInductiveConstructor{refl}\<%
\\
\>[0]\AgdaField{ty-cong}\AgdaSpace{}%
\AgdaSymbol{(}\AgdaFunction{▻}\AgdaSpace{}%
\AgdaBound{T}\AgdaSymbol{)}\AgdaSpace{}%
\AgdaSymbol{\{}\AgdaArgument{f}\AgdaSpace{}%
\AgdaSymbol{=}\AgdaSpace{}%
\AgdaInductiveConstructor{s≤s}\AgdaSpace{}%
\AgdaBound{m≤n}\AgdaSymbol{\}}\AgdaSpace{}%
\AgdaInductiveConstructor{refl}\AgdaSpace{}%
\AgdaSymbol{=}\AgdaSpace{}%
\AgdaField{ty-cong}\AgdaSpace{}%
\AgdaBound{T}\AgdaSpace{}%
\AgdaInductiveConstructor{refl}\<%
\\
\>[0]\AgdaField{ty-id}\AgdaSpace{}%
\AgdaSymbol{(}\AgdaFunction{▻}\AgdaSpace{}%
\AgdaBound{T}\AgdaSymbol{)}\AgdaSpace{}%
\AgdaSymbol{\{}\AgdaArgument{x}\AgdaSpace{}%
\AgdaSymbol{=}\AgdaSpace{}%
\AgdaInductiveConstructor{zero}\AgdaSymbol{\}}%
\>[24]\AgdaSymbol{=}\AgdaSpace{}%
\AgdaInductiveConstructor{refl}\<%
\\
\>[0]\AgdaField{ty-id}\AgdaSpace{}%
\AgdaSymbol{(}\AgdaFunction{▻}\AgdaSpace{}%
\AgdaBound{T}\AgdaSymbol{)}\AgdaSpace{}%
\AgdaSymbol{\{}\AgdaArgument{x}\AgdaSpace{}%
\AgdaSymbol{=}\AgdaSpace{}%
\AgdaInductiveConstructor{suc}\AgdaSpace{}%
\AgdaBound{n}\AgdaSymbol{\}}\AgdaSpace{}%
\AgdaSymbol{=}\AgdaSpace{}%
\AgdaFunction{strong-ty-id}\AgdaSpace{}%
\AgdaBound{T}\<%
\\
\>[0]\AgdaField{ty-comp}\AgdaSpace{}%
\AgdaSymbol{(}\AgdaFunction{▻}\AgdaSpace{}%
\AgdaBound{T}\AgdaSymbol{)}\AgdaSpace{}%
\AgdaSymbol{\{}\AgdaArgument{f}\AgdaSpace{}%
\AgdaSymbol{=}\AgdaSpace{}%
\AgdaInductiveConstructor{z≤n}\AgdaSymbol{\}}%
\>[42]\AgdaSymbol{=}\AgdaSpace{}%
\AgdaInductiveConstructor{refl}\<%
\\
\>[0]\AgdaField{ty-comp}\AgdaSpace{}%
\AgdaSymbol{(}\AgdaFunction{▻}\AgdaSpace{}%
\AgdaBound{T}\AgdaSymbol{)}\AgdaSpace{}%
\AgdaSymbol{\{}\AgdaArgument{f}\AgdaSpace{}%
\AgdaSymbol{=}\AgdaSpace{}%
\AgdaInductiveConstructor{s≤s}\AgdaSpace{}%
\AgdaBound{k≤m}\AgdaSymbol{\}}\AgdaSpace{}%
\AgdaSymbol{\{}\AgdaArgument{g}\AgdaSpace{}%
\AgdaSymbol{=}\AgdaSpace{}%
\AgdaInductiveConstructor{s≤s}\AgdaSpace{}%
\AgdaBound{m≤n}\AgdaSymbol{\}}\AgdaSpace{}%
\AgdaSymbol{=}\AgdaSpace{}%
\AgdaFunction{strong-ty-comp}\AgdaSpace{}%
\AgdaBound{T}\<%
\end{code}
Here we defined \later{\AB{T}}\AS\AFi{⟨}\AS\AB{n}\AS\AFi{,}\AS\AB{γ}\AS\AFi{⟩} by pattern matching on \AB{n} and \later{\AB{T}}\AS\AFi{⟪}\AS\AB{m≤n}\AS\AFi{,}\AS\AB{e}\AS\AFi{⟫}\AS\AB{t} by pattern matching on \AB{m≤n}.

Another intuition that we can make precise now is the representation of guarded streams as the diagram in \eqref{eq:diagram}.
\begin{code}[hide]%
\>[0]\AgdaSymbol{\{-\#}\AgdaSpace{}%
\AgdaKeyword{NON\AgdaUnderscore{}COVERING}\AgdaSpace{}%
\AgdaSymbol{\#-\}}\<%
\\
\>[0]\AgdaFunction{★}\AgdaSpace{}%
\AgdaSymbol{:}\AgdaSpace{}%
\AgdaRecord{BaseCategory}\<%
\\
\>[0]\AgdaField{BaseCategory.Ob}\AgdaSpace{}%
\AgdaFunction{★}\AgdaSpace{}%
\AgdaSymbol{=}\AgdaSpace{}%
\AgdaRecord{⊤}\<%
\\
\>[0]\AgdaField{BaseCategory.Hom}\AgdaSpace{}%
\AgdaFunction{★}\AgdaSpace{}%
\AgdaSymbol{=}\AgdaSpace{}%
\AgdaSymbol{λ}\AgdaSpace{}%
\AgdaBound{\AgdaUnderscore{}}\AgdaSpace{}%
\AgdaBound{\AgdaUnderscore{}}\AgdaSpace{}%
\AgdaSymbol{→}\AgdaSpace{}%
\AgdaRecord{⊤}\<%
\\
\>[0]\AgdaField{BaseCategory.hom-id}\AgdaSpace{}%
\AgdaFunction{★}\AgdaSpace{}%
\AgdaSymbol{=}\AgdaSpace{}%
\AgdaInductiveConstructor{tt}\<%
\\
\>[0]\AgdaOperator{\AgdaField{BaseCategory.\AgdaUnderscore{}∙\AgdaUnderscore{}}}\AgdaSpace{}%
\AgdaFunction{★}\AgdaSpace{}%
\AgdaSymbol{=}\AgdaSpace{}%
\AgdaSymbol{λ}\AgdaSpace{}%
\AgdaBound{\AgdaUnderscore{}}\AgdaSpace{}%
\AgdaBound{\AgdaUnderscore{}}\AgdaSpace{}%
\AgdaSymbol{→}\AgdaSpace{}%
\AgdaInductiveConstructor{tt}\<%
\\
\\[\AgdaEmptyExtraSkip]%
\>[0]\AgdaSymbol{\{-\#}\AgdaSpace{}%
\AgdaKeyword{NON\AgdaUnderscore{}COVERING}\AgdaSpace{}%
\AgdaSymbol{\#-\}}\<%
\\
\>[0]\AgdaFunction{constantly}\AgdaSpace{}%
\AgdaSymbol{:}\AgdaSpace{}%
\AgdaRecord{Modality}\AgdaSpace{}%
\AgdaFunction{★}\AgdaSpace{}%
\AgdaFunction{ω}\<%
\\
\>[0]\AgdaField{ctx-op}\AgdaSpace{}%
\AgdaSymbol{(}\AgdaField{ctx-functor}\AgdaSpace{}%
\AgdaFunction{constantly}\AgdaSymbol{)}\AgdaSpace{}%
\AgdaBound{Γ}\AgdaSpace{}%
\AgdaOperator{\AgdaField{⟨}}\AgdaSpace{}%
\AgdaInductiveConstructor{tt}\AgdaSpace{}%
\AgdaOperator{\AgdaField{⟩}}\AgdaSpace{}%
\AgdaSymbol{=}\AgdaSpace{}%
\AgdaBound{Γ}\AgdaSpace{}%
\AgdaOperator{\AgdaField{⟨}}\AgdaSpace{}%
\AgdaNumber{0}\AgdaSpace{}%
\AgdaOperator{\AgdaField{⟩}}\<%
\\
\>[0]\AgdaField{ctx-op}\AgdaSpace{}%
\AgdaSymbol{(}\AgdaField{ctx-functor}\AgdaSpace{}%
\AgdaFunction{constantly}\AgdaSymbol{)}\AgdaSpace{}%
\AgdaBound{Γ}\AgdaSpace{}%
\AgdaOperator{\AgdaField{⟪}}\AgdaSpace{}%
\AgdaInductiveConstructor{tt}\AgdaSpace{}%
\AgdaOperator{\AgdaField{⟫}}\AgdaSpace{}%
\AgdaBound{γ}\AgdaSpace{}%
\AgdaSymbol{=}\AgdaSpace{}%
\AgdaBound{γ}\<%
\\
\>[0]\AgdaOperator{\AgdaField{⟨\AgdaUnderscore{}∣\AgdaUnderscore{}⟩}}\AgdaSpace{}%
\AgdaFunction{constantly}\AgdaSpace{}%
\AgdaSymbol{\{}\AgdaBound{Γ}\AgdaSymbol{\}}\AgdaSpace{}%
\AgdaBound{T}\AgdaSpace{}%
\AgdaOperator{\AgdaField{⟨}}\AgdaSpace{}%
\AgdaBound{n}\AgdaSpace{}%
\AgdaOperator{\AgdaField{,}}\AgdaSpace{}%
\AgdaBound{γ}\AgdaSpace{}%
\AgdaOperator{\AgdaField{⟩}}\AgdaSpace{}%
\AgdaSymbol{=}\AgdaSpace{}%
\AgdaBound{T}\AgdaSpace{}%
\AgdaOperator{\AgdaField{⟨}}\AgdaSpace{}%
\AgdaInductiveConstructor{tt}\AgdaSpace{}%
\AgdaOperator{\AgdaField{,}}\AgdaSpace{}%
\AgdaBound{Γ}\AgdaSpace{}%
\AgdaOperator{\AgdaField{⟪}}\AgdaSpace{}%
\AgdaInductiveConstructor{z≤n}\AgdaSpace{}%
\AgdaOperator{\AgdaField{⟫}}\AgdaSpace{}%
\AgdaBound{γ}\AgdaSpace{}%
\AgdaOperator{\AgdaField{⟩}}\<%
\\
\\[\AgdaEmptyExtraSkip]%
\>[0]\AgdaSymbol{\{-\#}\AgdaSpace{}%
\AgdaKeyword{NON\AgdaUnderscore{}COVERING}\AgdaSpace{}%
\AgdaSymbol{\#-\}}\<%
\end{code}
\begin{code}%
\>[0]\AgdaFunction{GStream}\AgdaSpace{}%
\AgdaSymbol{:}\AgdaSpace{}%
\AgdaRecord{Ty}\AgdaSpace{}%
\AgdaSymbol{(}\AgdaFunction{lock}\AgdaSpace{}%
\AgdaFunction{constantly}\AgdaSpace{}%
\AgdaGeneralizable{Γ}\AgdaSymbol{)}\AgdaSpace{}%
\AgdaSymbol{→}\AgdaSpace{}%
\AgdaRecord{Ty}\AgdaSpace{}%
\AgdaGeneralizable{Γ}\<%
\\
\>[0]\AgdaFunction{GStream}\AgdaSpace{}%
\AgdaBound{A}\AgdaSpace{}%
\AgdaOperator{\AgdaField{⟨}}\AgdaSpace{}%
\AgdaBound{n}\AgdaSpace{}%
\AgdaOperator{\AgdaField{,}}\AgdaSpace{}%
\AgdaBound{γ}\AgdaSpace{}%
\AgdaOperator{\AgdaField{⟩}}\AgdaSpace{}%
\AgdaSymbol{=}\AgdaSpace{}%
\AgdaDatatype{Vec}\AgdaSpace{}%
\AgdaSymbol{(}\AgdaOperator{\AgdaField{⟨}}\AgdaSpace{}%
\AgdaFunction{constantly}\AgdaSpace{}%
\AgdaOperator{\AgdaField{∣}}\AgdaSpace{}%
\AgdaBound{A}\AgdaSpace{}%
\AgdaOperator{\AgdaField{⟩}}\AgdaSpace{}%
\AgdaOperator{\AgdaField{⟨}}\AgdaSpace{}%
\AgdaBound{n}\AgdaSpace{}%
\AgdaOperator{\AgdaField{,}}\AgdaSpace{}%
\AgdaBound{γ}\AgdaSpace{}%
\AgdaOperator{\AgdaField{⟩}}\AgdaSymbol{)}\AgdaSpace{}%
\AgdaSymbol{(}\AgdaInductiveConstructor{suc}\AgdaSpace{}%
\AgdaBound{n}\AgdaSymbol{)}\<%
\end{code}
\AF{GStream}'s input and output types live over a different base category, but the \AF{constantly} modality bridges this gap.
The implementation of the restriction maps and the \AF{constantly} modality can be found in the Agda code.